%% file: main.tex
\documentclass[sw,crcready]{iosart2x}

\usepackage[inline]{enumitem}
\usepackage{algorithm}
\usepackage[noend]{algpseudocode}
\usepackage{mathtools,amssymb}
\usepackage{listings}
\usepackage{color}
\usepackage{comment}
\usepackage{listings}
\usepackage{tikz}
\usepackage{breqn}
\usepackage{graphicx}
\usepackage{amsmath}
\usepackage[bf,small,tableposition=bottom]{caption}
\usepackage{amssymb}
\usepackage{pgfplots}
\usepackage{centernot}
\usepackage{mathtools}
\usepackage{algorithm}
\usepackage{subcaption}

\usepackage{makecell}

\algnewcommand{\IfThen}[2]{% \IfThenElse{<if>}{<then>}{<else>}
  \State \algorithmicif\ #1\ \algorithmicthen\ #2}
\algnewcommand{\ElseThen}[1]{% \IfThenElse{<if>}{<then>}{<else>}
  \State \algorithmicelse\ #1}
\algnewcommand{\ElseIfThen}[2]{% \IfThenElse{<if>}{<then>}{<else>}
  \State \algorithmicelse\ \algorithmicif\ #1\ \algorithmicthen\ #2}

\makeatletter
\newcommand*{\algrule}[1][\algorithmicindent]{%
  \makebox[#1][l]{%
    \hspace*{.2em}% <------------- This is where the rule starts from
    \vrule height .75\baselineskip depth .25\baselineskip
  }
}

\newcount\ALG@printindent@tempcnta
\def\ALG@printindent{%
    \ifnum \theALG@nested>0% is there anything to print
    \ifx\ALG@text\ALG@x@notext% is this an end group without any text?
    \else
    \unskip
    \ALG@printindent@tempcnta=1
    \loop
    \algrule[\csname ALG@ind@\the\ALG@printindent@tempcnta\endcsname]%
    \advance \ALG@printindent@tempcnta 1
    \ifnum \ALG@printindent@tempcnta<\numexpr\theALG@nested+1\relax
    \repeat
    \fi
    \fi
}
\patchcmd{\ALG@doentity}{\noindent\hskip\ALG@tlm}{\ALG@printindent}{}{\errmessage{failed to patch}}
\patchcmd{\ALG@doentity}{\item[]\nointerlineskip}{}{}{} % no spurious vertical space
\makeatother

\newcommand{\piqnic}{\textsc{Piqnic}}
\newcommand{\colchain}{\textsc{ColChain}}
\newcommand{\system}{\textsc{Lothbrok}}

\DeclareMathOperator{\mbin}{\>\scalebox{1.1}[0.8]{\rotatebox[origin=c]{180}{$\exists$}}}

\newtheorem{definition}{Definition}

\pubyear{0000}
\volume{0}
\firstpage{1}
\lastpage{1}

\begin{document}

\begin{frontmatter}

%\pretitle{}
\title{The Lothbrok approach for SPARQL Query Optimization over Decentralized Knowledge Graphs}
\runtitle{The Lothbrok approach for SPARQL Query Optimization over Decentralized Knowledge Graphs}
%\subtitle{}

% For one author:
%\author{\inits{N.}\fnms{Name1} \snm{Surname1}\ead[label=e1]{first@somewhere.com}}
%\address{Department first, \orgname{University or Company name},
%Abbreviate US states, \cny{Country}\printead[presep={\\}]{e1}}
%\runningauthor{N. Surname1}

% Two or more authors:
\author[First]{\inits{C.}\fnms{Christian} \snm{Aebeloe}\ead[label=e1]{caebel@cs.aau.dk}%
\thanks{Corresponding author. \printead{e1}.}},
\author[First]{\inits{G.}\fnms{Gabriela} \snm{Montoya}\ead[label=e2]{gmontoya@cs.aau.dk}}
and
\author[First]{\inits{K.}\fnms{Katja} \snm{Hose}\ead[label=e3]{khose@cs.aau.dk}}
\runauthor{C. Aebeloe et al.}
\address[First]{Department of Computer Science, \orgname{Aalborg University}, Selma Lagerl\"{o}fs Vej 300, DK-9220 Aalborg Ø, \cny{Denmark}\\ E-mails: \ caebel@cs.aau.dk, gmontoya@cs.aau.dk, khose@cs.aau.dk}

%\begin{review}{editor}
%\reviewer{\fnms{First} \snm{Editor}\address{\orgname{University or Company name}, \cny{Country}}}
%\reviewer{\fnms{Second} \snm{Editor}\address{\orgname{First University or Company name}, \cny{Country}
%    and \orgname{Second University or Company name}, \cny{Country}}}
%\end{review}
%\begin{review}{solicited}
%\reviewer{\fnms{First} \snm{Solicited reviewer}\address{\orgname{University or Company name}, \cny{Country}}}
%\reviewer{\snm{anonymous reviewer}}
%\end{review}
%\begin{review}{open}
%\reviewer{\fnms{First} \snm{Open Reviewer}\address{\orgname{University or Company name}, \cny{Country}}}
%\end{review}

\begin{abstract}
%The promise of a vast and readily accessible Web of Data has increased the popularity of the Semantic Web over recent years.
While the Web of Data in principle offers access to a wide range of interlinked data, the architecture of the Semantic Web today relies mostly on the data providers to maintain access to their data through SPARQL endpoints.
Several studies, however, have shown that such endpoints often experience downtime, meaning that the data they maintain becomes inaccessible.
While decentralized systems based on Peer-to-Peer (P2P) technology have previously shown to increase the availability of knowledge graphs, even when a large proportion of the nodes fail, processing queries in such a setup can be an expensive task since data necessary to answer a single query might be distributed over multiple nodes.
%relevant to a single query can be split out over several nodes.
%This means that nodes end up transferring a large amount of data when processing queries, resulting in a significant network overhead.
In this paper, we therefore propose an approach to optimizing SPARQL queries over decentralized knowledge graphs, called \system{}.
While there are potentially many aspects to consider when optimizing such queries, we focus on three aspects: cardinality estimation, locality awareness, and data fragmentation.
%(i) data fragmentation based on characteristic sets (i.e., predicate families),
%(ii) decentralized indexes that lets \system{} obtain more accurate cardinality estimations, and
%(iii) a query processing strategy that lets nodes consider the locality of data when processing queries and delegate subqueries to other, neighboring nodes in the network.
%While \system{} could be implemented on top of any P2P system, we implemented \system{} on top of two recent such approaches.
We empirically show that \system{} is able to achieve significantly faster query processing performance compared to the state of the art when processing challenging queries %using a well-known benchmarking suite for federated systems, 
as well as when the network is under high load. %using large-scale synthetic datasets and queries.
\end{abstract}


\end{frontmatter}

%%%%%%%%%%% The article body starts:

\section{Introduction}\label{sec:introduction}
\input{introduction}

\section{Related Work}\label{sec:relatedwork}
\input{relatedwork}

\section{Background}\label{sec:preliminaries}
\input{background}

\section{The \system{} Approach}\label{sec:system}
\input{system}

\section{Query Optimization}\label{sec:queryoptimization}
\input{queryoptimization}

\section{Query Execution}\label{sec:queryprocessing}
\input{queryprocessing}

\section{Experimental Evaluation}\label{sec:experiments}
\input{experiments}

\section{Conclusions}\label{sec:conclusions}
\input{conclusion}

\vspace*{2ex}
\noindent
\textbf{Acknowledgments.}
This research was partially funded by the Danish Council for Independent Research (DFF) under grant agreement no. DFF-8048-00051B, Aalborg University's Talent Programme, and the Poul Due Jensen Foundation.
\vspace*{-.2ex}

%%%%%%%%%%% The bibliography starts:

%%  ios1.bst will be used to                               %%

\nocite{*} 
% if your bibliography is in bibtex format, use those commands:
\bibliographystyle{ios1}           % Style BST file.
\bibliography{references}        % Bibliography file (usually '*.bib')

% or include bibliography directly:
%\begin{thebibliography}{0}
%\bibitem{r1} F. Author, Information about cited object.
%
%\bibitem{r2} S. Author and T. Author, Information about cited object.
%\end{thebibliography}

%\appendix
%\section{Additional LargeRDFBench Results}\label{app:largerdfbench}
%\input{appendix}

%\section{Additional WatDiv Results}\label{app:watdiv}
%\input{appendixb}

\end{document}

%% file: introduction.tex
Due to the popularity of decentralized knowledge graphs on the Web, more and increasingly large knowledge graphs encoded in RDF are becoming available~\cite{hose2021knowledge}.
%The ever increasing popularity of decentralized knowledge graphs on the Web means that the number of available knowledge graphs encoded using RDF is increasing significantly.
Furthermore, RDF knowledge graphs made available today are becoming exceedingly large.
For instance, Wikidata~\cite{DBLP:journals/cacm/VrandecicK14} and Bio2RDF~\cite{DBLP:conf/semweb/DumontierCCAEBD14} contain more than 14 billion triples each.
As a result, data providers experience an increasing burden of maintaining access to the datasets; and without any monetary incentives to do so, datasets often end up becoming unavailable~\cite{DBLP:conf/semweb/ArandaHUV13,DBLP:journals/semweb/VandenbusscheUM17,DBLP:conf/esws/AebeloeMH19} and outdated~\cite{DBLP:conf/www/AebeloeMH21}.

In recent years, several decentralized systems~\cite{DBLP:conf/esws/AebeloeMH19,DBLP:conf/www/AebeloeMH21,DBLP:conf/icde/KarnstedtSRMHSJ07,DBLP:journals/ws/VerborghSHHVMHC16,DBLP:conf/www/AzzamAMKPH21,DBLP:journals/corr/abs-2002-09172} have been proposed to alleviate the aforementioned burden from the data providers by reducing the computational load required to keep the data available, albeit using different methods to do so.
For instance, Linked Data Fragments (LDF)-based approaches~\cite{DBLP:journals/ws/VerborghSHHVMHC16,DBLP:conf/www/AzzamAMKPH21,DBLP:journals/corr/abs-2002-09172,smartkg,DBLP:conf/semweb/HelingAMS18} reduce the computational load on the server by distributing some of the query processing effort to the client, ensuring that the server only processes requests with low time complexity.
On the other hand, Peer-to-Peer (P2P) systems~\cite{DBLP:conf/esws/AebeloeMH19,DBLP:conf/www/AebeloeMH21,DBLP:conf/icde/KarnstedtSRMHSJ07} remove the centralized point of failure that a server represents and replicate the data across several nodes in a decentralized fashion, ensuring that even if the uploading node fails, the data is still accessible.
For instance, RDFPeers~\cite{DBLP:conf/www/CaiF04} uses a structured overlay over a P2P network that relies on Dynamic Hash Tables (DHTs) to determine where to replicate certain data.
However, in situations where nodes frequently leave or join the network (i.e., churn), and data is often uploaded to the network, nodes have to go through a costly adjustment process to update the overlay and redistribute the data.
Instead, systems like \piqnic~\cite{DBLP:conf/esws/AebeloeMH19} and \colchain~\cite{DBLP:conf/www/AebeloeMH21} use unstructured P2P systems as foundation, where there is no global control over where data is replicated, %giving the nodes the autonomy to decide which data to store and 
making the network more stable under churn.

\colchain{} builds upon \piqnic{} and divides the entire network into communities of nodes that not only replicate the same data, but also collaborate on keeping certain data (fragments) up-to-date.
This is done by using blockchain technology~\cite{bitcoin,DBLP:conf/i-semantics/GrauxSJLSMP18,DBLP:conf/www/SopekGKKTT18,DBLP:journals/ijwgs/ZhengXDCW18} where \emph{chains} of updates maintain the history of changes to the data fragments.
%Nevertheless, decentralized approaches often focus specifically on the availability problem and largely ignore the updatability problem.
%In fact, one of the only approaches that addresses the updatability as well as the availability is \colchain~\cite{DBLP:conf/www/AebeloeMH21}.
%\colchain{} is a P2P system that divides the entire network into communities of nodes that collaborate on keeping certain data (fragments) up-to-date and uses \emph{chains} of updates to maintain the history of the data fragments, building on blockchains~\cite{bitcoin,DBLP:conf/i-semantics/GrauxSJLSMP18,DBLP:conf/www/SopekGKKTT18,DBLP:journals/ijwgs/ZhengXDCW18}.
%Blockchains are chains of data blocks that represent global ledgers and rely on a majority consensus when adding a new block to the chain; when a new block is added, it is linked to the latest existing block on the chain using hashes to prevent changes without first reaching a consensus.
By linking such update chains to the data fragments in a community, \colchain{} allows community participants to collaborate on keeping the data up-to-date while using consensus to make malicious updates less likely and allowing users to roll-back updates to an earlier version on request.
Furthermore, the decentralized nature of \colchain{} also increases the availability of the uploaded data by replicating the data on nodes within the community.

Nevertheless, while \piqnic{} and \colchain{} already use decentralized indexes~\cite{ppbfs} to determine where data is located during query time, 
subgraphs needed to answer a query are usually scattered across multiple nodes.
Furthermore, the indexes
provide limited information that prevents the nodes from considering locality and accurately estimating join cardinalities when optimizing queries.
As a result, such systems often experience an unnecessarily large amount of intermediate results when processing a query.
This problem is exacerbated by the decentralized nature of the systems, since the intermediate results have to be transferred between nodes, causing a significant communication overhead.

%Nevertheless, while \colchain~\cite{DBLP:conf/www/AebeloeMH21} already uses   it does not improve upon query processing efficiency compared to approaches such as \piqnic{}~\cite{DBLP:conf/esws/AebeloeMH19} using the same decentralized indexes.
%As such, it suffers from the same shortcomings as the state of the art.
%,like many other P2P systems~\cite{DBLP:conf/esws/AebeloeMH19,DBLP:conf/icde/KarnstedtSRMHSJ07,DBLP:conf/www/CaiF04},
%experiences an unnecessarily large amount of intermediate results when processing a query.

While there are potentially many aspects to consider when optimizing queries in a decentralized setup, we will focus on three such aspects: cardinality estimation, locality awareness, and data fragmentation.
Suboptimal solutions to any of these three aspects can lead to an increased communication overhead and lower performance.
For instance, while fragmenting large knowledge graphs into smaller fragments ensures that nodes do not have to replicate entire knowledge graphs, using a fragmentation technique that spreads out the data relevant to a single (sub)query across several fragments can increase the communication overhead since nodes might have to send an excessive number of requests to obtain all relevant data to answer a particular query~\cite{DBLP:conf/vldb/AilamakiDHW99,kcap,DBLP:conf/icde/HoseS13,DBLP:conf/www/GalarragaHS14}.
On the other hand, inaccurate cardinality estimations can lead to a suboptimal join strategy that increases the amount of intermediate results and therefore runtime~\cite{DBLP:conf/icde/NeumannM11,DBLP:conf/semweb/MontoyaSH17}.
And while several approaches have proposed reasonably accurate cardinality estimation techniques~\cite{DBLP:conf/icde/NeumannM11,DBLP:conf/semweb/MontoyaSH17,DBLP:conf/sigmod/ParkKBKHH20} over knowledge graphs, and for federated engines in particular~\cite{DBLP:conf/semweb/MontoyaSH17,DBLP:conf/sigmod/HoseS12,DBLP:conf/www/HarthHKPSU10,DBLP:journals/www/UmbrichHKHP11}, such approaches cannot easily be transferred to a decentralized setup since nodes in a decentralized setup lack a global overview of the network and the data is scattered across multiple nodes.
Finally, considering locality of the data when processing queries can help ensure that larger subqueries are delegated to nodes that can process them without communicating with other nodes, lowering the data transfer overall.

Nevertheless, while an optimization approach that maximizes the degree to which entire queries can be processed by a single node could decrease the communication overhead, a study~\cite{DBLP:conf/vldb/AilamakiDHW99} found that processing entire queries on one node can actually decrease the overall performance when the network is under heavy load, and that it is equally important to balance out the query load between nodes.
As such, there is a need for a more holistic approach to query optimization that is able to delegate the processing of subqueries to other nodes in the network, thus reducing the communication overhead to the extent possible.
For instance, query optimization techniques that are based on star-shaped subqueries have previously been shown to increase performance by at least an order of magnitude~\cite{DBLP:conf/www/AzzamAMKPH21,smartkg,DBLP:journals/corr/abs-2002-09172,DBLP:conf/esws/VidalRLMSP10}.
This, and the fact that conjunctive subqueries are relatively efficient to process~\cite{DBLP:journals/tods/PerezAG09}, means that decomposing and processing queries based on star-shaped subqueries can significantly reduce the communication overhead in decentralized systems.
%For instance, several recent studies~\cite{DBLP:conf/www/AzzamAMKPH21,smartkg,DBLP:journals/corr/abs-2002-09172} found that, in a client-server architecture, conjunctive subqueries, such as star-shaped subqueries, can decrease the communication overhead significantly while not overloading the server.

In this paper, we therefore extend our work on \piqnic~\cite{DBLP:conf/esws/AebeloeMH19} and \colchain~\cite{DBLP:conf/www/AebeloeMH21} in three aspects that work together to reduce the communication overhead when processing SPARQL queries, and in doing so, improve query processing performance in an approach that we call \system{}.
\system{} adapts Characteristic Sets~\cite{DBLP:conf/icde/NeumannM11,DBLP:conf/www/AzzamAMKPH21,smartkg,DBLP:journals/corr/abs-2002-09172} to fragment data in decentralized P2P systems. %, and, based on those subgraphs determine the partitions based on $n$-hop replication~\cite{DBLP:conf/icde/HoseS13}
Furthermore, \system{} builds upon Prefix-Partitioned Bloom Filters (PPBFs)~\cite{ppbfs} and proposes a new indexing scheme called Semantically Partitioned Bloom Filters (SPBFs) to obtain more accurate cardinality estimations.
%\system{} adapts , such that the structure of the index lets \system{} obtain more accurate cardinality estimations and consider the locality of data when processing queries called .
Lastly, \system{} also introduces a locality-aware query optimization strategy that takes advantage of the SPBF indexes and is able to delegate the processing of (sub)queries to neighboring nodes in the network holding relevant data.
%Lastly, \system{} adapts Characteristic Sets~\cite{DBLP:conf/icde/NeumannM11} to the decentralized setup of \colchain{}, called \dcslong, improving the accuracy of the cardinality estimations in \colchain.
%\system{} thus reduces the communication overhead significantly while still distributing the query processing across multiple nodes.
%We combine the main contributions described above into a single approach for optimizing SPARQL queries over decentralized RDF peers that we call the \system{} approach to optimizing decentralized SPARQL queries.
We evaluate \system{} thoroughly using LargeRDFBench~\cite{largerdfbench}, a benchmark suite for federated RDF systems that comprises 13 datasets with over a billion triples and includes 40 queries of varying complexity and sizes of intermediate results.
Furthermore, we evaluate \system{} using synthetic data and queries from WatDiv~\cite{watdiv} to test the scalability of \system{} under load. %and query logs posed by real users from LSQ~\cite{DBLP:conf/semweb/SaleemAHMN15}.
In summary, we make the following contributions:

\begin{itemize}
\item A data fragmentation technique that builds on Characteristic Sets~\cite{DBLP:conf/icde/NeumannM11}
\item SPBF indexes adapted to the characteristic set fragmentation technique
\item A cardinality estimation approach over decentralized RDF fragments using the SPBF indexes to provide more accurate cardinality estimations
\item A locality-aware query optimization algorithm that uses SPBF indexes to delegate subqueries to neighboring nodes and reduce the communication overhead
\item A thorough experimental evaluation of the impact of the presented techniques on query processing performance using real-world data from a well-known benchmark suite, and large-scale synthetic datasets
\end{itemize}

\noindent
The paper is structured as follows:
Section~\ref{sec:relatedwork} discusses related work
while Section~\ref{sec:preliminaries} describes background information.
Then, Section~\ref{sec:system} presents \system,
Section~\ref{sec:queryoptimization} details how \system{} optimizes queries,
and Section~\ref{sec:queryprocessing} describes the query execution approach,
while Section~\ref{sec:experiments} presents our experimental evaluation.
Lastly, Section~\ref{sec:conclusions} concludes the paper with an outlook to future work.

%% file: relatedwork.tex
The availability problem has prompted significant amount of research in the areas of decentralized query processing and decentralized architectures for knowledge graphs.
In this section, we therefore discuss existing approaches related to \system{};
client-server architectures, federated systems, and P2P systems.

\subsection{Client-Server Architectures}
SPARQL endpoints are Web services providing an HTTP interface that accepts SPARQL queries and remain some of the most popular interfaces for querying RDF data on the Web.
However, several studies~\cite{DBLP:conf/semweb/ArandaHUV13,DBLP:journals/semweb/VandenbusscheUM17} have found that such endpoints are often unavailable and experience downtime.

Linked Data Fragment (LDF) interfaces, such as Triple Pattern Fragments (TPF)~\cite{DBLP:journals/ws/VerborghSHHVMHC16}, attempt to increase the availability of the server by shifting some of the query processing load towards the client while the server only processes requests with low time complexity.
For instance, TPF servers only process individual triple patterns while the TPF clients process joins and other expensive operations.
Today, several TPF clients exist that rely on either a greedy algorithm~\cite{DBLP:journals/ws/VerborghSHHVMHC16}, a metadata based strategy~\cite{DBLP:conf/esws/HerwegenVMW15}, or star-shaped query decomposition combined with adaptive query processing techniques~\cite{DBLP:conf/semweb/AcostaV15} to determine the join order of the triple patterns in a query. %, or process skyline queries~\cite{DBLP:conf/semweb/KelesH19}.
However, while in all these approaches the server can handle more concurrent requests in comparison to SPARQL endpoints without becoming unresponsive, TPF naturally incurs a large network overhead when processing queries since intermediate bindings from previously evaluated triple patterns are transferred along with subsequently evaluated triple patterns to limit the amount of intermediate results, one by one.
Furthermore, studies found that the performance of TPF is heavily affected by the type of triple pattern (i.e., the position of variables in the triple pattern)~\cite{DBLP:conf/semweb/HelingAMS18} and the shape of the query~\cite{DBLP:conf/semweb/MontoyaKH19,DBLP:journals/corr/abs-1912-08010}.

Several different systems have since been proposed to lower the network overhead.
For instance, Bindings-Restricted TPF (brTPF)~\cite{brtpf} bulks bindings from previously evaluated triple patterns such that multiple bindings can be attached to a single request.
While this reduces the number of requests made for a triple pattern, it still incurs a somewhat large data transfer overhead, since each request still evaluates a single triple pattern.
hybridSE~\cite{DBLP:conf/semweb/MontoyaAH18} combines a brTPF server with a SPARQL endpoint and takes advantage of the strengths of each approach; subqueries with large numbers of intermediate results are sent to the SPARQL endpoint to overcome the limitations posed by LDF systems.
However, hybridSE often answers complex queries using the SPARQL endpoint and is thus vulnerable to server failure. %as SPARQL endpoints often experience downtime~\cite{DBLP:conf/semweb/ArandaHUV13,DBLP:journals/semweb/VandenbusscheUM17}.

To further limit the network overhead, Star Pattern Fragments (SPF)~\cite{DBLP:journals/corr/abs-2002-09172} clients send conjunctive subqueries in the shape of stars (star patterns) to the server and process more complex patterns locally on the client.
Such conjunctive subqueries can be processed relatively efficiently by the server~\cite{DBLP:journals/tods/PerezAG09}, which results in the transfer of significantly fewer intermediate results than in systems like TPF and brTPF.
On the other hand, Smart-KG~\cite{smartkg} ships predicate-family partitions (i.e., characteristic sets) to the client and processes the entire query locally; however, triple patterns with infrequent predicate values (according to a certain threshold) are sent to and evaluated by the server.
While this takes advantage of the distributed resources that the clients possess, Smart-KG often ends up transferring excessive amounts of data unnecessarily since entire partitions of a dataset are transferred regardless of any bindings from previously evaluated star patterns.
WiseKG~\cite{DBLP:conf/www/AzzamAMKPH21} combines SPF and Smart-KG and uses a cost model to determine which strategy (SPF or Smart-KG) is the most cost-effective to process a given star-shaped subquery.
Like SPF and Smart-KG, WiseKG processes more complex patterns on the client.
%While this addresses some of the major issues of LDF systems, WiseKG still relies on a centralized server that is subject to failure
Nevertheless, all the aforementioned LDF approaches rely on a centralized server or a fixed set of servers that are subject to failure.

Lastly, different from LDF approaches, SaGe~\cite{DBLP:conf/www/MinierSM19} decreases the load on the server by suspending queries after a fixed time quantum to prevent long-running queries from exhausting server resources; the queries can then be restarted by making a new request to the server.
However, SaGe processes entire, and possibly complex, queries on the server, and as stated above, such servers are subject to failure.

\subsection{Federated Systems}
Federated systems enable answering queries over data spread out across multiple independent SPARQL endpoints~\cite{DBLP:conf/semweb/AcostaVLCR11,DBLP:conf/i-semantics/CharalambidisTK15,DBLP:conf/semweb/GorlitzS11,DBLP:conf/semweb/SchwarteHHSS11,IbragimovHPZ15} or LDF servers~\cite{DBLP:journals/corr/abs-2102-03269} offering access to different datasets.
While such approaches spread out query processing over several servers, lowering the load on each individual server, they sometimes generate suboptimal query execution plans that increase the number of intermediate results and the load on individual servers~\cite{DBLP:conf/esws/JakobsenMH19}.
As such, several approaches~\cite{DBLP:conf/semweb/MontoyaSH17,DBLP:conf/semweb/MontoyaVA12,DBLP:conf/i-semantics/0002PSHN18,DBLP:conf/sigmod/HoseS12,DBLP:conf/www/HarthHKPSU10,DBLP:journals/www/UmbrichHKHP11} have attempted to optimize federated queries in different ways.
For instance, \cite{DBLP:conf/semweb/SchwarteHHSS11} builds an index over time by remembering which endpoints in the federation can provide answers to which triple patterns.
Furthermore,
\cite{DBLP:conf/semweb/MontoyaVA12} decomposes queries into subqueries that can be evaluated by a single endpoint.
While \cite{DBLP:conf/semweb/MontoyaVA12} uses a similar query decomposition strategy as \system{}, they target federations over SPARQL endpoints, and as previously mentioned, such endpoints suffer from availability issues.
On the other hand, \cite{DBLP:conf/semweb/MontoyaSH17,DBLP:conf/i-semantics/0002PSHN18} estimate the selectivity of joins to produce more efficient join plans.
For instance, \cite{DBLP:conf/semweb/MontoyaSH17} uses characteristic sets~\cite{DBLP:conf/icde/NeumannM11} and pairs~\cite{DBLP:conf/edbt/Gubichev014} to index the data in the federation and combines this with Dynamic Programming (DP) to optimize query execution plans.
Furthermore, \cite{DBLP:journals/corr/abs-2102-03269} proposes an interface for processing federated queries over heterogeneous LDF interfaces.
To achieve this, the query optimizer is adapted to the characteristics of the different interfaces as well as the locality of the data, i.e., knowledge of which nodes hold which data.
%While \system{} networks generally consist of homogeneous nodes, the query optimizer in \system{} considers the locality of the data in a similar manner as the approach presented in~\cite{DBLP:journals/corr/abs-2102-03269}.
%While these approaches were proposed for contexts with a fixed set of servers, they could be applied to a decentralized setup such as the one we are targeting as well.
%Specifically, by fragmenting the data based on characteristic sets, \system{} builds on the approach presented in~\cite{DBLP:conf/semweb/MontoyaSH17} and uses a similar cardinality estimation technique and DP to optimize join plans based on data locality in a network.
Inspired by these approaches, \system{} fragments knowledge graphs based on characteristic sets and uses a similar cardinality estimation technique to optimize join plans in consideration of data locality in the network.

\subsection{Peer-to-Peer Systems}
Peer-to-Peer (P2P) systems~\cite{DBLP:conf/esws/AebeloeMH19,DBLP:conf/www/AebeloeMH21,DBLP:conf/www/CaiF04,DBLP:conf/icde/KarnstedtSRMHSJ07,DBLP:journals/ws/KaoudiKKMMP10,DBLP:conf/www/MansourSHZCGAB16,DBLP:conf/www/GalarragaHS14} tackle the availability issue from a different perspective: by removing the central point of failure completely and replicating the data across multiple nodes in a P2P network, they can ensure the data remains available even if the original node that uploaded the data fails.
As such, they consist of a set of nodes (often resource limited) that act both as servers and clients, maintaining a limited local datastore.
The structure of the network, i.e., connections between the nodes, as well as data placement (data allocation), varies from system to system.
For instance, some systems~\cite{DBLP:conf/www/CaiF04,DBLP:conf/icde/KarnstedtSRMHSJ07,DBLP:journals/ws/KaoudiKKMMP10} enforce data placement by applying a structured overlay over the network, such as Dynamic Hash Tables (DHTs)~\cite{DBLP:journals/cacm/Larson88}.
On the other hand, \piqnic~\cite{DBLP:conf/esws/AebeloeMH19} imposes no structure on top the network; nodes are connected randomly to a set of neighbors that are shuffled periodically with another node's neighbors to increase the degree of joinability between the fragments of neighboring nodes.
Lastly, \colchain~\cite{DBLP:conf/www/AebeloeMH21} extends \piqnic{} and divides the entire network into smaller communities of nodes that collaborate on keeping certain data available and up-to-date.
By applying community-based ledgers of updates and relying on a consensus protocol within a community, \colchain{} lets users actively participate in keeping the data up-to-date.

Each P2P system has different ways of processing queries.
For instance, due to the lack of global knowledge over the network, basic P2P systems have to flood the network with requests for a given horizon to increase the likelyhood of receiving complete query results.
To counteract this, distributed indexes~\cite{ppbfs,DBLP:conf/icdcs/CrespoG02,DBLP:journals/www/UmbrichHKHP11} like Prefix-Partitioned Bloom Filter (PPBF) indexes~\cite{ppbfs} determine which nodes may include relevant data for a given query and thus allow the system to prune nodes from consideration during query optimization.
%On the other hand, \colchain{} nodes process queries based on the communities they are a member of.
Yet, the aforementioned systems still experience a significant overhead partly caused by inaccurate cardinality estimations, query optimization that does not consider the locality of data, as well as data fragmentation that splits up closely related data.
For instance, \piqnic{} and \colchain{} both use a predicate-based fragmentation strategy that creates a fragment for each predicate.
This, together with the replication and allocation strategy used, means that data relevant to a single query is distributed over a significant number of fragments and nodes.

However, while an approach that maximizes the degree to which entire queries can be processed by one node can lower the communication overhead, distributing some of the query processing load across multiple nodes is equally important when optimizing queries in a decentralized context~\cite{DBLP:conf/vldb/AilamakiDHW99} to avoid overloading individual nodes.
As such, \system{} limits the communication overhead by fragmenting data based on characteristic sets and introducing a new indexing scheme that lets nodes take advantage of the fragmentation to more accurately estimate subquery cardinality and distribute the processing of subqueries to nodes in the network based on data locality.
Furthermore, since fragments are created based on characteristic sets, entire star patterns can be processed efficiently by single nodes, further distributing the query processing load, lowering the communication overhead at the same time, and increasing the query throughput.
%While in this paper, we focus mostly on \piqnic{} and \colchain, \system{} could be implemented on top of any decentralized architecture to also provide the benefits described in this paper.

%% file: background.tex
%\subsection{Knowledge Graphs}
A commonly used format for storing semantic data is the Resource Description Framework (RDF)~\cite{brickley2014rdf}.
RDF structures data as triples, defined as follows.

\begin{definition}[RDF Triple]
Let $I$, $B$, and $L$ be the disjoint sets of IRIs, blank nodes, and literals.
An RDF triple is a triple $t$ of the form $t=(s,p,o)\in (I\cup B)\times I\times (I\cup B\cup L)$, where $s$, $p$, and $o$ are called subject, predicate, and object.
\end{definition}

Given the definition of an RDF triple, a \emph{knowledge graph} $\mathcal{G}$ is a finite set of RDF triples.
%An RDF~\cite{brickley2014rdf} graph (knowledge graph) is a set of triples $t=(s,p,o)\in (I\cup B)\times I\times (I\cup B\cup L)$ where $I$, $B$, and $L$ are the (infinite) disjoint sets of IRIs, blank nodes, and literals.
%To ease the presentation of our approach, we henceforth, we define a knowledge graph as a directed labeled graph as follows.
%
%\begin{definition}[Knowledge Graph]
%A knowledge graph $\mathcal{G}$ is a directed labeled graph, i.e., $\mathcal{G}=(\Lambda,E,\xrightarrow{})$, where $\Lambda$ is the set of vertices (i.e., subjects and objects) and $\Lambda\subseteq I\cup B\cup L$, $E$ is the set of labels (i.e., predicates) and $E\subseteq I$, and $\xrightarrow{}$ is the set of labeled edges (i.e., $\xrightarrow{}\,\,\subseteq\Lambda\times E\times\Lambda$).
%\end{definition}
%
The most popular language to query knowledge graphs is SPARQL~\cite{world2013sparql}.
A SPARQL query consists of one or more \emph{triple patterns}.
A triple pattern $t$ is a triple of the form $t=(s,p,o)\in (I\cup B\cup V)\times (I\cup V)\times (I\cup B\cup L\cup V)$ where $V$ is the set of all variables.
%We say that a triple pattern $p$ \emph{matches} a knowledge graph $\mathcal{G}$ iff there exists a mapping from the variables in $p$ to a subject, predicate, or object in any triple in $\mathcal{G}$, such that substituting the variables in $p$ according to the mapping yields a triple in $\mathcal{G}$.
A Basic Graph Pattern (BGP) is a set of conjunctive triple patterns.
Without loss of generality, we focus our discussion in the main part of this paper on BGPs and describe in Section~\ref{sec:queryoptimization} how our approach can support other operators, such as \texttt{UNION} and \texttt{OPTIONAL}; our experimental evaluation in Section~\ref{sec:experiments} includes queries with a variety of SPARQL operators including \texttt{UNION} and \texttt{OPTIONAL}.

A complex BGP $P$ can be decomposed into a set of \emph{star patterns}.
A star pattern $P'$ is a set of triple patterns that share the same subject, i.e., $\forall t_1=(s_1,p_1,o_1),t_2=(s_2,p_2,o_2)$ such that $t_1,t_2\in P'$, it is the case that $s_1=s_2$.
Note that while star patterns can be defined as both subject-based and object-based star patterns, for ease of presentation, we focus on subject-based star patterns only since subject-subject joins are much more common in real query loads~\cite{DBLP:conf/www/0001SCBMN19}; \system{} can trivially be adapted to object-based star patterns by using the same principles presented in this paper for object-object joins rather than subject-subject joins.

\begin{definition}[Star Decomposition~\cite{DBLP:journals/corr/abs-2002-09172}]
Given a BGP $P=\{t_1,\dots,t_n\}$ with subjects $S_P=\{s_1,\dots,s_m\}$, the \emph{star decomposition} of $P$, $\mathcal{S}(P)=\{P_s(P)\mid s\in S_P\}$, is a set of star patterns $P_s(P)$ for each $s\in S_P$, such that $P=\cup_{s\in S_P}P_s(P)$ where $P_s(P)=\{(s',p',o')\mid (s',p',o')\in P\wedge s'=s\}$.
\end{definition}

%Since BGPs can be seen as a labeled directed graph like a knowledge graph but where any position can be a variable, we define a query graph as follows.
%
%\begin{definition}[Query Graph]
%A query graph $\mathcal{Q}$ is a directed labeled graph, i.e., $\mathcal{Q}=(\Lambda,E,\xrightarrow{})$, where $\Lambda\subseteq I\cup B\cup L\cup V$ is the set of vertices (i.e., subjects and objects), $E\subseteq I\cup V$ is the set of labels (i.e., predicates), and $\xrightarrow{}$ is the set of labeled edges (i.e., $\xrightarrow{}\,\,\subseteq\Lambda\times E\times\Lambda$).
%\end{definition}
%
%The empty knowledge/query graph $\mathcal{G}^{\emptyset}$ is defined as the graph where the sets of vertices, edges, and transitions are empty, i.e., $\mathcal{G}^{\emptyset}=(\emptyset,\emptyset,\emptyset)$.
%Furthermore, given two graphs $\mathcal{G}_1=(\Lambda_1,E_1,\xrightarrow{}_1)$ and $\mathcal{G}_2=(\Lambda_2,E_2,\xrightarrow{}_2)$, the union between $\mathcal{G}_1$ and $\mathcal{G}_2$, denoted $\mathcal{G}_1\cup\mathcal{G}_2$, is defined as $\mathcal{G}_1\cup\mathcal{G}_2=(\Lambda_1\cup\Lambda_2,E_1\cup E_2,\xrightarrow{}_1\cup\xrightarrow{}_2)$
The answer to a BGP $P$ over a knowledge graph $\mathcal{G}$ is a set of \emph{solution mappings}, defined as follows.

\begin{definition}[Solution mapping~\cite{DBLP:conf/www/AebeloeMH21,DBLP:journals/ws/VerborghSHHVMHC16}]
Given a BGP $P$ and a knowledge graph $\mathcal{G}$, the sets $I_{\mathcal{G}}$, $B_{\mathcal{G}}$, and $L_{\mathcal{G}}$ are the sets of IRIs, blank nodes, and literals in $\mathcal{G}$, and $V_P$ is the set of variables in $P$,
a \emph{solution mapping} $\mu$ is a partial mapping $\mu:{V_P}\mapsto({U_{\mathcal{G}}}\cup{B_{\mathcal{G}}}\cup{L_{\mathcal{G}}})$.
\end{definition}

Given a BGP $P$ and a solution mapping $\mu$, the notation $\mu[P]$ denotes the triple (patterns) obtained by replacing variables in $P$ according to the bindings in $\mu$.
Furthermore, given a knowledge graph $\mathcal{G}$ and BGP $P$, $[[P]]_{\mathcal{G}}$ denotes the set of solution mappings that constitute the answer to $P$ over $\mathcal{G}$, i.e., $\forall\mu\in [[P]]_{\mathcal{G}}$, $\mu[P]\in\mathcal{G}$, and $\forall T\in\mu[P]$, $T$ is a set of \emph{matching} triples to $P$, denoted $T[P]$.
%the answer to $P$ over $\mathcal{G}$.
%A set of triples $T$ is said to be \emph{matching} triples to $P$, denoted $T[P]$, if there exists a solution mapping $\mu\in [[P]]_{\mathcal{G}}$ such that $T=\mu[P]$.
Furthermore, $dom(\mu)$ returns the \emph{domain} of $\mu$, i.e., the set of variables that are bound in $\mu$ and $vars(P)$ returns the variables in $P$.

\subsection{Peer-to-Peer}\label{subsec:unp2p}
%In this section, we define unstructured P2P systems in the context of knowledge graphs that \system{} builds upon.´
In its simplest form, an unstructured P2P system consists of a set of interconnected nodes that all maintain a local datastore managing a set of (partial) knowledge graphs, where each node maintains a local view over the network, i.e., a set of \emph{neighboring} nodes (nodes within the local view over the network). %that they can interact with.
%Given a node $n$, a neighboring node to $n$ is a node that is included within $n$'s local view over the network.
%that maps triples to fragments available within a certain horizon; 
%The structure of the network (i.e., the connections between nodes) can vary from system to system.
%For instance, \piqnic~\cite{DBLP:conf/esws/AebeloeMH19} imposes no structure over the network and connections are random while periodic shuffles (i.e., where nodes exchange neighbors) continuously update the network structure.
%On the other hand, connections in \colchain~\cite{DBLP:conf/www/AebeloeMH21} are determined by the communities in which nodes participate.
%In \piqnic~\cite{DBLP:conf/esws/AebeloeMH19}, for instance, nodes maintain a set of neighboring nodes that are periodically shuffled between neighbors.
%Furthermore, knowledge graphs in \piqnic{} are replicated in chains, i.e., the uploading node sends the knowledge graph to one of its neighbors, which replicates the knowledge graph locally and passes it on to one of its neighbors, and so on, for a certain number of steps.
%On the other hand, in \colchain~\cite{DBLP:conf/www/AebeloeMH21}, nodes that participate in or observe the same communities are neighbors, and knowledge graphs are replicated across all nodes within the community they are uploaded to.
%However, to emphasize the generality of our approach, and to allow \system{} to theoretically be implemented on top of any P2P system, we base our approach on a more general definition of P2P networks than the ones presented in~\cite{DBLP:conf/esws/AebeloeMH19,DBLP:conf/www/AebeloeMH21}.
%

Formally, we define a P2P network $N$ as a set of interconnected nodes $N=\{n_1,\dots,n_n\}$ where each node maintains a local datastore and a local view over the network.
The data uploaded to a node in $N$ is replicated throughout the network.
Furthermore, in line with previous work~\cite{ppbfs,DBLP:conf/www/AebeloeMH21}, each node maintains a distributed index describing the knowledge graphs reachable within a certain number of steps (also known as hops), called the \emph{horizon} of a node.
A node $n$ is defined as follows:

\begin{definition}[Node~\cite{DBLP:conf/esws/AebeloeMH19,ppbfs}]
A node $n$ is a triple $n=(G,I,N_n)$ where:
\begin{itemize}
\item $G$ is the set of knowledge graphs in $n$'s local datastore
\item $I$ is $n$'s distributed index
\item $N_n$ is a set of neighboring nodes
\end{itemize}
\end{definition}

While maintaining the structure of the network is important for P2P systems, it is not relevant for the data and query processing techniques that this paper is focusing on.
As such, we do not go into detail on network topology, data replication and allocation, and periodic shuffles.
Instead, we refer the interested reader to related work such as~\cite{DBLP:conf/esws/AebeloeMH19,DBLP:conf/www/AebeloeMH21} for more details.
In the following, we define data fragmentation and introduce a running example.

%However, 
%Given the sizes of knowledge graphs today, replicating full datasets on several nodes is not feasible.
%As such, 
In line with previous work~\cite{DBLP:conf/esws/AebeloeMH19,DBLP:conf/www/AebeloeMH21}, and to avoid having to replicate large knowledge graphs throughout the network, \system{} divides knowledge graphs into smaller disjoint \emph{fragments}, i.e., partial knowledge graphs, which can be replicated more easily.
Fragments can be obtained using a \emph{fragmentation} function.
A fragmentation function is a function that, given a knowledge graph, returns a set of disjoint fragments, and is formally defined as follows:

\begin{definition}[Fragmentation Function~\cite{DBLP:conf/esws/AebeloeMH19,DBLP:conf/www/AebeloeMH21}]\label{def:partitioning}
A \emph{fragmentation} function $\mathcal{F}$ is a function that maps a knowledge graph $\mathcal{G}$ to a set of \emph{knowledge graph fragments}, i.e., $\mathcal{F}:\mathcal{G}\mapsto 2^{\mathcal{G}}$.% such that $\forall\mathcal{G}_1,\mathcal{G}_2\in \mathcal{P}(\mathcal{G}):\mathcal{G}_1\cap\mathcal{G}_2=\emptyset$.
\end{definition}

Different fragmentation functions can have different granularities.
For instance, the most coarse-granular fragmentation function is $\mathcal{F}_C(\mathcal{G})=\{\mathcal{G}\}$, i.e., the fragmentation function does not split up the original knowledge graph.
\colchain~\cite{DBLP:conf/www/AebeloeMH21} as well as \piqnic~\cite{DBLP:conf/esws/AebeloeMH19} use a \emph{predicate-based} fragmentation function for $\mathcal{G}$, i.e., $\mathcal{F}_P(\mathcal{G})=\{\{(s',p',o')\mid (s',p',o')\in\mathcal{G}\wedge p'=p\}\mid\exists s,o:(s,p,o)\in\mathcal{G}\}$, which creates a fragment for each unique predicate in $\mathcal{G}$.
\system{} uses a fragmentation function based on characteristic sets~\cite{DBLP:conf/icde/NeumannM11} (i.e., predicate families) that is detailed in Section~\ref{subsec:partitioning}.
%
%Similar to the network structure, the replication strategy differs for each system.
%For instance, once a dataset is uploaded to a \piqnic{} node, the uploading node selects one of its neighbors to replicate a fragment on, which chooses one of its neighbors and so on, forming a chain of neighbors to replicate each fragment on.
%On the other hand, \colchain{} uses community-based replication and replicates each fragment to all participants in the community.

The fragments created by the fragmentation function are replicated and allocated at multiple nodes in the network to ensure availability in case the original provider of the knowledge graph becomes unavailable and to enable load balancing.
The replication and allocation factor are parameters of the underlying network; for instance, in \piqnic~\cite{DBLP:conf/esws/AebeloeMH19}, fragments are replicated and allocated across the node's neighbors, and nodes index all fragments available within a certain horizon.
On the other hand, \colchain~\cite{DBLP:conf/www/AebeloeMH21} replicates and allocates fragments at nodes that participate within the same communities.
Since this paper focuses on data fragmentation and query optimization, we omit details on data replication and allocation and refer the interested reader to related work~\cite{DBLP:conf/esws/AebeloeMH19,DBLP:conf/www/AebeloeMH21} for details.

%The structure of the network, both in terms of the connections between nodes, the replication strategy, and indexing strategy, can vary from system to system.
%For instance, \piqnic~\cite{DBLP:conf/esws/AebeloeMH19} imposes no structure over the network and as such connections are random, while periodic shuffles (i.e., where nodes exchange neighbors) continuously update the network structure; fragments are replicated across the node's neighbors and nodes index all fragments available within a certain horizon.
%%Furthermore, fragments are replicated in chains starting from the uploading node, while indexes are compiled based on a horizon.
%On the other hand, connections, replication, and indexing in \colchain~\cite{DBLP:conf/www/AebeloeMH21} are determined by the communities in which nodes participate.
%To emphasize the generality of our approach, and to allow \system{} to be implemented on top of any P2P network, we will not go into details with the structural aspects of P2P systems, such as neighborhood management, periodic updates, communities, and so on, but focus our discussion on data fragmentation, data indexing, and query optimization.
%However, to emphasize the generality of our approach, and to allow \system{} to be implemented on top of any P2P network, we define
%base our approach on a more general definition of P2P networks than the ones presented in~\cite{DBLP:conf/esws/AebeloeMH19,DBLP:conf/www/AebeloeMH21}.

\begin{figure*}[htb!]
\centering
\begin{subfigure}[b]{0.5\textwidth}
  \centering
  \input{figures/running_network.tex}
  \caption{Example network $N$}\label{subfig:running_network}
\end{subfigure}
\begin{subfigure}[b]{0.3\textwidth}
  \centering
  \input{figures/running_node.tex}
  \caption{Node $n_5$}\label{subfig:running_node}
\end{subfigure}
\caption{(a) Example of an unstructured P2P network $N=\{n_1,\dots,n_5\}$ and (b) architecture of a single node $n_5$ that indexes data within a horizon of 2 nodes.}\label{fig:running_intro}
\end{figure*}
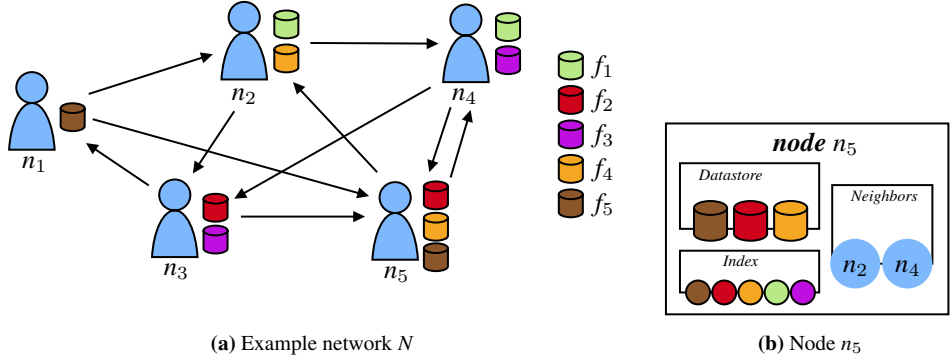

Consider, as a running example, the unstructured P2P network in Figure~\ref{subfig:running_network} consisting of five nodes ($N=\{n_1,\dots,n_5\}$) that replicate a total of five fragments ($f_1,\dots,f_5$).
In this example, each node maintains a set of two neighbors and each fragment is replicated across two nodes.
For instance, node $n_5$ has $\{n_2,n_4\}$ as its set of neighbors, and replicates the fragments $\{f_2,f_4,f_5\}$ in its local datastore.
While the running example is based on an unstructured network, such as the one presented in~\cite{DBLP:conf/esws/AebeloeMH19}, \system{} could be adapted to more structured setups, such as the one presented in~\cite{DBLP:conf/www/AebeloeMH21}.
%To emphasize the generality of our approach, and to allow \system{} to be implemented on top of any P2P network, we focus our discussion on a 
%we base our approach on a more general definition of P2P networks than the ones presented in~\cite{DBLP:conf/esws/AebeloeMH19,DBLP:conf/www/AebeloeMH21}.

\subsection{Distributed Indexes}\label{subsec:dis_index}
%In this section, we reiterate the definition of a distributed index as defined in~\cite{ppbfs,DBLP:conf/www/AebeloeMH21}.
To speed up query processing performance, systems like \piqnic~\cite{DBLP:conf/esws/AebeloeMH19} and \colchain~\cite{DBLP:conf/www/AebeloeMH21} use distributed indexes~\cite{ppbfs,DBLP:conf/icdcs/CrespoG02} to efficiently identify nodes holding relevant data for a given SPARQL query.
%This is done by pruning fragments that would not contribute to the overall query result.
The indexes capture information about the fragments stored locally at the node itself as well as information about fragments that can be accessed via its neighbors.
%Nodes in P2P systems furthermore index fragments that are not within their local datastore; this is done to include the data within such fragments in the query processing strategy.

%A distributed index is a structure that specifies which fragment contain relevant data to which subquery and which nodes store the fragments in their local datastore.
In~\cite{DBLP:conf/www/AebeloeMH21,ppbfs}, a distributed index is formally defined as consisting of two mappings; (1) from a triple pattern to the set of fragments containing relevant data to the triple pattern, and (2) from a fragment to the set of nodes that store the fragment.
Furthermore, to build the indexes for a node's local view over the network, nodes share partial indexes, i.e., partial mappings, for the fragments that they have access to, called index \emph{slices}.
%Note that since this definition maps from triple patterns to fragments, it only works due to the predicate-based fragmentation used by the approaches; in Section~\ref{subsec:indexing} we extend this definition to map from star patterns to fragments rather than triple patterns to work with the fragmentation approach presented in Section~\ref{subsec:partitioning}.
%However, first, and 
In line with \cite{DBLP:conf/www/AebeloeMH21,ppbfs}, we define distributed indexes and index slices in the following.

\begin{definition}[Distributed Index~\cite{DBLP:conf/www/AebeloeMH21,ppbfs}]\label{def:index}
Let $\mathcal{N}$ be the set of nodes within a network, $n$ be a node such that $n\in\mathcal{N}$, $\mathcal{T}$ be the set of all possible triple patterns, and $\mathcal{F}$ be the set of fragments that $n$ has access to within its local view over the network.
A \emph{distributed index} on $n$ is a tuple $I_n=(\nu,\eta)$ with $\nu:\mathcal{T}\mapsto 2^{\mathcal{F}}$ and $\eta:\mathcal{F}\mapsto 2^{\mathcal{N}}$.
For a triple pattern $t$, $\nu(t)$ returns the set of fragments in $\mathcal{F}$ that $t$ matches.
For a fragment $f\in\mathcal{F}$, $\eta(f)$ returns the nodes on which $f$ is located.
\end{definition}

Given a node $n$, $n$'s distributed index is denoted $I_n$.
%Given a distributed index $I$, the function $range(I.\nu)$ returns all the fragments mapped to by some triple pattern in $I.\nu$, i.e., given $\mathcal{T}$ as the (infinite) set of all triple patterns, $range(I.\nu)=\{I.\nu(t)\mid t\in\mathcal{T}\}$.
Given the definition of a distributed index, we define a \emph{node mapping} as a mapping from a triple pattern $t$ in a BGP $P$ to a set of nodes that contain relevant fragments to $t$, as follows:

\begin{definition}[Node Mapping~\cite{ppbfs,DBLP:conf/www/AebeloeMH21}]\label{def:nmatch}
For any BPG $P$ and distributed index $I$, there exists a function $match(P,I)$ that returns a \emph{node mapping} $M:P\mapsto 2^{\mathcal{N}}$, such that $\forall t\in P$, $M(t)$ returns the indexed nodes that have fragments holding data matching the triple $t$.
\end{definition}

An index slice for a fragment is a partial mapping from triple patterns to the fragments that contain relevant triples to the triple patterns, as well as a mapping from the fragment to the nodes that replicate it, and is defined as follows:

\begin{definition}[Index Slice~\cite{ppbfs,DBLP:conf/www/AebeloeMH21}]\label{def:slice}
Let $f$ be a fragment.
The index \emph{slice} of $f$, $s_f$, is a tuple $s_{f}=(\nu',\eta')$, where $\nu'(t)$ returns $\{f\}$ if there exists a triple in $f$ that matches $t$, and $\eta'(f)$ returns the set of all nodes that contain $f$ in their local datastore.
The function $s(f)$ returns the index slice describing $f$.
\end{definition}

Index slices for the fragments that a node has access to are combined into a distributed index for that particular node using the $\oplus$ operator\footnote{$\oplus$ is defined in~\cite{DBLP:conf/www/AebeloeMH21,ppbfs} as $(f\oplus g)(x) = f(x)\cup g(x)$ if $f$ and $g$ are defined at $x$; 
$(f\oplus g)(x) = f(x)$  if $f$ is defined at $x$; $(f\oplus g)(x) = g(x)$ if $g$ is defined at $x$.}.
The distributed index is then used to check the relevancy and overlap of fragments during query time to optimize the query. 
Given a set of slices $S$, the index obtained by combining the slices in $S$, $I(S)$, can be computed using the formula in Equation~\ref{eq:combine}~\cite{ppbfs,DBLP:conf/www/AebeloeMH21}.

\begin{equation}\label{eq:combine}
I(S)=\left(\bigoplus_{s\in S}s.\nu',\bigoplus_{s\in S}s.\eta'\right)
\end{equation}

While the definition of distributed indexes allows for several different types of indexes, the index slices used in \piqnic~\cite{DBLP:conf/esws/AebeloeMH19} and \colchain~\cite{DBLP:conf/www/AebeloeMH21} correspond to Prefix-Partitioned Bloom Filters (PPBFs)~\cite{ppbfs},
which extend regular Bloom filters~\cite{Bloom}.
A Bloom filter $\mathcal{B}$ for a set $S$ of IRIs such that $|S|=n$ is a tuple $\mathcal{B}=(\hat{b},H)$ where $\hat{b}$ is a bitvector of size $m$ and $H$ is a set of $k$ hash functions~\cite{ppbfs}.
Each hash function in $H$ maps the elements from $S$ (i.e., IRIs) to a position in $\hat{b}$; these positions are thus set to $1$ whereas the positions not mapped to by a function in $H$ are $0$.
In other words, \cite{ppbfs} represents the combined set of subjects and objects in a fragment in a prefix-partitioned bitvector.
Looking up whether an element $e$ is in $S$ using the Bloom filter for $S$ is done by hashing $e$ using the hash functions in $H$ and checking the value of each position in $\hat{b}$.
If at least one of those positions is set to $0$, it is certain that $e\not\in S$.
However, if all corresponding bits are set to $1$, it is not certain that $e\in S$, since it could be a false positive caused by hash collisions, i.e., different values are mapped to the same positions in the underlying bitvector.
In this case, we say that $e$ \textit{may} be in $S$, denoted $e\mbin S$.

To check the compatibility of two fragments relevant for conjunctive triple patterns, we check whether or not they produce any join results.
To do this, we could check whether or not the intersection of the bitvectors describing the subjects and objects of the fragments is empty (i.e., if they have some IRI in common).
Given two Bloom filters $\mathcal{B}_1=(\hat{b}_1,H)$ and $\mathcal{B}_2=(\hat{b}_2,H)$, the intersection of $\mathcal{B}_1$ and $\mathcal{B}_2$ is approximated by the logic \texttt{AND} operation between $\hat{b}_1$ and $\hat{b}_2$, $\mathcal{B}_1\cap \mathcal{B}_2\approx\hat{b}_1\&\hat{b}_2$.

To avoid exceedingly large bitvectors, PPBFs partition the bitvector based on the prefix of the IRIs.
A PPBF is formally defined in~\cite{ppbfs} as follows.

\begin{definition}[Prefix-Partitioned Bloom Filter~\cite{ppbfs}]\label{def:ppbf}
A PPBF $\mathcal{B}^P$ is a 4-tuple $\mathcal{B}^P=(P,\hat{B},\theta,H)$ where
\begin{itemize}
\item $P$ a set of prefixes
\item $\hat{B}$ is a set of bitvectors such that $\forall\hat{b}_1,\hat{b}_2\in\hat{B}:|\hat{b}_1|=|\hat{b}_2|$
\item $\theta:P\rightarrow \hat{B}$ is a prefix-mapping function
\item $H$ is a set of hash functions
\end{itemize}
For each $p_i \in P$, $\mathcal{B}_i=(\theta(p_i),H)$ is the Bloom Filter that encodes the names of the IRIs with prefix $p_i$ and is called a partition of $\mathcal{B}^P$.
\end{definition}

\begin{figure*}[htb!]
\centering
\begin{subfigure}[b]{0.5\textwidth}
  \centering
  \input{figures/ppbfex.tex}
  \caption{PPBF $\mathcal{B}^P_1$}\label{subfig:ppbfex}
\end{subfigure}
\begin{subfigure}[b]{0.45\textwidth}
  \centering
  \input{figures/bsint.tex}
  \caption{$\mathcal{B}^P_1\cap\mathcal{B}^P_2$}\label{subfig:bsint}
\end{subfigure}
\caption{Example of (a) inserting an IRI into a PPBF $\mathcal{B}^P_1$ and (b) intersection between two PPBFs $\mathcal{B}^P_1\cap\mathcal{B}^P_2$~\cite{ppbfs}.}
\label{fig:ppbfex}
\end{figure*}
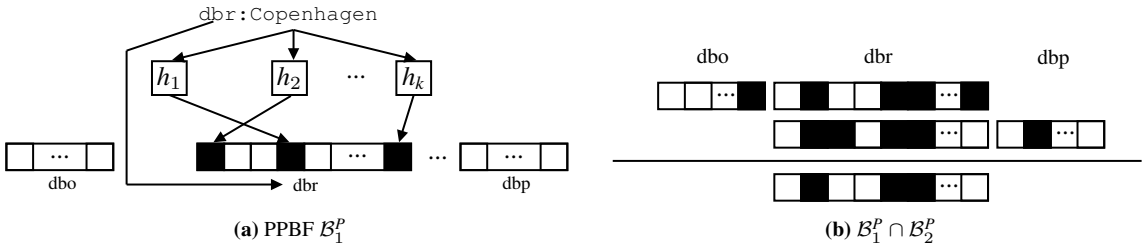

Consider the example where the IRI \texttt{dbr:Copenhagen} is inserted into a PPBF, visualized in Figure~\ref{subfig:ppbfex}.
In this case, the IRI is matched to the prefix \texttt{dbr}, and the IRI is hashed using each hash function in the PPBF;
each corresponding bit in the bitvector for the \texttt{dbr} prefix is thus set to 1.

Like for regular Bloom filters, we say that an IRI $i$ with prefix $p$ \textit{may} be in a PPBF $\mathcal{B}^P$, denoted $i\mbin\mathcal{B}^P$, if and only if all positions given by $h(i)$ such that $h\in H$ are set to $1$ in the bitvector $\theta(p)$.
PPBFs are used by \piqnic{} and \colchain{} to prune non-overlapping fragments of joining triple patterns from the query execution plan (i.e., the $match(P,I)$ function in Definition~\ref{def:nmatch}).
This is done by finding the intersection of the two PPBFs to check whether or not they overlap; if the intersection of the two PPBFs is empty, the corresponding fragments do not produce any join results.
The PPBF intersection is defined in~\cite{ppbfs} as follows.

\begin{definition}[Prefix-Partitioned Bloom Filter Intersection~\cite{ppbfs}]
The intersection of two PPBFs with the same set of hash functions H and bitvectors of the same size, denoted $\mathcal{B}^P_1\cap \mathcal{B}^P_2$, is
$\mathcal{B}^P_1\cap \mathcal{B}^P_2=(P_\cap,\hat{B}_\cap,\theta_\cap,H)$, where $P_\cap=\mathcal{B}^P_1.P\cap\mathcal{B}^P_2.P$, $\hat{B}_\cap=\{\mathcal{B}^P_1.\theta(p)\,\&\,\mathcal{B}^P_2.\theta(p)\mid p\in P_\cap\}$, and $\theta_\cap:P_\cap\rightarrow \hat{B}_\cap$.
\end{definition}

Consider the example intersection visualized in Figure~\ref{subfig:bsint}.
As described above, the intersection of two PPBFs is the bitwise \texttt{AND} operation on the bitvectors for the prefixes that $\mathcal{B}^P_1$ and $\mathcal{B}^P_2$ have in common.
In this example, $\mathcal{B}^P_2$ does not have a bitvector with the prefix \texttt{dbp}, thus this partition is omitted from the intersection.
Similarly, the bitvector partition with the \texttt{dbo} prefix is omitted.
Since both PPBFs have bitvectors for the \texttt{dbr} prefix, the resulting PPBF has one partition for the \texttt{dbr} prefix that is a result of the bitwise \texttt{AND} operation between the two corresponding partitions in $\mathcal{B}^P_1$ and $\mathcal{B}^P_2$.

%% file: figures/running_network.tex
\tikzset{every picture/.style={line width=0.75pt}} %set default line width to 0.75pt        

\begin{tikzpicture}[x=0.5pt,y=0.5pt,yscale=-1,xscale=1]

%Flowchart: Magnetic Disk [id:dp6352431530638349] 
\draw  [fill={rgb, 255:red, 208; green, 2; blue, 27 }  ,fill opacity=1 ] (284,194) -- (284,207.65) .. controls (284,209.68) and (279.97,211.33) .. (275,211.33) .. controls (270.03,211.33) and (266,209.68) .. (266,207.65) -- (266,194)(284,194) .. controls (284,196.03) and (279.97,197.68) .. (275,197.68) .. controls (270.03,197.68) and (266,196.03) .. (266,194) .. controls (266,191.97) and (270.03,190.33) .. (275,190.33) .. controls (279.97,190.33) and (284,191.97) .. (284,194) -- cycle ;
%Flowchart: Magnetic Disk [id:dp5174783359184019] 
\draw  [fill={rgb, 255:red, 208; green, 2; blue, 27 }  ,fill opacity=1 ] (449,184.68) -- (449,198.33) .. controls (449,200.35) and (444.97,202) .. (440,202) .. controls (435.03,202) and (431,200.35) .. (431,198.33) -- (431,184.68)(449,184.68) .. controls (449,186.7) and (444.97,188.35) .. (440,188.35) .. controls (435.03,188.35) and (431,186.7) .. (431,184.68) .. controls (431,182.65) and (435.03,181) .. (440,181) .. controls (444.97,181) and (449,182.65) .. (449,184.68) -- cycle ;
%Straight Lines [id:da8850526574661792] 
\draw    (180,115) -- (270.14,85.92) ;
\draw [shift={(273,85)}, rotate = 162.12] [fill={rgb, 255:red, 0; green, 0; blue, 0 }  ][line width=0.08]  [draw opacity=0] (8.93,-4.29) -- (0,0) -- (8.93,4.29) -- cycle    ;
%Straight Lines [id:da39307299957818476] 
\draw    (183,134) -- (387.11,191.19) ;
\draw [shift={(390,192)}, rotate = 195.65] [fill={rgb, 255:red, 0; green, 0; blue, 0 }  ][line width=0.08]  [draw opacity=0] (8.93,-4.29) -- (0,0) -- (8.93,4.29) -- cycle    ;
%Straight Lines [id:da004801653920722848] 
\draw    (289,128) -- (259.65,172.5) ;
\draw [shift={(258,175)}, rotate = 303.41] [fill={rgb, 255:red, 0; green, 0; blue, 0 }  ][line width=0.08]  [draw opacity=0] (8.93,-4.29) -- (0,0) -- (8.93,4.29) -- cycle    ;
%Straight Lines [id:da6187310185209809] 
\draw    (346,76.5) -- (439,76.98) ;
\draw [shift={(442,77)}, rotate = 180.3] [fill={rgb, 255:red, 0; green, 0; blue, 0 }  ][line width=0.08]  [draw opacity=0] (8.93,-4.29) -- (0,0) -- (8.93,4.29) -- cycle    ;
%Straight Lines [id:da8836735963449948] 
\draw    (225,184) -- (180.48,153.69) ;
\draw [shift={(178,152)}, rotate = 34.25] [fill={rgb, 255:red, 0; green, 0; blue, 0 }  ][line width=0.08]  [draw opacity=0] (8.93,-4.29) -- (0,0) -- (8.93,4.29) -- cycle    ;
%Straight Lines [id:da8738031944664053] 
\draw    (396,174) -- (335.04,108.2) ;
\draw [shift={(333,106)}, rotate = 47.19] [fill={rgb, 255:red, 0; green, 0; blue, 0 }  ][line width=0.08]  [draw opacity=0] (8.93,-4.29) -- (0,0) -- (8.93,4.29) -- cycle    ;
%Straight Lines [id:da7101019251571523] 
\draw    (436,110) -- (291.6,192.51) ;
\draw [shift={(289,194)}, rotate = 330.26] [fill={rgb, 255:red, 0; green, 0; blue, 0 }  ][line width=0.08]  [draw opacity=0] (8.93,-4.29) -- (0,0) -- (8.93,4.29) -- cycle    ;
%Straight Lines [id:da5112941509625328] 
\draw    (451,125) -- (435.93,171.15) ;
\draw [shift={(435,174)}, rotate = 288.08] [fill={rgb, 255:red, 0; green, 0; blue, 0 }  ][line width=0.08]  [draw opacity=0] (8.93,-4.29) -- (0,0) -- (8.93,4.29) -- cycle    ;
%Straight Lines [id:da12679212930590267] 
\draw    (294,207.5) -- (384,207.5) ;
\draw [shift={(387,207.5)}, rotate = 180] [fill={rgb, 255:red, 0; green, 0; blue, 0 }  ][line width=0.08]  [draw opacity=0] (8.93,-4.29) -- (0,0) -- (8.93,4.29) -- cycle    ;
%Straight Lines [id:da7537555635927422] 
\draw    (451,178) -- (465.15,129.88) ;
\draw [shift={(466,127)}, rotate = 106.39] [fill={rgb, 255:red, 0; green, 0; blue, 0 }  ][line width=0.08]  [draw opacity=0] (8.93,-4.29) -- (0,0) -- (8.93,4.29) -- cycle    ;
%Flowchart: Magnetic Disk [id:dp4793084370933033] 
\draw  [fill={rgb, 255:red, 184; green, 233; blue, 134 }  ,fill opacity=1 ] (550,86.68) -- (550,100.33) .. controls (550,102.35) and (545.97,104) .. (541,104) .. controls (536.03,104) and (532,102.35) .. (532,100.33) -- (532,86.68)(550,86.68) .. controls (550,88.7) and (545.97,90.35) .. (541,90.35) .. controls (536.03,90.35) and (532,88.7) .. (532,86.68) .. controls (532,84.65) and (536.03,83) .. (541,83) .. controls (545.97,83) and (550,84.65) .. (550,86.68) -- cycle ;
%Flowchart: Magnetic Disk [id:dp6468030958162768] 
\draw  [fill={rgb, 255:red, 208; green, 2; blue, 27 }  ,fill opacity=1 ] (550,112.68) -- (550,126.33) .. controls (550,128.35) and (545.97,130) .. (541,130) .. controls (536.03,130) and (532,128.35) .. (532,126.33) -- (532,112.68)(550,112.68) .. controls (550,114.7) and (545.97,116.35) .. (541,116.35) .. controls (536.03,116.35) and (532,114.7) .. (532,112.68) .. controls (532,110.65) and (536.03,109) .. (541,109) .. controls (545.97,109) and (550,110.65) .. (550,112.68) -- cycle ;
%Flowchart: Magnetic Disk [id:dp5716763281142162] 
\draw  [fill={rgb, 255:red, 189; green, 16; blue, 224 }  ,fill opacity=1 ] (550,138.68) -- (550,152.32) .. controls (550,154.35) and (545.97,156) .. (541,156) .. controls (536.03,156) and (532,154.35) .. (532,152.32) -- (532,138.68)(550,138.68) .. controls (550,140.7) and (545.97,142.35) .. (541,142.35) .. controls (536.03,142.35) and (532,140.7) .. (532,138.68) .. controls (532,136.65) and (536.03,135) .. (541,135) .. controls (545.97,135) and (550,136.65) .. (550,138.68) -- cycle ;
%Flowchart: Magnetic Disk [id:dp6286074456205583] 
\draw  [fill={rgb, 255:red, 189; green, 16; blue, 224 }  ,fill opacity=1 ] (503,82.68) -- (503,96.33) .. controls (503,98.35) and (498.97,100) .. (494,100) .. controls (489.03,100) and (485,98.35) .. (485,96.33) -- (485,82.68)(503,82.68) .. controls (503,84.7) and (498.97,86.35) .. (494,86.35) .. controls (489.03,86.35) and (485,84.7) .. (485,82.68) .. controls (485,80.65) and (489.03,79) .. (494,79) .. controls (498.97,79) and (503,80.65) .. (503,82.68) -- cycle ;
%Flowchart: Magnetic Disk [id:dp9890972643050008] 
\draw  [fill={rgb, 255:red, 245; green, 166; blue, 35 }  ,fill opacity=1 ] (550,164.68) -- (550,178.33) .. controls (550,180.35) and (545.97,182) .. (541,182) .. controls (536.03,182) and (532,180.35) .. (532,178.33) -- (532,164.68)(550,164.68) .. controls (550,166.7) and (545.97,168.35) .. (541,168.35) .. controls (536.03,168.35) and (532,166.7) .. (532,164.68) .. controls (532,162.65) and (536.03,161) .. (541,161) .. controls (545.97,161) and (550,162.65) .. (550,164.68) -- cycle ;
%Flowchart: Magnetic Disk [id:dp17978437030553518] 
\draw  [fill={rgb, 255:red, 139; green, 87; blue, 42 }  ,fill opacity=1 ] (550,190.68) -- (550,204.33) .. controls (550,206.35) and (545.97,208) .. (541,208) .. controls (536.03,208) and (532,206.35) .. (532,204.33) -- (532,190.68)(550,190.68) .. controls (550,192.7) and (545.97,194.35) .. (541,194.35) .. controls (536.03,194.35) and (532,192.7) .. (532,190.68) .. controls (532,188.65) and (536.03,187) .. (541,187) .. controls (545.97,187) and (550,188.65) .. (550,190.68) -- cycle ;
%Flowchart: Magnetic Disk [id:dp3741147063849106] 
\draw  [fill={rgb, 255:red, 139; green, 87; blue, 42 }  ,fill opacity=1 ] (177,125.68) -- (177,139.32) .. controls (177,141.35) and (172.97,143) .. (168,143) .. controls (163.03,143) and (159,141.35) .. (159,139.32) -- (159,125.68)(177,125.68) .. controls (177,127.7) and (172.97,129.35) .. (168,129.35) .. controls (163.03,129.35) and (159,127.7) .. (159,125.68) .. controls (159,123.65) and (163.03,122) .. (168,122) .. controls (172.97,122) and (177,123.65) .. (177,125.68) -- cycle ;
%Flowchart: Magnetic Disk [id:dp019000640976979066] 
\draw  [fill={rgb, 255:red, 184; green, 233; blue, 134 }  ,fill opacity=1 ] (337,55.68) -- (337,69.33) .. controls (337,71.35) and (332.97,73) .. (328,73) .. controls (323.03,73) and (319,71.35) .. (319,69.33) -- (319,55.68)(337,55.68) .. controls (337,57.7) and (332.97,59.35) .. (328,59.35) .. controls (323.03,59.35) and (319,57.7) .. (319,55.68) .. controls (319,53.65) and (323.03,52) .. (328,52) .. controls (332.97,52) and (337,53.65) .. (337,55.68) -- cycle ;
%Flowchart: Magnetic Disk [id:dp9318868802697046] 
\draw  [fill={rgb, 255:red, 245; green, 166; blue, 35 }  ,fill opacity=1 ] (337,81.33) -- (337,94.98) .. controls (337,97) and (332.97,98.65) .. (328,98.65) .. controls (323.03,98.65) and (319,97) .. (319,94.98) -- (319,81.33)(337,81.33) .. controls (337,83.35) and (332.97,85) .. (328,85) .. controls (323.03,85) and (319,83.35) .. (319,81.33) .. controls (319,79.3) and (323.03,77.65) .. (328,77.65) .. controls (332.97,77.65) and (337,79.3) .. (337,81.33) -- cycle ;
%Flowchart: Magnetic Disk [id:dp7769952530172142] 
\draw  [fill={rgb, 255:red, 189; green, 16; blue, 224 }  ,fill opacity=1 ] (284,218) -- (284,231.65) .. controls (284,233.68) and (279.97,235.33) .. (275,235.33) .. controls (270.03,235.33) and (266,233.68) .. (266,231.65) -- (266,218)(284,218) .. controls (284,220.03) and (279.97,221.68) .. (275,221.68) .. controls (270.03,221.68) and (266,220.03) .. (266,218) .. controls (266,215.97) and (270.03,214.33) .. (275,214.33) .. controls (279.97,214.33) and (284,215.97) .. (284,218) -- cycle ;
%Flowchart: Magnetic Disk [id:dp9419660020949427] 
\draw  [fill={rgb, 255:red, 139; green, 87; blue, 42 }  ,fill opacity=1 ] (449,231.33) -- (449,244.97) .. controls (449,247) and (444.97,248.65) .. (440,248.65) .. controls (435.03,248.65) and (431,247) .. (431,244.97) -- (431,231.33)(449,231.33) .. controls (449,233.35) and (444.97,235) .. (440,235) .. controls (435.03,235) and (431,233.35) .. (431,231.33) .. controls (431,229.3) and (435.03,227.65) .. (440,227.65) .. controls (444.97,227.65) and (449,229.3) .. (449,231.33) -- cycle ;
%Flowchart: Magnetic Disk [id:dp24350919550825156] 
\draw  [fill={rgb, 255:red, 184; green, 233; blue, 134 }  ,fill opacity=1 ] (503,57.68) -- (503,71.33) .. controls (503,73.35) and (498.97,75) .. (494,75) .. controls (489.03,75) and (485,73.35) .. (485,71.33) -- (485,57.68)(503,57.68) .. controls (503,59.7) and (498.97,61.35) .. (494,61.35) .. controls (489.03,61.35) and (485,59.7) .. (485,57.68) .. controls (485,55.65) and (489.03,54) .. (494,54) .. controls (498.97,54) and (503,55.65) .. (503,57.68) -- cycle ;
%Flowchart: Magnetic Disk [id:dp22121121049134418] 
\draw  [fill={rgb, 255:red, 245; green, 166; blue, 35 }  ,fill opacity=1 ] (449,208.33) -- (449,221.97) .. controls (449,224) and (444.97,225.65) .. (440,225.65) .. controls (435.03,225.65) and (431,224) .. (431,221.97) -- (431,208.33)(449,208.33) .. controls (449,210.35) and (444.97,212) .. (440,212) .. controls (435.03,212) and (431,210.35) .. (431,208.33) .. controls (431,206.3) and (435.03,204.65) .. (440,204.65) .. controls (444.97,204.65) and (449,206.3) .. (449,208.33) -- cycle ;
%Shape: Path Data [id:dp7586317961955396] 
\draw  [fill={rgb, 255:red, 125; green, 184; blue, 253 }  ,fill opacity=1 ] (137.47,121.81) .. controls (146.18,121.81) and (153.39,137.16) .. (154.5,157.04) -- (120.43,157.04) .. controls (121.54,137.16) and (128.75,121.81) .. (137.47,121.81) -- cycle (137.47,98.5) .. controls (143.9,98.5) and (149.12,103.72) .. (149.12,110.16) .. controls (149.12,116.59) and (143.9,121.81) .. (137.47,121.81) .. controls (131.03,121.81) and (125.81,116.59) .. (125.81,110.16) .. controls (125.81,103.72) and (131.03,98.5) .. (137.47,98.5) -- cycle ;
%Shape: Path Data [id:dp3855727592911915] 
\draw  [fill={rgb, 255:red, 125; green, 184; blue, 253 }  ,fill opacity=1 ] (244.47,202.81) .. controls (253.18,202.81) and (260.39,218.16) .. (261.5,238.04) -- (227.43,238.04) .. controls (228.54,218.16) and (235.75,202.81) .. (244.47,202.81) -- cycle (244.47,179.5) .. controls (250.9,179.5) and (256.12,184.72) .. (256.12,191.16) .. controls (256.12,197.59) and (250.9,202.81) .. (244.47,202.81) .. controls (238.03,202.81) and (232.81,197.59) .. (232.81,191.16) .. controls (232.81,184.72) and (238.03,179.5) .. (244.47,179.5) -- cycle ;
%Shape: Path Data [id:dp8607063020087472] 
\draw  [fill={rgb, 255:red, 125; green, 184; blue, 253 }  ,fill opacity=1 ] (296.47,68.81) .. controls (305.18,68.81) and (312.39,84.16) .. (313.5,104.04) -- (279.43,104.04) .. controls (280.54,84.16) and (287.75,68.81) .. (296.47,68.81) -- cycle (296.47,45.5) .. controls (302.9,45.5) and (308.12,50.72) .. (308.12,57.16) .. controls (308.12,63.59) and (302.9,68.81) .. (296.47,68.81) .. controls (290.03,68.81) and (284.81,63.59) .. (284.81,57.16) .. controls (284.81,50.72) and (290.03,45.5) .. (296.47,45.5) -- cycle ;
%Shape: Path Data [id:dp1521604148935043] 
\draw  [fill={rgb, 255:red, 125; green, 184; blue, 253 }  ,fill opacity=1 ] (462.47,70.81) .. controls (471.18,70.81) and (478.39,86.16) .. (479.5,106.04) -- (445.43,106.04) .. controls (446.54,86.16) and (453.75,70.81) .. (462.47,70.81) -- cycle (462.47,47.5) .. controls (468.9,47.5) and (474.12,52.72) .. (474.12,59.16) .. controls (474.12,65.59) and (468.9,70.81) .. (462.47,70.81) .. controls (456.03,70.81) and (450.81,65.59) .. (450.81,59.16) .. controls (450.81,52.72) and (456.03,47.5) .. (462.47,47.5) -- cycle ;
%Shape: Path Data [id:dp7882406703964574] 
\draw  [fill={rgb, 255:red, 125; green, 184; blue, 253 }  ,fill opacity=1 ] (409.47,204.81) .. controls (418.18,204.81) and (425.39,220.16) .. (426.5,240.04) -- (392.43,240.04) .. controls (393.54,220.16) and (400.75,204.81) .. (409.47,204.81) -- cycle (409.47,181.5) .. controls (415.9,181.5) and (421.12,186.72) .. (421.12,193.16) .. controls (421.12,199.59) and (415.9,204.81) .. (409.47,204.81) .. controls (403.03,204.81) and (397.81,199.59) .. (397.81,193.16) .. controls (397.81,186.72) and (403.03,181.5) .. (409.47,181.5) -- cycle ;

% Text Node
\draw (128,159.4) node [anchor=north west][inner sep=0.75pt]    {$n_{1}$};
% Text Node
\draw (285,108.4) node [anchor=north west][inner sep=0.75pt]    {$n_{2}$};
% Text Node
\draw (234,240.4) node [anchor=north west][inner sep=0.75pt]    {$n_{3}$};
% Text Node
\draw (450,108.4) node [anchor=north west][inner sep=0.75pt]    {$n_{4}$};
% Text Node
\draw (399,242.4) node [anchor=north west][inner sep=0.75pt]    {$n_{5}$};
% Text Node
\draw (554,83.4) node [anchor=north west][inner sep=0.75pt]    {$f_{1}$};
% Text Node
\draw (554,109.4) node [anchor=north west][inner sep=0.75pt]    {$f_{2}$};
% Text Node
\draw (554,135.4) node [anchor=north west][inner sep=0.75pt]    {$f_{3}$};
% Text Node
\draw (554,161.4) node [anchor=north west][inner sep=0.75pt]    {$f_{4}$};
% Text Node
\draw (554,187.4) node [anchor=north west][inner sep=0.75pt]    {$f_{5}$};

\end{tikzpicture}

%% file: figures/running_node.tex
\tikzset{every picture/.style={line width=0.75pt}} %set default line width to 0.75pt        

\begin{tikzpicture}[x=0.75pt,y=0.75pt,yscale=-1,xscale=1]

%Shape: Rectangle [id:dp26914463233082064] 
\draw  [line width=0.75]  (153,116) -- (294,116) -- (294,212) -- (153,212) -- cycle ;
%Shape: Rectangle [id:dp7303055948220758] 
\draw   (160,136) -- (229.5,136) -- (229.5,169) -- (160,169) -- cycle ;
%Flowchart: Magnetic Disk [id:dp17575695748455855] 
\draw  [fill={rgb, 255:red, 139; green, 87; blue, 42 }  ,fill opacity=1 ] (183.5,158.5) -- (183.5,171.5) .. controls (183.5,173.43) and (179.81,175) .. (175.25,175) .. controls (170.69,175) and (167,173.43) .. (167,171.5) -- (167,158.5)(183.5,158.5) .. controls (183.5,160.43) and (179.81,162) .. (175.25,162) .. controls (170.69,162) and (167,160.43) .. (167,158.5) .. controls (167,156.57) and (170.69,155) .. (175.25,155) .. controls (179.81,155) and (183.5,156.57) .. (183.5,158.5) -- cycle ;
%Flowchart: Magnetic Disk [id:dp79552616451014] 
\draw  [fill={rgb, 255:red, 208; green, 2; blue, 27 }  ,fill opacity=1 ] (203.5,158.5) -- (203.5,171.5) .. controls (203.5,173.43) and (199.81,175) .. (195.25,175) .. controls (190.69,175) and (187,173.43) .. (187,171.5) -- (187,158.5)(203.5,158.5) .. controls (203.5,160.43) and (199.81,162) .. (195.25,162) .. controls (190.69,162) and (187,160.43) .. (187,158.5) .. controls (187,156.57) and (190.69,155) .. (195.25,155) .. controls (199.81,155) and (203.5,156.57) .. (203.5,158.5) -- cycle ;
%Flowchart: Magnetic Disk [id:dp8417565938329913] 
\draw  [fill={rgb, 255:red, 245; green, 166; blue, 35 }  ,fill opacity=1 ] (223.5,158.5) -- (223.5,171.5) .. controls (223.5,173.43) and (219.81,175) .. (215.25,175) .. controls (210.69,175) and (207,173.43) .. (207,171.5) -- (207,158.5)(223.5,158.5) .. controls (223.5,160.43) and (219.81,162) .. (215.25,162) .. controls (210.69,162) and (207,160.43) .. (207,158.5) .. controls (207,156.57) and (210.69,155) .. (215.25,155) .. controls (219.81,155) and (223.5,156.57) .. (223.5,158.5) -- cycle ;
%Shape: Rectangle [id:dp3369668731365222] 
\draw   (160,180) -- (229.5,180) -- (229.5,201) -- (160,201) -- cycle ;
%Shape: Circle [id:dp6023874245865927] 
\draw  [fill={rgb, 255:red, 139; green, 87; blue, 42 }  ,fill opacity=1 ] (163,201) .. controls (163,197.69) and (165.69,195) .. (169,195) .. controls (172.31,195) and (175,197.69) .. (175,201) .. controls (175,204.31) and (172.31,207) .. (169,207) .. controls (165.69,207) and (163,204.31) .. (163,201) -- cycle ;
%Shape: Circle [id:dp25567600448433936] 
\draw  [fill={rgb, 255:red, 208; green, 2; blue, 27 }  ,fill opacity=1 ] (176.09,201) .. controls (176.09,197.69) and (178.78,195) .. (182.09,195) .. controls (185.4,195) and (188.09,197.69) .. (188.09,201) .. controls (188.09,204.31) and (185.4,207) .. (182.09,207) .. controls (178.78,207) and (176.09,204.31) .. (176.09,201) -- cycle ;
%Shape: Circle [id:dp6929761000356008] 
\draw  [fill={rgb, 255:red, 245; green, 166; blue, 35 }  ,fill opacity=1 ] (189.18,201) .. controls (189.18,197.69) and (191.87,195) .. (195.18,195) .. controls (198.5,195) and (201.18,197.69) .. (201.18,201) .. controls (201.18,204.31) and (198.5,207) .. (195.18,207) .. controls (191.87,207) and (189.18,204.31) .. (189.18,201) -- cycle ;
%Shape: Circle [id:dp15373061466689975] 
\draw  [fill={rgb, 255:red, 184; green, 233; blue, 134 }  ,fill opacity=1 ] (202.27,201) .. controls (202.27,197.69) and (204.96,195) .. (208.27,195) .. controls (211.59,195) and (214.27,197.69) .. (214.27,201) .. controls (214.27,204.31) and (211.59,207) .. (208.27,207) .. controls (204.96,207) and (202.27,204.31) .. (202.27,201) -- cycle ;
%Shape: Circle [id:dp2949803749055241] 
\draw  [fill={rgb, 255:red, 189; green, 16; blue, 224 }  ,fill opacity=1 ] (215.36,201) .. controls (215.36,197.69) and (218.05,195) .. (221.36,195) .. controls (224.68,195) and (227.36,197.69) .. (227.36,201) .. controls (227.36,204.31) and (224.68,207) .. (221.36,207) .. controls (218.05,207) and (215.36,204.31) .. (215.36,201) -- cycle ;
%Shape: Rectangle [id:dp26643268784095964] 
\draw   (236,147) -- (286,147) -- (286,186.5) -- (236,186.5) -- cycle ;
%Shape: Circle [id:dp2347478513917549] 
\draw  [draw opacity=0][fill={rgb, 255:red, 125; green, 184; blue, 253 }  ,fill opacity=1 ] (235,186.5) .. controls (235,179.6) and (240.6,174) .. (247.5,174) .. controls (254.4,174) and (260,179.6) .. (260,186.5) .. controls (260,193.4) and (254.4,199) .. (247.5,199) .. controls (240.6,199) and (235,193.4) .. (235,186.5) -- cycle ;
%Shape: Circle [id:dp019439928672374318] 
\draw  [draw opacity=0][fill={rgb, 255:red, 125; green, 184; blue, 253 }  ,fill opacity=1 ] (261,186.5) .. controls (261,179.6) and (266.6,174) .. (273.5,174) .. controls (280.4,174) and (286,179.6) .. (286,186.5) .. controls (286,193.4) and (280.4,199) .. (273.5,199) .. controls (266.6,199) and (261,193.4) .. (261,186.5) -- cycle ;

% Text Node
\draw (168,138) node [anchor=north west][inner sep=0.75pt]   [align=left] {\textit{{\tiny Datastore}}};
% Text Node
\draw (180,181) node [anchor=north west][inner sep=0.75pt]   [align=left] {{\tiny \textit{Index}}};
% Text Node
\draw (205,119) node [anchor=north west][inner sep=0.75pt]   [align=left] {\textbf{\textit{node }}$\displaystyle n_{5}$};
% Text Node
\draw (244,148) node [anchor=north west][inner sep=0.75pt]   [align=left] {\textit{{\tiny Neighbors}}};
% Text Node
\draw (240,183.4) node [anchor=north west][inner sep=0.75pt]    {$n_{2}$};
% Text Node
\draw (266,183.4) node [anchor=north west][inner sep=0.75pt]    {$n_{4}$};

\end{tikzpicture}

%% file: figures/ppbfex.tex
\tikzset{every picture/.style={line width=0.75pt}} %set default line width to 0.75pt        

\begin{tikzpicture}[x=0.5pt,y=0.5pt,yscale=-1,xscale=1]

%Shape: Rectangle [id:dp4142055042198082] 
\draw  [color={rgb, 255:red, 0; green, 0; blue, 0 }  ,draw opacity=1 ] (129,186.13) -- (209,186.13) -- (209,206.13) -- (129,206.13) -- cycle ;
%Shape: Rectangle [id:dp3014560934421402] 
\draw   (129,186.13) -- (149,186.13) -- (149,206.13) -- (129,206.13) -- cycle ;
%Shape: Rectangle [id:dp05608647522341936] 
\draw   (189,186.13) -- (209,186.13) -- (209,206.13) -- (189,206.13) -- cycle ;
%Shape: Rectangle [id:dp7619783496833951] 
\draw   (238,123.96) -- (264,123.96) -- (264,149.45) -- (238,149.45) -- cycle ;
%Shape: Rectangle [id:dp7241819916328813] 
\draw   (328,123.96) -- (354,123.96) -- (354,149.45) -- (328,149.45) -- cycle ;
%Shape: Rectangle [id:dp2664736710704614] 
\draw   (421,123.96) -- (447,123.96) -- (447,149.45) -- (421,149.45) -- cycle ;
%Straight Lines [id:da8952391467953862] 
\draw    (251,149.45) -- (338.21,184.03) ;
\draw [shift={(341,185.13)}, rotate = 201.63] [fill={rgb, 255:red, 0; green, 0; blue, 0 }  ][line width=0.08]  [draw opacity=0] (8.93,-4.29) -- (0,0) -- (8.93,4.29) -- cycle    ;
%Straight Lines [id:da28227429578312435] 
\draw    (342,150.18) -- (286.57,183.59) ;
\draw [shift={(284,185.13)}, rotate = 328.92] [fill={rgb, 255:red, 0; green, 0; blue, 0 }  ][line width=0.08]  [draw opacity=0] (8.93,-4.29) -- (0,0) -- (8.93,4.29) -- cycle    ;
%Straight Lines [id:da611355384943438] 
\draw    (434,149.45) -- (423.88,182.27) ;
\draw [shift={(423,185.13)}, rotate = 287.13] [fill={rgb, 255:red, 0; green, 0; blue, 0 }  ][line width=0.08]  [draw opacity=0] (8.93,-4.29) -- (0,0) -- (8.93,4.29) -- cycle    ;
%Straight Lines [id:da1718312958854884] 
\draw    (344,100.36) -- (255.91,122.36) ;
\draw [shift={(253,123.08)}, rotate = 345.98] [fill={rgb, 255:red, 0; green, 0; blue, 0 }  ][line width=0.08]  [draw opacity=0] (8.93,-4.29) -- (0,0) -- (8.93,4.29) -- cycle    ;
%Straight Lines [id:da7412977463352984] 
\draw    (344,100.36) -- (344,120.08) ;
\draw [shift={(344,123.08)}, rotate = 270] [fill={rgb, 255:red, 0; green, 0; blue, 0 }  ][line width=0.08]  [draw opacity=0] (8.93,-4.29) -- (0,0) -- (8.93,4.29) -- cycle    ;
%Straight Lines [id:da13544282813615116] 
\draw    (344,100.36) -- (433.09,122.36) ;
\draw [shift={(436,123.08)}, rotate = 193.87] [fill={rgb, 255:red, 0; green, 0; blue, 0 }  ][line width=0.08]  [draw opacity=0] (8.93,-4.29) -- (0,0) -- (8.93,4.29) -- cycle    ;
%Shape: Rectangle [id:dp600639296089989] 
\draw  [color={rgb, 255:red, 0; green, 0; blue, 0 }  ,draw opacity=1 ] (469,186.13) -- (549,186.13) -- (549,206.13) -- (469,206.13) -- cycle ;
%Shape: Rectangle [id:dp18262642410916097] 
\draw   (469,186.13) -- (489,186.13) -- (489,206.13) -- (469,206.13) -- cycle ;
%Shape: Rectangle [id:dp9741199482240703] 
\draw   (529,186.13) -- (549,186.13) -- (549,206.13) -- (529,206.13) -- cycle ;
%Shape: Rectangle [id:dp2798055867890199] 
\draw  [color={rgb, 255:red, 0; green, 0; blue, 0 }  ,draw opacity=1 ] (272,186.13) -- (432,186.13) -- (432,206.13) -- (272,206.13) -- cycle ;
%Shape: Rectangle [id:dp9791567375717689] 
\draw  [fill={rgb, 255:red, 0; green, 0; blue, 0 }  ,fill opacity=1 ] (272,186.13) -- (292,186.13) -- (292,206.13) -- (272,206.13) -- cycle ;
%Shape: Rectangle [id:dp07242934968327275] 
\draw   (292,186.13) -- (312,186.13) -- (312,206.13) -- (292,206.13) -- cycle ;
%Shape: Rectangle [id:dp19740163073839445] 
\draw   (312,186.13) -- (332,186.13) -- (332,206.13) -- (312,206.13) -- cycle ;
%Shape: Rectangle [id:dp5871584719365329] 
\draw   (352,186.13) -- (372,186.13) -- (372,206.13) -- (352,206.13) -- cycle ;
%Shape: Rectangle [id:dp49543172598594365] 
\draw  [color={rgb, 255:red, 0; green, 0; blue, 0 }  ,draw opacity=1 ][fill={rgb, 255:red, 0; green, 0; blue, 0 }  ,fill opacity=1 ] (332,186.13) -- (352,186.13) -- (352,206.13) -- (332,206.13) -- cycle ;
%Shape: Rectangle [id:dp6595857774806794] 
\draw  [fill={rgb, 255:red, 0; green, 0; blue, 0 }  ,fill opacity=1 ] (412,186.13) -- (432,186.13) -- (432,206.13) -- (412,206.13) -- cycle ;
%Straight Lines [id:da6990514601420407] 
\draw    (219,217) -- (333,217) ;
\draw [shift={(336,217)}, rotate = 180] [fill={rgb, 255:red, 0; green, 0; blue, 0 }  ][line width=0.08]  [draw opacity=0] (8.93,-4.29) -- (0,0) -- (8.93,4.29) -- cycle    ;
%Straight Lines [id:da395491638806845] 
\draw    (219,217) -- (219,116) ;
%Straight Lines [id:da5281058895737583] 
\draw    (284,94) -- (219,116) ;

% Text Node
\draw (169,196.13) node   [align=left] {...};
% Text Node
\draw (251,136.7) node    {$h_{1}$};
% Text Node
\draw (341,136.7) node    {$h_{2}$};
% Text Node
\draw (434,136.7) node    {$h_{k}$};
% Text Node
\draw (340,89) node   [align=left] {\footnotesize\texttt{dbr:Copenhagen}};
% Text Node
\draw (391,136.07) node   [align=left] {...};
% Text Node
\draw (171,216) node  [font=\scriptsize] [align=left] {dbo};
% Text Node
\draw (509,196.13) node   [align=left] {...};
% Text Node
\draw (511,217) node  [font=\scriptsize] [align=left] {dbp};
% Text Node
\draw (353,217) node  [font=\scriptsize] [align=left] {dbr};
% Text Node
\draw (452,197.07) node   [align=left] {...};
% Text Node
\draw (392,195.13) node   [align=left] {...};

\end{tikzpicture}

%% file: figures/bsint.tex
\tikzset{every picture/.style={line width=0.75pt}} %set default line width to 0.75pt        

\begin{tikzpicture}[x=0.5pt,y=0.5pt,yscale=-1,xscale=1]

%Straight Lines [id:da18352266510628046] 
\draw    (139,181) -- (535,181) ;
%Shape: Rectangle [id:dp08734995192856154] 
\draw  [color={rgb, 255:red, 0; green, 0; blue, 0 }  ,draw opacity=1 ] (340,151) -- (420,151) -- (420,171) -- (340,171) -- cycle ;
%Shape: Rectangle [id:dp9660264488846108] 
\draw   (360,151) -- (380,151) -- (380,171) -- (360,171) -- cycle ;
%Shape: Rectangle [id:dp38578573432036267] 
\draw   (400,151) -- (420,151) -- (420,171) -- (400,171) -- cycle ;
%Shape: Rectangle [id:dp47091388239638854] 
\draw  [color={rgb, 255:red, 0; green, 0; blue, 0 }  ,draw opacity=1 ] (260,151) -- (340,151) -- (340,171) -- (260,171) -- cycle ;
%Shape: Rectangle [id:dp7932710485988459] 
\draw   (280,151) -- (300,151) -- (300,171) -- (280,171) -- cycle ;
%Shape: Rectangle [id:dp30027514151301027] 
\draw   (320,151) -- (340,151) -- (340,171) -- (320,171) -- cycle ;
%Shape: Rectangle [id:dp3592306917578667] 
\draw  [fill={rgb, 255:red, 0; green, 0; blue, 0 }  ,fill opacity=1 ] (340,151) -- (360,151) -- (360,171) -- (340,171) -- cycle ;
%Shape: Rectangle [id:dp3672965904428305] 
\draw  [fill={rgb, 255:red, 0; green, 0; blue, 0 }  ,fill opacity=1 ] (280,151) -- (300,151) -- (300,171) -- (280,171) -- cycle ;
%Shape: Rectangle [id:dp8716096276369809] 
\draw   (487,151) -- (507,151) -- (507,171) -- (487,171) -- cycle ;
%Shape: Rectangle [id:dp1748144033824922] 
\draw  [color={rgb, 255:red, 0; green, 0; blue, 0 }  ,draw opacity=1 ] (427,151) -- (507,151) -- (507,171) -- (427,171) -- cycle ;
%Shape: Rectangle [id:dp7965680899664451] 
\draw   (447,151) -- (467,151) -- (467,171) -- (447,171) -- cycle ;
%Shape: Rectangle [id:dp6732853782792414] 
\draw  [fill={rgb, 255:red, 0; green, 0; blue, 0 }  ,fill opacity=1 ] (447,151) -- (467,151) -- (467,171) -- (447,171) -- cycle ;
%Shape: Rectangle [id:dp31086567334225723] 
\draw  [fill={rgb, 255:red, 0; green, 0; blue, 0 }  ,fill opacity=1 ] (360,151) -- (380,151) -- (380,171) -- (360,171) -- cycle ;
%Shape: Rectangle [id:dp765818424697233] 
\draw   (233,122) -- (253,122) -- (253,142) -- (233,142) -- cycle ;
%Shape: Rectangle [id:dp32736130608748004] 
\draw  [color={rgb, 255:red, 0; green, 0; blue, 0 }  ,draw opacity=1 ] (173,122) -- (253,122) -- (253,142) -- (173,142) -- cycle ;
%Shape: Rectangle [id:dp36108894287606075] 
\draw   (193,122) -- (213,122) -- (213,142) -- (193,142) -- cycle ;
%Shape: Rectangle [id:dp8745644454828159] 
\draw  [fill={rgb, 255:red, 0; green, 0; blue, 0 }  ,fill opacity=1 ] (233,122) -- (253,122) -- (253,142) -- (233,142) -- cycle ;
%Shape: Rectangle [id:dp712010928785983] 
\draw  [color={rgb, 255:red, 0; green, 0; blue, 0 }  ,draw opacity=1 ] (340,122) -- (420,122) -- (420,142) -- (340,142) -- cycle ;
%Shape: Rectangle [id:dp5435167957882441] 
\draw   (360,122) -- (380,122) -- (380,142) -- (360,142) -- cycle ;
%Shape: Rectangle [id:dp13943735222839804] 
\draw   (400,122) -- (420,122) -- (420,142) -- (400,142) -- cycle ;
%Shape: Rectangle [id:dp04756711032179195] 
\draw  [color={rgb, 255:red, 0; green, 0; blue, 0 }  ,draw opacity=1 ] (260,122) -- (340,122) -- (340,142) -- (260,142) -- cycle ;
%Shape: Rectangle [id:dp5690016889981447] 
\draw   (280,122) -- (300,122) -- (300,142) -- (280,142) -- cycle ;
%Shape: Rectangle [id:dp6811615696592271] 
\draw   (320,122) -- (340,122) -- (340,142) -- (320,142) -- cycle ;
%Shape: Rectangle [id:dp6461082644228638] 
\draw  [fill={rgb, 255:red, 0; green, 0; blue, 0 }  ,fill opacity=1 ] (340,122) -- (360,122) -- (360,142) -- (340,142) -- cycle ;
%Shape: Rectangle [id:dp3361534813539174] 
\draw  [fill={rgb, 255:red, 0; green, 0; blue, 0 }  ,fill opacity=1 ] (280,122) -- (300,122) -- (300,142) -- (280,142) -- cycle ;
%Shape: Rectangle [id:dp000926792175268254] 
\draw  [fill={rgb, 255:red, 0; green, 0; blue, 0 }  ,fill opacity=1 ] (360,122) -- (380,122) -- (380,142) -- (360,142) -- cycle ;
%Shape: Rectangle [id:dp45251462448401714] 
\draw  [color={rgb, 255:red, 0; green, 0; blue, 0 }  ,draw opacity=1 ] (340,191) -- (420,191) -- (420,211) -- (340,211) -- cycle ;
%Shape: Rectangle [id:dp5496614004896594] 
\draw   (360,191) -- (380,191) -- (380,211) -- (360,211) -- cycle ;
%Shape: Rectangle [id:dp28722857105587984] 
\draw   (400,191) -- (420,191) -- (420,211) -- (400,211) -- cycle ;
%Shape: Rectangle [id:dp766908463030603] 
\draw  [color={rgb, 255:red, 0; green, 0; blue, 0 }  ,draw opacity=1 ] (260,191) -- (340,191) -- (340,211) -- (260,211) -- cycle ;
%Shape: Rectangle [id:dp6273650723583887] 
\draw   (280,191) -- (300,191) -- (300,211) -- (280,211) -- cycle ;
%Shape: Rectangle [id:dp4020898256702735] 
\draw   (320,191) -- (340,191) -- (340,211) -- (320,211) -- cycle ;
%Shape: Rectangle [id:dp3161806951473476] 
\draw  [fill={rgb, 255:red, 0; green, 0; blue, 0 }  ,fill opacity=1 ] (340,191) -- (360,191) -- (360,211) -- (340,211) -- cycle ;
%Shape: Rectangle [id:dp3902297522134792] 
\draw  [fill={rgb, 255:red, 0; green, 0; blue, 0 }  ,fill opacity=1 ] (280,191) -- (300,191) -- (300,211) -- (280,211) -- cycle ;
%Shape: Rectangle [id:dp671125046650777] 
\draw  [fill={rgb, 255:red, 0; green, 0; blue, 0 }  ,fill opacity=1 ] (360,191) -- (380,191) -- (380,211) -- (360,211) -- cycle ;
%Shape: Rectangle [id:dp2077830505409899] 
\draw  [fill={rgb, 255:red, 0; green, 0; blue, 0 }  ,fill opacity=1 ] (300,151) -- (320,151) -- (320,171) -- (300,171) -- cycle ;
%Shape: Rectangle [id:dp7463438184908047] 
\draw  [fill={rgb, 255:red, 0; green, 0; blue, 0 }  ,fill opacity=1 ] (400,122) -- (420,122) -- (420,142) -- (400,142) -- cycle ;

% Text Node
\draw (391,160) node   [align=left] {...};
% Text Node
\draw (338,102) node  [font=\footnotesize] [align=left] {dbr};
% Text Node
\draw (478,160) node   [align=left] {...};
% Text Node
\draw (469,104) node  [font=\footnotesize] [align=left] {dbp};
% Text Node
\draw (224,131) node   [align=left] {...};
% Text Node
\draw (214,102) node  [font=\footnotesize] [align=left] {dbo};
% Text Node
\draw (391,131) node   [align=left] {...};
% Text Node
\draw (391,200) node   [align=left] {...};

\end{tikzpicture}

%% file: system.tex
%\system{} reduces the communication overhead when  and in doing so increases performance when processing queries in a decentralized setup.
%Differently from in \colchain, however, with \system{}, community participants do not replicate all the data within the community; instead, replication of the data throughout a community is based on the representative workload of each participants.
Differently from \piqnic{} and \colchain, \system{} uses a fragmentation strategy based on characteristic sets.
To accommodate efficient query processing over such fragments, as well as to enable locality-awareness and more accurate cardinality estimation, \system{} introduces an indexing scheme that maps star patterns to fragments rather than triple patterns.
%uses a fragmentation and decentralized indexing scheme that supports multiple predicates; note that \piqnic{} and \colchain{} use a fragmentation scheme limited to one predicate per fragment.
%changes the decentralized indexing schema used in \colchain{} to support multiple predicates in each fragment while maintaining the link between the subjects, predicates, and objects.
%\system{} uses a more accurate cardinality estimations that lets the query processor create more optimal join strategies.
In the remainder of this section, we provide a brief overview of the \system{} architecture and how \system{} optimizes SPARQL queries over decentralized knowledge graphs, followed by a formal definition of the fragmentation and indexing approach.
% indexing,
%and cardinality estimations, 
%and how \system{} reduces the network overhead.
Query optimization with details on how to exploit locality-awareness and join ordering are explained in Section~\ref{sec:queryoptimization}.
%For details on locality-awareness and join ordering, refer to Section~\ref{sec:queryprocessing}.

\subsection{Design and Overview}\label{subsec:design}
%As stated in Section~\ref{sec:preliminaries}, since \system{} could be implemented on top of any P2P system, we do not go into details with the structural aspects of P2P systems, such as neighborhood management and periodic shuffles, but base our discussion on a more generalized P2P setup.
%Instead, we focus our discussion on the contributions of \system{} and how it optimizes query processing performance in such P2P systems.
\system{} introduces three contributions, that altogether decrease the communication overhead and in doing so increases query processing performance.
First, \system{} creates fragments based on characteristic sets such that entire star patterns can be answered by a single fragment.
This is beneficial since, as we discussed in Section~\ref{sec:introduction}, such star patterns are relatively efficiently processed by the nodes~\cite{DBLP:journals/tods/PerezAG09} and reduce the communication overhead.
The characteristic set of a subject value (entity) is the set of predicates that occur in triples with that subject.
As such, \system{} creates one fragment per unique characteristic set and each fragment thus contains all the triples with the subjects that match the characteristic set of the fragment.
Consider, for instance, the example network in Figure~\ref{fig:running_intro} and query $Q$ shown in Figure~\ref{subfig:running_q}.
Table~\ref{subfig:running_cs} shows the characteristic sets of each fragment in the network. %; each fragment contains the triples for all subjects in the uploaded knowledge graph that match the fragment's characteristic set.
Using this fragmentation method, each fragment can provide answers to entire star patterns; for instance, $P_3\in\mathcal{S(Q)}$ can be processed over just $f_5$, since it is the only fragment containing triples with both predicates present in $P_3$.
%Assume that $Q$ is issued on node $n_1$.
%Each fragment that is replicated throughout the network contains the triples relevant for a characteristic set in the uploaded dataset.
%Note that in this example, some characteristic sets are subsets of others (e.g., for $f_1$ and $f_2$); this can happen when subjects that share similar characteristic sets have slightly different predicate combinations.
The formal definition of the fragmentation approach is presented in Section~\ref{subsec:partitioning}.

\begin{figure*}[htb!]
\centering
\begin{subfigure}[b]{0.38\textwidth}
  \centering
  \begin{lstlisting}[basicstyle=\sffamily\footnotesize,language=SPARQL, firstnumber=0, numbersep=5pt, numberstyle=\tiny, columns=flexible,  label=examplequery, commentstyle=\color{gray}\scriptsize]

select * where {
  ?person dbo:nationality ?country . # tp1 (P1)
  ?person dbo:author ?publication . # tp2 (P1)
  ?country dbo:capital ?capital . # tp3 (P2)
  ?country dbo:currency ?currency . # tp4 (P2)
  ?publication dbo:publisher ?publisher . #tp5 (P3)
  ?publication dbp:language ?language . #tp6 (P3)
}
\end{lstlisting}
\vspace{-3ex}
  \caption{Query $Q$}\label{subfig:running_q}
\end{subfigure}
\begin{subfigure}[b]{0.61\textwidth}
\scalebox{0.9}{
  \begin{tabular}{c|c}
\hline \textbf{Fragment} & \textbf{CS}  \\ \hline
$f_1$ & $\{$\texttt{dbo:nationality},\texttt{dbo:author},\texttt{dbo:deathDate}$\}$ \\
$f_2$ & $\{$\texttt{dbo:nationality},\texttt{dbo:author}$\}$ \\
$f_3$ & $\{$\texttt{dbo:capital},\texttt{dbo:currency},\texttt{dbo:population}$\}$ \\
$f_4$ & $\{$\texttt{dbo:capital},\texttt{dbo:currency}$\}$ \\
$f_5$ & $\{$\texttt{dbo:publisher},\texttt{dbo:language}$\}$ \\ \hline \hline
\end{tabular}}
  \caption{CSs of each fragment in the running example}\label{subfig:running_cs}
\end{subfigure}
\caption{(a) Example SPARQL query $Q$ and (b) corresponding characteristic sets in the example network.}
\label{fig:running_csq}
\end{figure*}

Second, to accommodate processing entire star patterns over individual fragments, and to encode structural information that can be used for cardinality estimation and locality awareness, \system{} introduces a novel indexing scheme, called Semantically Partitioned Blooms Filter (SPBF) Indexes, that builds upon the Prefix-Partitioned Bloom Filter (PPBF) indexes presented in~\cite{ppbfs}.
%PPBF indexes encode the set of subject and object values as bitvectors partitioned based on the prefix of the values (e.g., \texttt{dbr:} and \texttt{dbo:}).
In particular, SPBFs partition the bitvectors based on the IRI's position in the fragment, i.e., whether it is a subject, predicate, or object.
For instance, in  the running example, the SPBF for $f_5$ contains a partition encoding all the subjects with the characteristic set $\{$\texttt{dbo:publisher},\texttt{dbo:language}$\}$, as well as partitions encoding all the objects in $f_5$ that occur in a triple with each predicate.
The formal definition of SPBF indexes is discussed in Section~\ref{subsec:indexing}.

Third, \system{} proposes a query optimization technique that takes advantage of the fragmentation based on characteristic sets and the SPBF indexes to estimate cardinalities and consider data locality while optimizing the query execution plan.
First, \system{} builds a \emph{compatibility graph} using the SPBF indexes that describes, for a given query, which fragments are compatible with one another for each star join in the query (i.e., which fragments may produce results for the joins).
Then, \system{} builds a query execution plan using a Dynamic Programming (DP) algorithm that considers the compatibility of fragments in the compatibility graph and the locality of the fragments in the index.
%In particular, the DP algorithm estimates the cost of processing a star join on each node 
%as the amount of intermediate results that processing the join on them would incur; 
%and delegates each join to the node that can process the join while incurring the least amount of intermediate results to be transferred over the networks.
%Furthermore, the DP algorithm takes into account that disconnected branches can be processed in parallel.
%The cost of processing a join on a specific node is computed by estimating the join cardinality and considering which fragments the nodes store locally.
%Given a query $Q$, processing $Q$ in \system{} follows the following steps:
%
%\begin{enumerate}
%\item Build the compatibility graph $G^C$ for $Q$ by checking the overlap of the corresponding partitions in the SPBFs that contain relevant data to joining star patterns.
%\item Build a query execution plan using the Dynamic Programming algorithm on $Q$ and $G^C$ to compute the cost of processing each subquery on each node with relevant data.
%The DP algorithm uses information about data locality and cardinality estimations.
%\item Process the strategy by delegating star joins to the nodes specified in the plan.
%\end{enumerate}

In the remainder of this section, we detail data fragmentation (Section~\ref{subsec:partitioning}) and indexing (Section~\ref{subsec:indexing}) in \system.
Section~\ref{sec:queryoptimization} details the query optimization approach used by \system.

\subsection{Data Fragmentation}\label{subsec:partitioning}
As discussed in Section~\ref{sec:introduction}, star-shaped subqueries can be processed relatively efficiently over a fragment~\cite{DBLP:journals/tods/PerezAG09}, thus they can also help achieving a better balance between reducing the communication overhead and distributing the query processing load~\cite{DBLP:conf/www/AzzamAMKPH21,smartkg,DBLP:journals/corr/abs-2002-09172}.
To facilitate processing such star patterns on single nodes, we propose to fragment the uploaded knowledge graphs based on \emph{characteristic sets}~\cite{DBLP:conf/icde/NeumannM11,DBLP:conf/www/AzzamAMKPH21,smartkg}.
Formally, a characteristic set is defined as follows:

\begin{definition}[Characteristic Set~\cite{DBLP:conf/www/AzzamAMKPH21,smartkg,DBLP:conf/icde/NeumannM11}]\label{def:cs}
The characteristic set for a subject $s$ in a given knowledge graph $\mathcal{G}$, $C_{\mathcal{G}}(s)$, is the set of predicates associated with $s$, i.e., $C_{\mathcal{G}}(s)=\{p\mid (s,p,o)\in {\mathcal{G}}\}$.
The set of characteristic sets of a knowledge graph ${\mathcal{G}}$ is $C({\mathcal{G}})=\{C_{\mathcal{G}}(s)\mid (s,p,o)\in {\mathcal{G}}\}$.
\end{definition}

In other words, the characteristic set of a subject is the set of predicates (i.e., predicate combination) used to describe the subject, i.e., that occur in the same triples as the subject.
For instance, if the triples (\texttt{dbr:Denmark},\texttt{dbo:capital,}\texttt{dbr:Copenhagen}) and (\texttt{dbr:Denmark},\texttt{dbo:currency},\texttt{dbr:Danish\_Krone}) are the only ones with subject \texttt{dbr:Denmark}, then this subject is described by the characteristic set $\{$\texttt{dbo:capital}$,$\texttt{dbo:currency}$\}$.

Characteristic sets were first introduced in~\cite{DBLP:conf/icde/NeumannM11}, used for cardinality estimation and, in extension of that, join ordering.
WiseKG~\cite{DBLP:conf/www/AzzamAMKPH21} and Smart-KG~\cite{smartkg} used the notion of characteristic sets for fragmentation of knowledge graphs in LDF systems to balance the query load between clients and servers.
In this paper, we use characteristic set based fragments as an alternative to the purely predicate-based fragmentation used by for example \piqnic.
We define the characteristic set based fragmentation function as follows:

%\vspace{-2ex}
\begin{definition}[Characteristic Set Fragmentation Function]
Let $\mathcal{G}$ be a knowledge graph,
then the \emph{characteristic set fragmentation function} of $\mathcal{G}$, $\mathcal{F}_C(\mathcal{G})$, is defined using the notation introduced in Definition~\ref{def:cs}, as:
\begin{equation}\label{eq:partition}
%\mathcal{F}_C(\mathcal{G})=\bigcup_{C_i\in C(\mathcal{G})}\{\{(s,p,o)\mid (s,p,o)\in\mathcal{G}\wedge C_{\mathcal{G}}(s)=C_i\}\}
\mathcal{F}_C(\mathcal{G})=\{\{(s,p,o)\mid (s,p,o)\in\mathcal{G}\wedge C_{\mathcal{G}}(s)=C_i\}\mid C_i\in C(\mathcal{G})\}
\end{equation}
\end{definition}

That is, the characteristic set fragmentation function creates a fragment for each characteristic set in the knowledge graph.
In the characteristic sets shown in Figure~\ref{subfig:running_cs}, $f_4$ thus contains all triples of all subjects that are described by the characteristic set $\{$\texttt{dbo:capital}$,$\texttt{dbo:currency}$\}$.

\system{} nodes can then use these fragments to process entire star patterns.
However, for relatively unstructured knowledge graphs, %such as parts of LargeRDFBench, 
using fragmentation purely based on characteristic sets can lead to an unwieldy number of fragments.
For instance, in our experimental evaluation in Section~\ref{sec:experiments}, fragmenting the data from LargeRDFBench~\cite{largerdfbench} using Equation~\ref{eq:partition} led to 181,859 distinct fragments, most of which contain very few subjects.
Usually, these fragments are created for subjects that are unique due to one or two predicates, while the remaining predicates could fit into larger fragments.

Consider, for instance, in the running example, the situation where the following five characteristic sets are found in the uploaded knowledge graph; for illustration purposes we have extended the notation with the number of subjects covered by each characteristic set:

\vspace{-2ex}
\begin{equation*}
\begin{split}
CS_1=(\{\mathtt{dbo:nationality},\mathtt{dbo:author},\mathtt{dbo:deathDate}\},500)\\
CS_2=(\{\mathtt{dbo:nationality},\mathtt{dbo:author}\},500)\\
CS_3=(\{\mathtt{dbo:publisher},\mathtt{dbo:language}\},1000)\\
CS_4=(\{\mathtt{dbo:nationality},\mathtt{dbo:author},\mathtt{dbo:language}\},2)\\
CS_5=(\{\mathtt{dbo:nationality}\},1)
\end{split}
\end{equation*}

In this case, a separate fragment is created for $CS_4$ even though it does not carry very much information because it describes only two subjects.
While this is not such a big issue in terms of space, it affects the lookup time when optimizing the join order and estimating the cardinalities, since the query processor has to consider potentially thousands of such small fragments.
As such, and similar to~\cite{DBLP:conf/icde/NeumannM11}, we merge infrequent characteristic sets into fragments with a larger number of subjects.
After fragmenting datasets using Equation~\ref{eq:partition}, we apply a strategy with two sequential steps for fragments with infrequent characteristic sets.

First, we merge a fragment $f_1$ with characteristic set $CS_1$ into a fragment $f_2$ with characteristic set $CS_2$ if $CS_1\subseteq CS_2$ by adding the triples of $f_1$ to $f_2$; if there are multiple candidates for $f_2$, we select the one with the smallest set of predicates.
In the example above, for instance, we merge $CS_5$ into $CS_2$ by adding the subject from the fragment with $CS_5$ to the fragment with $CS_2$.

Second, we split fragments $f$ with infrequent characteristic sets into two separate fragments, $f_1$ and $f_2$, such that $f_1$ and $f_2$ can be merged into other fragments with more frequent characteristic sets.
In the example above, we thus split the fragment with $CS_4$ into two smaller fragments $f_4'$ and $f_4''$ such that $f_4'$ has the characteristic set $\{$\texttt{dbo:nationality},\texttt{dbo:author}$\}$ and $f_4''$ has the characteristic set $\{$\texttt{dbo:language}$\}$; $f_4'$ is then merged into the fragment with $CS_2$ and $f_4''$ is merged into the fragment with $CS_3$.
%
%Each fragment with an infrequent predicate combination (i.e., with the number of entities being below a certain threshold) is thus either merged into a fragment containing a superset, or split into smaller fragments that can be merged individually, following the approach presented in~\cite{DBLP:conf/icde/NeumannM11}.
%This significantly reduces the number of fragments since the infrequent characteristic sets are the most common ones.
%Specifically, for LargeRDFBench, merging infrequent fragments into supersets (i.e., step 1 above), results in 12,392 fragments, and splitting up and merging the remaining infrequent fragments (i.e., step 2 above), results in a total of 1,160 remaining fragments.
For example, in the example above, we end up with the following fragments:

\begin{equation*}
\begin{split}
CS_1=(\{\mathtt{dbo:nationality},\mathtt{dbo:author},\mathtt{dbo:deathDate}\},500)\\
CS_2=(\{\mathtt{dbo:nationality},\mathtt{dbo:author}\},503)\\
CS_3=(\{\mathtt{dbo:publisher},\mathtt{dbo:language}\},1002)
\end{split}
\end{equation*}

\subsection{Semantically Partitioned Bloom Filter Indexes}\label{subsec:indexing}
The indexing schema presented in~\cite{ppbfs} (Definition~\ref{def:index}) represents the set of subject and object values as prefix-partitioned bitvectors based on Bloom filters~\cite{Bloom} called Prefix-Partitioned Bloom Filters (PPBFs).
However, PPBFs encode the entire set of subjects and objects in a fragment as a single set and ignore the position (subject or object) of the IRIs in the triples; as such, in a situation where two fragments, for instance, use the same IRIs in the object position, the intersection of the two PPBFs is non-empty.
Then, if the corresponding triple patterns in the query are joined with a subject-object join, the fragments are not pruned since the PPBFs overlap; however, since we are looking for a subject-object join rather than an object-object join, these fragments could have been pruned without affecting the query completeness.
Furthermore, PPBF indexes do not include the predicate values in the index slices, rather they associate the predicate value with the index slice itself (Definition~\ref{def:ppbf}), thus maintaining information about the links between the subjects, predicates, and objects.
This is possible since the implementations of \piqnic{} and \colchain{} use the predicate-based fragmentation function; however, \system{} allows for fragments with several distinct predicates.

Hence, to efficiently estimate whether or not fragments join for a particular query and to maintain the connection between the subjets, predicates, and objects for fragments with multiple predicates, we propose an indexing schema called \emph{Semantically Partitioned Bloom Filters} (SPBFs), which builds upon PPBF as baseline.
%SPBFs provide another layer of partitioning, on top of the prefix partitioning provided by PPBFs, based on the elements' position in the triple (subject, predicate, object) relative to the predicates.
%In other words, SPBFs partition the object values by the predicates they are connected with (i.e., one object partition per predicate).
As the triples contained in fragments defined based on characteristic sets (Section~\ref{subsec:partitioning}) share the same subjects, SPBFs encode the subject values in a single prefix-partitioned bitvector, while there is one prefix-partitioned bitvector for each predicate in the fragment that encodes the objects occurring in triples with that predicate.
%However, since fragments in \system{} correspond to characteristic sets, the subjects within a fragment are connected to each predicate.
%As such, to avoid duplication, the subjects are described by just one (subject) partition.
%This is similar to the \emph{distinct} value in Characteristic Sets.
For instance, in the running example, each subject within $f_2$ occurs in triples with both \texttt{dbo:nationality} and \texttt{dbo:author} as predicates.
The SPBF for $f_2$ contains one partition describing the subject values, one partition describing the object values connected with the \texttt{dbo:nationality} predicate, and one partition describing the object values connected with the \texttt{dbo:author} predicate.
Formally, an SPBF is defined as follows:

\begin{definition}[Semantically Partitioned Bloom Filter]
An SPBF $\mathcal{B}^S$ is a 5-tuple $\mathcal{B}^S=(P,\mathcal{B}_s,B_o,\Phi,H)$ where:
\begin{itemize}
\item $P$ is a set of distinct predicate values
\item $\mathcal{B}_s$ is the prefix-partitioned bitvector that summarizes the subjects
\item $B_o$ is the set of prefix-partitioned bitvectors that summarize the objects
\item $\forall\mathcal{B}_i\in \{\mathcal{B}_s\}\cup B_o$ , $\mathcal{B}_i=(P_i,\hat{B}_i,\theta_i)$ where:
\begin{itemize}
\item $P_i$ is a set of prefixes
\item $\hat{B}_i$ is a set of bitvectors such that $\forall\hat{b}_1,\hat{b}_2\in\hat{B}_i:|\hat{b}_1|=|\hat{b}_2|$
\item $\theta_i:P_i\rightarrow\hat{B}_i$ is a prefix-mapping function 
\end{itemize}
\item $\Phi:P\rightarrow B_o$ is a predicate-mapping function such that $\forall p\in P:\Phi(p)\in B_o$
\item $H$ is a set of hash functions
\end{itemize}
\end{definition}

Similarly to prefix-partitioned bitvectors, we say that an IRI $i$ at position $\rho\in\{s,p,o\}$ \textit{may} be in an SPBF $\mathcal{B}^S$, denoted $i\mbin^{\rho}\mathcal{B}^S$, if and only if $i\mbin\mathcal{B}^S.\mathcal{B}_s$ if $\rho=s$, $\exists p\in\mathcal{B}^S.P:i\mbin\mathcal{B}^S.\Phi(p)$ if $\rho=o$, or $i\in\mathcal{B}^S.P$ if $\rho = p$.
Furthermore, $\mathcal{B}_p(\mathcal{B}^S)$ denotes a function that computes and returns the prefix-partitioned bitvector that contains all predicates in $\mathcal{B}^S.P$.
Given a fragment $f$, $\mathcal{B}^S(f)$ describes the SPBF for $f$.

\begin{figure*}[htb!]
\centering
\begin{subfigure}[b]{0.45\textwidth}
  \centering
  \input{figures/bs1.tex}
  \caption{SPBF $\mathcal{B}^S(f_1)$}\label{subfig:bs1}
\end{subfigure}
\begin{subfigure}[b]{0.45\textwidth}
  \centering
  \input{figures/bs2.tex}
  \caption{SPBF $\mathcal{B}^S(f_4)$}\label{subfig:bs2}
\end{subfigure}
%\vspace{-2ex}
\caption{SPBFs of $f_1$, $\mathcal{B}^S(f_1)$ (a) and $f_4$, $\mathcal{B}^S(f_4)$ (b) in the running example.}
\label{fig:sppbfex2}
\end{figure*}
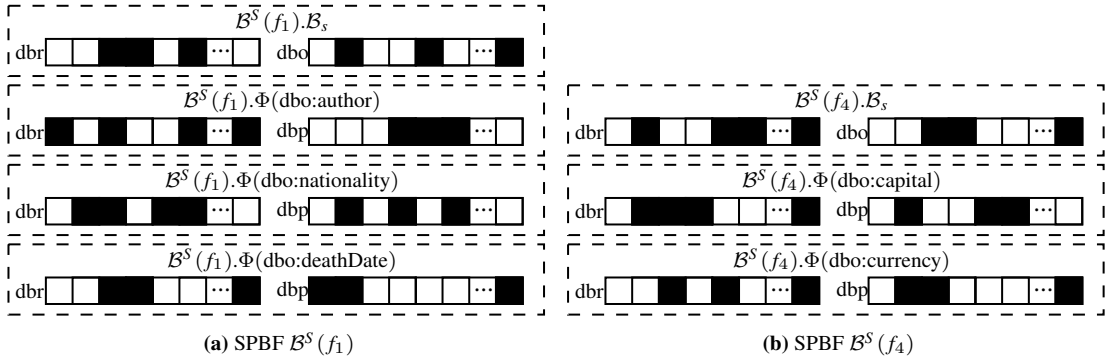

Consider again the running example from Figure~\ref{fig:running_intro}.
Figure~\ref{fig:sppbfex2} shows the SPBFs of fragments $f_1$ (Figure~\ref{subfig:bs1}) and $f_4$ (Figure~\ref{subfig:bs2}).
The SPBF for $f_1$ contains a prefix-partitioned bitvector that encodes all the subject values in $f_1$, $\mathcal{B}^S(f_1).\mathcal{B}_s$, as well as a prefix-partitioned bitvector for each predicate that encodes the object values that are connected with the predicates, i.e., the partition $\mathcal{B}^S(f_1).\Phi($\texttt{dbo:author}$)$ that describes the objects that are connected with the \texttt{dbo:author} predicate, and so on.
Similar for the SPBF for $f_4$, $\mathcal{B}^S(f_4)$.

%This is also visualized in Figure~\ref{fig:sppbfex1}.
%In this case, the IRI \texttt{dbr:Denmark} is added to the PPBF describing the subjects in the partition with the \texttt{dbr} prefix, and the IRI \texttt{dbr:Copenhagen} is added to the bit vector describing the objects associated with the \texttt{dbo:capital} predicate in the partition with the \texttt{dbr} prefix.
%This is done by hashing both IRIs using the hash functions in the SPBF and setting the corresponding bits to 1 in the corresponding PPBF partitions.
%
%\begin{figure*}[htb!]
%\centering
%\input{figures/sppbfex1.tex}
%\caption{Example of updating an SPBF $\mathcal{B}^S(f_4)$ when adding an RDF triple (adapted from~\cite{ppbfs}).}\label{fig:sppbfex1}
%\end{figure*}

A distributed index as defined in Definition~\ref{def:index} and~\cite{DBLP:conf/www/AebeloeMH21,ppbfs} associates triple patterns in the query with fragments that contain relevant data to the triple patterns.
However, since \system{} partitions data based on characteristic sets, 
%associating triples with relevant fragment could yield inaccurate estimations of which fragments are relevant to which star patterns in the query and lead to an excacerbated communication overhead.
%As such, 
we adapt the definition of a distributed index to the fragmentation based on characteristic sets and SPBF indexes.
Let $relevantFragment(P,f)$ be a function that returns \texttt{true} if $\forall t=(s,p,o)\in P$, $s\in V$ or $s\mbin^s\mathcal{B}^S(f)$, $p\in V$ or $p\mbin^p\mathcal{B}^S(f)$, and $o\in V$ or $o\mbin\mathcal{B}^S(f).\Phi(p)$, or \texttt{false} otherwise.
We define an SPBF index as follows:

\begin{definition}[Semantically Partitioned Bloom Filter Index~\cite{DBLP:conf/www/AebeloeMH21,ppbfs}]\label{def:spbfindex}
Let $n$ be a node and $\mathcal{N}$ be the set of nodes within $n$'s local view of the network, $\mathcal{P}$ be the set of all possible star patterns, and $\mathcal{F}$ be the set of fragments stored by at least one node in $\mathcal{N}$.
The \emph{SPBF index} on $n$ is a tuple $I^S_n=(\upsilon,\eta)$ with $\upsilon:\mathcal{P}\mapsto 2^{\mathcal{F}}$ and $\eta:\mathcal{F}\mapsto 2^{\mathcal{N}}$.
$\upsilon(P)$ returns the set of fragments $F$ such that $\forall f\in F$, $relevantFragment(P,f)=$ \texttt{true}.
$\eta(f)$ returns the set of nodes $N$ such that $f\in n_i.G$, $\forall n_i\in N$ and $n_i\in\mathcal{N}$.
%For a star pattern $P\in\mathcal{P}$, $\upsilon(P)$ returns the set of fragments in $\mathcal{F}$ that $S$ matches.
%For a fragment $f\in\mathcal{F}$, $\eta(f)$ returns the nodes on which $f$ is located.
\end{definition}

In other words, an SPBF index maps a star pattern to the fragments that may contain all the constants within the star pattern, and the fragments to the nodes that store them.
Furthermore, since \system, like \piqnic{} and \colchain, builds partial indexes, i.e., slices, for each fragment that are combined to form the node's distributed index, we define an SPBF index slice as follows:

\begin{definition}[SPBF Slice]
Let $f$ be a fragment.
The \emph{SPBF slice} describing $f$ is a tuple $s^S_{f}=(\upsilon',\eta')$ where $\upsilon'(P)$ returns $\{f\}$ if and only if $relevantFragment(P,f)=$ \texttt{true}, and $\eta'(f)$ returns the set of all nodes that contain $f$ in its local datastore.
\end{definition}

The function $s^S(f)$ finds the SPBF slice describing $f$.
%\todo{Algorithm that describes matching a BGP to S-PPBFs (maybe first in next section).}
The SPBF slice describing a fragment is the SPBF obtained from the respective fragment.
For instance, in the running example, the SPBF slice of $f_1$ corresponds to the SPBF obtained from $f_1$, i.e., the one in Figure~\ref{subfig:bs1}.
In Section~\ref{sec:queryoptimization}, we detail how SPBF indexes are used to optimize queries using cardinality estimations and the locality of the data.

%% file: figures/bs1.tex
\tikzset{every picture/.style={line width=0.75pt}} %set default line width to 0.75pt        

\begin{tikzpicture}[x=0.5pt,y=0.5pt,yscale=-1,xscale=1]
%uncomment if require: \path (0,502); %set diagram left start at 0, and has height of 502

%Shape: Rectangle [id:dp6322100235411865] 
\draw  [color={rgb, 255:red, 0; green, 0; blue, 0 }  ,draw opacity=1 ] (255,203) -- (335,203) -- (335,223) -- (255,223) -- cycle ;
%Shape: Rectangle [id:dp7540631355440066] 
\draw   (275,203) -- (295,203) -- (295,223) -- (275,223) -- cycle ;
%Shape: Rectangle [id:dp06331769632615503] 
\draw   (315,203) -- (335,203) -- (335,223) -- (315,223) -- cycle ;
%Shape: Rectangle [id:dp6765989676942908] 
\draw  [color={rgb, 255:red, 0; green, 0; blue, 0 }  ,draw opacity=1 ] (175,203) -- (255,203) -- (255,223) -- (175,223) -- cycle ;
%Shape: Rectangle [id:dp882550539809469] 
\draw   (195,203) -- (215,203) -- (215,223) -- (195,223) -- cycle ;
%Shape: Rectangle [id:dp306507946149404] 
\draw   (235,203) -- (255,203) -- (255,223) -- (235,223) -- cycle ;
%Shape: Rectangle [id:dp1211979876345225] 
\draw  [fill={rgb, 255:red, 0; green, 0; blue, 0 }  ,fill opacity=1 ] (235,203) -- (255,203) -- (255,223) -- (235,223) -- cycle ;
%Shape: Rectangle [id:dp921362332075942] 
\draw  [dash pattern={on 4.5pt off 4.5pt}] (147,178) -- (548,178) -- (548,231) -- (147,231) -- cycle ;
%Shape: Rectangle [id:dp5713292353010847] 
\draw  [fill={rgb, 255:red, 0; green, 0; blue, 0 }  ,fill opacity=1 ] (215,203) -- (235,203) -- (235,223) -- (215,223) -- cycle ;
%Shape: Rectangle [id:dp6348399713471977] 
\draw  [color={rgb, 255:red, 0; green, 0; blue, 0 }  ,draw opacity=1 ] (452,203) -- (532,203) -- (532,223) -- (452,223) -- cycle ;
%Shape: Rectangle [id:dp22552828054525764] 
\draw   (472,203) -- (492,203) -- (492,223) -- (472,223) -- cycle ;
%Shape: Rectangle [id:dp5210841239488082] 
\draw   (512,203) -- (532,203) -- (532,223) -- (512,223) -- cycle ;
%Shape: Rectangle [id:dp9470143027573386] 
\draw  [color={rgb, 255:red, 0; green, 0; blue, 0 }  ,draw opacity=1 ] (372,203) -- (452,203) -- (452,223) -- (372,223) -- cycle ;
%Shape: Rectangle [id:dp9714826736953572] 
\draw   (392,203) -- (412,203) -- (412,223) -- (392,223) -- cycle ;
%Shape: Rectangle [id:dp8243807030589893] 
\draw   (432,203) -- (452,203) -- (452,223) -- (432,223) -- cycle ;
%Shape: Rectangle [id:dp514684396179907] 
\draw  [fill={rgb, 255:red, 0; green, 0; blue, 0 }  ,fill opacity=1 ] (452,203) -- (472,203) -- (472,223) -- (452,223) -- cycle ;
%Shape: Rectangle [id:dp12580379581954437] 
\draw  [fill={rgb, 255:red, 0; green, 0; blue, 0 }  ,fill opacity=1 ] (512,203) -- (532,203) -- (532,223) -- (512,223) -- cycle ;
%Shape: Rectangle [id:dp7085852242231774] 
\draw  [fill={rgb, 255:red, 0; green, 0; blue, 0 }  ,fill opacity=1 ] (392,203) -- (412,203) -- (412,223) -- (392,223) -- cycle ;
%Shape: Rectangle [id:dp9484762808169802] 
\draw  [color={rgb, 255:red, 0; green, 0; blue, 0 }  ,draw opacity=1 ] (255,263) -- (335,263) -- (335,283) -- (255,283) -- cycle ;
%Shape: Rectangle [id:dp8908142827282336] 
\draw   (275,263) -- (295,263) -- (295,283) -- (275,283) -- cycle ;
%Shape: Rectangle [id:dp2217206162649099] 
\draw   (315,263) -- (335,263) -- (335,283) -- (315,283) -- cycle ;
%Shape: Rectangle [id:dp007690613332826279] 
\draw  [color={rgb, 255:red, 0; green, 0; blue, 0 }  ,draw opacity=1 ] (175,263) -- (255,263) -- (255,283) -- (175,283) -- cycle ;
%Shape: Rectangle [id:dp3707199362410306] 
\draw   (195,263) -- (215,263) -- (215,283) -- (195,283) -- cycle ;
%Shape: Rectangle [id:dp8175078950703853] 
\draw   (235,263) -- (255,263) -- (255,283) -- (235,283) -- cycle ;
%Shape: Rectangle [id:dp07638557327295914] 
\draw  [fill={rgb, 255:red, 0; green, 0; blue, 0 }  ,fill opacity=1 ] (215,263) -- (235,263) -- (235,283) -- (215,283) -- cycle ;
%Shape: Rectangle [id:dp5218795258430653] 
\draw  [dash pattern={on 4.5pt off 4.5pt}] (147,238) -- (548,238) -- (548,291) -- (147,291) -- cycle ;
%Shape: Rectangle [id:dp8318031035639365] 
\draw  [fill={rgb, 255:red, 0; green, 0; blue, 0 }  ,fill opacity=1 ] (315,263) -- (335,263) -- (335,283) -- (315,283) -- cycle ;
%Shape: Rectangle [id:dp008197027505760235] 
\draw  [fill={rgb, 255:red, 0; green, 0; blue, 0 }  ,fill opacity=1 ] (175,263) -- (195,263) -- (195,283) -- (175,283) -- cycle ;
%Shape: Rectangle [id:dp006709639315284011] 
\draw  [color={rgb, 255:red, 0; green, 0; blue, 0 }  ,draw opacity=1 ] (452,263) -- (532,263) -- (532,283) -- (452,283) -- cycle ;
%Shape: Rectangle [id:dp34848015585468306] 
\draw   (472,263) -- (492,263) -- (492,283) -- (472,283) -- cycle ;
%Shape: Rectangle [id:dp1297765468643618] 
\draw   (512,263) -- (532,263) -- (532,283) -- (512,283) -- cycle ;
%Shape: Rectangle [id:dp35847854848129235] 
\draw  [color={rgb, 255:red, 0; green, 0; blue, 0 }  ,draw opacity=1 ] (372,263) -- (452,263) -- (452,283) -- (372,283) -- cycle ;
%Shape: Rectangle [id:dp9260732788710374] 
\draw   (392,263) -- (412,263) -- (412,283) -- (392,283) -- cycle ;
%Shape: Rectangle [id:dp5611088995640918] 
\draw   (432,263) -- (452,263) -- (452,283) -- (432,283) -- cycle ;
%Shape: Rectangle [id:dp02955144484061123] 
\draw  [fill={rgb, 255:red, 0; green, 0; blue, 0 }  ,fill opacity=1 ] (452,263) -- (472,263) -- (472,283) -- (452,283) -- cycle ;
%Shape: Rectangle [id:dp04216722465841771] 
\draw  [fill={rgb, 255:red, 0; green, 0; blue, 0 }  ,fill opacity=1 ] (472,263) -- (492,263) -- (492,283) -- (472,283) -- cycle ;
%Shape: Rectangle [id:dp6830963193755012] 
\draw  [fill={rgb, 255:red, 0; green, 0; blue, 0 }  ,fill opacity=1 ] (432,263) -- (452,263) -- (452,283) -- (432,283) -- cycle ;
%Shape: Rectangle [id:dp035229013634325734] 
\draw  [fill={rgb, 255:red, 0; green, 0; blue, 0 }  ,fill opacity=1 ] (275,263) -- (295,263) -- (295,283) -- (275,283) -- cycle ;
%Shape: Rectangle [id:dp7049643555186892] 
\draw  [color={rgb, 255:red, 0; green, 0; blue, 0 }  ,draw opacity=1 ] (255,323) -- (335,323) -- (335,343) -- (255,343) -- cycle ;
%Shape: Rectangle [id:dp5984048135063407] 
\draw   (275,323) -- (295,323) -- (295,343) -- (275,343) -- cycle ;
%Shape: Rectangle [id:dp517314232429618] 
\draw   (315,323) -- (335,323) -- (335,343) -- (315,343) -- cycle ;
%Shape: Rectangle [id:dp5736297122427622] 
\draw  [color={rgb, 255:red, 0; green, 0; blue, 0 }  ,draw opacity=1 ] (175,323) -- (255,323) -- (255,343) -- (175,343) -- cycle ;
%Shape: Rectangle [id:dp611653936280701] 
\draw   (195,323) -- (215,323) -- (215,343) -- (195,343) -- cycle ;
%Shape: Rectangle [id:dp9906582435882691] 
\draw   (235,323) -- (255,323) -- (255,343) -- (235,343) -- cycle ;
%Shape: Rectangle [id:dp26783087825769425] 
\draw  [fill={rgb, 255:red, 0; green, 0; blue, 0 }  ,fill opacity=1 ] (195,323) -- (215,323) -- (215,343) -- (195,343) -- cycle ;
%Shape: Rectangle [id:dp5698258242649619] 
\draw  [dash pattern={on 4.5pt off 4.5pt}] (147,298) -- (548,298) -- (548,351) -- (147,351) -- cycle ;
%Shape: Rectangle [id:dp2744262928619675] 
\draw  [color={rgb, 255:red, 0; green, 0; blue, 0 }  ,draw opacity=1 ] (452,323) -- (532,323) -- (532,343) -- (452,343) -- cycle ;
%Shape: Rectangle [id:dp4913820823794346] 
\draw   (472,323) -- (492,323) -- (492,343) -- (472,343) -- cycle ;
%Shape: Rectangle [id:dp6195947029132618] 
\draw   (512,323) -- (532,323) -- (532,343) -- (512,343) -- cycle ;
%Shape: Rectangle [id:dp4704317367705386] 
\draw  [color={rgb, 255:red, 0; green, 0; blue, 0 }  ,draw opacity=1 ] (372,323) -- (452,323) -- (452,343) -- (372,343) -- cycle ;
%Shape: Rectangle [id:dp9627237971590218] 
\draw   (392,323) -- (412,323) -- (412,343) -- (392,343) -- cycle ;
%Shape: Rectangle [id:dp9307377539977749] 
\draw   (432,323) -- (452,323) -- (452,343) -- (432,343) -- cycle ;
%Shape: Rectangle [id:dp5176725187442326] 
\draw  [fill={rgb, 255:red, 0; green, 0; blue, 0 }  ,fill opacity=1 ] (432,323) -- (452,323) -- (452,343) -- (432,343) -- cycle ;
%Shape: Rectangle [id:dp45967929589279566] 
\draw  [fill={rgb, 255:red, 0; green, 0; blue, 0 }  ,fill opacity=1 ] (472,323) -- (492,323) -- (492,343) -- (472,343) -- cycle ;
%Shape: Rectangle [id:dp226299075834548] 
\draw  [fill={rgb, 255:red, 0; green, 0; blue, 0 }  ,fill opacity=1 ] (392,323) -- (412,323) -- (412,343) -- (392,343) -- cycle ;
%Shape: Rectangle [id:dp8367910581897061] 
\draw  [fill={rgb, 255:red, 0; green, 0; blue, 0 }  ,fill opacity=1 ] (215,323) -- (235,323) -- (235,343) -- (215,343) -- cycle ;
%Shape: Rectangle [id:dp11323288066483439] 
\draw  [fill={rgb, 255:red, 0; green, 0; blue, 0 }  ,fill opacity=1 ] (275,203) -- (295,203) -- (295,223) -- (275,223) -- cycle ;
%Shape: Rectangle [id:dp9767153696955705] 
\draw  [color={rgb, 255:red, 0; green, 0; blue, 0 }  ,draw opacity=1 ] (255,383) -- (335,383) -- (335,403) -- (255,403) -- cycle ;
%Shape: Rectangle [id:dp8384321164375254] 
\draw   (275,383) -- (295,383) -- (295,403) -- (275,403) -- cycle ;
%Shape: Rectangle [id:dp7172897775505861] 
\draw   (315,383) -- (335,383) -- (335,403) -- (315,403) -- cycle ;
%Shape: Rectangle [id:dp023282652620890132] 
\draw  [color={rgb, 255:red, 0; green, 0; blue, 0 }  ,draw opacity=1 ] (175,383) -- (255,383) -- (255,403) -- (175,403) -- cycle ;
%Shape: Rectangle [id:dp3550210814506808] 
\draw   (195,383) -- (215,383) -- (215,403) -- (195,403) -- cycle ;
%Shape: Rectangle [id:dp491822060059267] 
\draw   (235,383) -- (255,383) -- (255,403) -- (235,403) -- cycle ;
%Shape: Rectangle [id:dp4418710645032531] 
\draw  [fill={rgb, 255:red, 0; green, 0; blue, 0 }  ,fill opacity=1 ] (215,383) -- (235,383) -- (235,403) -- (215,403) -- cycle ;
%Shape: Rectangle [id:dp4195309034438275] 
\draw  [dash pattern={on 4.5pt off 4.5pt}] (147,358) -- (548,358) -- (548,411) -- (147,411) -- cycle ;
%Shape: Rectangle [id:dp17661321782085748] 
\draw  [fill={rgb, 255:red, 0; green, 0; blue, 0 }  ,fill opacity=1 ] (315,383) -- (335,383) -- (335,403) -- (315,403) -- cycle ;
%Shape: Rectangle [id:dp5994480142323826] 
\draw  [color={rgb, 255:red, 0; green, 0; blue, 0 }  ,draw opacity=1 ] (452,383) -- (532,383) -- (532,403) -- (452,403) -- cycle ;
%Shape: Rectangle [id:dp4383122998402087] 
\draw   (472,383) -- (492,383) -- (492,403) -- (472,403) -- cycle ;
%Shape: Rectangle [id:dp8133092573410704] 
\draw   (512,383) -- (532,383) -- (532,403) -- (512,403) -- cycle ;
%Shape: Rectangle [id:dp4333716619547786] 
\draw  [color={rgb, 255:red, 0; green, 0; blue, 0 }  ,draw opacity=1 ] (372,383) -- (452,383) -- (452,403) -- (372,403) -- cycle ;
%Shape: Rectangle [id:dp07050175016358917] 
\draw   (392,383) -- (412,383) -- (412,403) -- (392,403) -- cycle ;
%Shape: Rectangle [id:dp6345754394737093] 
\draw   (432,383) -- (452,383) -- (452,403) -- (432,403) -- cycle ;
%Shape: Rectangle [id:dp09847703178533918] 
\draw  [fill={rgb, 255:red, 0; green, 0; blue, 0 }  ,fill opacity=1 ] (372,383) -- (392,383) -- (392,403) -- (372,403) -- cycle ;
%Shape: Rectangle [id:dp07207391600225277] 
\draw  [fill={rgb, 255:red, 0; green, 0; blue, 0 }  ,fill opacity=1 ] (512,383) -- (532,383) -- (532,403) -- (512,403) -- cycle ;
%Shape: Rectangle [id:dp6822200428478957] 
\draw  [fill={rgb, 255:red, 0; green, 0; blue, 0 }  ,fill opacity=1 ] (392,383) -- (412,383) -- (412,403) -- (392,403) -- cycle ;
%Shape: Rectangle [id:dp9088037967552351] 
\draw  [fill={rgb, 255:red, 0; green, 0; blue, 0 }  ,fill opacity=1 ] (235,383) -- (255,383) -- (255,403) -- (235,403) -- cycle ;
%Shape: Rectangle [id:dp8798021005436084] 
\draw  [fill={rgb, 255:red, 0; green, 0; blue, 0 }  ,fill opacity=1 ] (255,323) -- (275,323) -- (275,343) -- (255,343) -- cycle ;
%Shape: Rectangle [id:dp09701981143983285] 
\draw  [fill={rgb, 255:red, 0; green, 0; blue, 0 }  ,fill opacity=1 ] (275,323) -- (295,323) -- (295,343) -- (275,343) -- cycle ;

% Text Node
\draw (306,212) node   [align=left] {...};
% Text Node
\draw (352,190) node  [font=\footnotesize] [align=left] {$\mathcal{B}^S(f_1).\mathcal{B}_s$};
% Text Node
\draw (163,213) node  [font=\footnotesize] [align=left] {dbr};
% Text Node
\draw (503,212) node   [align=left] {...};
% Text Node
\draw (360,213) node  [font=\footnotesize] [align=left] {dbo};
% Text Node
\draw (306,272) node   [align=left] {...};
% Text Node
\draw (352,250) node  [font=\footnotesize] [align=left] {$\mathcal{B}^S(f_1).\Phi($dbo:author$)$};
% Text Node
\draw (163,273) node  [font=\footnotesize] [align=left] {dbr};
% Text Node
\draw (503,272) node   [align=left] {...};
% Text Node
\draw (360,273) node  [font=\footnotesize] [align=left] {dbp};
% Text Node
\draw (306,332) node   [align=left] {...};
% Text Node
\draw (352,310) node  [font=\footnotesize] [align=left] {$\mathcal{B}^S(f_1).\Phi($dbo:nationality$)$};
% Text Node
\draw (163,333) node  [font=\footnotesize] [align=left] {dbr};
% Text Node
\draw (503,332) node   [align=left] {...};
% Text Node
\draw (360,333) node  [font=\footnotesize] [align=left] {dbp};
% Text Node
\draw (306,392) node   [align=left] {...};
% Text Node
\draw (352,370) node  [font=\footnotesize] [align=left] {$\mathcal{B}^S(f_1).\Phi($dbo:deathDate$)$};
% Text Node
\draw (163,393) node  [font=\footnotesize] [align=left] {dbr};
% Text Node
\draw (503,392) node   [align=left] {...};
% Text Node
\draw (360,393) node  [font=\footnotesize] [align=left] {dbp};

\end{tikzpicture}

%% file: figures/bs2.tex
\tikzset{every picture/.style={line width=0.75pt}} %set default line width to 0.75pt        

\begin{tikzpicture}[x=0.5pt,y=0.5pt,yscale=-1,xscale=1]
%uncomment if require: \path (0,502); %set diagram left start at 0, and has height of 502

%Shape: Rectangle [id:dp6322100235411865] 
\draw  [color={rgb, 255:red, 0; green, 0; blue, 0 }  ,draw opacity=1 ] (255,203) -- (335,203) -- (335,223) -- (255,223) -- cycle ;
%Shape: Rectangle [id:dp7540631355440066] 
\draw   (275,203) -- (295,203) -- (295,223) -- (275,223) -- cycle ;
%Shape: Rectangle [id:dp06331769632615503] 
\draw   (315,203) -- (335,203) -- (335,223) -- (315,223) -- cycle ;
%Shape: Rectangle [id:dp6765989676942908] 
\draw  [color={rgb, 255:red, 0; green, 0; blue, 0 }  ,draw opacity=1 ] (175,203) -- (255,203) -- (255,223) -- (175,223) -- cycle ;
%Shape: Rectangle [id:dp882550539809469] 
\draw   (195,203) -- (215,203) -- (215,223) -- (195,223) -- cycle ;
%Shape: Rectangle [id:dp306507946149404] 
\draw   (235,203) -- (255,203) -- (255,223) -- (235,223) -- cycle ;
%Shape: Rectangle [id:dp1211979876345225] 
\draw  [fill={rgb, 255:red, 0; green, 0; blue, 0 }  ,fill opacity=1 ] (255,203) -- (275,203) -- (275,223) -- (255,223) -- cycle ;
%Shape: Rectangle [id:dp921362332075942] 
\draw  [dash pattern={on 4.5pt off 4.5pt}] (147,178) -- (548,178) -- (548,231) -- (147,231) -- cycle ;
%Shape: Rectangle [id:dp012470301125139138] 
\draw  [fill={rgb, 255:red, 0; green, 0; blue, 0 }  ,fill opacity=1 ] (315,203) -- (335,203) -- (335,223) -- (315,223) -- cycle ;
%Shape: Rectangle [id:dp5713292353010847] 
\draw  [fill={rgb, 255:red, 0; green, 0; blue, 0 }  ,fill opacity=1 ] (195,203) -- (215,203) -- (215,223) -- (195,223) -- cycle ;
%Shape: Rectangle [id:dp6348399713471977] 
\draw  [color={rgb, 255:red, 0; green, 0; blue, 0 }  ,draw opacity=1 ] (452,203) -- (532,203) -- (532,223) -- (452,223) -- cycle ;
%Shape: Rectangle [id:dp22552828054525764] 
\draw   (472,203) -- (492,203) -- (492,223) -- (472,223) -- cycle ;
%Shape: Rectangle [id:dp5210841239488082] 
\draw   (512,203) -- (532,203) -- (532,223) -- (512,223) -- cycle ;
%Shape: Rectangle [id:dp9470143027573386] 
\draw  [color={rgb, 255:red, 0; green, 0; blue, 0 }  ,draw opacity=1 ] (372,203) -- (452,203) -- (452,223) -- (372,223) -- cycle ;
%Shape: Rectangle [id:dp9714826736953572] 
\draw   (392,203) -- (412,203) -- (412,223) -- (392,223) -- cycle ;
%Shape: Rectangle [id:dp8243807030589893] 
\draw   (432,203) -- (452,203) -- (452,223) -- (432,223) -- cycle ;
%Shape: Rectangle [id:dp514684396179907] 
\draw  [fill={rgb, 255:red, 0; green, 0; blue, 0 }  ,fill opacity=1 ] (432,203) -- (452,203) -- (452,223) -- (432,223) -- cycle ;
%Shape: Rectangle [id:dp12580379581954437] 
\draw  [fill={rgb, 255:red, 0; green, 0; blue, 0 }  ,fill opacity=1 ] (512,203) -- (532,203) -- (532,223) -- (512,223) -- cycle ;
%Shape: Rectangle [id:dp7085852242231774] 
\draw  [fill={rgb, 255:red, 0; green, 0; blue, 0 }  ,fill opacity=1 ] (412,203) -- (432,203) -- (432,223) -- (412,223) -- cycle ;
%Shape: Rectangle [id:dp9484762808169802] 
\draw  [color={rgb, 255:red, 0; green, 0; blue, 0 }  ,draw opacity=1 ] (255,263) -- (335,263) -- (335,283) -- (255,283) -- cycle ;
%Shape: Rectangle [id:dp8908142827282336] 
\draw   (275,263) -- (295,263) -- (295,283) -- (275,283) -- cycle ;
%Shape: Rectangle [id:dp2217206162649099] 
\draw   (315,263) -- (335,263) -- (335,283) -- (315,283) -- cycle ;
%Shape: Rectangle [id:dp007690613332826279] 
\draw  [color={rgb, 255:red, 0; green, 0; blue, 0 }  ,draw opacity=1 ] (175,263) -- (255,263) -- (255,283) -- (175,283) -- cycle ;
%Shape: Rectangle [id:dp3707199362410306] 
\draw   (195,263) -- (215,263) -- (215,283) -- (195,283) -- cycle ;
%Shape: Rectangle [id:dp8175078950703853] 
\draw   (235,263) -- (255,263) -- (255,283) -- (235,283) -- cycle ;
%Shape: Rectangle [id:dp07638557327295914] 
\draw  [fill={rgb, 255:red, 0; green, 0; blue, 0 }  ,fill opacity=1 ] (215,263) -- (235,263) -- (235,283) -- (215,283) -- cycle ;
%Shape: Rectangle [id:dp5218795258430653] 
\draw  [dash pattern={on 4.5pt off 4.5pt}] (147,238) -- (548,238) -- (548,291) -- (147,291) -- cycle ;
%Shape: Rectangle [id:dp8318031035639365] 
\draw  [fill={rgb, 255:red, 0; green, 0; blue, 0 }  ,fill opacity=1 ] (315,263) -- (335,263) -- (335,283) -- (315,283) -- cycle ;
%Shape: Rectangle [id:dp008197027505760235] 
\draw  [fill={rgb, 255:red, 0; green, 0; blue, 0 }  ,fill opacity=1 ] (195,263) -- (215,263) -- (215,283) -- (195,283) -- cycle ;
%Shape: Rectangle [id:dp006709639315284011] 
\draw  [color={rgb, 255:red, 0; green, 0; blue, 0 }  ,draw opacity=1 ] (452,263) -- (532,263) -- (532,283) -- (452,283) -- cycle ;
%Shape: Rectangle [id:dp34848015585468306] 
\draw   (472,263) -- (492,263) -- (492,283) -- (472,283) -- cycle ;
%Shape: Rectangle [id:dp1297765468643618] 
\draw   (512,263) -- (532,263) -- (532,283) -- (512,283) -- cycle ;
%Shape: Rectangle [id:dp35847854848129235] 
\draw  [color={rgb, 255:red, 0; green, 0; blue, 0 }  ,draw opacity=1 ] (372,263) -- (452,263) -- (452,283) -- (372,283) -- cycle ;
%Shape: Rectangle [id:dp9260732788710374] 
\draw   (392,263) -- (412,263) -- (412,283) -- (392,283) -- cycle ;
%Shape: Rectangle [id:dp5611088995640918] 
\draw   (432,263) -- (452,263) -- (452,283) -- (432,283) -- cycle ;
%Shape: Rectangle [id:dp02955144484061123] 
\draw  [fill={rgb, 255:red, 0; green, 0; blue, 0 }  ,fill opacity=1 ] (452,263) -- (472,263) -- (472,283) -- (452,283) -- cycle ;
%Shape: Rectangle [id:dp04216722465841771] 
\draw  [fill={rgb, 255:red, 0; green, 0; blue, 0 }  ,fill opacity=1 ] (472,263) -- (492,263) -- (492,283) -- (472,283) -- cycle ;
%Shape: Rectangle [id:dp6830963193755012] 
\draw  [fill={rgb, 255:red, 0; green, 0; blue, 0 }  ,fill opacity=1 ] (392,263) -- (412,263) -- (412,283) -- (392,283) -- cycle ;
%Shape: Rectangle [id:dp035229013634325734] 
\draw  [fill={rgb, 255:red, 0; green, 0; blue, 0 }  ,fill opacity=1 ] (235,263) -- (255,263) -- (255,283) -- (235,283) -- cycle ;
%Shape: Rectangle [id:dp7049643555186892] 
\draw  [color={rgb, 255:red, 0; green, 0; blue, 0 }  ,draw opacity=1 ] (255,323) -- (335,323) -- (335,343) -- (255,343) -- cycle ;
%Shape: Rectangle [id:dp5984048135063407] 
\draw   (275,323) -- (295,323) -- (295,343) -- (275,343) -- cycle ;
%Shape: Rectangle [id:dp517314232429618] 
\draw   (315,323) -- (335,323) -- (335,343) -- (315,343) -- cycle ;
%Shape: Rectangle [id:dp5736297122427622] 
\draw  [color={rgb, 255:red, 0; green, 0; blue, 0 }  ,draw opacity=1 ] (175,323) -- (255,323) -- (255,343) -- (175,343) -- cycle ;
%Shape: Rectangle [id:dp611653936280701] 
\draw   (195,323) -- (215,323) -- (215,343) -- (195,343) -- cycle ;
%Shape: Rectangle [id:dp9906582435882691] 
\draw   (235,323) -- (255,323) -- (255,343) -- (235,343) -- cycle ;
%Shape: Rectangle [id:dp26783087825769425] 
\draw  [fill={rgb, 255:red, 0; green, 0; blue, 0 }  ,fill opacity=1 ] (215,323) -- (235,323) -- (235,343) -- (215,343) -- cycle ;
%Shape: Rectangle [id:dp5698258242649619] 
\draw  [dash pattern={on 4.5pt off 4.5pt}] (147,298) -- (548,298) -- (548,351) -- (147,351) -- cycle ;
%Shape: Rectangle [id:dp5550400770631237] 
\draw  [fill={rgb, 255:red, 0; green, 0; blue, 0 }  ,fill opacity=1 ] (315,323) -- (335,323) -- (335,343) -- (315,343) -- cycle ;
%Shape: Rectangle [id:dp2744262928619675] 
\draw  [color={rgb, 255:red, 0; green, 0; blue, 0 }  ,draw opacity=1 ] (452,323) -- (532,323) -- (532,343) -- (452,343) -- cycle ;
%Shape: Rectangle [id:dp4913820823794346] 
\draw   (472,323) -- (492,323) -- (492,343) -- (472,343) -- cycle ;
%Shape: Rectangle [id:dp6195947029132618] 
\draw   (512,323) -- (532,323) -- (532,343) -- (512,343) -- cycle ;
%Shape: Rectangle [id:dp4704317367705386] 
\draw  [color={rgb, 255:red, 0; green, 0; blue, 0 }  ,draw opacity=1 ] (372,323) -- (452,323) -- (452,343) -- (372,343) -- cycle ;
%Shape: Rectangle [id:dp9627237971590218] 
\draw   (392,323) -- (412,323) -- (412,343) -- (392,343) -- cycle ;
%Shape: Rectangle [id:dp9307377539977749] 
\draw   (432,323) -- (452,323) -- (452,343) -- (432,343) -- cycle ;
%Shape: Rectangle [id:dp5176725187442326] 
\draw  [fill={rgb, 255:red, 0; green, 0; blue, 0 }  ,fill opacity=1 ] (412,323) -- (432,323) -- (432,343) -- (412,343) -- cycle ;
%Shape: Rectangle [id:dp45967929589279566] 
\draw  [fill={rgb, 255:red, 0; green, 0; blue, 0 }  ,fill opacity=1 ] (512,323) -- (532,323) -- (532,343) -- (512,343) -- cycle ;
%Shape: Rectangle [id:dp226299075834548] 
\draw  [fill={rgb, 255:red, 0; green, 0; blue, 0 }  ,fill opacity=1 ] (392,323) -- (412,323) -- (412,343) -- (392,343) -- cycle ;
%Shape: Rectangle [id:dp8367910581897061] 
\draw  [fill={rgb, 255:red, 0; green, 0; blue, 0 }  ,fill opacity=1 ] (255,323) -- (275,323) -- (275,343) -- (255,343) -- cycle ;
%Shape: Rectangle [id:dp11323288066483439] 
\draw  [fill={rgb, 255:red, 0; green, 0; blue, 0 }  ,fill opacity=1 ] (275,203) -- (295,203) -- (295,223) -- (275,223) -- cycle ;

% Text Node
\draw (306,212) node   [align=left] {...};
% Text Node
\draw (352,190) node  [font=\footnotesize] [align=left] {$\mathcal{B}^S(f_4).\mathcal{B}_s$};
% Text Node
\draw (163,213) node  [font=\footnotesize] [align=left] {dbr};
% Text Node
\draw (503,212) node   [align=left] {...};
% Text Node
\draw (360,213) node  [font=\footnotesize] [align=left] {dbo};
% Text Node
\draw (306,272) node   [align=left] {...};
% Text Node
\draw (352,250) node  [font=\footnotesize] [align=left] {$\mathcal{B}^S(f_4).\Phi($dbo:capital$)$};
% Text Node
\draw (163,273) node  [font=\footnotesize] [align=left] {dbr};
% Text Node
\draw (503,272) node   [align=left] {...};
% Text Node
\draw (360,273) node  [font=\footnotesize] [align=left] {dbp};
% Text Node
\draw (306,332) node   [align=left] {...};
% Text Node
\draw (352,310) node  [font=\footnotesize] [align=left] {$\mathcal{B}^S(f_4).\Phi($dbo:currency$)$};
% Text Node
\draw (163,333) node  [font=\footnotesize] [align=left] {dbr};
% Text Node
\draw (503,332) node   [align=left] {...};
% Text Node
\draw (360,333) node  [font=\footnotesize] [align=left] {dbp};

\end{tikzpicture}

%% file: queryoptimization.tex
To optimize queries over the network, \system{} first determines which fragments are compatible i.e., produce join results for the given query.
This is to prune fragments that would not contribute to the overall query result. %, and (2) identify combinations of fragments and subqueries that can be processed in parallel.
To do this, \system{} builds a graph that includes the fragments that are compatible for star patterns in the given query, called a \emph{compatibility graph}.
%\emph{compatibility graph} for the query that describes which fragments are compatible for the given query.
In other words, the nodes in a compatibility graph are fragments, and the edges connect the compatible ones.
%This achieves both points above, since (1) fragments that are not compatible with any other fragment for the query are not included in the compatibility graph, and (2) unconnected branches of compatible fragment for the same subqueries can be processed in parallel.
%

Compatibility graphs encapsulate two things.
First, the fragments within a compatibility graph are the fragments that contribute to the overall query result, i.e., fragments that do not contribute to the result are pruned.
Second, different branches of a compatibility graph for the same subqueries can be processed in parallel.
Take, for instance, again query $Q$ in Figure~\ref{subfig:running_q}.
In this case, assuming the join order $P_2\bowtie P_1\bowtie P_3$ (details on join order optimization in Section~\ref{subsec:join}) and the compatibility of the fragments given in Figure~\ref{subfig:cgraph} (details on compatibility graphs in Section~\ref{subsec:sourceselection}), then subquery $P_2\bowtie P_1$ could be processed concurrently over $\{f_1,f_4\}$ and $\{f_2,f_3\}$ since $f_1$ only depends on the intermediate results from $f_4$ and $f_2$ only depends on the intermediate results from $f_3$.
Hence, it could be beneficial to process $P_1\cup P_2$ by delegating the subquery to nodes $n_2$ and $n_3$ concurrently such that $n_2$ processes $[[P_2]]_{f_4}\bowtie [[P_1]]_{f_1}$ locally and $n_3$ processes $[[P_2]]_{f_3}\bowtie [[P_1]]_{f_2}$ locally, and using the combined (by union) results as intermediate bindings when processing $[[P_3]]_{f_5}$ on node $n_1$.
%In other words, when $n_1$ processes $Q'$, it 
%delegates $P_2\cup P_1\cup P_3$ to a node containing at least one fragment in the compatibility graph such that the number of intermediate results to be transferred is minimized (itself in this example).
%Then, it 
%delegates $P_2\cup P_1$ concurrently to both $n_2$ and $n_3$ (since they contain $\{f_1,f_4\}$ and $\{f_2,f_3\}$ respectively) as described above and computes the query answer locally by joining the intermediate results from $n_2$ and $n_3$ locally, 
%Then, $n_2$ and $n_3$ process $P_2\bowtie P_1$ locally, the results of which are joined with $P_3$ on $n_1$.
%This is determined based on the join plan the issuing node computes (described later in this section), 

To this end, \system{} applies Dynamic Programming (DP) similar to~\cite{DBLP:conf/semweb/MontoyaSH17,DBLP:conf/icde/NeumannM11} to build a query execution plan specifying join delegations and parallel processing of subqueries. % based on the lowest estimated cost (i.e., the amount of data to be transferred over the network) of the execution plans.
To further decrease the network overhead, we adapt the cost function in the DP algorithm to consider data \emph{locality} and cardinality estimations available using the SPBF indexes.
In other words, the cost function estimates, given a query execution plan, how many intermediate results processing the join on a particular node incurs, and selects the execution plan that incurs the least data transfer overhead.
%Furthermore, the DP algorithm considers the compatibility graph to determine which (sub)plans can be processed in parallel.
%Description of cardinality estimation / DP algorithm that outputs an execution plan

In summary, given a BGP $P$, \system{} optimizes $P$ by applying the following steps:
\begin{enumerate}
%\item Obtain the compatibility graph for $Q$ using the SPBF index
%\item Iteratively build the query execution plan using Dynamic Programming (DP) as follows:
%\begin{enumerate}
%\item Compute the query execution plan and cost of processing each star pattern over the fragments specified in the compatibility graph and insert the result into the DP table
%\item In each iteration of the DP algorithm compute the execution plans for subqueries of increasing size by combining the (sub)plans already in the DP table for the smaller subqueries in consideration of the edges in the compatibility graph
%\item For each subquery in the current step, compute the cheapest join delegation by considering the locality in the SPBF index (i.e., the join delegation that incurs the least network traffic)
%\end{enumerate}
%\item Repeat steps a-c above for increasingly sized subqueries until the execution plan for the entire query is found
%\end{enumerate}
%\begin{enumerate}
\item Select the relevant fragments for each star pattern in $P$ using the SPBF index.
\item Build the compatibility graph $G^C$ for $P$ (Section~\ref{subsec:sourceselection}) by checking the overlap of the corresponding bitvector partitions in the SPBF index.
\item Build a query execution plan using Dynamic Programming (DP) on $P$ and $G^C$ in consideration of cardinality estimations (Section~\ref{subsec:card}) and data locality (Section~\ref{subsec:join}).
%to compute the cost of processing each subquery on each node with relevant data.
%The DP algorithm uses information about data locality and cardinality estimations.
%\item Process the strategy by delegating star joins to the nodes specified in the plan.
\end{enumerate}

\noindent The output of the above steps is a query execution plan.
In the remainder of this section, we go into details with source selection using compatibility graphs, cardinality estimation, and the query optimization strategy using Dynamic Programming.
%In other words, we describe how, given a query, a query execution plan is obtained.
In Section~\ref{sec:queryprocessing}, we describe how a query execution plan is processed.

\subsection{Fragment and Source Selection}\label{subsec:sourceselection}
As mentioned above, query optimization in \system{} exploits fragment \emph{compatibility}. %, i.e., whether or not two fragments produce join results for a given query. %, and locality awareness.
%To do this, the nodes build a query execution plan based on the compatibility of the fragments based on the intersection of the relevant PPBF partitions in their SPBF indexes.
%When processing queries in \system{}, nodes attempt to minimize the number of requests they have to make, while also distributing some of the query processing load to neighboring nodes, by using the cardinality estimations presented in Section~\ref{subsec:card} and locality awareness.
%
To achieve this, nodes build a \emph{compatibility graph} describing which fragments are compatible for a given query.
%This compatibility graph is then used in the Dynamic Programming (DP) algorithm detailed in Section~\ref{subsec:join} to optimize the join order of star patterns in the query.
%In the remainder of this section, we detail source selection using compatibility graphs.
Two fragments are said to be compatible for a given query if the intersection of the corresponding SPBF partitions is non-empty.
A compatibility graph is thus an undirected graph where nodes are the relevant fragments for the star patterns in the query (determined using the SPBF index) and edges describe the compatible ones.
%To build a compatibility graph, nodes use their SPBF index to (1) find the relevant fragments for the star patterns in the query, and (2) find sets of compatible fragments, thus building the compatibility graph.
%As stated above, the compatibility graph is then used by the DP algorithm in Section~\ref{subsec:join} to compute the query execution plan.
%In the following, we formally define compatibility graphs.
%
%Consider again the running example with the fragments specified on Figure~\ref{subfig:running_cs}.
%In this situation, node $n_1$ uses its SPBF index to determine that $\{f_1,f_2\}$ are relevant fragments to $P_1$, $\{f_3,f_4\}$ to $P_2$, and $f_5$ to $P_3$.
%Consider the situation where $f_1$ only produces join results with $f_4$ and $f_2$ with $f_3$.
%In this case, by checking the overlap of the corresponding SPBF partitions, $n_1$ builds the compatibility graph that connects $f_1$ with $f_4$ as well as $f_2$ with $f_3$.

Recall the function $\mathcal{B}^S(f)$ that returns the SPBF for a fragment $f$, and let $vars(P)$ be a function that returns all the variables in a star pattern $P$.
Furthermore, given an SPBF $\mathcal{B}^S$, a star pattern $P$, and a variable $v$, let $\mathcal{B}(\mathcal{B}^S,P,v)$ denote a function that returns %the PPBF partition in $\mathcal{B}^S$ that corresponds to the position of $v$ in $P$ 
(assuming $v$ can only occur once in $P$) $\mathcal{B}^S.\mathcal{B}_s$ if $v$ is the subject in $P$, $\mathcal{B}^S.\Phi(p)$ if $v$ is the object with predicate $p$, i.e., $(s,p,v)\in P$, or $\mathcal{B}_p(\mathcal{B}^S)$ if $v$ is a predicate in $P$.
Then, a compatibility graph of a BGP $P$ and SPBF index $I^S$ is formally defined as follows.

\begin{definition}[Compatibility Graph]
Given an SPBF index $I^S$ and a BGP $P$, the \emph{compatibility graph} $G^C$ of $P$ over $I^S$ is a tuple $G^C(P,I^S)=(F,C)$ such that $\forall P_1,P_2\in\mathcal{S}(P)$ where $vars(P_1)\cap vars(P_2)\neq\emptyset$ and $\forall v\in vars(P_1)\cap vars(P_2)$, it is the case that $\forall f_1\in I^S.\upsilon(P_1),f_2\in I^S.\upsilon(P_2)$ where $\mathcal{B}(\mathcal{B}^S(f_1),P_1,v)\cap\mathcal{B}(\mathcal{B}^S(f_2),P_2,v)\neq\emptyset$, $(f_1,f_2)\in C$ and $f_1,f_2\in F$.
Furthermore, $\forall P'\subseteq P$ where $vars(P')\cap vars(P-P')=\emptyset$ (i.e., for Cartesian products), it is the case that $\forall f_1\in F$ such that $f_1\in I^S.\upsilon(P_1)$ for some $P_1\in\mathcal{S}(P')$ and $\forall f_2\in F$ such that $f_2\in I^S.\upsilon(P_2)$ for some $P_2\in\mathcal{S}(P-P')$, $(f_1,f_2)\in C$.
\end{definition}

For instance, in the running example, let $\mathcal{B}^S(f_1).\Phi(\mathtt{dbo:nationality})\cap\mathcal{B}^S(f_4).\mathcal{B}_s\neq\emptyset$, i.e., $f_1$ and $f_4$ produce join results, and $\mathcal{B}^S(f_1).\Phi(\mathtt{dbo:nationality})\cap\mathcal{B}^S(f_3).\mathcal{B}_s=\emptyset$, i.e., $f_1$ and $f_3$ do not overlap.
Then, the compatibility graph for query $Q$ in Figure~\ref{subfig:running_q} contains an edge between $f_1$ and $f_4$, but no edge between $f_1$ and $f_3$.
We denote the empty compatibility graph (i.e., where $F$ and $C$ are empty sets) as $G^C_{\emptyset}$.
Algorithm~\ref{algo:compatibility} defines the $G^C(P,I^S)$ function in lines \ref{ln:gcfunc}-\ref{ln:gcfuncend} that computes a compatibility graph given a BGP $P$ and SPBF index $I^S$. % by recursively checking the overlap of fragments that are relevant for joining star patterns using PPBF intersections on the fragment's SPBF (described in Section~\ref{subsec:indexing}).

\begin{algorithm*}[htb]
\caption{Compute the Compatibility Graph of a BGP over an SPBF index}
\label{algo:compatibility}
\begin{algorithmic}[1]
\Statex \textbf{Input:} A BGP $P=P_1\cup\dots\cup P_n$; an SPBF index $I^S=(\upsilon,\eta)$
\Statex \textbf{Output:} A compatibility graph $G^C$
\Function{G$^C$}{$P$,$I^S$}\label{ln:gcfunc}
\State $P'\leftarrow P_k$ where $P_k\in\mathcal{S}(P)$ and $card_B(P_k)\leq card_B(P_j)\forall P_j\in\mathcal{S}(P)$;\label{ln:selfirst}
\State $P_{\epsilon}\leftarrow P'$;\label{ln:accumdef}
\State $F,C\leftarrow\emptyset$;
\ForAll{$f\in I^S.\upsilon(P')$}\label{ln:fafragstart}
\State $G^C_{\epsilon}\leftarrow$ \texttt{buildBranch}$(P-P',I^S,f,P',P_{\epsilon})$;\label{ln:bbcall}
%\If{$G^C_{\epsilon}\neq G^C_{\emptyset}$}
\State $F\leftarrow F\cup G^C_{\epsilon}.F$;\label{ln:mergef}
\State $C\leftarrow C\cup G^C_{\epsilon}.C$;\label{ln:mergec}
%\EndIf
\EndFor\label{ln:fafragend}
\If{$P-P_{\epsilon}\neq\emptyset$}\label{ln:carttstart}
\State $G^C_{\epsilon}\leftarrow$ \texttt{G}$^C(P-P_{\epsilon},I^S)$;\label{ln:gcrecurs}
\IfThen{$G^C_{\epsilon}=G^C_{\emptyset}$}{\Return $G^C_{\emptyset}$}\label{ln:operandempty}
%\If{$G^C_{\epsilon}\neq G^C_{\emptyset}$}
\ForAll{$f_1\in F,f_2\in G^C_{\epsilon}.F$}\label{ln:combinstart}
\State $C\leftarrow C\cup \{(f_1,f_2)\}$;\label{ln:combinend}
\EndFor
\State $F\leftarrow F\cup G^C_{\epsilon}.F$;\label{ln:mmergef}
\State $C\leftarrow C\cup G^C_{\epsilon}.C$;\label{ln:mmergec}
%\EndIf
\EndIf\label{ln:carttend}
\State \Return $(F,C)$;\label{ln:retgc}
\EndFunction\label{ln:gcfuncend}
\Function{buildBranch}{$P$,$I^S$,$f$,$P'$,$P_{\epsilon}$}\label{ln:bbfunc}
\If{$P=\emptyset$ \textbf{or} $\forall P''\in\mathcal{S}(P):vars(P')\cap vars(P'')=\emptyset$}\label{ln:retemptyif}
\State \Return $(\{f\},\emptyset)$;\label{ln:retempty}
\EndIf
\State $F,C\leftarrow\emptyset$;
\ForAll{$P''\in\mathcal{S}(P)$ s.t. $vars(P')\cap vars(P'')\neq\emptyset$}\label{ln:joinnstart}
\State $P_{\epsilon}'\leftarrow P_{\epsilon}\cup P''$;
\State $V\leftarrow vars(P')\cap vars(P'')$;
\ForAll{$f'\in I^S.\upsilon(P'')$ s.t. $\forall v\in V:\mathcal{B}(\mathcal{B}^S(f),P',v)\cap\mathcal{B}(\mathcal{B}^S(f'),P'',v)\neq\emptyset$}\label{ln:compat}
\State $G^C_{\epsilon}\leftarrow$ \texttt{buildBranch}$(P-P'',I^S,f',P'',P_{\epsilon}')$;\label{ln:bbrecurs}
\If{$G^C_{\epsilon}\neq G^C_{\emptyset}$}\label{ln:nonemptystart}
\State $F\leftarrow F\cup G^C_{\epsilon}.F\cup\{f\}$;\label{ln:addf}
\State $C\leftarrow C\cup G^C_{\epsilon}.C\cup\{(f,f')\}$;
\EndIf\label{ln:nonemptyend}
\EndFor
\State $P_{\epsilon}\leftarrow P_{\epsilon}\cup P_{\epsilon}'$;\label{ln:accumincr}
\EndFor\label{ln:joinend}
\State \Return $(F,C)$;\label{ln:retbb}
\EndFunction\label{ln:bbfuncend}
\end{algorithmic}
\end{algorithm*}

\begin{figure*}[tb!]
\centering
\begin{subfigure}[b]{0.12\textwidth}
  \centering
  \input{figures/cgex1.tex}
  \caption{}\label{subfig:cgex1}
\end{subfigure}
\begin{subfigure}[b]{0.12\textwidth}
  \centering
  \input{figures/cgex2.tex}
  \caption{}\label{subfig:cgex2}
\end{subfigure}
\begin{subfigure}[b]{0.12\textwidth}
  \centering
  \input{figures/cgex3.tex}
  \caption{}\label{subfig:cgex3}
\end{subfigure}
\begin{subfigure}[b]{0.12\textwidth}
  \centering
  \input{figures/cgex1.tex}
  \caption{}\label{subfig:cgex4}
\end{subfigure}
\begin{subfigure}[b]{0.12\textwidth}
  \centering
  \input{figures/cgex5.tex}
  \caption{}\label{subfig:cgex5}
\end{subfigure}
\begin{subfigure}[b]{0.12\textwidth}
  \centering
  \input{figures/cgex4.tex}
  \caption{}\label{subfig:cgex6}
\end{subfigure}
\begin{subfigure}[b]{0.18\textwidth}
  \centering
  \input{figures/compatgraphex.tex}
  \caption{}\label{subfig:cgraph}
\end{subfigure}
\caption{Recursively building the compatibility graph for the query in Figure~\ref{subfig:running_q} by applying Algorithm~\ref{algo:compatibility} resulting in $G^C(Q,I^S_{n_1})$. Yellow nodes denote the fragments relevant for $P_2$, blue nodes the fragments relevant for $P_1$, and the green nodes the fragments relevant for $P_3$.}
\label{fig:cgex}
\end{figure*}
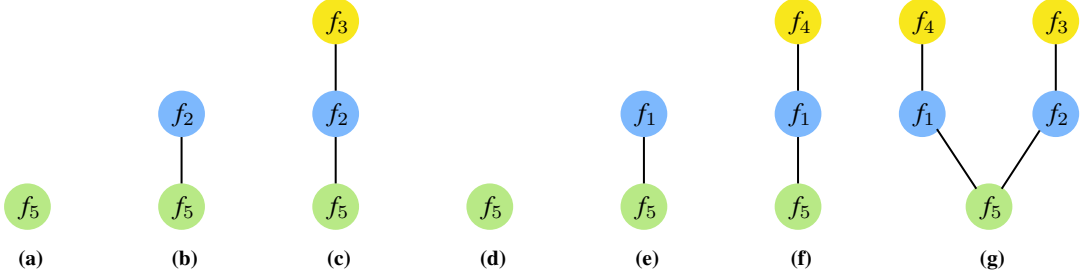

Figure~\ref{fig:cgex} shows how Algorithm~\ref{algo:compatibility} builds the compatibility graph for query $Q$ in Figure~\ref{subfig:running_q}.
In the following, we go through each intermediate step of the algorithm, describing the intermediate compatibility graphs built in the process.
First, the $G^C$ function selects the star pattern in $\mathcal{S}(P)$ with the lowest estimated cardinality in line~\ref{ln:selfirst} (cardinality estimation is detailed in Section~\ref{subsec:card}).
%Assume in this example that $card^P(P_3)=6000$, then 
Assume in the running example, that $P_2$ is the star pattern with the lowest estimated cardinality (Section~\ref{subsec:card}), and that it is therefore selected in line~\ref{ln:selfirst} as the first star pattern.
Furthermore, assume that $f_1$ is only compatible with $f_4$ and $f_2$ is compatible with $f_3$.

%The function $G^C(P,I^S)$ on lines \ref{ln:gcfunc}-\ref{ln:gcfuncend} first selects the star pattern in $\mathcal{S}(P)$ with the lowest estimated cardinality (on line~\ref{ln:selfirst}).
Then, the relevant fragments for the selected star pattern are found using the $I^S.\upsilon$ function from the SPBF index (Definition~\ref{def:spbfindex}) and iterated over in the for loop in lines~\ref{ln:fafragstart}-\ref{ln:fafragend};
for each of these fragments, the function calls the \texttt{buildBranch}$(P,I^S,f,P',P_{\epsilon})$ function in lines~\ref{ln:bbfunc}-\ref{ln:bbfuncend} that builds the (sub)graph starting from the current fragment.
In the example, the loop in lines~\ref{ln:fafragstart}-\ref{ln:fafragend} iterates over $\{f_3,f_4\}$, since these are the fragments relevant for $P_2$.

The \texttt{buildBranch}$(P,I^S,f,P',P_{\epsilon})$ function defines a recursive function that builds a sub-graph starting from a specific fragment and star pattern.
In the first iteration in the running example (i.e., for $f_3$), \texttt{buildBranch} is called with $P=P_1\cup P_3$, $f=f_3$, and $P'=P_2$ as parameters. %, which builds the subgraph for $f_3$. %, and with $f=f_4$ as in the second iteration.
%While, in this example, there are two iterations of the for loops, and thus two calls to \texttt{buildBranch}, to give an overview over the algorihtm, we will focus our description on the first iteration (for $f_3$) and briefly state what the output of the second iteration is when it is used.
First, if $P$ does not contain any star patterns that join with $P'$, i.e., if $P'$ is the outer-most star pattern in the join tree or for a Cartesian product, the function returns the compatibility graph just containing $f$ without any edges (lines~\ref{ln:retemptyif}-\ref{ln:retempty}).
In the example, since $P_1$ joins with $P_2$, the algorithm does not enter the if statement in line~\ref{ln:retempty}.

Instead, the for loop in lines~\ref{ln:joinnstart}-\ref{ln:joinend} iterates through the star patterns $P''\in P$ that join with $P'$, i.e., star patterns that have at least one variable in common.
For each fragment $f'$ relevant for $P''$ (again found using the SPBF index), the function checks the compatibility of $f$ and $f'$ for each join variable $v$ in line~\ref{ln:compat}, i.e., whether or not $f$ and $f'$ may produce join results for each join variable, by intersecting the corresponding partitioned bitvectors in $\mathcal{B}^S(f)$ and $\mathcal{B}^S(f')$.
If the fragments may produce join results, a recursive call is made in line~\ref{ln:bbrecurs} with the $P=P-P''$, $f=f'$, and $P'=P''$ as parameters. %, resulting in the sub-graph that describes the compatibility of $f'$ with fragments for the remaining star patterns in the query.
In the example, the for loop in line~\ref{ln:joinnstart} has only one iteration for $P''=P_1$, i.e., the only star pattern in $\mathcal{S}(P)$ that joins with $P_2$.
Hence, the for loop in line~\ref{ln:compat} checks the compatibility of each fragment relevant for $P_1$ ($f_1$ and $f_2$) with $f_3$ (since $f=f_3$ in this call to the function).
Since $f_2$ is compatible with $f_3$ (cf. the join cardinalities in Table~\ref{tab:cardinalities}), a recursive call is made in line~\ref{ln:bbrecurs} with $P=P_3$, $f=f_2$, and $P'=P_1$.

Since $P_3$ joins with $P_1$, the for loop in line~\ref{ln:compat} checks the compatibility of $f_5$ and $f_2$ and makes another recursive call to the function in line~\ref{ln:bbrecurs} with $P=\emptyset$, $f=f_5$, and $P'=P_3$.
In this iteration of the function, $P$ is empty, thus the graph $(\{f_5\},\emptyset)$ is returned in line~\ref{ln:retempty}.
This graph is visualized in Figure~\ref{subfig:cgex1} and contains only $f_5$ with no edges.
%This resulting subgraph is then merged with the intermediate compatibility graph (currently empty) on lines~\ref{ln:mergef} and~\ref{ln:mergec}.
Since this compatibility graph is non-empty, it is added to the output graph in lines~\ref{ln:nonemptystart}-\ref{ln:nonemptyend} together with $f_2$ (since $f=f_2$ in this iteration of \texttt{buildBranch}) and the edge between $f_5$ and $f_2$.
%In this case, the function adds $f_2$  and an edge between $f_5$ and $f_2$ to the aforementioned compatibility graph on lines~\ref{ln:nonemptystart}-\ref{ln:nonemptyend}.
This graph is visualized in Figure~\ref{subfig:cgex2} and returned by the current iteration of the \texttt{buildBranch} function.
Upon receiving the graph in Figure~\ref{subfig:cgex2}, the function adds $f_3$ (since $f=f_3$ in the current iteration) and an edge between $f_2$ and $f_3$ in lines~\ref{ln:nonemptystart}-\ref{ln:nonemptyend}, resulting in the compatibility graph shown on Figure~\ref{subfig:cgex3} that is returned in line~\ref{ln:retbb}.

%In the next iteration of the for loop on line~\ref{ln:joinnstart}, the function considers the relevant fragments for $P_3$; in this case, only $f_5$.
%Assume in the example that $f_5$ produces join results with both $f_3$ and $f_4$.
%Then, on line~\ref{ln:bbrecurs}, the function makes a recursive call with $P=P_1$, $f=f_5$, and $P'=P_3$.
%Since $P_3$ does not join with $P_1$, the graph $(\{f_5\},\emptyset)$ is returned and added to the compatibility graph along with an edge between $f_5$ and $f_3$.
%This is visualized on Figure~\ref{subfig:cgex3}.
%This graph is then returned by the \texttt{buildBranch} function on line~\ref{ln:retbb}.

In the next iteration of the for loop in line~\ref{ln:fafragstart}, the \texttt{buildBranch} is called with $P=P_1\cup P_3$, $f=f_4$, and $P'=P_2$.
Following the same procedure as described above for $f_3$, we first build the subgraph containing only $f_5$ shown in Figure~\ref{subfig:cgex4}.
Then, $f_1$ is added to the graph along with an edge between $f_1$ and $f_5$ (since they produce join results), resulting in the subgraph shown in Figure~\ref{subfig:cgex5}.
Next, $f_4$ is added along with an edge between $f_4$ and $f_1$, resulting in the compatibility graph for $f_4$ shown in Figure~\ref{subfig:cgex6}.
After merging this in lines~\ref{ln:mergef}-\ref{ln:mergec} with the compatibility graph in Figure~\ref{subfig:cgex3}, the resulting compatibility graph can be seen in Figure~\ref{subfig:cgraph}.

The if statement in lines~\ref{ln:cartstart}-\ref{ln:cartend} ensures that subqueries with star patterns that do not join (i.e., in the case of Cartesian products) are included in the compatibility graph.
This is done by keeping track of the considered star patterns in $P$ using the accumulator $P_{\epsilon}$ defined in line~\ref{ln:accumdef} and updated in line~\ref{ln:accumincr}.
%If at least one such subquery does not have any relevant fragment (i.e., returns an empty compatibility graph), the result likewise is empty, and the empty compatibility graph is returned on line~\ref{ln:operandempty}.
%In the case that all subqueries have non-empty compatibility graphs, each combination of fragments in the subgraphs is added as compatible fragments in the for loop on lines~\ref{ln:combinstart}-\ref{ln:combinend} since they are compatible regardless of their joinability.
%Lastly, the two subgraphs are merged on lines~\ref{ln:mmergef}-\ref{ln:mmergec}.
The example query contains no Cartesian products and so the compatibility graph on Figure~\ref{subfig:cgraph} is returned by the algorithm.

The output of Algorithm~\ref{algo:compatibility} in the example is the compatibility graph shown in Figure~\ref{subfig:cgraph}, specifying that $f_1$ is compatible with $\{f_4,f_5\}$ and $f_2$ is compatible with $\{f_3,f_5\}$.

\subsection{Cardinality Estimation}\label{subsec:card}
In Section~\ref{subsec:partitioning} we have described how \system{} fragments knowledge graphs based on characteristic sets.
Furthermore, in Section~\ref{subsec:indexing} we described how SPBF indexes connect the objects in a fragment to the predicates they occur in triples with.
Since the SPBF of a fragment includes partitioned bitvectors describing the subjects and objects (Definition~\ref{def:spbfindex}), we can estimate the number of values within these partitioned bitvectors and use those estimations to obtain cardinality estimations in a similar way as~\cite{DBLP:conf/icde/NeumannM11,DBLP:conf/semweb/MontoyaSH17}.
%As such, instead of aiming for a cardinality estimation approach that stores the number of object values for each predicate in the fragments like in~\cite{DBLP:conf/icde/NeumannM11,DBLP:conf/semweb/MontoyaSH17}, we estimate the number of values within the corresponding PPBF partitions in the SPBFs for a fragment and use those estimations as the values describing the number of occurrences of subjects and objects in the characteristic set to obtain cardinality estimations in a similar way as~\cite{DBLP:conf/icde/NeumannM11,DBLP:conf/semweb/MontoyaSH17}.
%
%Since \system{} indexes object IRIs based on the predicates they are associated with, and thus that the fragments created by \system{} are Characteristic Sets, the estimated size of the PPBFs can be used to estimate the cardinality of the star patterns.
To achieve this, we first define the estimated number of values in a partitioned bitvector.

Given a partitioned bitvector $\mathcal{B}$ and $\hat{b}\in\mathcal{B}.\hat{B}$, let $t(\hat{b})$ be a function that returns the number of bits in $\hat{b}$ that are set.
Then, the estimated cardinality of a partitioned bitvector $\mathcal{B}$, denoted $card^P(\mathcal{B})$, 
is the sum of the estimated cardinality for all bitvector partitions in $\mathcal{B}.\hat{B}$~\cite{ppbfs,DBLP:journals/dpd/PapapetrouSN10} and is formally defined as follows:

\begin{equation}\label{eq:ppbfcard}
card^P(\mathcal{B})=\sum_{\hat{b}\in\mathcal{B}.\hat{B}} \dfrac{ln(1-t(\hat{b})/|\hat{b}|)}{|\mathcal{B}.H|\cdot ln(1-1/|\hat{b}|)}
\end{equation}

\noindent Consider, for instance, again the running example introduced in Section~\ref{subsec:unp2p} and the SPBF for $f_4$, $\mathcal{B}^S(f_4)$, in Figure~\ref{subfig:bs2}.
Assume that $|\mathcal{B}^S(f_4).H|=5$ and that $|\hat{b}|=20000$ for all $\hat{b}\in\mathcal{B}^S(f_4).\hat{B}$.
Since the partitioned bitvector for the predicate \texttt{dbo:capital} in $f_4$ (Figure~\ref{subfig:bs2}) has two partitions, \texttt{dbr} and \texttt{dbp}, obtaining the estimated cardinality for $\mathcal{B}^S(f_4).\Phi($\texttt{dbo:capital}$)$ is the sum of estimating the cardinality of both prefix partitions.
Let the number of set bits in the bitvector for the \texttt{dbr} prefix be $736$ and the number of set bits in the bitvector for the \texttt{dbp} prefix be $249$.
Then, the estimated cardinality using Equation~\ref{eq:ppbfcard} is:
\begin{multline*}
card^P(\mathcal{B}^S(f_4).\Phi(\text{\texttt{dbo:capital}}))=\\\dfrac{ln(1-736/20000)}{5\cdot ln(1-1/20000)}+\dfrac{ln(1-249/20000)}{5\cdot ln(1-1/20000)}\approx\dfrac{-0.0375}{-0.00025}+\dfrac{-0.0125}{-0.00025}\approx 150+50\approx 200
\end{multline*}

\begin{table}[h]
\caption{Estimated cardinalities for the SPBFs $\mathcal{B}^S(f_1)$, $\mathcal{B}^S(f_2)$, $\mathcal{B}^S(f_3)$, and $\mathcal{B}^S(f_4)$ for the running example in Figure~\ref{fig:running_intro}}
\label{tab:cardinalities}
\begin{tabular}{ll|ll}
\hline
\multicolumn{1}{c}{\textbf{Partitioned Bitvector}} & \multicolumn{1}{c|}{$\mathbf{card^P}$} & \multicolumn{1}{c}{\textbf{Partitioned Bitvector}} & \multicolumn{1}{c}{$\mathbf{card^P}$} \\ \hline
$\mathcal{B}^S(f_1).\mathcal{B}_s$ & 1000  & $\mathcal{B}^S(f_3).\mathcal{B}_s$ &  100 \\
$\mathcal{B}^S(f_1).\Phi($\texttt{dbo:author}$)$ & 5000 & $\mathcal{B}^S(f_3).\Phi($\texttt{dbo:capital}$)$ & 100 \\
$\mathcal{B}^S(f_1).\Phi($\texttt{dbo:nationality}$)$ & 1000 &  $\mathcal{B}^S(f_3).\Phi($\texttt{dbo:currency}$)$ &  150 \\ 
$\mathcal{B}^S(f_1).\Phi($\texttt{dbo:deathDate}$)$ & 1000 &  $\mathcal{B}^S(f_3).\Phi($\texttt{dbo:population}$)$ &  100 \\ \hline
%\multicolumn{1}{c}{\textbf{PPBF}} & \multicolumn{1}{c|}{\textbf{Cardinality}} & \multicolumn{1}{c}{\textbf{PPBF}} & \multicolumn{1}{c}{\textbf{Cardinality}} \\ \hline
$\mathcal{B}^S(f_2).\mathcal{B}_s$ & 2000  & $\mathcal{B}^S(f_4).\mathcal{B}_s$ &  200 \\
$\mathcal{B}^S(f_2).\Phi($\texttt{dbo:author}$)$ & 3000 & $\mathcal{B}^S(f_4).\Phi($\texttt{dbo:capital}$)$ & 200 \\
$\mathcal{B}^S(f_2).\Phi($\texttt{dbo:nationality}$)$ & 2000 &  $\mathcal{B}^S(f_4).\Phi($\texttt{dbo:currency}$)$ &  500 \\ \hline 
$\mathcal{B}^S(f_1).\Phi($\texttt{dbo:nationality}$)\cap\mathcal{B}^S(f_3).\mathcal{B}_s$ & 0 &  $\mathcal{B}^S(f_2).\Phi($\texttt{dbo:nationality}$)\cap\mathcal{B}^S(f_3).\mathcal{B}_s$ &  100 \\
$\mathcal{B}^S(f_1).\Phi($\texttt{dbo:nationality}$)\cap\mathcal{B}^S(f_4).\mathcal{B}_s$ & 50 &  $\mathcal{B}^S(f_2).\Phi($\texttt{dbo:nationality}$)\cap\mathcal{B}^S(f_4).\mathcal{B}_s$ &  0 \\ \hline 
$\mathcal{B}^S(f_5).\mathcal{B}_s$ & 8000  & $\mathcal{B}^S(f_1).\Phi($\texttt{dbo:author}$)\cap\mathcal{B}^S(f_5).\mathcal{B}_s$ &  500  \\
$\mathcal{B}^S(f_5).\Phi($\texttt{dbo:publisher}$)$ & 8000 &  $\mathcal{B}^S(f_2).\Phi($\texttt{dbo:author}$)\cap\mathcal{B}^S(f_5).\mathcal{B}_s$ &  1000 \\
$\mathcal{B}^S(f_5).\Phi($\texttt{dbo:language}$)$ & 9000 &  &  \\ \hline 
\hline
\end{tabular}
\end{table}

\noindent Table~\ref{tab:cardinalities} shows the estimated cardinalities of each partitioned bitvector in the running example.

To estimate the cardinality of star-shaped subqueries, we utilize the fact that the subjects are described by a single partitioned bitvector.
For a star-shaped subquery asking for the set of unique subject values described by a given set of predicates (i.e., queries with the \texttt{DISTINCT} keyword), the cardinality can be estimated as the sum of the number of subjects in each fragment that includes all the predicates in the query.
For instance, the cardinality of $P_1$ in the query in Figure~\ref{subfig:running_q} is the number of distinct subject values in $f_1$ and $f_2$.

Given a star pattern $P$ and a fragment $f$, the cardinality of $P$ over $f$, assuming that $f$ is a relevant fragment for $P$, is the number of values in the partitioned bitvector on the subject position in $\mathcal{B}^S(f)$, and is formally defined as:

%\vspace{-1ex}
\begin{equation}\label{eq:carddistinctsingle}
card_D(P,f)=card^P(\mathcal{B}^S(f).\mathcal{B}_s)
\end{equation}

\noindent For queries not including the \texttt{DISTINCT} keyword, we need to account for duplicates by considering, on average, the number of triples for each non-variable predicate value in $P$ that each subject value is associated with.
Given a star pattern $P$ and fragment $f$, let $preds(P)$ denote the non-variable predicate values in $P$ (in the case of a variable on the predicate position in $P$, we consider the average number of predicate occurences in the characteristic set).
The cardinality of $P$ is thus estimated as follows~\cite{DBLP:conf/semweb/MontoyaSH17,DBLP:conf/icde/NeumannM11}:

%\vspace{-1ex}
\begin{equation}\label{eq:cardnondistinctsingle}
card_S(P,f)=card_D(P,f) \cdot \prod_{p_i\in preds(P)} \dfrac{card^P(\mathcal{B}^S(f).\Phi(p_i))}{card^P(\mathcal{B}^S(f).\mathcal{B}_s)}
\end{equation}

\noindent Henceforth, we will refer to the more generalized function $card$ rather than $card_D$ and $card_S$ to be equivalent to $card_D$ for queries with the \texttt{DISTINCT} modifier and $card_S$ for queries without.
Using Equations~\ref{eq:carddistinctsingle} or~\ref{eq:cardnondistinctsingle}, the cardinality of a star pattern $P$ over a node $n$'s SPBF index is, for all queries (both with and without the \texttt{DISTINCT} keyword), the aggregated cardinality over each relevant fragment to $P$, and is formally defined as follows:
%we compute the cardinality for each SPBF that corresponds to a relevant fragment for $P$ as the cardinality of the PPBF on the subject position.
%The cardinality of $P$ is the sum of cardinalities for each relevant SPBF.
%This is formally defined, for a set $B$ of SPBFs, as follows~\cite{DBLP:conf/semweb/MontoyaSH17,DBLP:conf/icde/NeumannM11}:

\vspace{-1ex}
\begin{equation}\label{eq:card}
card_n(P)=\sum_{f\in I^S_n.\eta(P)} card(P,f)
\end{equation}

\begin{figure*}[htb!]
\centering
\input{figures/cardex1.tex}
\caption{Estimating the cardinality of $P_1$ with the \texttt{DISTINCT} modifier as the number of subjects in $f_1$ and $f_2$ found using Equation~\ref{eq:carddistinctsingle}.}
\label{fig:cardex1}
\end{figure*}
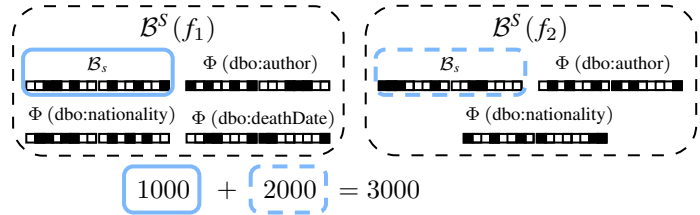

Consider, for instance, in the running example, the star-shaped BGP $P_1$ in Figure~\ref{subfig:running_q} and the estimated cardinalities of the partitioned bitvectors for each fragment in Table~\ref{tab:cardinalities}.
Assume in this case that the \texttt{DISTINCT} keyword is given in the query.
%Let $B$ be the set of SPBFs for all fragments in the example.
Then, $card_{n_1}(P_1)$ is computed as the aggregated estimation of subject values in $f_1$ and $f_2$, $card_{n_1}(P_1)=1000+2000=3000$.
This is visualized in Figure~\ref{fig:cardex1}.

%Formally, this is defined as follows
%In this case, we formally define the estimated cardinality of a star-shaped subquery, for each relevant fragment, as the number of subject values 
%As such, we consider the estimated cardinality of the PPBFs on the object positions for each predicate $p_i\in P$. % to account for multiple triples with the same predicates.
%In this case, the estimated cardinality for a star-shaped BGP $P$ is, for each relevant fragment, the number of subject values in the fragment (i.e., the cardinality of the PPBF at the subject position) multiplied by the ratio of duplicate object values for each predicate $p_i\in P$, i.e., the cardinality of the PPBF on the object position for each $p_i$ divided by the cardinality of the PPBF on the subject position. %, aggregated over the SPBF for each relevant fragment.
%This is formalized as follows
%\cite{DBLP:conf/semweb/MontoyaSH17,DBLP:conf/icde/NeumannM11}:

%\begin{equation}\label{eq:cardnondistinct}
%card_B(P)=\sum_{\mathcal{B}^S\in\{\mathcal{B}^S\in B\mid P\subseteq\mathcal{B}^S.P\}}\left( card^P(\mathcal{B}^S.\mathcal{B}_s) \cdot \prod_{p_i\in P} \dfrac{card^P(\mathcal{B}^S.\Phi(p_i))}{card^P(\mathcal{B}^S.\mathcal{B}_s)}\right)
%\end{equation}

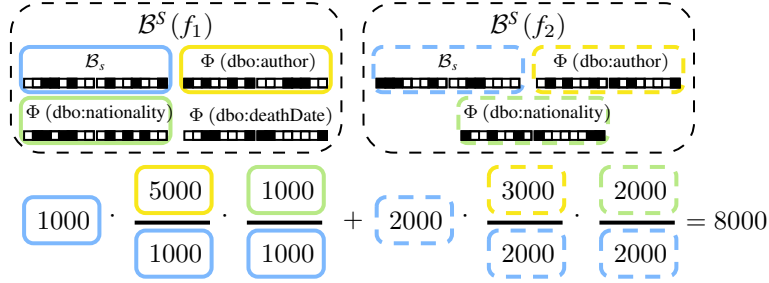
\begin{figure*}[htb!]
\centering
\input{figures/cardex2.tex}
\caption{Estimating the cardinality of $P_1$ without the \texttt{DISTINCT} modifier. Outlines show which bitvector each value is computed from.}
\label{fig:cardex2}
\end{figure*}

If, instead, the \texttt{DISTINCT} keyword was not included in the query, the cardinality $card_{n_1}(P_1)$ is, for each relevant fragment ($f_1$ and $f_2$), the number of subject values within the fragment multiplied with the average number of triples with each predicate $p_i\in preds(P_1)$ that each subject value is associated with, $card_{n_1}(P_1)=1000\cdot(5000/1000)\cdot(1000/1000)+2000\cdot(3000/2000)\cdot(2000/2000)=5000+3000=8000$.
Figure~\ref{fig:cardex2} visualizes the above computations and shows which bitvector each value is computed from.

Until now, the cardinality estimations presented in this section are useful for estimating the cardinality of individual star patterns in a query~\cite{DBLP:conf/semweb/MontoyaSH17,DBLP:conf/icde/NeumannM11}.
%However, since \system{} fragments knowledge graphs based on characteristic sets, such star patterns can be viewed as meta-nodes that does not require any internal optimization; the join order of the triple patterns within each star pattern is determined by the selector on the node that processes the star.
%This is detailed in Section~\ref{sec:queryprocessing}.
%While Equations~\ref{eq:carddistinct} and~\ref{eq:cardnondistinct} estimate the cardinality of star patterns, more complex queries require considering the connections between URIs in different fragments (i.e., different characteristic sets).
%
However, to estimate the cardinality of arbitrary BGPs, \cite{DBLP:conf/edbt/Gubichev014} introduced characteristic pairs that describe the connections between IRIs described by different characteristic sets.
%That is, given two characteristic sets $CS_1$ and $CS_2$, characteristic pairs capture the predicates that connect them as well as how many objects in $CS_1$ correspond to subjects in $CS_2$ for the specific predicate.
In our case, however, we rely on the SPBFs of the relevant fragments to compute characteristic pairs without storing additional information; %no additional information on top of the SPBF indexes is needed.
by intersecting the partitioned bitvectors on the positions corresponding to the join variable, we can estimate the selectivity of a given join and use that to estimate the cardinality of the join.
%
%In other words, given a query with two star patterns with predicates $P_k$ and $P_l$ on a join predicate $p$ (e.g., \texttt{dbo:nationality} for the join between $P_1$ and $P_2$ in Figure~\ref{subfig:running_q}), the join cardinality can, for each pair of fragments $f_k$ and $f_l$ relevant for the star patterns, be computed as the estimated cardinality of the intersection between the PPBF on the object position for the predicate value $p$ in $\mathcal{B}^S(f_k)$ and the PPBF on the subject position in $\mathcal{B}^S(f_l)$.

Formally, for queries including the \texttt{DISTINCT} keyword, given two star patterns $P_k$ and $P_l$ that join on a variable $v$ such that $(s,p,v)\in P_k$ and $v$ is the subject of all triple patterns in $P_l$, and two fragments $f_k$ and $f_l$ that are relevant for $P_k$ and $P_l$ respectively, the cardinality of the join is estimated as the number of IRIs on the subject position in $f_k$ multiplied by the selectivity of the join, i.e., the chance that each subject in the right side corresponds to a value in the join.
This is defined as follows:

\begin{equation}\label{eq:cardpairdistinctsingle}
card_D(P_k,P_l,p,f_k,f_l)=card^P(\mathcal{B}^S(f_k).\mathcal{B}_s)\cdot\bigg(\dfrac{card^P(\mathcal{B}^S(f_k).\Phi(p)\cap\mathcal{B}^S(f_l).\mathcal{B}_s)}{card^P(\mathcal{B}^S(f_k).\Phi(p))}\bigg)
\end{equation}

\noindent For queries that do not include the \texttt{DISTINCT} keyword, we again consider the average predicate occurrences for each triple pattern in both $P_k$ and $P_l$, similar to Equation~\ref{eq:cardnondistinctsingle}
%However, given a query with $P_k$ and $P_l$ and join predicate $p$ \emph{without} the \texttt{DISTINCT} keyword, we have to consider multiple triples with the same subject and predicate values.
%As such, building on Equation~\ref{eq:cardnondistinct}, the estimated join cardinality between $P_k$ and $P_l$ on $p$ is the cardinality of the PPBF intersection from Equation~\ref{eq:cardpairdistinct} multiplied with the ratio of duplicate object values for each predicate in $P_k$ and $P_l$.
%This is formalized as follows~
\cite{DBLP:conf/semweb/MontoyaSH17,DBLP:conf/edbt/Gubichev014}:

%\vspace*{-2ex}
\begin{multline}\label{eq:cardpairnondistinctsingle}
card_S(P_k,P_l,p,f_k,f_l)=card_D(P_k,P_l,p,f_k,f_l)\\
\cdot \prod_{p_k\in P_k-\{p\}} \bigg(\dfrac{card^P(\mathcal{B}^S(f_k).\Phi(p_k))}{card^P(\mathcal{B}^S(f_k).\mathcal{B}_s)}\bigg)
\cdot \prod_{p_l\in P_l} \bigg(\dfrac{card^P(\mathcal{B}^S(f_l).\Phi(p_l))}{card^P(\mathcal{B}^S(f_l).\mathcal{B}_s)}\bigg)
\end{multline}

\noindent Processing a join between two star patterns over a node $n$'s SPBF is the aggregated cardinality over each pair of relevant fragments to the two star patterns, and is for all queries formally defined as follows:

\begin{equation}\label{eq:cardpair}
card_n(P_k,P_l,p)=\sum_{f_k\in I^S_n.\eta(P_k)\wedge f_l\in I^S_n.\eta(P_l)} card(P_k,P_l,p,f_k,f_l)
\end{equation}

\begin{figure*}[htb!]
\centering
\input{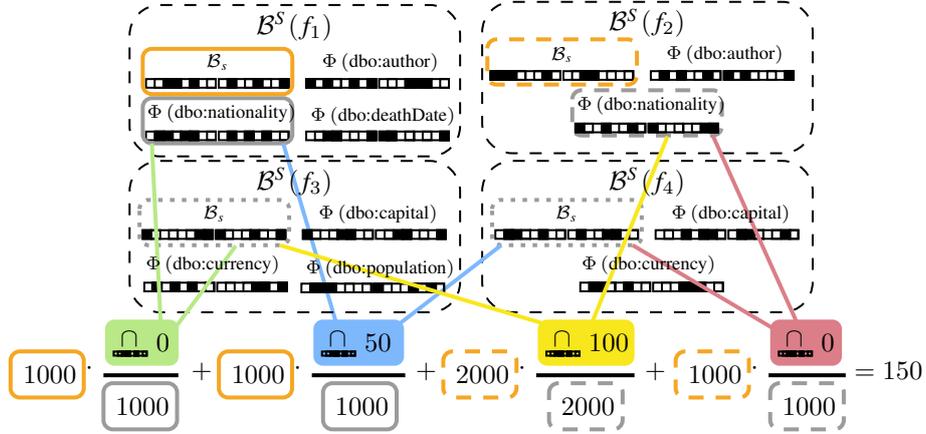}
\caption{Estimating the cardinality of $P_1\bowtie P_2$ with the \texttt{DISTINCT} modifier.}
\label{fig:cardex3}
\end{figure*}

For instance, consider the join between $P_1$ and $P_2$ in Figure~\ref{subfig:running_q} where the \texttt{DISTINCT} keyword is given in the query.
Here, the cardinality $card_{n_1}(P_1,P_2,$\texttt{dbo:nationality}$)$ is the aggregated cardinality of the partitioned bitvectors obtained by intersecting the partitioned bitvector on the object position for the \texttt{dbo:nationality} for each $f_k\in\{f_1,f_2\}$ with the partitioned bitvector on the subject position for each $f_l\in\{f_3,f_4\}$.
That is, given the cardinalities of the intersections shown in Table~\ref{tab:cardinalities}, $card_{n_1}(P_1,P_2,$\texttt{dbo:nationality}$)=1000\cdot(0/1000)+1000\cdot(50/1000)+2000\cdot(100/2000)+2000\cdot(0/1000)=150$.
We have visualized this computation in Figure~\ref{fig:cardex3}.

%For queries that do not include the \texttt{DISTINCT} keyword, we consider the average predicate occurrences for each triple pattern in both $P_k$ and $P_l$, similar to Equation~\ref{eq:cardnondistinct}
%%However, given a query with $P_k$ and $P_l$ and join predicate $p$ \emph{without} the \texttt{DISTINCT} keyword, we have to consider multiple triples with the same subject and predicate values.
%%As such, building on Equation~\ref{eq:cardnondistinct}, the estimated join cardinality between $P_k$ and $P_l$ on $p$ is the cardinality of the PPBF intersection from Equation~\ref{eq:cardpairdistinct} multiplied with the ratio of duplicate object values for each predicate in $P_k$ and $P_l$.
%%This is formalized as follows~
%\cite{DBLP:conf/semweb/MontoyaSH17,DBLP:conf/edbt/Gubichev014}:
%
%\begin{multline}\label{eq:cardpairnondistinct}
%card_B(P_k,P_l,p)=\sum_{\mathcal{B}^S_i\in\{\mathcal{B}^S_i\in B\mid P_k\subseteq\mathcal{B}^S_i.P\}\wedge\mathcal{B}^S_j\in\{\mathcal{B}^S_j\in B\mid P_l\subseteq\mathcal{B}^S_j.P\}}\bigg( card^P(\mathcal{B}^S_i.\Phi(p)\cap\mathcal{B}^S_j.\mathcal{B}_s) \\
%\cdot \prod_{p_k\in P_k-\{p\}} \big(\dfrac{card^P(\mathcal{B}^S_i.\Phi(p_k))}{card^P(\mathcal{B}^S_i.\mathcal{B}_s)}\big)
%\cdot \prod_{p_l\in P_l} \big(\dfrac{card^P(\mathcal{B}^S_j.\Phi(p_l))}{card^P(\mathcal{B}^S_j.\mathcal{B}_s)}\big) \bigg)
%\end{multline}

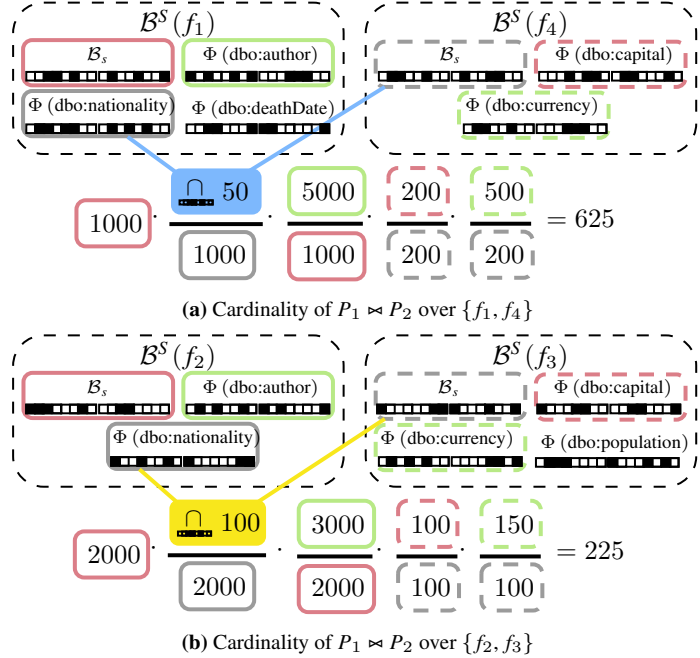
\begin{figure*}[htb!]
\centering
\begin{subfigure}[b]{\textwidth}
  \centering
  \input{figures/cardex4.tex}
  \caption{Cardinality of $P_1\bowtie P_2$ over $\{f_1,f_4\}$}\label{subfig:cardex41}
\end{subfigure}
\begin{subfigure}[b]{\textwidth}
  \centering
  \input{figures/cardex5.tex}
  \caption{Cardinality of $P_1\bowtie P_2$ over $\{f_2,f_3\}$}\label{subfig:cardex42}
\end{subfigure}
\caption{Estimating the cardinality of $P_1\bowtie P_2$ without the \texttt{DISTINCT} modifier over (a) $\{f_1,f_4\}$ and (b) $\{f_2,f_3\}$.
The output of Equation~\ref{eq:cardpair} is thus the sum of the two formulas ($625+225=850$).}
\label{fig:cardex4}
\end{figure*}

In the case where the \texttt{DISTINCT} keyword is not included in the query, the join between $P_1$ and $P_2$ in Figure~\ref{subfig:running_q} given the partitioned bitvector cardinalities in Table~\ref{tab:cardinalities} yields the following equation: $card_{n_1}(P_1,P_2,$\texttt{dbo:nationality}$)=1000\cdot(50/1000)\cdot(5000/1000)\cdot(200/200)\cdot(500/200)+2000\cdot(100/2000)\cdot(3000/2000)\cdot(100/100)\cdot(150/100)=625+225=850$.
%Using Equation~\ref{eq:cardnondistinct} to estimate the cardinalities of $P_1$ and $P_2$ yields the results $card^P(P_1)=1000\cdot (5000/1000)\cdot (1000/1000)=5000$ and $card^P(P_2)=200\cdot (200/200)\cdot (500/200)=500$.
%As such, the estimated cardinality of $P_1$ is 5000, $P_2$ is 500 , and the entire query is 625.
Figure~\ref{fig:cardex4} visualizes the above computation; Figure~\ref{subfig:cardex41} shows the computation over $\{f_1,f_4\}$ and Figure~\ref{subfig:cardex42} over $\{f_2,f_3\}$.
The outlines show which partitioned bitvector is used to compute each number, e.g., the value $50$ is found by computing the cardinality of the bitvector intersection $\mathcal{B}^S(f_1).\Phi(\mathtt{dbo:nationality})\cap\mathcal{B}^S(f_4).\mathcal{B}_s$.
The result is the sum of the formulas.

%Given the cardinality estimation of the joins between star patterns in the query coupled with locality information available in the indexes and compatibility graphs, we use Dynamic Programming (DP) to determine the optimal join order that minimizes the number of %intermediate results that have to be transferred over the network for a given query.
%This is detailed in Section~\ref{subsec:join}.
%First, we detail source selection using compatibility graphs in Section~\ref{subsec:sourceselection}.

\subsection{Optimizing Query Execution Plans}\label{subsec:join}
%Before going into details with how \system{} optimizes queries, we first formally define a query execution plan.
Based on the compatibility graph (Section~\ref{subsec:sourceselection}), the locality of fragments, and cardinality estimations (Section~\ref{subsec:card}), \system{} builds a \emph{query execution plan} that specifies which subqueries can be processed in parallel and which joins are delegated to which nodes as well as the join order.
A query execution plan is defined as follows:

\begin{definition}[Query Execution Plan]\label{def:qep}
A \emph{query execution plan} $\Pi$ consists of the execution plan and the node that processes the plan, called a \emph{delegation}.
A query execution plan can be one of four types:
\begin{itemize}
\item \emph{Join} $\Pi=\Pi_1\bowtie^{n}\Pi_2$ where $\Pi_1$ and $\Pi_2$ are two (sub)plans and $n$ is the node the join is delegated to.
\item \emph{Cartesian product} $\Pi=\Pi_1\times^{n}\Pi_2$ where $\Pi_1$ and $\Pi_2$ are two (sub)plans and $n$ is the node the Cartesian product is delegated to.
\item \emph{Union} $\Pi=\Pi_1\cup^{n}\Pi_2$ where $\Pi_1$ and $\Pi_2$ are two (sub)plans and $n$ is the node the union is delegated to.
\item \emph{Selection} $\Pi=[[P]]_f^n$ where $P$ is a star pattern, $f$ is the fragment that $P$ is processed over, and $n$ is the node the selection is delegated to.
\end{itemize}
\end{definition}

\begin{figure*}[tb!]
\centering
\begin{subfigure}[b]{0.2\textwidth}
  \centering
  \input{figures/pex1.tex}
  \caption{$P_1$}\label{subfig:pex1}
\end{subfigure}
\begin{subfigure}[b]{0.2\textwidth}
  \centering
  \input{figures/pex2.tex}
  \caption{$P_2$}\label{subfig:pex2}
\end{subfigure}
\begin{subfigure}[b]{0.12\textwidth}
  \centering
  \input{figures/pex3.tex}
  \caption{$P_3$}\label{subfig:pex3}
\end{subfigure}
\begin{subfigure}[b]{0.38\textwidth}
  \centering
  \input{figures/pex4.tex}
  \caption{$P_2\bowtie P_1$}\label{subfig:pex4}
\end{subfigure}
\begin{subfigure}[b]{0.28\textwidth}
  \centering
  \input{figures/pex5.tex}
  \caption{$P_2\bowtie P_3$}\label{subfig:pex5}
\end{subfigure}
\begin{subfigure}[b]{0.28\textwidth}
  \centering
  \input{figures/pex6.tex}
  \caption{$P_1\bowtie P_3$}\label{subfig:pex6}
\end{subfigure}
\begin{subfigure}[b]{0.38\textwidth}
  \centering
  \input{figures/plan.tex}
  \caption{$P_1\bowtie P_2\bowtie P_3$}\label{subfig:planex}
\end{subfigure}
\vspace{-1ex}
\caption{Best query execution plan for each subquery in the DP table (Table~\ref{tbl:dp}).}
\label{fig:qpex}
%\vspace{-1ex}
\end{figure*}
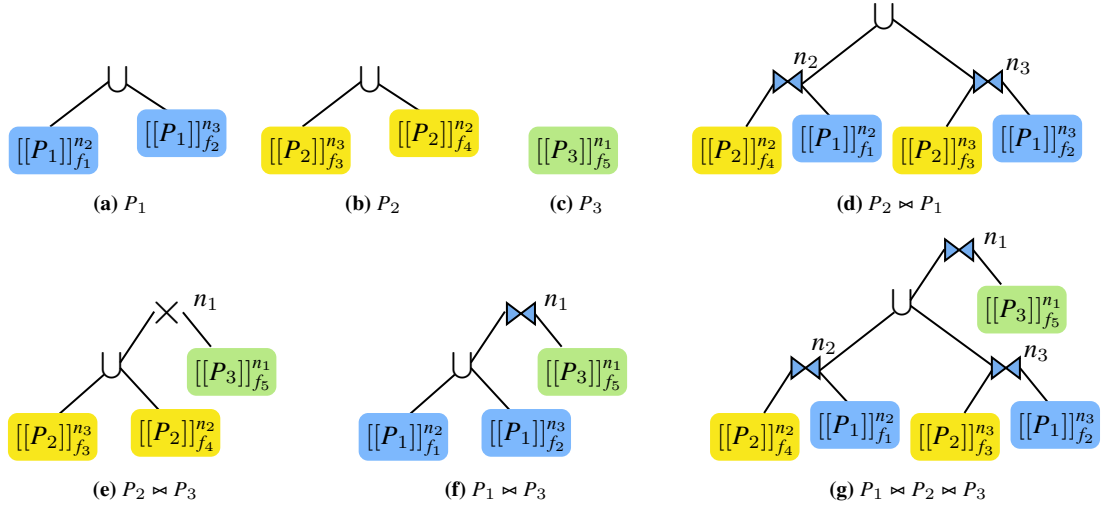

Since unions are not explicitly executed by any node, instead the partial results of each subplan in the union are transferred to the nodes that use those intermediate results, we simply omit the specification of delegations for unions from the description below.
%Since delegating unions to an independant node (i.e., a node that holds neither of the partial results that should be combined) causes the additional cost fir transferring the partial results, we always only consider the involved nodes for delegation and therefore simply omit this case from the specification below.
Furthermore, we assume that query execution plans are always left-deep, i.e., the right side of a join can only consist of a selection or a union of selections.
For instance, the execution plan for query $Q$, $\Pi=(([[P_2]]_{f_4}^{n_2}\bowtie^{n_2} [[P_1]]_{f_1}^{n_2})\cup ([[P_2]]_{f_3}^{n_3}\bowtie^{n_3} [[P_1]]_{f_2}^{n_3}))\bowtie^{n_1} [[P_3]]_{f_5}^{n_1}$ (Figure~\ref{subfig:planex}) specifies that the join $[[P_2]]_{f_4}\bowtie [[P_1]]_{f_1}$ is delegated to $n_2$ and processed in parallel with $[[P_2]]_{f_3}\bowtie [[P_1]]_{f_2}$ on $n_3$ (specified by the union), the result of which is transferred to $n_1$ and joined with $[[P_3]]_{f_5}$.

Since our cost function includes the estimated cardinality of a particular subplan, we first extend the framework for cardinality estimation described in Section~\ref{subsec:card} to enable cardinality estimation of an entire query execution plan.
This is straightforward for Cartesian products, unions, and selections; for Cartesian products it is the multiplication of the cardinality of the operands, for unions it is the sum of the cardinality of the operands, and for selections it is the cardinality of the star pattern over a specific fragment defined in Equations~\ref{eq:carddistinctsingle} and~\ref{eq:cardnondistinctsingle}.
Given the reasoning above, we define the cardinality of a query execution plan $\Pi$, $card(\Pi)$, covering all types of $\Pi$, as follows:

\begin{equation}\label{eq:cardPlan}
card(\Pi)=
\begin{cases}
card(\Pi_1)\cdot card(\Pi_2), & \text{if } \Pi=\Pi_1\times^n\Pi_2 \\
card(\Pi_1)+card(\Pi_2), & \text{if } \Pi=\Pi_1\cup\Pi_2 \\
card(P,f), & \text{if } \Pi=[[P]]_f^{n} \\
card(\Pi_1\bowtie^n\Pi_2), & \text{if } \Pi=\Pi_1\bowtie^n\Pi_2 
\end{cases}
\end{equation}

\noindent To generalize Equation~\ref{eq:cardpair} such that we can compute the cardinality of any join $\Pi=\Pi_1\bowtie^n\Pi_2$ (e.g., including joins between a BGP with multiple star patterns and a star pattern), we consider two cases: 
%(1) where $\Pi_1$ is a union $\Pi_1=\Pi_1'\cup\Pi_1''$, 
(1) where $\Pi_2$ is a union $\Pi_2=\Pi_2'\cup\Pi_2''$, and
(2) where $\Pi_2$ is a selection $\Pi_2=[[P]]_f^{n_1}$.
%Note, that while $\Pi_1$ could be another join (e.g., for the plan in Figure~\ref{subfig:planex}), we do not consider this as a special case.
%Recall that $\Pi_2$ can only be a selection or a union of selections.
The cardinality of the join can thus be estimated using the following formula:

\begin{equation}\label{eq:cardjoin}
card(\Pi_1\bowtie^n\Pi_2)=
\begin{cases}
%card^{\bowtie}(\Pi_1'\bowtie^n\Pi_2)+card(\Pi_1''\bowtie^n\Pi_2), & \text{if } \Pi_1=\Pi_1'\cup^{n_1}\Pi_1'' \\
card(\Pi_1\bowtie^n\Pi_2')+card(\Pi_1\bowtie^n\Pi_2''), & \text{if } \Pi_2=\Pi_2'\cup\Pi_2'' \\
card^{\bowtie}(\Pi_1,P,f), & \text{if } \Pi_2=[[P]]_f^{n_1}
\end{cases}
\end{equation}

\noindent The function $card^{\bowtie}(\Pi,P,f)$ in the second case of Equation~\ref{eq:cardjoin} computes the cardinality of the join for a particular selection on the right side of the join, $[[P]]_f$.
To achieve this estimation, we consider the estimated cardinality of $\Pi$ and the selectivity of the join similar to Equation~\ref{eq:cardpairdistinctsingle}.
To avoid a significant overestimation due to the possible correlation between multiple join variables in the same join, we only consider the most selective join variable for any specific join.
%; for queries including the \texttt{DISTINCT} keyword, this is the number of values in the join divided by the number of subject values in the joining fragments in $\Pi$ (since this is the number of possible distinct values on the left side).
%
Recall the $\mathcal{B}(\mathcal{B}^S,P,v)$ function that returns the partitioned bitvector in $\mathcal{B}^S$ that corresponds to $v$'s position in $P$, and let $S(\Pi,P)$ denote the set of star patterns in $\Pi$ that join with $P$ and $F(\Pi,f)$ denote the set of fragments in $\Pi$ that join with $f$. 
For instance, for the execution plan in Figure~\ref{subfig:pex4} and the compatibility graph in Figure~\ref{subfig:cgraph}, $S(\Pi,P_3)=\{P_1\}$ and $F(\Pi,f_5)=\{f_1,f_2\}$.
Furthermore, given two star patterns $P_1$ and $P_2$, let $v(P_1,P_2)=\{v\mid v\in vars(P_1)\cap vars(P_2)\}$, i.e., the set of join variables.
The cardinality of the join between a plan $\Pi$ and a selection $[[P]]_f$ is, given the \texttt{DISTINCT} keyword, generalized from Equation~\ref{eq:cardpairdistinctsingle} as follows:

\begin{equation}\label{eq:cardjoindistinct}
card^{\bowtie}_D(\Pi,P,f)=
card(\Pi)\cdot
\min_{P'\in S(\Pi,P)\wedge v\in v(P,P')}\bigg(\dfrac{\sum_{f'\in F(\Pi,f)}card^P(\mathcal{B}(\mathcal{B}^S(f),P,v)\cap\mathcal{B}(\mathcal{B}^S(f'),P',v))}{\sum_{f'\in F(\Pi,f)}card^P(\mathcal{B}(\mathcal{B}^S(f'),P',v))}\bigg)
\end{equation}

\begin{figure*}[htb!]
\centering
\input{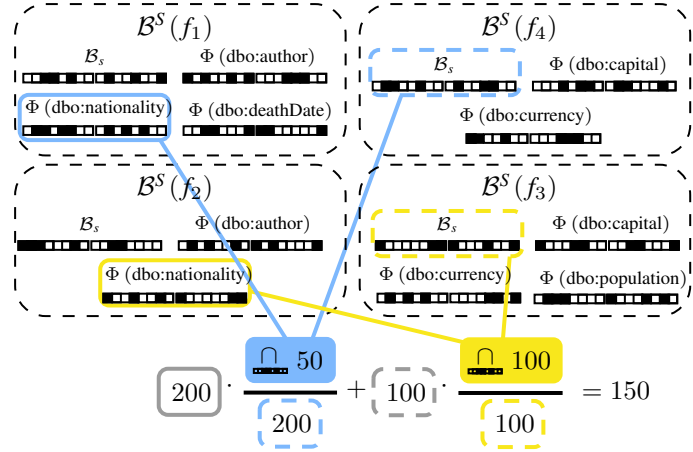}
\caption{Estimating the cardinality of $\Pi=([[P_2]]_{f_4}^{n_2}\bowtie^{n_2}[[P_1]]_{f_1}^{n_2})\cup([[P_2]]_{f_3}^{n_3}\bowtie^{n_3}[[P_1]]_{f_2}^{n_3})$ with the \texttt{DISTINCT} keyword using the cardinalities from Table~\ref{tab:cardinalities} and Equation~\ref{eq:cardjoindistinct}.}
\label{fig:cardex6}
\end{figure*}

%For instance, in the join $[[P_2]]_{f_4}\bowtie [[P_1]]_{f_1}$, the selectivity is $50/200$, since each of the 500 results to $[[P_2]]_{f_4}$ has a $50$ in $200$ chance, on average, to be one of the values within the join.
As an example, consider computing the cardinality $card(\Pi)$ of the plan $\Pi$ visualized in Figure~\ref{subfig:pex4} using the \texttt{DISTINCT} keyword.
Since $\Pi$ is a union, we compute the cardinality of $\Pi_1=[[P_2]]_{f_4}^{n_2}\bowtie^{n_2}[[P_1]]_{f_1}^{n_2}$ and $\Pi_2=[[P_2]]_{f_3}^{n_3}\bowtie^{n_3}[[P_1]]_{f_2}^{n_3}$ and let $card(\Pi)=card(\Pi_1)+card(\Pi_2)$.
Using Equation~\ref{eq:cardjoindistinct} on $\Pi_1$ and $\Pi_2$, we get the formula $card(\Pi)=200\cdot(50/200)+100\cdot(100/100)=150$ as visualized in Figure~\ref{fig:cardex6} (the gray values are the cardinalities of the left selections in each join obtained using Equation~\ref{eq:carddistinctsingle}).

For queries without the \texttt{DISTINCT} keyword, we once again consider the average predicate occurences.
However, since the predicate occurrences in $\Pi$ are already considered in $card(\Pi)$ in Equation~\ref{eq:cardjoindistinct}, we only consider the average number of occurrences in $f$ for each triple pattern in $P$ that does not join with $\Pi$ on the object.
%we compute the selectivity as the number of values in the join divided by the number of values in the corresponding SPBF partitions for the fragments in $\Pi$ (since this is the number of possible non-distinct values on the left side of the join), and consider the average number of predicate occurrences in $f$ for each triple pattern in $P$ that does not  (similar to Equation~\ref{eq:cardpairnondistinctsingle}).
%
The cardinality of the join between a plan $\Pi$ and selection $[[P]]_f$, without the \texttt{DISTINCT} keyword, is computed as:
% $\Pi$ that constitutes , e.g., applying $\Psi(\Pi)$ to the plan $([[P_2]]_{f_4}\bowtie^{n_2}[[P_1]]_{f_1})\cup([[P_2]]_{f_3}\bowtie^{n_3}[[P_1]]_{f_2})$ in Figure~\ref{subfig:pex4} results in the plans $[[P_2]]_{f_4}\bowtie^{n_2}[[P_1]]_{f_1}$ and $[[P_2]]_{f_3}\bowtie^{n_3}[[P_1]]_{f_2}$.

\begin{equation}\label{eq:cardjoinnondistinct}
card^{\bowtie}_S(\Pi,P,f)=
card^{\bowtie}_D(\Pi,P,f)\cdot
\prod_{p\in preds(P): (s,p,o)\in P\wedge o\not\in v(P,P')\forall P'\in S(\Pi,P)}\bigg(\dfrac{card^P(\mathcal{B}^S(f).\Phi(p))}{card^P(\mathcal{B}^S(f).\mathcal{B}_s)}\bigg)
\end{equation}

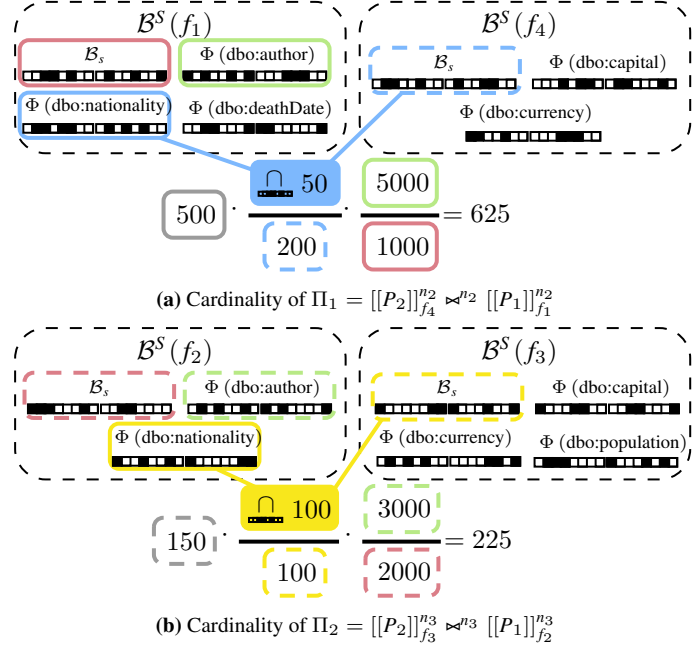
\begin{figure*}[htb!]
\centering
\begin{subfigure}[b]{\textwidth}
  \centering
  \input{figures/cardex7.tex}
  \caption{Cardinality of $\Pi_1=[[P_2]]_{f_4}^{n_2}\bowtie^{n_2}[[P_1]]_{f_1}^{n_2}$}\label{subfig:cardex71}
\end{subfigure}
\begin{subfigure}[b]{\textwidth}
  \centering
  \input{figures/cardex8.tex}
  \caption{Cardinality of $\Pi_2=[[P_2]]_{f_3}^{n_3}\bowtie^{n_3}[[P_1]]_{f_2}^{n_3}$}\label{subfig:cardex72}
\end{subfigure}
\vspace{-4ex}
\caption{Estimating the cardinality of $\Pi$ in Figure~\ref{subfig:pex4} without the \texttt{DISTINCT} modifier for (a) $\Pi_1=[[P_2]]_{f_4}^{n_2}\bowtie^{n_2}[[P_1]]_{f_1}^{n_2}$ and (b) $\Pi_2=[[P_2]]_{f_3}^{n_3}\bowtie^{n_3}[[P_1]]_{f_2}^{n_3}$.
The output of Equation~\ref{eq:cardPlan} is thus the sum of the two formulas ($625+225=850$).}
\label{fig:cardex7}
\vspace{-1ex}
\end{figure*}

\begin{algorithm*}[b!]
\caption{Compute the transfer cost of a query execution plan}
\label{algo:transfercost}
\begin{algorithmic}[1]
\Statex \textbf{Input:} A query execution plan $\Pi$; a node $n$
\Statex \textbf{Output:} The estimated transfer cost $cost$
\Function{transferCost}{$\Pi$,$n$}
\State $cost\leftarrow 0$;
\If{$\Pi=[[P]]_f^{n_i}$}\label{ln:costSelStart}
\IfThen{$n\neq n_i$}{$cost\leftarrow card(P,f)$;}\label{ln:costSelEnd}
\ElsIf{$\Pi=\Pi_1\cup\Pi_2$}\label{ln:costUnionStart}
\State $cost\leftarrow \mathtt{transferCost}(\Pi_1,n)+\mathtt{transferCost}(\Pi_2,n)$;\label{ln:costUnionEnd}
\ElsIf{$\Pi=\Pi_1\times^{n_i}\Pi_2$}\label{ln:costCartStart}
\State $cost\leftarrow \mathtt{transferCost}(\Pi_1,n_i)+\mathtt{transferCost}(\Pi_2,n_i)$;\label{ln:costCartMid}
\IfThen{$n_i\neq n$}{$cost\leftarrow cost + card(\Pi)$;}\label{ln:costCartEnd}
\ElsIf{$\Pi=\Pi_1\bowtie^{n_i}\Pi_2$}\label{ln:costJoinStart}
\If{$\Pi_2=\Pi_2'\cup\Pi_2''$}
\State $cost\leftarrow \mathtt{transferCost}(\Pi_1\bowtie^{n_i}\Pi_2',n) + \mathtt{transferCost}(\Pi_1\bowtie^{n_i}\Pi_2'',n)$;\label{ln:costJoinUn}
\ElsIf{$\Pi_2=[[P]]_f^{n_j}$}
\State $cost\leftarrow \mathtt{transferCost}(\Pi_1,n_i)$;\label{ln:costJoinMid1}
\IfThen{$n_i\neq n_j$}{$cost\leftarrow cost+card^{\bowtie}_S(\Pi_1,P,f)$;}\label{ln:costJoinMid2}
\EndIf
%\State $cost\leftarrow \mathtt{transferCost}(\Pi_1,n_i)+\mathtt{transferCost}(\Pi_2,n_i)$;
\IfThen{$n\neq n_i$}{$cost\leftarrow cost + card(\Pi)$;}\label{ln:costJoinEnd}
\EndIf
\State \Return $cost$;
\EndFunction
\end{algorithmic}
\end{algorithm*}

%As an example, consider computing the cardinality $card(\Pi)$ of the plan $\Pi$ visualized in Figure~\ref{subfig:pex4}.
%Since $\Pi$ is a union, we compute the cardinality of $\Pi_1=[[P_2]]_{f_4}^{n_2}\bowtie^{n_2}[[P_1]]_{f_1}^{n_2}$ and $\Pi_2=[[P_2]]_{f_3}^{n_3}\bowtie^{n_3}[[P_1]]_{f_2}^{n_3}$ and let $card(\Pi)=card(\Pi_1)+card(\Pi_2)$.
%For $\Pi_1$, we apply 
%Since the left side of the top-most join is a union, we compute the first case of Equation~\ref{eq:cardjoin}, i.e., the aggregated cardinality of the (sub)plans $\Pi_1=([[P_2]]_{f_4}\bowtie^{n_2}[[P_1]]_{f_1})\bowtie^{n_1}[[P_3]]_{f_5}$ and $\Pi_2=([[P_2]]_{f_3}\bowtie^{n_3}[[P_1]]_{f_2})\bowtie^{n_1}[[P_3]]_{f_5}$, i.e., $card(\Pi)=card(\Pi_1)+card(\Pi_2)$.
%For each of these computations, we apply Equation~\ref{eq:cardjoinsingle}.
%Considering the SPBF cardinalities in Table~\ref{tab:cardinalities} and the cardinalities of the joins given in Figure~\ref{fig:cardex4}\footnote{Computing the cardinality of $\Pi_1=[[P_2]]_{f_4}\bowtie^{n_2}[[P_1]]_{f_1}$ and $\Pi_2=[[P_2]]_{f_3}\bowtie^{n_3}[[P_1]]_{f_2}$ using Equation~\ref{eq:cardjoinsingle} results in the same cardinalities as in Figure~\ref{fig:cardex4}; $card(\Pi_1)=500\cdot(50/200)\cdot(5000/1000)=625$, and $card(\Pi_2)=150\cdot(100/100)\cdot(3000/2000)=225$.} yields the following equation: $card(\Pi)=625\cdot(500/5000)\cdot(9000/8000)+225\cdot(1000/3000)\cdot(9000/8000)\approx 70+84\approx 154$.
%This is visualized in Figure~\ref{fig:cardex6} where the source of each value in the equation is shown.
\noindent Once again, computing the cardinality of $\Pi$ in Figure~\ref{subfig:pex4} not including the \texttt{DISTINCT} keyword is $card(\Pi)=card(\Pi_1)+card(\Pi_2)$.
Using Equation~\ref{eq:cardjoinnondistinct} on each of these yields the equation $card(\Pi)=500\cdot(50/200)\cdot(5000/1000)+150\cdot(100/100)\cdot(3000/2000)=625+225=850$.
Figure~\ref{fig:cardex7} visualizes this computation. %; the cardinality of $\Pi_1$ is shown on Figure~\ref{subfig:cardex71} and $\Pi_2$ on Figure~\ref{subfig:cardex72}.

Using the cardinality estimation shown in Equation~\ref{eq:cardPlan}, Algorithm~\ref{algo:transfercost} shows how the transfer cost of a query execution plan $\Pi$ on a node $n$ is computed taking into account the locality of the fragments.
First, if $\Pi=[[P]]_f^{n_i}$, i.e., $\Pi$ is a selection, the algorithm checks whether $n=n_i$ (line~\ref{ln:costSelEnd}); if they are equal it is 0 (since it incurs no transfer cost), otherwise the transfer cost of $\Pi$ is equal to the cardinality of the selection (Equation~\ref{eq:cardnondistinctsingle}).
For instance, the transfer cost of the execution plan shown in Figure~\ref{subfig:pex3} ($[[P_3]]_{f_5}^{n_1}$) on $n_1$ is $0$ since $f_5$ is available on $n_1$.

If, instead, $\Pi=\Pi_1\cup\Pi_2$, i.e., $\Pi$ is a union, the transfer cost is the sum of the transfer costs for $\Pi_1$ and $\Pi_2$ (line~\ref{ln:costUnionEnd}).
For instance, the transfer cost of the execution plan shown in Figure~\ref{subfig:pex1} ($[[P_1]]_{f_1}^{n_2}\cup [[P_1]]_{f_2}^{n_3}$) on $n_1$ is $5000+3000=8000$, since neither $f_1$ or $f_2$ is available on $n_1$.

Otherwise, if $\Pi=\Pi_1\times^{n_i}\Pi_2$, i.e., $\Pi$ is a Cartesian product, the transfer cost is the sum of the transfer costs for $\Pi_1$ and $\Pi_2$ (line~\ref{ln:costUnionEnd}), plus the cardinality of the Cartesian product if it is delegated to a different node than the one processing the (sub)plan, i.e., if $n\neq n_i$ (since they have to be transferred from $n_i$ to $n$).
%For instance, the transfer cost of the execution plan shown in Figure~\ref{subfig:pex1} ($[[P_1]]_{f_1}\cup [[P_1]]_{f_2}$) on $n_1$ is $5000+3000=8000$, since neither $f_1$ or $f_2$ is available on $n_1$.

Finally, if $\Pi=\Pi_1\bowtie^{n_i}\Pi_2$, i.e., $\Pi$ is a join, we once again take advantage of the fact that the right side of a join is always either a selection or a union of selections; in the latter case, we aggregate the transfer cost over each subplan in the union (line~\ref{ln:costJoinUn}).
However, if the right side of the join is a selection $\Pi_2=[[P]]_f^{n_j}$, we start by estimating the transfer cost of the left side of the join (line~\ref{ln:costJoinMid1}); if $n_j\neq n_i$, we further add in line~\ref{ln:costJoinMid2} %the cardinality of the left side (assuming the intermediate results from the left operand always have to be transferred to the location of the right operand) and 
the cardinality of the join (since these results should have to be sent back to $n_i$).
Furthermore, if $n_i\neq n$, we add in line~\ref{ln:costJoinEnd} the cardinality of the execution plan to the cost, since the results have to be transferred from $n_i$ to $n$.

%Consider as an example the plan $\Pi$ shown in Figure~\ref{subfig:planex} and calling $transferCost(\Pi,n_1)$.
%First, we compute the cost of the left side of the join (line~\ref{ln:costJoinMid1}).
%Since this (sub)plan (Figure~\ref{subfig:pex4}) is a union, we compute each part of the union separately and aggregate the results (line~\ref{ln:costUnionEnd}).
%For the left side, i.e., $[[P_2]]_{f_4}^{n_2}\bowtie^{n_2}[[P_1]]_{f_1}^{n_2}$, since both $f_4$ and $f_1$ are available on $n_2$, the cost of the join itself is $0$ (lines~\ref{ln:costSelEnd} and~\ref{ln:costJoinMid2}), while the cost of the plan is $625$, since $n_1\neq n_2$ (line~\ref{ln:costJoinEnd}), i.e., the results have to be transferred from $n_2$ to $n_1$.
%Similarly for the right side of the union, we get a cost of $625+225=850$ for processing the execution plan in Figure~\ref{subfig:pex4} on $n_1$.
%Since $f_5$ is available on $n_1$, $\Pi$ incurs no additional transfer cost (line~\ref{ln:costJoinEnd}), resulting in $transferCost(\Pi,n_1)=850$.

Given the transfer cost in Algorithm~\ref{algo:transfercost}, the cost of processing a query execution plan $\Pi$ on a node $n$ is the transfer cost plus the cardinality of $\Pi$.
This is formally defined as follows:

\vspace{-1ex}
\begin{equation}\label{eq:costPlan}
cost_n(\Pi)=transferCost(\Pi,n)+card(\Pi)
\end{equation}
\vspace{-2ex}

\noindent Given the cost function in Equation~\ref{eq:costPlan}, we compute the cost of each possible join delegation and apply Dynamic Programming (DP) to achieve the execution plan with the lowest cost.
Table~\ref{tbl:dp} shows the best execution plan in the DP table for each (sub)plan for processing query $Q$ (Figure~\ref{subfig:running_q}) on node $n_1$ in the running example; Figure~\ref{fig:qpex} visualizes the execution plans in Table~\ref{tbl:dp}.
%For instance, the cost of processing $\Pi$ in Figure~\ref{subfig:planex} on $n_1$ is $transferCost(\Pi,n_1)+card(\Pi)=850+154=1004$. %cf. the transfer cost computation above and cardinality estimation in Figure~\ref{fig:cardex6}.

\begin{table}[!htb]
\caption{Entries in the DP table for query $Q$ (Figure~\ref{subfig:running_q})}\label{tbl:dp}
\begin{tabular}{cccc}
 \textbf{Subquery} & \textbf{Execution Plan} & \textbf{Cardinality} & \textbf{Cost}  \\ \hline
 $P_1$ & $[[P_1]]_{f_1}^{n_2}\cup [[P_1]]_{f_2}^{n_3}$ & 8,000 & 8,000 \\
 $P_2$ & $[[P_2]]_{f_3}^{n_3}\cup [[P_2]]_{f_4}^{n_2}$ & 650 & 650 \\
 $P_3$ & $[[P_3]]_{f_5}^{n_1}$ & 9,000 & 9,000 \\
 $P_2\bowtie P_1$ & $([[P_2]]_{f_4}^{n_2}\bowtie^{n_2} [[P_1]]_{f_1}^{n_2})\cup ([[P_2]]_{f_3}^{n_3}\bowtie^{n_3} [[P_1]]_{f_2}^{n_3})$ & \textbf{850} & 1,700 \\
 $P_2\bowtie P_3$ & $([[P_2]]_{f_3}^{n_3}\cup [[P_2]]_{f_4}^{n_2})\times^{n_1}[[P_3]]_{f_5}^{n_1}$ & 5,850,000 & 5,850,650 \\
 $P_1\bowtie P_3$ & $([[P_1]]_{f_1}^{n_2}\cup [[P_1]]_{f_2}^{n_3})\bowtie^{n_1} [[P_3]]_{f_5}^{n_1}$ & 1,688 & 9,688 \\
 $P_2\bowtie P_1\bowtie P_3$ & $(([[P_2]]_{f_4}^{n_2}\bowtie^{n_2} [[P_1]]_{f_1}^{n_2})\cup ([[P_2]]_{f_3}^{n_3}\bowtie^{n_3} [[P_1]]_{f_2}^{n_3}))\bowtie^{n_1}[[P_3]]_{f_5}^{n_1}$ & 154 & 1,004
\end{tabular}
\end{table}

%% file: figures/cgex1.tex
\tikzset{every picture/.style={line width=0.75pt}} %set default line width to 0.75pt        

\begin{tikzpicture}[x=0.5pt,y=0.5pt,yscale=-1,xscale=1]
%uncomment if require: \path (0,436); %set diagram left start at 0, and has height of 436

%Shape: Circle [id:dp3280439895815911] 
\draw  [draw opacity=0][fill={rgb, 255:red, 184; green, 233; blue, 134 } ,fill opacity=1 ] (324,252.5) .. controls (324,242.84) and (331.84,235) .. (341.5,235) .. controls (351.16,235) and (359,242.84) .. (359,252.5) .. controls (359,262.16) and (351.16,270) .. (341.5,270) .. controls (331.84,270) and (324,262.16) .. (324,252.5) -- cycle ;

% Text Node
\draw (332,242.4) node [anchor=north west][inner sep=0.75pt]    {$f_{5}$};

\end{tikzpicture}

%% file: figures/cgex2.tex
\tikzset{every picture/.style={line width=0.75pt}} %set default line width to 0.75pt        

\begin{tikzpicture}[x=0.5pt,y=0.5pt,yscale=-1,xscale=1]
%uncomment if require: \path (0,436); %set diagram left start at 0, and has height of 436

%Shape: Circle [id:dp1358178850418491] 
\draw  [draw opacity=0][fill={rgb, 255:red, 125; green, 184; blue, 253 }  ,fill opacity=1 ] (324,182.5) .. controls (324,172.84) and (331.84,165) .. (341.5,165) .. controls (351.16,165) and (359,172.84) .. (359,182.5) .. controls (359,192.16) and (351.16,200) .. (341.5,200) .. controls (331.84,200) and (324,192.16) .. (324,182.5) -- cycle ;
%Shape: Circle [id:dp3280439895815911] 
\draw  [draw opacity=0][fill={rgb, 255:red, 184; green, 233; blue, 134 }  ,fill opacity=1 ] (324,252.5) .. controls (324,242.84) and (331.84,235) .. (341.5,235) .. controls (351.16,235) and (359,242.84) .. (359,252.5) .. controls (359,262.16) and (351.16,270) .. (341.5,270) .. controls (331.84,270) and (324,262.16) .. (324,252.5) -- cycle ;
%Straight Lines [id:da10576094690019244] 
\draw    (341.5,200) -- (341.5,235) ;

% Text Node
\draw (332,172.4) node [anchor=north west][inner sep=0.75pt]    {$f_{2}$};
% Text Node
\draw (332,242.4) node [anchor=north west][inner sep=0.75pt]    {$f_{5}$};

\end{tikzpicture}

%% file: figures/cgex3.tex
\tikzset{every picture/.style={line width=0.75pt}} %set default line width to 0.75pt        

\begin{tikzpicture}[x=0.5pt,y=0.5pt,yscale=-1,xscale=1]
%uncomment if require: \path (0,436); %set diagram left start at 0, and has height of 436

%Shape: Circle [id:dp1358178850418491] 
\draw  [draw opacity=0][fill={rgb, 255:red, 248; green, 231; blue, 28 }  ,fill opacity=1 ] (324,182.5) .. controls (324,172.84) and (331.84,165) .. (341.5,165) .. controls (351.16,165) and (359,172.84) .. (359,182.5) .. controls (359,192.16) and (351.16,200) .. (341.5,200) .. controls (331.84,200) and (324,192.16) .. (324,182.5) -- cycle ;
%Shape: Circle [id:dp3280439895815911] 
\draw  [draw opacity=0][fill={rgb, 255:red, 125; green, 184; blue, 253 }  ,fill opacity=1 ] (324,252.5) .. controls (324,242.84) and (331.84,235) .. (341.5,235) .. controls (351.16,235) and (359,242.84) .. (359,252.5) .. controls (359,262.16) and (351.16,270) .. (341.5,270) .. controls (331.84,270) and (324,262.16) .. (324,252.5) -- cycle ;
%Shape: Circle [id:dp8128701237473113] 
\draw  [draw opacity=0][fill={rgb, 255:red, 184; green, 233; blue, 134 }  ,fill opacity=1 ] (324,322.5) .. controls (324,312.84) and (331.84,305) .. (341.5,305) .. controls (351.16,305) and (359,312.84) .. (359,322.5) .. controls (359,332.16) and (351.16,340) .. (341.5,340) .. controls (331.84,340) and (324,332.16) .. (324,322.5) -- cycle ;
%Straight Lines [id:da10576094690019244] 
\draw    (341.5,200) -- (341.5,235) ;
%Straight Lines [id:da7326800482493573] 
\draw    (341.5,305) -- (341.5,270) ;

% Text Node
\draw (332,172.4) node [anchor=north west][inner sep=0.75pt]    {$f_{3}$};
% Text Node
\draw (332,242.4) node [anchor=north west][inner sep=0.75pt]    {$f_{2}$};
% Text Node
\draw (332,312.4) node [anchor=north west][inner sep=0.75pt]    {$f_{5}$};

\end{tikzpicture}

%% file: figures/cgex5.tex
\tikzset{every picture/.style={line width=0.75pt}} %set default line width to 0.75pt        

\begin{tikzpicture}[x=0.5pt,y=0.5pt,yscale=-1,xscale=1]
%uncomment if require: \path (0,436); %set diagram left start at 0, and has height of 436

%Shape: Circle [id:dp1358178850418491] 
\draw  [draw opacity=0][fill={rgb, 255:red, 125; green, 184; blue, 253 }  ,fill opacity=1 ] (324,182.5) .. controls (324,172.84) and (331.84,165) .. (341.5,165) .. controls (351.16,165) and (359,172.84) .. (359,182.5) .. controls (359,192.16) and (351.16,200) .. (341.5,200) .. controls (331.84,200) and (324,192.16) .. (324,182.5) -- cycle ;
%Shape: Circle [id:dp3280439895815911] 
\draw  [draw opacity=0][fill={rgb, 255:red, 184; green, 233; blue, 134 }  ,fill opacity=1 ] (324,252.5) .. controls (324,242.84) and (331.84,235) .. (341.5,235) .. controls (351.16,235) and (359,242.84) .. (359,252.5) .. controls (359,262.16) and (351.16,270) .. (341.5,270) .. controls (331.84,270) and (324,262.16) .. (324,252.5) -- cycle ;
%Straight Lines [id:da10576094690019244] 
\draw    (341.5,200) -- (341.5,235) ;

% Text Node
\draw (332,172.4) node [anchor=north west][inner sep=0.75pt]    {$f_{1}$};
% Text Node
\draw (332,242.4) node [anchor=north west][inner sep=0.75pt]    {$f_{5}$};

\end{tikzpicture}

%% file: figures/cgex4.tex
\tikzset{every picture/.style={line width=0.75pt}} %set default line width to 0.75pt        

\begin{tikzpicture}[x=0.5pt,y=0.5pt,yscale=-1,xscale=1]
%uncomment if require: \path (0,436); %set diagram left start at 0, and has height of 436

%Shape: Circle [id:dp47750815641295064] 
\draw  [draw opacity=0][fill={rgb, 255:red, 248; green, 231; blue, 28 }  ,fill opacity=1 ] (224,182.5) .. controls (224,172.84) and (231.84,165) .. (241.5,165) .. controls (251.16,165) and (259,172.84) .. (259,182.5) .. controls (259,192.16) and (251.16,200) .. (241.5,200) .. controls (231.84,200) and (224,192.16) .. (224,182.5) -- cycle ;
%Shape: Circle [id:dp703699277162218] 
\draw  [draw opacity=0][fill={rgb, 255:red, 125; green, 184; blue, 253 }  ,fill opacity=1 ] (224,252.5) .. controls (224,242.84) and (231.84,235) .. (241.5,235) .. controls (251.16,235) and (259,242.84) .. (259,252.5) .. controls (259,262.16) and (251.16,270) .. (241.5,270) .. controls (231.84,270) and (224,262.16) .. (224,252.5) -- cycle ;
%Shape: Circle [id:dp8128701237473113] 
\draw  [draw opacity=0][fill={rgb, 255:red, 184; green, 233; blue, 134 }  ,fill opacity=1 ] (224,322.5) .. controls (224,312.84) and (231.84,305) .. (241.5,305) .. controls (251.16,305) and (259,312.84) .. (259,322.5) .. controls (259,332.16) and (251.16,340) .. (241.5,340) .. controls (231.84,340) and (224,332.16) .. (224,322.5) -- cycle ;
%Straight Lines [id:da5948486508667451] 
\draw    (241.5,200) -- (241.5,235) ;
%Straight Lines [id:da3505352749144771] 
\draw    (241.5,270) -- (241.5,305) ;

% Text Node
\draw (232,172.4) node [anchor=north west][inner sep=0.75pt]    {$f_{4}$};
% Text Node
\draw (232,242.4) node [anchor=north west][inner sep=0.75pt]    {$f_{1}$};
% Text Node
\draw (232,312.4) node [anchor=north west][inner sep=0.75pt]    {$f_{5}$};

\end{tikzpicture}

%% file: figures/compatgraphex.tex
\tikzset{every picture/.style={line width=0.75pt}} %set default line width to 0.75pt        

\begin{tikzpicture}[x=0.5pt,y=0.5pt,yscale=-1,xscale=1]

%Shape: Circle [id:dp47750815641295064] 
\draw  [draw opacity=0][fill={rgb, 255:red, 248; green, 231; blue, 28 }  ,fill opacity=1 ] (224,182.5) .. controls (224,172.84) and (231.84,165) .. (241.5,165) .. controls (251.16,165) and (259,172.84) .. (259,182.5) .. controls (259,192.16) and (251.16,200) .. (241.5,200) .. controls (231.84,200) and (224,192.16) .. (224,182.5) -- cycle ;
%Shape: Circle [id:dp1358178850418491] 
\draw  [draw opacity=0][fill={rgb, 255:red, 248; green, 231; blue, 28 }  ,fill opacity=1 ] (324,182.5) .. controls (324,172.84) and (331.84,165) .. (341.5,165) .. controls (351.16,165) and (359,172.84) .. (359,182.5) .. controls (359,192.16) and (351.16,200) .. (341.5,200) .. controls (331.84,200) and (324,192.16) .. (324,182.5) -- cycle ;
%Shape: Circle [id:dp703699277162218] 
\draw  [draw opacity=0][fill={rgb, 255:red, 125; green, 184; blue, 253 }  ,fill opacity=1 ] (224,252.5) .. controls (224,242.84) and (231.84,235) .. (241.5,235) .. controls (251.16,235) and (259,242.84) .. (259,252.5) .. controls (259,262.16) and (251.16,270) .. (241.5,270) .. controls (231.84,270) and (224,262.16) .. (224,252.5) -- cycle ;
%Shape: Circle [id:dp3280439895815911] 
\draw  [draw opacity=0][fill={rgb, 255:red, 125; green, 184; blue, 253 }  ,fill opacity=1 ] (324,252.5) .. controls (324,242.84) and (331.84,235) .. (341.5,235) .. controls (351.16,235) and (359,242.84) .. (359,252.5) .. controls (359,262.16) and (351.16,270) .. (341.5,270) .. controls (331.84,270) and (324,262.16) .. (324,252.5) -- cycle ;
%Shape: Circle [id:dp8128701237473113] 
\draw  [draw opacity=0][fill={rgb, 255:red, 184; green, 233; blue, 134 }  ,fill opacity=1 ] (274,322.5) .. controls (274,312.84) and (281.84,305) .. (291.5,305) .. controls (301.16,305) and (309,312.84) .. (309,322.5) .. controls (309,332.16) and (301.16,340) .. (291.5,340) .. controls (281.84,340) and (274,332.16) .. (274,322.5) -- cycle ;
%Straight Lines [id:da5948486508667451] 
\draw    (241.5,200) -- (241.5,235) ;
%Straight Lines [id:da10576094690019244] 
\draw    (341.5,200) -- (341.5,235) ;
%Straight Lines [id:da3505352749144771] 
\draw    (252.5,264) -- (282,309) ;
%Straight Lines [id:da7326800482493573] 
\draw    (301,309) -- (329.5,265) ;

% Text Node
\draw (232,172.4) node [anchor=north west][inner sep=0.75pt]    {$f_{4}$};
% Text Node
\draw (332,172.4) node [anchor=north west][inner sep=0.75pt]    {$f_{3}$};
% Text Node
\draw (232,242.4) node [anchor=north west][inner sep=0.75pt]    {$f_{1}$};
% Text Node
\draw (332,242.4) node [anchor=north west][inner sep=0.75pt]    {$f_{2}$};
% Text Node
\draw (282,312.4) node [anchor=north west][inner sep=0.75pt]    {$f_{5}$};

\end{tikzpicture}

%% file: figures/cardex1.tex
\tikzset{every picture/.style={line width=0.75pt}} %set default line width to 0.75pt        

\begin{tikzpicture}[x=0.6pt,y=0.6pt,yscale=-1,xscale=1]
%uncomment if require: \path (0,762); %set diagram left start at 0, and has height of 762

%Shape: Rectangle [id:dp8387399614999044] 
\draw  [color={rgb, 255:red, 0; green, 0; blue, 0 }  ,draw opacity=1 ] (119.77,110.56) -- (141.55,110.56) -- (141.55,116) -- (119.77,116) -- cycle ;
%Shape: Rectangle [id:dp4295953583775852] 
\draw   (125.22,110.56) -- (130.66,110.56) -- (130.66,116) -- (125.22,116) -- cycle ;
%Shape: Rectangle [id:dp3306481358194292] 
\draw   (136.1,110.56) -- (141.55,110.56) -- (141.55,116) -- (136.1,116) -- cycle ;
%Shape: Rectangle [id:dp9615378257249877] 
\draw  [color={rgb, 255:red, 0; green, 0; blue, 0 }  ,draw opacity=1 ] (98,110.56) -- (119.77,110.56) -- (119.77,116) -- (98,116) -- cycle ;
%Shape: Rectangle [id:dp22195807349685726] 
\draw   (103.44,110.56) -- (108.89,110.56) -- (108.89,116) -- (103.44,116) -- cycle ;
%Shape: Rectangle [id:dp11140650792547868] 
\draw   (114.33,110.56) -- (119.77,110.56) -- (119.77,116) -- (114.33,116) -- cycle ;
%Shape: Rectangle [id:dp1622219004134312] 
\draw  [fill={rgb, 255:red, 0; green, 0; blue, 0 }  ,fill opacity=1 ] (114.33,110.56) -- (119.77,110.56) -- (119.77,116) -- (114.33,116) -- cycle ;
%Shape: Rectangle [id:dp536903437731283] 
\draw  [fill={rgb, 255:red, 0; green, 0; blue, 0 }  ,fill opacity=1 ] (108.89,110.56) -- (114.33,110.56) -- (114.33,116) -- (108.89,116) -- cycle ;
%Shape: Rectangle [id:dp2679143543226373] 
\draw  [color={rgb, 255:red, 0; green, 0; blue, 0 }  ,draw opacity=1 ] (165.23,110.56) -- (187,110.56) -- (187,116) -- (165.23,116) -- cycle ;
%Shape: Rectangle [id:dp20074758169683027] 
\draw   (170.67,110.56) -- (176.11,110.56) -- (176.11,116) -- (170.67,116) -- cycle ;
%Shape: Rectangle [id:dp4389221057731434] 
\draw   (181.56,110.56) -- (187,110.56) -- (187,116) -- (181.56,116) -- cycle ;
%Shape: Rectangle [id:dp47104130800868793] 
\draw  [color={rgb, 255:red, 0; green, 0; blue, 0 }  ,draw opacity=1 ] (143.45,110.56) -- (165.23,110.56) -- (165.23,116) -- (143.45,116) -- cycle ;
%Shape: Rectangle [id:dp4795901042169357] 
\draw   (148.9,110.56) -- (154.34,110.56) -- (154.34,116) -- (148.9,116) -- cycle ;
%Shape: Rectangle [id:dp1586933010210959] 
\draw   (159.78,110.56) -- (165.23,110.56) -- (165.23,116) -- (159.78,116) -- cycle ;
%Shape: Rectangle [id:dp7819588186789117] 
\draw  [fill={rgb, 255:red, 0; green, 0; blue, 0 }  ,fill opacity=1 ] (165.23,110.56) -- (170.67,110.56) -- (170.67,116) -- (165.23,116) -- cycle ;
%Shape: Rectangle [id:dp986435151196106] 
\draw  [fill={rgb, 255:red, 0; green, 0; blue, 0 }  ,fill opacity=1 ] (181.56,110.56) -- (187,110.56) -- (187,116) -- (181.56,116) -- cycle ;
%Shape: Rectangle [id:dp2069179774909723] 
\draw  [fill={rgb, 255:red, 0; green, 0; blue, 0 }  ,fill opacity=1 ] (148.9,110.56) -- (154.34,110.56) -- (154.34,116) -- (148.9,116) -- cycle ;
%Shape: Rectangle [id:dp03362018400189082] 
\draw  [fill={rgb, 255:red, 0; green, 0; blue, 0 }  ,fill opacity=1 ] (125.22,110.56) -- (130.66,110.56) -- (130.66,116) -- (125.22,116) -- cycle ;
%Shape: Rectangle [id:dp20839781982453176] 
\draw  [color={rgb, 255:red, 0; green, 0; blue, 0 }  ,draw opacity=1 ] (219.77,110.56) -- (241.55,110.56) -- (241.55,116) -- (219.77,116) -- cycle ;
%Shape: Rectangle [id:dp4052886258298046] 
\draw   (225.22,110.56) -- (230.66,110.56) -- (230.66,116) -- (225.22,116) -- cycle ;
%Shape: Rectangle [id:dp8361603104937438] 
\draw   (236.1,110.56) -- (241.55,110.56) -- (241.55,116) -- (236.1,116) -- cycle ;
%Shape: Rectangle [id:dp017731928437753064] 
\draw  [color={rgb, 255:red, 0; green, 0; blue, 0 }  ,draw opacity=1 ] (198,110.56) -- (219.77,110.56) -- (219.77,116) -- (198,116) -- cycle ;
%Shape: Rectangle [id:dp14418179362908856] 
\draw   (203.44,110.56) -- (208.89,110.56) -- (208.89,116) -- (203.44,116) -- cycle ;
%Shape: Rectangle [id:dp9621023136240734] 
\draw   (214.33,110.56) -- (219.77,110.56) -- (219.77,116) -- (214.33,116) -- cycle ;
%Shape: Rectangle [id:dp30482821150372497] 
\draw  [fill={rgb, 255:red, 0; green, 0; blue, 0 }  ,fill opacity=1 ] (208.89,110.56) -- (213.75,110.56) -- (213.75,116) -- (208.89,116) -- cycle ;
%Shape: Rectangle [id:dp7681276143438305] 
\draw  [fill={rgb, 255:red, 0; green, 0; blue, 0 }  ,fill opacity=1 ] (197.89,110.56) -- (203.33,110.56) -- (203.33,116) -- (197.89,116) -- cycle ;
%Shape: Rectangle [id:dp10153164614094523] 
\draw  [color={rgb, 255:red, 0; green, 0; blue, 0 }  ,draw opacity=1 ] (265.23,110.56) -- (287,110.56) -- (287,116) -- (265.23,116) -- cycle ;
%Shape: Rectangle [id:dp8226845049423511] 
\draw   (270.67,110.56) -- (276.11,110.56) -- (276.11,116) -- (270.67,116) -- cycle ;
%Shape: Rectangle [id:dp9046826303812424] 
\draw   (281.56,110.56) -- (287,110.56) -- (287,116) -- (281.56,116) -- cycle ;
%Shape: Rectangle [id:dp13555959261068196] 
\draw  [color={rgb, 255:red, 0; green, 0; blue, 0 }  ,draw opacity=1 ] (243.45,110.56) -- (265.23,110.56) -- (265.23,116) -- (243.45,116) -- cycle ;
%Shape: Rectangle [id:dp17981641982785557] 
\draw   (248.9,110.56) -- (254.34,110.56) -- (254.34,116) -- (248.9,116) -- cycle ;
%Shape: Rectangle [id:dp25851216796396403] 
\draw   (259.78,110.56) -- (265.23,110.56) -- (265.23,116) -- (259.78,116) -- cycle ;
%Shape: Rectangle [id:dp6158785262016018] 
\draw  [fill={rgb, 255:red, 0; green, 0; blue, 0 }  ,fill opacity=1 ] (265.23,110.56) -- (270.67,110.56) -- (270.67,116) -- (265.23,116) -- cycle ;
%Shape: Rectangle [id:dp0929053601627543] 
\draw  [fill={rgb, 255:red, 0; green, 0; blue, 0 }  ,fill opacity=1 ] (270.56,110.56) -- (276,110.56) -- (276,116) -- (270.56,116) -- cycle ;
%Shape: Rectangle [id:dp12906292886672255] 
\draw  [fill={rgb, 255:red, 0; green, 0; blue, 0 }  ,fill opacity=1 ] (259.9,110.56) -- (265.34,110.56) -- (265.34,116) -- (259.9,116) -- cycle ;
%Shape: Rectangle [id:dp6796836313605346] 
\draw  [fill={rgb, 255:red, 0; green, 0; blue, 0 }  ,fill opacity=1 ] (225.22,110.56) -- (230.66,110.56) -- (230.66,116) -- (225.22,116) -- cycle ;
%Shape: Rectangle [id:dp29191588566752824] 
\draw  [color={rgb, 255:red, 0; green, 0; blue, 0 }  ,draw opacity=1 ] (119.77,143.56) -- (141.55,143.56) -- (141.55,149) -- (119.77,149) -- cycle ;
%Shape: Rectangle [id:dp7995013223208742] 
\draw   (125.22,143.56) -- (130.66,143.56) -- (130.66,149) -- (125.22,149) -- cycle ;
%Shape: Rectangle [id:dp43961461402111923] 
\draw   (136.1,143.56) -- (141.55,143.56) -- (141.55,149) -- (136.1,149) -- cycle ;
%Shape: Rectangle [id:dp32503408510653464] 
\draw  [color={rgb, 255:red, 0; green, 0; blue, 0 }  ,draw opacity=1 ] (98,143.56) -- (119.77,143.56) -- (119.77,149) -- (98,149) -- cycle ;
%Shape: Rectangle [id:dp5642845524256369] 
\draw   (103.44,143.56) -- (108.89,143.56) -- (108.89,149) -- (103.44,149) -- cycle ;
%Shape: Rectangle [id:dp8407641095100401] 
\draw   (114.33,143.56) -- (119.77,143.56) -- (119.77,149) -- (114.33,149) -- cycle ;
%Shape: Rectangle [id:dp18769626116117144] 
\draw  [fill={rgb, 255:red, 0; green, 0; blue, 0 }  ,fill opacity=1 ] (109.33,143.56) -- (114.77,143.56) -- (114.77,149) -- (109.33,149) -- cycle ;
%Shape: Rectangle [id:dp1798374344094137] 
\draw  [fill={rgb, 255:red, 0; green, 0; blue, 0 }  ,fill opacity=1 ] (103.89,143.56) -- (109.33,143.56) -- (109.33,149) -- (103.89,149) -- cycle ;
%Shape: Rectangle [id:dp6420601793307151] 
\draw  [color={rgb, 255:red, 0; green, 0; blue, 0 }  ,draw opacity=1 ] (165.23,143.56) -- (187,143.56) -- (187,149) -- (165.23,149) -- cycle ;
%Shape: Rectangle [id:dp692813838692275] 
\draw   (170.67,143.56) -- (176.11,143.56) -- (176.11,149) -- (170.67,149) -- cycle ;
%Shape: Rectangle [id:dp7628182988115574] 
\draw   (181.56,143.56) -- (187,143.56) -- (187,149) -- (181.56,149) -- cycle ;
%Shape: Rectangle [id:dp792474379535553] 
\draw  [color={rgb, 255:red, 0; green, 0; blue, 0 }  ,draw opacity=1 ] (143.45,143.56) -- (165.23,143.56) -- (165.23,149) -- (143.45,149) -- cycle ;
%Shape: Rectangle [id:dp7118756909159807] 
\draw   (148.9,143.56) -- (154.34,143.56) -- (154.34,149) -- (148.9,149) -- cycle ;
%Shape: Rectangle [id:dp552693571576886] 
\draw   (159.78,143.56) -- (165.23,143.56) -- (165.23,149) -- (159.78,149) -- cycle ;
%Shape: Rectangle [id:dp7097537153255002] 
\draw  [fill={rgb, 255:red, 0; green, 0; blue, 0 }  ,fill opacity=1 ] (160.23,143.56) -- (165.67,143.56) -- (165.67,149) -- (160.23,149) -- cycle ;
%Shape: Rectangle [id:dp8742587691997274] 
\draw  [fill={rgb, 255:red, 0; green, 0; blue, 0 }  ,fill opacity=1 ] (170.56,143.56) -- (176,143.56) -- (176,149) -- (170.56,149) -- cycle ;
%Shape: Rectangle [id:dp47544132574853715] 
\draw  [fill={rgb, 255:red, 0; green, 0; blue, 0 }  ,fill opacity=1 ] (148.9,143.56) -- (154.34,143.56) -- (154.34,149) -- (148.9,149) -- cycle ;
%Shape: Rectangle [id:dp567979468100233] 
\draw  [fill={rgb, 255:red, 0; green, 0; blue, 0 }  ,fill opacity=1 ] (120.22,143.56) -- (125.66,143.56) -- (125.66,149) -- (120.22,149) -- cycle ;
%Shape: Rectangle [id:dp8848452249709213] 
\draw  [color={rgb, 255:red, 0; green, 0; blue, 0 }  ,draw opacity=1 ] (219.77,143.56) -- (241.55,143.56) -- (241.55,149) -- (219.77,149) -- cycle ;
%Shape: Rectangle [id:dp885581562887734] 
\draw   (225.22,143.56) -- (230.66,143.56) -- (230.66,149) -- (225.22,149) -- cycle ;
%Shape: Rectangle [id:dp7745804793634135] 
\draw   (236.1,143.56) -- (241.55,143.56) -- (241.55,149) -- (236.1,149) -- cycle ;
%Shape: Rectangle [id:dp3526756619968049] 
\draw  [color={rgb, 255:red, 0; green, 0; blue, 0 }  ,draw opacity=1 ] (198,143.56) -- (219.77,143.56) -- (219.77,149) -- (198,149) -- cycle ;
%Shape: Rectangle [id:dp5815759867935808] 
\draw   (203.44,143.56) -- (208.89,143.56) -- (208.89,149) -- (203.44,149) -- cycle ;
%Shape: Rectangle [id:dp11313135632431193] 
\draw   (214.33,143.56) -- (219.77,143.56) -- (219.77,149) -- (214.33,149) -- cycle ;
%Shape: Rectangle [id:dp377051158556549] 
\draw  [fill={rgb, 255:red, 0; green, 0; blue, 0 }  ,fill opacity=1 ] (214.33,143.56) -- (219.77,143.56) -- (219.77,149) -- (214.33,149) -- cycle ;
%Shape: Rectangle [id:dp6737920549428935] 
\draw  [fill={rgb, 255:red, 0; green, 0; blue, 0 }  ,fill opacity=1 ] (208.89,143.56) -- (214.33,143.56) -- (214.33,149) -- (208.89,149) -- cycle ;
%Shape: Rectangle [id:dp5898140163359423] 
\draw  [color={rgb, 255:red, 0; green, 0; blue, 0 }  ,draw opacity=1 ] (265.23,143.56) -- (287,143.56) -- (287,149) -- (265.23,149) -- cycle ;
%Shape: Rectangle [id:dp3080511376590268] 
\draw   (270.67,143.56) -- (276.11,143.56) -- (276.11,149) -- (270.67,149) -- cycle ;
%Shape: Rectangle [id:dp5258739039682021] 
\draw   (281.56,143.56) -- (287,143.56) -- (287,149) -- (281.56,149) -- cycle ;
%Shape: Rectangle [id:dp9400974144378738] 
\draw  [color={rgb, 255:red, 0; green, 0; blue, 0 }  ,draw opacity=1 ] (243.45,143.56) -- (265.23,143.56) -- (265.23,149) -- (243.45,149) -- cycle ;
%Shape: Rectangle [id:dp5173366579293642] 
\draw   (248.9,143.56) -- (254.34,143.56) -- (254.34,149) -- (248.9,149) -- cycle ;
%Shape: Rectangle [id:dp8914470092699311] 
\draw   (259.78,143.56) -- (265.23,143.56) -- (265.23,149) -- (259.78,149) -- cycle ;
%Shape: Rectangle [id:dp7310004388484735] 
\draw  [fill={rgb, 255:red, 0; green, 0; blue, 0 }  ,fill opacity=1 ] (248.23,143.56) -- (253.67,143.56) -- (253.67,149) -- (248.23,149) -- cycle ;
%Shape: Rectangle [id:dp35355785095784564] 
\draw  [fill={rgb, 255:red, 0; green, 0; blue, 0 }  ,fill opacity=1 ] (281.56,143.56) -- (287,143.56) -- (287,149) -- (281.56,149) -- cycle ;
%Shape: Rectangle [id:dp746611566755886] 
\draw  [fill={rgb, 255:red, 0; green, 0; blue, 0 }  ,fill opacity=1 ] (243.9,143.56) -- (249.34,143.56) -- (249.34,149) -- (243.9,149) -- cycle ;
%Shape: Rectangle [id:dp6712474035845276] 
\draw  [fill={rgb, 255:red, 0; green, 0; blue, 0 }  ,fill opacity=1 ] (236.22,143.56) -- (241.66,143.56) -- (241.66,149) -- (236.22,149) -- cycle ;
%Rounded Rect [id:dp6527217395443599] 
\draw  [dash pattern={on 4.5pt off 4.5pt}] (89.78,82.05) .. controls (89.78,71.67) and (98.2,63.25) .. (108.58,63.25) -- (277.2,63.25) .. controls (287.58,63.25) and (296,71.67) .. (296,82.05) -- (296,138.45) .. controls (296,148.83) and (287.58,157.25) .. (277.2,157.25) -- (108.58,157.25) .. controls (98.2,157.25) and (89.78,148.83) .. (89.78,138.45) -- cycle ;
%Shape: Rectangle [id:dp5861704792273547] 
\draw  [color={rgb, 255:red, 0; green, 0; blue, 0 }  ,draw opacity=1 ] (339.77,110.56) -- (361.55,110.56) -- (361.55,116) -- (339.77,116) -- cycle ;
%Shape: Rectangle [id:dp8652286188539163] 
\draw   (345.22,110.56) -- (350.66,110.56) -- (350.66,116) -- (345.22,116) -- cycle ;
%Shape: Rectangle [id:dp6553686219851063] 
\draw   (356.1,110.56) -- (361.55,110.56) -- (361.55,116) -- (356.1,116) -- cycle ;
%Shape: Rectangle [id:dp33096734613493217] 
\draw  [color={rgb, 255:red, 0; green, 0; blue, 0 }  ,draw opacity=1 ] (318,110.56) -- (339.77,110.56) -- (339.77,116) -- (318,116) -- cycle ;
%Shape: Rectangle [id:dp12349908140137411] 
\draw   (323.44,110.56) -- (328.89,110.56) -- (328.89,116) -- (323.44,116) -- cycle ;
%Shape: Rectangle [id:dp9635156330233046] 
\draw   (334.33,110.56) -- (339.77,110.56) -- (339.77,116) -- (334.33,116) -- cycle ;
%Shape: Rectangle [id:dp453467002224598] 
\draw  [fill={rgb, 255:red, 0; green, 0; blue, 0 }  ,fill opacity=1 ] (323.33,110.56) -- (328.77,110.56) -- (328.77,116) -- (323.33,116) -- cycle ;
%Shape: Rectangle [id:dp5667661563867294] 
\draw  [fill={rgb, 255:red, 0; green, 0; blue, 0 }  ,fill opacity=1 ] (317.89,110.56) -- (323.33,110.56) -- (323.33,116) -- (317.89,116) -- cycle ;
%Shape: Rectangle [id:dp18368381226393837] 
\draw  [color={rgb, 255:red, 0; green, 0; blue, 0 }  ,draw opacity=1 ] (385.23,110.56) -- (407,110.56) -- (407,116) -- (385.23,116) -- cycle ;
%Shape: Rectangle [id:dp6384956850046604] 
\draw   (390.67,110.56) -- (396.11,110.56) -- (396.11,116) -- (390.67,116) -- cycle ;
%Shape: Rectangle [id:dp9607004976167075] 
\draw   (401.56,110.56) -- (407,110.56) -- (407,116) -- (401.56,116) -- cycle ;
%Shape: Rectangle [id:dp2106746094458557] 
\draw  [color={rgb, 255:red, 0; green, 0; blue, 0 }  ,draw opacity=1 ] (363.45,110.56) -- (385.23,110.56) -- (385.23,116) -- (363.45,116) -- cycle ;
%Shape: Rectangle [id:dp3831807817615869] 
\draw   (368.9,110.56) -- (374.34,110.56) -- (374.34,116) -- (368.9,116) -- cycle ;
%Shape: Rectangle [id:dp7247383825467211] 
\draw   (379.78,110.56) -- (385.23,110.56) -- (385.23,116) -- (379.78,116) -- cycle ;
%Shape: Rectangle [id:dp9246418474353738] 
\draw  [fill={rgb, 255:red, 0; green, 0; blue, 0 }  ,fill opacity=1 ] (379.23,110.56) -- (384.67,110.56) -- (384.67,116) -- (379.23,116) -- cycle ;
%Shape: Rectangle [id:dp9645990451010458] 
\draw  [fill={rgb, 255:red, 0; green, 0; blue, 0 }  ,fill opacity=1 ] (373.9,110.56) -- (379.34,110.56) -- (379.34,116) -- (373.9,116) -- cycle ;
%Shape: Rectangle [id:dp599618835804869] 
\draw  [fill={rgb, 255:red, 0; green, 0; blue, 0 }  ,fill opacity=1 ] (328.22,110.56) -- (333.66,110.56) -- (333.66,116) -- (328.22,116) -- cycle ;
%Shape: Rectangle [id:dp8900152210843846] 
\draw  [color={rgb, 255:red, 0; green, 0; blue, 0 }  ,draw opacity=1 ] (439.77,110.56) -- (461.55,110.56) -- (461.55,116) -- (439.77,116) -- cycle ;
%Shape: Rectangle [id:dp6048879739770341] 
\draw   (445.22,110.56) -- (450.66,110.56) -- (450.66,116) -- (445.22,116) -- cycle ;
%Shape: Rectangle [id:dp40613933690081616] 
\draw   (456.1,110.56) -- (461.55,110.56) -- (461.55,116) -- (456.1,116) -- cycle ;
%Shape: Rectangle [id:dp3694720086065709] 
\draw  [color={rgb, 255:red, 0; green, 0; blue, 0 }  ,draw opacity=1 ] (418,110.56) -- (439.77,110.56) -- (439.77,116) -- (418,116) -- cycle ;
%Shape: Rectangle [id:dp2544005334397377] 
\draw   (423.44,110.56) -- (428.89,110.56) -- (428.89,116) -- (423.44,116) -- cycle ;
%Shape: Rectangle [id:dp49521746441525005] 
\draw   (434.33,110.56) -- (439.77,110.56) -- (439.77,116) -- (434.33,116) -- cycle ;
%Shape: Rectangle [id:dp882234857926115] 
\draw  [fill={rgb, 255:red, 0; green, 0; blue, 0 }  ,fill opacity=1 ] (434.33,110.56) -- (439.77,110.56) -- (439.77,116) -- (434.33,116) -- cycle ;
%Shape: Rectangle [id:dp1977449502547841] 
\draw  [fill={rgb, 255:red, 0; green, 0; blue, 0 }  ,fill opacity=1 ] (423.89,110.56) -- (429.33,110.56) -- (429.33,116) -- (423.89,116) -- cycle ;
%Shape: Rectangle [id:dp7957773701366023] 
\draw  [color={rgb, 255:red, 0; green, 0; blue, 0 }  ,draw opacity=1 ] (485.23,110.56) -- (507,110.56) -- (507,116) -- (485.23,116) -- cycle ;
%Shape: Rectangle [id:dp22389356835277985] 
\draw   (490.67,110.56) -- (496.11,110.56) -- (496.11,116) -- (490.67,116) -- cycle ;
%Shape: Rectangle [id:dp592588007009627] 
\draw   (501.56,110.56) -- (507,110.56) -- (507,116) -- (501.56,116) -- cycle ;
%Shape: Rectangle [id:dp7321831273379731] 
\draw  [color={rgb, 255:red, 0; green, 0; blue, 0 }  ,draw opacity=1 ] (463.45,110.56) -- (485.23,110.56) -- (485.23,116) -- (463.45,116) -- cycle ;
%Shape: Rectangle [id:dp6409032174076374] 
\draw   (468.9,110.56) -- (474.34,110.56) -- (474.34,116) -- (468.9,116) -- cycle ;
%Shape: Rectangle [id:dp6428171177972177] 
\draw   (479.78,110.56) -- (485.23,110.56) -- (485.23,116) -- (479.78,116) -- cycle ;
%Shape: Rectangle [id:dp34391960056492255] 
\draw  [fill={rgb, 255:red, 0; green, 0; blue, 0 }  ,fill opacity=1 ] (474.23,110.56) -- (479.67,110.56) -- (479.67,116) -- (474.23,116) -- cycle ;
%Shape: Rectangle [id:dp6480114712835973] 
\draw  [fill={rgb, 255:red, 0; green, 0; blue, 0 }  ,fill opacity=1 ] (501.56,110.56) -- (507,110.56) -- (507,116) -- (501.56,116) -- cycle ;
%Shape: Rectangle [id:dp5798495527142489] 
\draw  [fill={rgb, 255:red, 0; green, 0; blue, 0 }  ,fill opacity=1 ] (463.9,110.56) -- (469.34,110.56) -- (469.34,116) -- (463.9,116) -- cycle ;
%Shape: Rectangle [id:dp5989897309740557] 
\draw  [fill={rgb, 255:red, 0; green, 0; blue, 0 }  ,fill opacity=1 ] (451.22,110.56) -- (456.66,110.56) -- (456.66,116) -- (451.22,116) -- cycle ;
%Shape: Rectangle [id:dp9574636901151966] 
\draw  [color={rgb, 255:red, 0; green, 0; blue, 0 }  ,draw opacity=1 ] (392.77,143.56) -- (414.55,143.56) -- (414.55,149) -- (392.77,149) -- cycle ;
%Shape: Rectangle [id:dp4437195918570437] 
\draw   (398.22,143.56) -- (403.66,143.56) -- (403.66,149) -- (398.22,149) -- cycle ;
%Shape: Rectangle [id:dp8092864778088349] 
\draw   (409.1,143.56) -- (414.55,143.56) -- (414.55,149) -- (409.1,149) -- cycle ;
%Shape: Rectangle [id:dp6458826468181249] 
\draw  [color={rgb, 255:red, 0; green, 0; blue, 0 }  ,draw opacity=1 ] (371,143.56) -- (392.77,143.56) -- (392.77,149) -- (371,149) -- cycle ;
%Shape: Rectangle [id:dp7408635501037414] 
\draw   (376.44,143.56) -- (381.89,143.56) -- (381.89,149) -- (376.44,149) -- cycle ;
%Shape: Rectangle [id:dp25101181226921354] 
\draw   (387.33,143.56) -- (392.77,143.56) -- (392.77,149) -- (387.33,149) -- cycle ;
%Shape: Rectangle [id:dp5981989849043672] 
\draw  [fill={rgb, 255:red, 0; green, 0; blue, 0 }  ,fill opacity=1 ] (387.33,143.56) -- (392.77,143.56) -- (392.77,149) -- (387.33,149) -- cycle ;
%Shape: Rectangle [id:dp5500673298742877] 
\draw  [fill={rgb, 255:red, 0; green, 0; blue, 0 }  ,fill opacity=1 ] (370.89,143.56) -- (376.33,143.56) -- (376.33,149) -- (370.89,149) -- cycle ;
%Shape: Rectangle [id:dp2357994140066244] 
\draw  [color={rgb, 255:red, 0; green, 0; blue, 0 }  ,draw opacity=1 ] (438.23,143.56) -- (460,143.56) -- (460,149) -- (438.23,149) -- cycle ;
%Shape: Rectangle [id:dp1382302217730672] 
\draw   (443.67,143.56) -- (449.11,143.56) -- (449.11,149) -- (443.67,149) -- cycle ;
%Shape: Rectangle [id:dp8095057171250256] 
\draw   (454.56,143.56) -- (460,143.56) -- (460,149) -- (454.56,149) -- cycle ;
%Shape: Rectangle [id:dp8924074634471413] 
\draw  [color={rgb, 255:red, 0; green, 0; blue, 0 }  ,draw opacity=1 ] (416.45,143.56) -- (438.23,143.56) -- (438.23,149) -- (416.45,149) -- cycle ;
%Shape: Rectangle [id:dp8849073919046266] 
\draw   (421.9,143.56) -- (427.34,143.56) -- (427.34,149) -- (421.9,149) -- cycle ;
%Shape: Rectangle [id:dp5685875937148861] 
\draw   (432.78,143.56) -- (438.23,143.56) -- (438.23,149) -- (432.78,149) -- cycle ;
%Shape: Rectangle [id:dp4627308800662172] 
\draw  [fill={rgb, 255:red, 0; green, 0; blue, 0 }  ,fill opacity=1 ] (449.23,143.56) -- (454.67,143.56) -- (454.67,149) -- (449.23,149) -- cycle ;
%Shape: Rectangle [id:dp33324795874184165] 
\draw  [fill={rgb, 255:red, 0; green, 0; blue, 0 }  ,fill opacity=1 ] (454.56,143.56) -- (460,143.56) -- (460,149) -- (454.56,149) -- cycle ;
%Shape: Rectangle [id:dp09942177184906287] 
\draw  [fill={rgb, 255:red, 0; green, 0; blue, 0 }  ,fill opacity=1 ] (416.9,143.56) -- (422.34,143.56) -- (422.34,149) -- (416.9,149) -- cycle ;
%Shape: Rectangle [id:dp26475216678831326] 
\draw  [fill={rgb, 255:red, 0; green, 0; blue, 0 }  ,fill opacity=1 ] (404.22,143.56) -- (409.66,143.56) -- (409.66,149) -- (404.22,149) -- cycle ;
%Rounded Rect [id:dp38860855528006655] 
\draw  [dash pattern={on 4.5pt off 4.5pt}] (309.78,81.8) .. controls (309.78,71.42) and (318.2,63) .. (328.58,63) -- (497.2,63) .. controls (507.58,63) and (516,71.42) .. (516,81.8) -- (516,138.2) .. controls (516,148.58) and (507.58,157) .. (497.2,157) -- (328.58,157) .. controls (318.2,157) and (309.78,148.58) .. (309.78,138.2) -- cycle ;
%Rounded Rect [id:dp2651870404432606] 
\draw  [color={rgb, 255:red, 125; green, 184; blue, 253 }  ,draw opacity=1 ][line width=1.5]  (96,96.25) .. controls (96,93.1) and (98.55,90.56) .. (101.69,90.56) -- (183.65,90.56) .. controls (186.79,90.56) and (189.34,93.1) .. (189.34,96.25) -- (189.34,113.31) .. controls (189.34,116.45) and (186.79,119) .. (183.65,119) -- (101.69,119) .. controls (98.55,119) and (96,116.45) .. (96,113.31) -- cycle ;
%Rounded Rect [id:dp5766246136104448] 
\draw  [color={rgb, 255:red, 125; green, 184; blue, 253 }  ,draw opacity=1 ][line width=1.5]  (159,169.8) .. controls (159,166.6) and (161.6,164) .. (164.8,164) -- (200.2,164) .. controls (203.4,164) and (206,166.6) .. (206,169.8) -- (206,187.2) .. controls (206,190.4) and (203.4,193) .. (200.2,193) -- (164.8,193) .. controls (161.6,193) and (159,190.4) .. (159,187.2) -- cycle ;
%Rounded Rect [id:dp8340954198059106] 
\draw  [color={rgb, 255:red, 125; green, 184; blue, 253 }  ,draw opacity=1 ][dash pattern={on 5.63pt off 4.5pt}][line width=1.5]  (238,169.8) .. controls (238,166.6) and (240.6,164) .. (243.8,164) -- (279.2,164) .. controls (282.4,164) and (285,166.6) .. (285,169.8) -- (285,187.2) .. controls (285,190.4) and (282.4,193) .. (279.2,193) -- (243.8,193) .. controls (240.6,193) and (238,190.4) .. (238,187.2) -- cycle ;
%Rounded Rect [id:dp9867617922591404] 
\draw  [color={rgb, 255:red, 125; green, 184; blue, 253 }  ,draw opacity=1 ][dash pattern={on 5.63pt off 4.5pt}][line width=1.5]  (316,96.25) .. controls (316,93.1) and (318.55,90.56) .. (321.69,90.56) -- (403.65,90.56) .. controls (406.79,90.56) and (409.34,93.1) .. (409.34,96.25) -- (409.34,113.31) .. controls (409.34,116.45) and (406.79,119) .. (403.65,119) -- (321.69,119) .. controls (318.55,119) and (316,116.45) .. (316,113.31) -- cycle ;
%Shape: Rectangle [id:dp8450992048901718] 
\draw  [fill={rgb, 255:red, 0; green, 0; blue, 0 }  ,fill opacity=1 ] (236.22,110.56) -- (241.66,110.56) -- (241.66,116) -- (236.22,116) -- cycle ;
%Shape: Rectangle [id:dp053136509283846056] 
\draw  [fill={rgb, 255:red, 0; green, 0; blue, 0 }  ,fill opacity=1 ] (125.22,143.56) -- (130.66,143.56) -- (130.66,149) -- (125.22,149) -- cycle ;
%Shape: Rectangle [id:dp10476102096166562] 
\draw  [fill={rgb, 255:red, 0; green, 0; blue, 0 }  ,fill opacity=1 ] (350.22,110.56) -- (355.66,110.56) -- (355.66,116) -- (350.22,116) -- cycle ;

% Text Node
\draw (134,93.4) node [anchor=north west][inner sep=0.75pt]  [font=\scriptsize]  {$\mathcal{B}_{s}$};
% Text Node
\draw (167,66.4) node [anchor=north west][inner sep=0.75pt]    {$\mathcal{B}^{S}( f_{1})$};
% Text Node
\draw (206,92.4) node [anchor=north west][inner sep=0.75pt]  [font=\scriptsize]  {$\Phi \left(\text{dbo:author}\right)$};
% Text Node
\draw (199,126.4) node [anchor=north west][inner sep=0.75pt]  [font=\scriptsize]  {$\Phi \left(\text{dbo:deathDate}\right)$};
% Text Node
\draw (97,124.4) node [anchor=north west][inner sep=0.75pt]  [font=\scriptsize]  {$\Phi \left(\text{dbo:nationality}\right)$};
% Text Node
\draw (354,93.4) node [anchor=north west][inner sep=0.75pt]  [font=\scriptsize]  {$\mathcal{B}_{s}$};
% Text Node
\draw (387,66.4) node [anchor=north west][inner sep=0.75pt]    {$\mathcal{B}^{S}( f_{2})$};
% Text Node
\draw (426,92.4) node [anchor=north west][inner sep=0.75pt]  [font=\scriptsize]  {$\Phi \left(\text{dbo:author}\right)$};
% Text Node
\draw (370,124.4) node [anchor=north west][inner sep=0.75pt]  [font=\scriptsize]  {$\Phi \left(\text{dbo:nationality}\right)$};
% Text Node
\draw (165,171.4) node [anchor=north west][inner sep=0.75pt]    {$1000$};
% Text Node
\draw (215,171.4) node [anchor=north west][inner sep=0.75pt]    {$+$};
% Text Node
\draw (291,171.4) node [anchor=north west][inner sep=0.75pt]    {$=3000$};
% Text Node
\draw (244,171.4) node [anchor=north west][inner sep=0.75pt]    {$2000$};

\end{tikzpicture}

%% file: figures/cardex2.tex
\tikzset{every picture/.style={line width=0.75pt}} %set default line width to 0.75pt        

\begin{tikzpicture}[x=0.6pt,y=0.6pt,yscale=-1,xscale=1]
%uncomment if require: \path (0,762); %set diagram left start at 0, and has height of 762

%Rounded Rect [id:dp3847368628501874] 
\draw  [dash pattern={on 4.5pt off 4.5pt}] (69.78,111.8) .. controls (69.78,101.42) and (78.2,93) .. (88.58,93) -- (257.2,93) .. controls (267.58,93) and (276,101.42) .. (276,111.8) -- (276,168.2) .. controls (276,178.58) and (267.58,187) .. (257.2,187) -- (88.58,187) .. controls (78.2,187) and (69.78,178.58) .. (69.78,168.2) -- cycle ;
%Rounded Rect [id:dp22717240962608265] 
\draw  [dash pattern={on 4.5pt off 4.5pt}] (289.78,111.8) .. controls (289.78,101.42) and (298.2,93) .. (308.58,93) -- (477.2,93) .. controls (487.58,93) and (496,101.42) .. (496,111.8) -- (496,168.2) .. controls (496,178.58) and (487.58,187) .. (477.2,187) -- (308.58,187) .. controls (298.2,187) and (289.78,178.58) .. (289.78,168.2) -- cycle ;
%Rounded Rect [id:dp7614243105545082] 
\draw  [color={rgb, 255:red, 125; green, 184; blue, 253 }  ,draw opacity=1 ][line width=1.5]  (76,126.25) .. controls (76,123.1) and (78.55,120.56) .. (81.69,120.56) -- (163.65,120.56) .. controls (166.79,120.56) and (169.34,123.1) .. (169.34,126.25) -- (169.34,143.31) .. controls (169.34,146.45) and (166.79,149) .. (163.65,149) -- (81.69,149) .. controls (78.55,149) and (76,146.45) .. (76,143.31) -- cycle ;
%Rounded Rect [id:dp5247188582853796] 
\draw  [color={rgb, 255:red, 125; green, 184; blue, 253 }  ,draw opacity=1 ][line width=1.5]  (78,220.8) .. controls (78,217.6) and (80.6,215) .. (83.8,215) -- (119.2,215) .. controls (122.4,215) and (125,217.6) .. (125,220.8) -- (125,238.2) .. controls (125,241.4) and (122.4,244) .. (119.2,244) -- (83.8,244) .. controls (80.6,244) and (78,241.4) .. (78,238.2) -- cycle ;
%Rounded Rect [id:dp7389819660506797] 
\draw  [color={rgb, 255:red, 248; green, 231; blue, 28 }  ,draw opacity=1 ][line width=1.5]  (176,126.25) .. controls (176,123.1) and (178.55,120.56) .. (181.69,120.56) -- (263.65,120.56) .. controls (266.79,120.56) and (269.34,123.1) .. (269.34,126.25) -- (269.34,143.31) .. controls (269.34,146.45) and (266.79,149) .. (263.65,149) -- (181.69,149) .. controls (178.55,149) and (176,146.45) .. (176,143.31) -- cycle ;
%Rounded Rect [id:dp13647792909611178] 
\draw  [color={rgb, 255:red, 184; green, 233; blue, 134 }  ,draw opacity=1 ][line width=1.5]  (76,158.25) .. controls (76,155.1) and (78.55,152.56) .. (81.69,152.56) -- (163.65,152.56) .. controls (166.79,152.56) and (169.34,155.1) .. (169.34,158.25) -- (169.34,175.31) .. controls (169.34,178.45) and (166.79,181) .. (163.65,181) -- (81.69,181) .. controls (78.55,181) and (76,178.45) .. (76,175.31) -- cycle ;
%Rounded Rect [id:dp5727189551915882] 
\draw  [color={rgb, 255:red, 248; green, 231; blue, 28 }  ,draw opacity=1 ][line width=1.5]  (148,201.8) .. controls (148,198.6) and (150.6,196) .. (153.8,196) -- (189.2,196) .. controls (192.4,196) and (195,198.6) .. (195,201.8) -- (195,219.2) .. controls (195,222.4) and (192.4,225) .. (189.2,225) -- (153.8,225) .. controls (150.6,225) and (148,222.4) .. (148,219.2) -- cycle ;
%Rounded Rect [id:dp2216752360382307] 
\draw  [color={rgb, 255:red, 125; green, 184; blue, 253 }  ,draw opacity=1 ][line width=1.5]  (148,241.8) .. controls (148,238.6) and (150.6,236) .. (153.8,236) -- (189.2,236) .. controls (192.4,236) and (195,238.6) .. (195,241.8) -- (195,259.2) .. controls (195,262.4) and (192.4,265) .. (189.2,265) -- (153.8,265) .. controls (150.6,265) and (148,262.4) .. (148,259.2) -- cycle ;
%Straight Lines [id:da013395931383922921] 
\draw [line width=1.5]    (147,231) -- (194,231) ;
%Rounded Rect [id:dp959385434532154] 
\draw  [color={rgb, 255:red, 184; green, 233; blue, 134 }  ,draw opacity=1 ][line width=1.5]  (218,201.8) .. controls (218,198.6) and (220.6,196) .. (223.8,196) -- (259.2,196) .. controls (262.4,196) and (265,198.6) .. (265,201.8) -- (265,219.2) .. controls (265,222.4) and (262.4,225) .. (259.2,225) -- (223.8,225) .. controls (220.6,225) and (218,222.4) .. (218,219.2) -- cycle ;
%Rounded Rect [id:dp13723202713641025] 
\draw  [color={rgb, 255:red, 125; green, 184; blue, 253 }  ,draw opacity=1 ][line width=1.5]  (218,241.8) .. controls (218,238.6) and (220.6,236) .. (223.8,236) -- (259.2,236) .. controls (262.4,236) and (265,238.6) .. (265,241.8) -- (265,259.2) .. controls (265,262.4) and (262.4,265) .. (259.2,265) -- (223.8,265) .. controls (220.6,265) and (218,262.4) .. (218,259.2) -- cycle ;
%Straight Lines [id:da5126441398963224] 
\draw [line width=1.5]    (217,231) -- (264,231) ;
%Rounded Rect [id:dp38794274986532673] 
\draw  [color={rgb, 255:red, 125; green, 184; blue, 253 }  ,draw opacity=1 ][dash pattern={on 5.63pt off 4.5pt}][line width=1.5]  (298,220.8) .. controls (298,217.6) and (300.6,215) .. (303.8,215) -- (339.2,215) .. controls (342.4,215) and (345,217.6) .. (345,220.8) -- (345,238.2) .. controls (345,241.4) and (342.4,244) .. (339.2,244) -- (303.8,244) .. controls (300.6,244) and (298,241.4) .. (298,238.2) -- cycle ;
%Rounded Rect [id:dp18058571659917544] 
\draw  [color={rgb, 255:red, 248; green, 231; blue, 28 }  ,draw opacity=1 ][dash pattern={on 5.63pt off 4.5pt}][line width=1.5]  (368,201.8) .. controls (368,198.6) and (370.6,196) .. (373.8,196) -- (409.2,196) .. controls (412.4,196) and (415,198.6) .. (415,201.8) -- (415,219.2) .. controls (415,222.4) and (412.4,225) .. (409.2,225) -- (373.8,225) .. controls (370.6,225) and (368,222.4) .. (368,219.2) -- cycle ;
%Rounded Rect [id:dp3949556520121368] 
\draw  [color={rgb, 255:red, 125; green, 184; blue, 253 }  ,draw opacity=1 ][dash pattern={on 5.63pt off 4.5pt}][line width=1.5]  (368,241.8) .. controls (368,238.6) and (370.6,236) .. (373.8,236) -- (409.2,236) .. controls (412.4,236) and (415,238.6) .. (415,241.8) -- (415,259.2) .. controls (415,262.4) and (412.4,265) .. (409.2,265) -- (373.8,265) .. controls (370.6,265) and (368,262.4) .. (368,259.2) -- cycle ;
%Straight Lines [id:da48942965556453366] 
\draw [line width=1.5]    (367,231) -- (414,231) ;
%Rounded Rect [id:dp9512643939745119] 
\draw  [color={rgb, 255:red, 184; green, 233; blue, 134 }  ,draw opacity=1 ][dash pattern={on 5.63pt off 4.5pt}][line width=1.5]  (438,201.8) .. controls (438,198.6) and (440.6,196) .. (443.8,196) -- (479.2,196) .. controls (482.4,196) and (485,198.6) .. (485,201.8) -- (485,219.2) .. controls (485,222.4) and (482.4,225) .. (479.2,225) -- (443.8,225) .. controls (440.6,225) and (438,222.4) .. (438,219.2) -- cycle ;
%Rounded Rect [id:dp9059760482427242] 
\draw  [color={rgb, 255:red, 125; green, 184; blue, 253 }  ,draw opacity=1 ][dash pattern={on 5.63pt off 4.5pt}][line width=1.5]  (438,241.8) .. controls (438,238.6) and (440.6,236) .. (443.8,236) -- (479.2,236) .. controls (482.4,236) and (485,238.6) .. (485,241.8) -- (485,259.2) .. controls (485,262.4) and (482.4,265) .. (479.2,265) -- (443.8,265) .. controls (440.6,265) and (438,262.4) .. (438,259.2) -- cycle ;
%Straight Lines [id:da7883848433869057] 
\draw [line width=1.5]    (437,231) -- (484,231) ;
%Rounded Rect [id:dp20659652149893049] 
\draw  [color={rgb, 255:red, 125; green, 184; blue, 253 }  ,draw opacity=1 ][dash pattern={on 5.63pt off 4.5pt}][line width=1.5]  (296,126.25) .. controls (296,123.1) and (298.55,120.56) .. (301.69,120.56) -- (383.65,120.56) .. controls (386.79,120.56) and (389.34,123.1) .. (389.34,126.25) -- (389.34,143.31) .. controls (389.34,146.45) and (386.79,149) .. (383.65,149) -- (301.69,149) .. controls (298.55,149) and (296,146.45) .. (296,143.31) -- cycle ;
%Rounded Rect [id:dp7779764299347182] 
\draw  [color={rgb, 255:red, 248; green, 231; blue, 28 }  ,draw opacity=1 ][dash pattern={on 5.63pt off 4.5pt}][line width=1.5]  (396,126.25) .. controls (396,123.1) and (398.55,120.56) .. (401.69,120.56) -- (483.65,120.56) .. controls (486.79,120.56) and (489.34,123.1) .. (489.34,126.25) -- (489.34,143.31) .. controls (489.34,146.45) and (486.79,149) .. (483.65,149) -- (401.69,149) .. controls (398.55,149) and (396,146.45) .. (396,143.31) -- cycle ;
%Rounded Rect [id:dp9309903698102842] 
\draw  [color={rgb, 255:red, 184; green, 233; blue, 134 }  ,draw opacity=1 ][dash pattern={on 5.63pt off 4.5pt}][line width=1.5]  (349,158.25) .. controls (349,155.1) and (351.55,152.56) .. (354.69,152.56) -- (436.65,152.56) .. controls (439.79,152.56) and (442.34,155.1) .. (442.34,158.25) -- (442.34,175.31) .. controls (442.34,178.45) and (439.79,181) .. (436.65,181) -- (354.69,181) .. controls (351.55,181) and (349,178.45) .. (349,175.31) -- cycle ;
%Shape: Rectangle [id:dp49121860072615775] 
\draw  [color={rgb, 255:red, 0; green, 0; blue, 0 }  ,draw opacity=1 ] (99.77,140.56) -- (121.55,140.56) -- (121.55,146) -- (99.77,146) -- cycle ;
%Shape: Rectangle [id:dp18404298184682077] 
\draw   (105.22,140.56) -- (110.66,140.56) -- (110.66,146) -- (105.22,146) -- cycle ;
%Shape: Rectangle [id:dp026232920674231552] 
\draw   (116.1,140.56) -- (121.55,140.56) -- (121.55,146) -- (116.1,146) -- cycle ;
%Shape: Rectangle [id:dp4042278155602148] 
\draw  [color={rgb, 255:red, 0; green, 0; blue, 0 }  ,draw opacity=1 ] (78,140.56) -- (99.77,140.56) -- (99.77,146) -- (78,146) -- cycle ;
%Shape: Rectangle [id:dp7297549400300131] 
\draw   (83.44,140.56) -- (88.89,140.56) -- (88.89,146) -- (83.44,146) -- cycle ;
%Shape: Rectangle [id:dp01017805021429874] 
\draw   (94.33,140.56) -- (99.77,140.56) -- (99.77,146) -- (94.33,146) -- cycle ;
%Shape: Rectangle [id:dp1533226024708001] 
\draw  [fill={rgb, 255:red, 0; green, 0; blue, 0 }  ,fill opacity=1 ] (94.33,140.56) -- (99.77,140.56) -- (99.77,146) -- (94.33,146) -- cycle ;
%Shape: Rectangle [id:dp5857867244166927] 
\draw  [fill={rgb, 255:red, 0; green, 0; blue, 0 }  ,fill opacity=1 ] (88.89,140.56) -- (94.33,140.56) -- (94.33,146) -- (88.89,146) -- cycle ;
%Shape: Rectangle [id:dp6721139472777592] 
\draw  [color={rgb, 255:red, 0; green, 0; blue, 0 }  ,draw opacity=1 ] (145.23,140.56) -- (167,140.56) -- (167,146) -- (145.23,146) -- cycle ;
%Shape: Rectangle [id:dp2887879026295417] 
\draw   (150.67,140.56) -- (156.11,140.56) -- (156.11,146) -- (150.67,146) -- cycle ;
%Shape: Rectangle [id:dp8066933810433047] 
\draw   (161.56,140.56) -- (167,140.56) -- (167,146) -- (161.56,146) -- cycle ;
%Shape: Rectangle [id:dp031901554956735745] 
\draw  [color={rgb, 255:red, 0; green, 0; blue, 0 }  ,draw opacity=1 ] (123.45,140.56) -- (145.23,140.56) -- (145.23,146) -- (123.45,146) -- cycle ;
%Shape: Rectangle [id:dp6234972419470488] 
\draw   (128.9,140.56) -- (134.34,140.56) -- (134.34,146) -- (128.9,146) -- cycle ;
%Shape: Rectangle [id:dp7203345457667116] 
\draw   (139.78,140.56) -- (145.23,140.56) -- (145.23,146) -- (139.78,146) -- cycle ;
%Shape: Rectangle [id:dp14599469561197476] 
\draw  [fill={rgb, 255:red, 0; green, 0; blue, 0 }  ,fill opacity=1 ] (145.23,140.56) -- (150.67,140.56) -- (150.67,146) -- (145.23,146) -- cycle ;
%Shape: Rectangle [id:dp8093796524683704] 
\draw  [fill={rgb, 255:red, 0; green, 0; blue, 0 }  ,fill opacity=1 ] (161.56,140.56) -- (167,140.56) -- (167,146) -- (161.56,146) -- cycle ;
%Shape: Rectangle [id:dp895646171747254] 
\draw  [fill={rgb, 255:red, 0; green, 0; blue, 0 }  ,fill opacity=1 ] (128.9,140.56) -- (134.34,140.56) -- (134.34,146) -- (128.9,146) -- cycle ;
%Shape: Rectangle [id:dp7032796937655139] 
\draw  [fill={rgb, 255:red, 0; green, 0; blue, 0 }  ,fill opacity=1 ] (105.22,140.56) -- (110.66,140.56) -- (110.66,146) -- (105.22,146) -- cycle ;
%Shape: Rectangle [id:dp4559419351064571] 
\draw  [color={rgb, 255:red, 0; green, 0; blue, 0 }  ,draw opacity=1 ] (199.77,140.56) -- (221.55,140.56) -- (221.55,146) -- (199.77,146) -- cycle ;
%Shape: Rectangle [id:dp35072607017160407] 
\draw   (205.22,140.56) -- (210.66,140.56) -- (210.66,146) -- (205.22,146) -- cycle ;
%Shape: Rectangle [id:dp056740080220742284] 
\draw   (216.1,140.56) -- (221.55,140.56) -- (221.55,146) -- (216.1,146) -- cycle ;
%Shape: Rectangle [id:dp22911780935323556] 
\draw  [color={rgb, 255:red, 0; green, 0; blue, 0 }  ,draw opacity=1 ] (178,140.56) -- (199.77,140.56) -- (199.77,146) -- (178,146) -- cycle ;
%Shape: Rectangle [id:dp6874493184912627] 
\draw   (183.44,140.56) -- (188.89,140.56) -- (188.89,146) -- (183.44,146) -- cycle ;
%Shape: Rectangle [id:dp9712222264144089] 
\draw   (194.33,140.56) -- (199.77,140.56) -- (199.77,146) -- (194.33,146) -- cycle ;
%Shape: Rectangle [id:dp4275099552569622] 
\draw  [fill={rgb, 255:red, 0; green, 0; blue, 0 }  ,fill opacity=1 ] (188.89,140.56) -- (193.75,140.56) -- (193.75,146) -- (188.89,146) -- cycle ;
%Shape: Rectangle [id:dp004622251036311753] 
\draw  [fill={rgb, 255:red, 0; green, 0; blue, 0 }  ,fill opacity=1 ] (177.89,140.56) -- (183.33,140.56) -- (183.33,146) -- (177.89,146) -- cycle ;
%Shape: Rectangle [id:dp14983610905195144] 
\draw  [color={rgb, 255:red, 0; green, 0; blue, 0 }  ,draw opacity=1 ] (245.23,140.56) -- (267,140.56) -- (267,146) -- (245.23,146) -- cycle ;
%Shape: Rectangle [id:dp8904001750472106] 
\draw   (250.67,140.56) -- (256.11,140.56) -- (256.11,146) -- (250.67,146) -- cycle ;
%Shape: Rectangle [id:dp12429242868714252] 
\draw   (261.56,140.56) -- (267,140.56) -- (267,146) -- (261.56,146) -- cycle ;
%Shape: Rectangle [id:dp45666142947665855] 
\draw  [color={rgb, 255:red, 0; green, 0; blue, 0 }  ,draw opacity=1 ] (223.45,140.56) -- (245.23,140.56) -- (245.23,146) -- (223.45,146) -- cycle ;
%Shape: Rectangle [id:dp854679153146026] 
\draw   (228.9,140.56) -- (234.34,140.56) -- (234.34,146) -- (228.9,146) -- cycle ;
%Shape: Rectangle [id:dp11619385093009704] 
\draw   (239.78,140.56) -- (245.23,140.56) -- (245.23,146) -- (239.78,146) -- cycle ;
%Shape: Rectangle [id:dp641209160803135] 
\draw  [fill={rgb, 255:red, 0; green, 0; blue, 0 }  ,fill opacity=1 ] (245.23,140.56) -- (250.67,140.56) -- (250.67,146) -- (245.23,146) -- cycle ;
%Shape: Rectangle [id:dp9453689371750449] 
\draw  [fill={rgb, 255:red, 0; green, 0; blue, 0 }  ,fill opacity=1 ] (250.56,140.56) -- (256,140.56) -- (256,146) -- (250.56,146) -- cycle ;
%Shape: Rectangle [id:dp005060003571495275] 
\draw  [fill={rgb, 255:red, 0; green, 0; blue, 0 }  ,fill opacity=1 ] (239.9,140.56) -- (245.34,140.56) -- (245.34,146) -- (239.9,146) -- cycle ;
%Shape: Rectangle [id:dp6389295597292995] 
\draw  [fill={rgb, 255:red, 0; green, 0; blue, 0 }  ,fill opacity=1 ] (205.22,140.56) -- (210.66,140.56) -- (210.66,146) -- (205.22,146) -- cycle ;
%Shape: Rectangle [id:dp027094630097064254] 
\draw  [color={rgb, 255:red, 0; green, 0; blue, 0 }  ,draw opacity=1 ] (99.77,173.56) -- (121.55,173.56) -- (121.55,179) -- (99.77,179) -- cycle ;
%Shape: Rectangle [id:dp7670146900869991] 
\draw   (105.22,173.56) -- (110.66,173.56) -- (110.66,179) -- (105.22,179) -- cycle ;
%Shape: Rectangle [id:dp9743642954976218] 
\draw   (116.1,173.56) -- (121.55,173.56) -- (121.55,179) -- (116.1,179) -- cycle ;
%Shape: Rectangle [id:dp7681470075223196] 
\draw  [color={rgb, 255:red, 0; green, 0; blue, 0 }  ,draw opacity=1 ] (78,173.56) -- (99.77,173.56) -- (99.77,179) -- (78,179) -- cycle ;
%Shape: Rectangle [id:dp16514919565305186] 
\draw   (83.44,173.56) -- (88.89,173.56) -- (88.89,179) -- (83.44,179) -- cycle ;
%Shape: Rectangle [id:dp10483416132407797] 
\draw   (94.33,173.56) -- (99.77,173.56) -- (99.77,179) -- (94.33,179) -- cycle ;
%Shape: Rectangle [id:dp31739211291159586] 
\draw  [fill={rgb, 255:red, 0; green, 0; blue, 0 }  ,fill opacity=1 ] (89.33,173.56) -- (94.77,173.56) -- (94.77,179) -- (89.33,179) -- cycle ;
%Shape: Rectangle [id:dp8901446807053793] 
\draw  [fill={rgb, 255:red, 0; green, 0; blue, 0 }  ,fill opacity=1 ] (83.89,173.56) -- (89.33,173.56) -- (89.33,179) -- (83.89,179) -- cycle ;
%Shape: Rectangle [id:dp5305461061298274] 
\draw  [color={rgb, 255:red, 0; green, 0; blue, 0 }  ,draw opacity=1 ] (145.23,173.56) -- (167,173.56) -- (167,179) -- (145.23,179) -- cycle ;
%Shape: Rectangle [id:dp8499679021492669] 
\draw   (150.67,173.56) -- (156.11,173.56) -- (156.11,179) -- (150.67,179) -- cycle ;
%Shape: Rectangle [id:dp12602747551802762] 
\draw   (161.56,173.56) -- (167,173.56) -- (167,179) -- (161.56,179) -- cycle ;
%Shape: Rectangle [id:dp732657952042898] 
\draw  [color={rgb, 255:red, 0; green, 0; blue, 0 }  ,draw opacity=1 ] (123.45,173.56) -- (145.23,173.56) -- (145.23,179) -- (123.45,179) -- cycle ;
%Shape: Rectangle [id:dp5159335449594321] 
\draw   (128.9,173.56) -- (134.34,173.56) -- (134.34,179) -- (128.9,179) -- cycle ;
%Shape: Rectangle [id:dp39537693423606446] 
\draw   (139.78,173.56) -- (145.23,173.56) -- (145.23,179) -- (139.78,179) -- cycle ;
%Shape: Rectangle [id:dp640101494634245] 
\draw  [fill={rgb, 255:red, 0; green, 0; blue, 0 }  ,fill opacity=1 ] (140.23,173.56) -- (145.67,173.56) -- (145.67,179) -- (140.23,179) -- cycle ;
%Shape: Rectangle [id:dp3583451050632922] 
\draw  [fill={rgb, 255:red, 0; green, 0; blue, 0 }  ,fill opacity=1 ] (150.56,173.56) -- (156,173.56) -- (156,179) -- (150.56,179) -- cycle ;
%Shape: Rectangle [id:dp1733565894485669] 
\draw  [fill={rgb, 255:red, 0; green, 0; blue, 0 }  ,fill opacity=1 ] (128.9,173.56) -- (134.34,173.56) -- (134.34,179) -- (128.9,179) -- cycle ;
%Shape: Rectangle [id:dp17585143307065643] 
\draw  [fill={rgb, 255:red, 0; green, 0; blue, 0 }  ,fill opacity=1 ] (100.22,173.56) -- (105.66,173.56) -- (105.66,179) -- (100.22,179) -- cycle ;
%Shape: Rectangle [id:dp41427800232588663] 
\draw  [color={rgb, 255:red, 0; green, 0; blue, 0 }  ,draw opacity=1 ] (199.77,173.56) -- (221.55,173.56) -- (221.55,179) -- (199.77,179) -- cycle ;
%Shape: Rectangle [id:dp43594540206891874] 
\draw   (205.22,173.56) -- (210.66,173.56) -- (210.66,179) -- (205.22,179) -- cycle ;
%Shape: Rectangle [id:dp2652837877676325] 
\draw   (216.1,173.56) -- (221.55,173.56) -- (221.55,179) -- (216.1,179) -- cycle ;
%Shape: Rectangle [id:dp17430784539695088] 
\draw  [color={rgb, 255:red, 0; green, 0; blue, 0 }  ,draw opacity=1 ] (178,173.56) -- (199.77,173.56) -- (199.77,179) -- (178,179) -- cycle ;
%Shape: Rectangle [id:dp1727866802177418] 
\draw   (183.44,173.56) -- (188.89,173.56) -- (188.89,179) -- (183.44,179) -- cycle ;
%Shape: Rectangle [id:dp19060482632262532] 
\draw   (194.33,173.56) -- (199.77,173.56) -- (199.77,179) -- (194.33,179) -- cycle ;
%Shape: Rectangle [id:dp6651131956694276] 
\draw  [fill={rgb, 255:red, 0; green, 0; blue, 0 }  ,fill opacity=1 ] (194.33,173.56) -- (199.77,173.56) -- (199.77,179) -- (194.33,179) -- cycle ;
%Shape: Rectangle [id:dp6122578481019029] 
\draw  [fill={rgb, 255:red, 0; green, 0; blue, 0 }  ,fill opacity=1 ] (188.89,173.56) -- (194.33,173.56) -- (194.33,179) -- (188.89,179) -- cycle ;
%Shape: Rectangle [id:dp32153243875885074] 
\draw  [color={rgb, 255:red, 0; green, 0; blue, 0 }  ,draw opacity=1 ] (245.23,173.56) -- (267,173.56) -- (267,179) -- (245.23,179) -- cycle ;
%Shape: Rectangle [id:dp8040132327590503] 
\draw   (250.67,173.56) -- (256.11,173.56) -- (256.11,179) -- (250.67,179) -- cycle ;
%Shape: Rectangle [id:dp13819823962736855] 
\draw   (261.56,173.56) -- (267,173.56) -- (267,179) -- (261.56,179) -- cycle ;
%Shape: Rectangle [id:dp7221406838015886] 
\draw  [color={rgb, 255:red, 0; green, 0; blue, 0 }  ,draw opacity=1 ] (223.45,173.56) -- (245.23,173.56) -- (245.23,179) -- (223.45,179) -- cycle ;
%Shape: Rectangle [id:dp9264199293868157] 
\draw   (228.9,173.56) -- (234.34,173.56) -- (234.34,179) -- (228.9,179) -- cycle ;
%Shape: Rectangle [id:dp3866711051993982] 
\draw   (239.78,173.56) -- (245.23,173.56) -- (245.23,179) -- (239.78,179) -- cycle ;
%Shape: Rectangle [id:dp09109876105025316] 
\draw  [fill={rgb, 255:red, 0; green, 0; blue, 0 }  ,fill opacity=1 ] (228.23,173.56) -- (233.67,173.56) -- (233.67,179) -- (228.23,179) -- cycle ;
%Shape: Rectangle [id:dp4720077648795926] 
\draw  [fill={rgb, 255:red, 0; green, 0; blue, 0 }  ,fill opacity=1 ] (261.56,173.56) -- (267,173.56) -- (267,179) -- (261.56,179) -- cycle ;
%Shape: Rectangle [id:dp5848425791016713] 
\draw  [fill={rgb, 255:red, 0; green, 0; blue, 0 }  ,fill opacity=1 ] (223.9,173.56) -- (229.34,173.56) -- (229.34,179) -- (223.9,179) -- cycle ;
%Shape: Rectangle [id:dp25781724688032714] 
\draw  [fill={rgb, 255:red, 0; green, 0; blue, 0 }  ,fill opacity=1 ] (216.22,173.56) -- (221.66,173.56) -- (221.66,179) -- (216.22,179) -- cycle ;
%Shape: Rectangle [id:dp4995181376517819] 
\draw  [fill={rgb, 255:red, 0; green, 0; blue, 0 }  ,fill opacity=1 ] (216.22,140.56) -- (221.66,140.56) -- (221.66,146) -- (216.22,146) -- cycle ;
%Shape: Rectangle [id:dp4286617629109051] 
\draw  [fill={rgb, 255:red, 0; green, 0; blue, 0 }  ,fill opacity=1 ] (105.22,173.56) -- (110.66,173.56) -- (110.66,179) -- (105.22,179) -- cycle ;
%Shape: Rectangle [id:dp9306464354230708] 
\draw  [color={rgb, 255:red, 0; green, 0; blue, 0 }  ,draw opacity=1 ] (319.77,140.56) -- (341.55,140.56) -- (341.55,146) -- (319.77,146) -- cycle ;
%Shape: Rectangle [id:dp29652437844937063] 
\draw   (325.22,140.56) -- (330.66,140.56) -- (330.66,146) -- (325.22,146) -- cycle ;
%Shape: Rectangle [id:dp5051862101948283] 
\draw   (336.1,140.56) -- (341.55,140.56) -- (341.55,146) -- (336.1,146) -- cycle ;
%Shape: Rectangle [id:dp4312961913397021] 
\draw  [color={rgb, 255:red, 0; green, 0; blue, 0 }  ,draw opacity=1 ] (298,140.56) -- (319.77,140.56) -- (319.77,146) -- (298,146) -- cycle ;
%Shape: Rectangle [id:dp019072242848628296] 
\draw   (303.44,140.56) -- (308.89,140.56) -- (308.89,146) -- (303.44,146) -- cycle ;
%Shape: Rectangle [id:dp9088875546103644] 
\draw   (314.33,140.56) -- (319.77,140.56) -- (319.77,146) -- (314.33,146) -- cycle ;
%Shape: Rectangle [id:dp4509146933483469] 
\draw  [fill={rgb, 255:red, 0; green, 0; blue, 0 }  ,fill opacity=1 ] (303.33,140.56) -- (308.77,140.56) -- (308.77,146) -- (303.33,146) -- cycle ;
%Shape: Rectangle [id:dp06659409548372286] 
\draw  [fill={rgb, 255:red, 0; green, 0; blue, 0 }  ,fill opacity=1 ] (297.89,140.56) -- (303.33,140.56) -- (303.33,146) -- (297.89,146) -- cycle ;
%Shape: Rectangle [id:dp2227934368178257] 
\draw  [color={rgb, 255:red, 0; green, 0; blue, 0 }  ,draw opacity=1 ] (365.23,140.56) -- (387,140.56) -- (387,146) -- (365.23,146) -- cycle ;
%Shape: Rectangle [id:dp8531328449737762] 
\draw   (370.67,140.56) -- (376.11,140.56) -- (376.11,146) -- (370.67,146) -- cycle ;
%Shape: Rectangle [id:dp5687142654583596] 
\draw   (381.56,140.56) -- (387,140.56) -- (387,146) -- (381.56,146) -- cycle ;
%Shape: Rectangle [id:dp6987612671621292] 
\draw  [color={rgb, 255:red, 0; green, 0; blue, 0 }  ,draw opacity=1 ] (343.45,140.56) -- (365.23,140.56) -- (365.23,146) -- (343.45,146) -- cycle ;
%Shape: Rectangle [id:dp10284306888798844] 
\draw   (348.9,140.56) -- (354.34,140.56) -- (354.34,146) -- (348.9,146) -- cycle ;
%Shape: Rectangle [id:dp8747461842929224] 
\draw   (359.78,140.56) -- (365.23,140.56) -- (365.23,146) -- (359.78,146) -- cycle ;
%Shape: Rectangle [id:dp5882489576603956] 
\draw  [fill={rgb, 255:red, 0; green, 0; blue, 0 }  ,fill opacity=1 ] (359.23,140.56) -- (364.67,140.56) -- (364.67,146) -- (359.23,146) -- cycle ;
%Shape: Rectangle [id:dp8153030099744215] 
\draw  [fill={rgb, 255:red, 0; green, 0; blue, 0 }  ,fill opacity=1 ] (353.9,140.56) -- (359.34,140.56) -- (359.34,146) -- (353.9,146) -- cycle ;
%Shape: Rectangle [id:dp9341116869975464] 
\draw  [fill={rgb, 255:red, 0; green, 0; blue, 0 }  ,fill opacity=1 ] (308.22,140.56) -- (313.66,140.56) -- (313.66,146) -- (308.22,146) -- cycle ;
%Shape: Rectangle [id:dp024476773899236304] 
\draw  [color={rgb, 255:red, 0; green, 0; blue, 0 }  ,draw opacity=1 ] (419.77,140.56) -- (441.55,140.56) -- (441.55,146) -- (419.77,146) -- cycle ;
%Shape: Rectangle [id:dp3281698375505394] 
\draw   (425.22,140.56) -- (430.66,140.56) -- (430.66,146) -- (425.22,146) -- cycle ;
%Shape: Rectangle [id:dp3570033608563904] 
\draw   (436.1,140.56) -- (441.55,140.56) -- (441.55,146) -- (436.1,146) -- cycle ;
%Shape: Rectangle [id:dp14989426958341678] 
\draw  [color={rgb, 255:red, 0; green, 0; blue, 0 }  ,draw opacity=1 ] (398,140.56) -- (419.77,140.56) -- (419.77,146) -- (398,146) -- cycle ;
%Shape: Rectangle [id:dp5356842599310161] 
\draw   (403.44,140.56) -- (408.89,140.56) -- (408.89,146) -- (403.44,146) -- cycle ;
%Shape: Rectangle [id:dp16282000395925067] 
\draw   (414.33,140.56) -- (419.77,140.56) -- (419.77,146) -- (414.33,146) -- cycle ;
%Shape: Rectangle [id:dp9088099045150276] 
\draw  [fill={rgb, 255:red, 0; green, 0; blue, 0 }  ,fill opacity=1 ] (414.33,140.56) -- (419.77,140.56) -- (419.77,146) -- (414.33,146) -- cycle ;
%Shape: Rectangle [id:dp4236857305268481] 
\draw  [fill={rgb, 255:red, 0; green, 0; blue, 0 }  ,fill opacity=1 ] (403.89,140.56) -- (409.33,140.56) -- (409.33,146) -- (403.89,146) -- cycle ;
%Shape: Rectangle [id:dp47471672391200725] 
\draw  [color={rgb, 255:red, 0; green, 0; blue, 0 }  ,draw opacity=1 ] (465.23,140.56) -- (487,140.56) -- (487,146) -- (465.23,146) -- cycle ;
%Shape: Rectangle [id:dp5046782397295588] 
\draw   (470.67,140.56) -- (476.11,140.56) -- (476.11,146) -- (470.67,146) -- cycle ;
%Shape: Rectangle [id:dp5398427840502985] 
\draw   (481.56,140.56) -- (487,140.56) -- (487,146) -- (481.56,146) -- cycle ;
%Shape: Rectangle [id:dp9530653156150113] 
\draw  [color={rgb, 255:red, 0; green, 0; blue, 0 }  ,draw opacity=1 ] (443.45,140.56) -- (465.23,140.56) -- (465.23,146) -- (443.45,146) -- cycle ;
%Shape: Rectangle [id:dp9018101808702101] 
\draw   (448.9,140.56) -- (454.34,140.56) -- (454.34,146) -- (448.9,146) -- cycle ;
%Shape: Rectangle [id:dp6236348541944832] 
\draw   (459.78,140.56) -- (465.23,140.56) -- (465.23,146) -- (459.78,146) -- cycle ;
%Shape: Rectangle [id:dp8733512750694163] 
\draw  [fill={rgb, 255:red, 0; green, 0; blue, 0 }  ,fill opacity=1 ] (454.23,140.56) -- (459.67,140.56) -- (459.67,146) -- (454.23,146) -- cycle ;
%Shape: Rectangle [id:dp657881340687533] 
\draw  [fill={rgb, 255:red, 0; green, 0; blue, 0 }  ,fill opacity=1 ] (481.56,140.56) -- (487,140.56) -- (487,146) -- (481.56,146) -- cycle ;
%Shape: Rectangle [id:dp28693516905497596] 
\draw  [fill={rgb, 255:red, 0; green, 0; blue, 0 }  ,fill opacity=1 ] (443.9,140.56) -- (449.34,140.56) -- (449.34,146) -- (443.9,146) -- cycle ;
%Shape: Rectangle [id:dp42932310452352407] 
\draw  [fill={rgb, 255:red, 0; green, 0; blue, 0 }  ,fill opacity=1 ] (431.22,140.56) -- (436.66,140.56) -- (436.66,146) -- (431.22,146) -- cycle ;
%Shape: Rectangle [id:dp8595557696680994] 
\draw  [color={rgb, 255:red, 0; green, 0; blue, 0 }  ,draw opacity=1 ] (372.77,173.56) -- (394.55,173.56) -- (394.55,179) -- (372.77,179) -- cycle ;
%Shape: Rectangle [id:dp7269145145904885] 
\draw   (378.22,173.56) -- (383.66,173.56) -- (383.66,179) -- (378.22,179) -- cycle ;
%Shape: Rectangle [id:dp5236811737509476] 
\draw   (389.1,173.56) -- (394.55,173.56) -- (394.55,179) -- (389.1,179) -- cycle ;
%Shape: Rectangle [id:dp8958306034501012] 
\draw  [color={rgb, 255:red, 0; green, 0; blue, 0 }  ,draw opacity=1 ] (351,173.56) -- (372.77,173.56) -- (372.77,179) -- (351,179) -- cycle ;
%Shape: Rectangle [id:dp32645091734453147] 
\draw   (356.44,173.56) -- (361.89,173.56) -- (361.89,179) -- (356.44,179) -- cycle ;
%Shape: Rectangle [id:dp1999196870922424] 
\draw   (367.33,173.56) -- (372.77,173.56) -- (372.77,179) -- (367.33,179) -- cycle ;
%Shape: Rectangle [id:dp9658463124738417] 
\draw  [fill={rgb, 255:red, 0; green, 0; blue, 0 }  ,fill opacity=1 ] (367.33,173.56) -- (372.77,173.56) -- (372.77,179) -- (367.33,179) -- cycle ;
%Shape: Rectangle [id:dp6785693865798004] 
\draw  [fill={rgb, 255:red, 0; green, 0; blue, 0 }  ,fill opacity=1 ] (350.89,173.56) -- (356.33,173.56) -- (356.33,179) -- (350.89,179) -- cycle ;
%Shape: Rectangle [id:dp6019195811705763] 
\draw  [color={rgb, 255:red, 0; green, 0; blue, 0 }  ,draw opacity=1 ] (418.23,173.56) -- (440,173.56) -- (440,179) -- (418.23,179) -- cycle ;
%Shape: Rectangle [id:dp38647774093214204] 
\draw   (423.67,173.56) -- (429.11,173.56) -- (429.11,179) -- (423.67,179) -- cycle ;
%Shape: Rectangle [id:dp13859708518878888] 
\draw   (434.56,173.56) -- (440,173.56) -- (440,179) -- (434.56,179) -- cycle ;
%Shape: Rectangle [id:dp016343302815948757] 
\draw  [color={rgb, 255:red, 0; green, 0; blue, 0 }  ,draw opacity=1 ] (396.45,173.56) -- (418.23,173.56) -- (418.23,179) -- (396.45,179) -- cycle ;
%Shape: Rectangle [id:dp2976263734201169] 
\draw   (401.9,173.56) -- (407.34,173.56) -- (407.34,179) -- (401.9,179) -- cycle ;
%Shape: Rectangle [id:dp725298638940311] 
\draw   (412.78,173.56) -- (418.23,173.56) -- (418.23,179) -- (412.78,179) -- cycle ;
%Shape: Rectangle [id:dp43716497545454147] 
\draw  [fill={rgb, 255:red, 0; green, 0; blue, 0 }  ,fill opacity=1 ] (429.23,173.56) -- (434.67,173.56) -- (434.67,179) -- (429.23,179) -- cycle ;
%Shape: Rectangle [id:dp7382209189284608] 
\draw  [fill={rgb, 255:red, 0; green, 0; blue, 0 }  ,fill opacity=1 ] (434.56,173.56) -- (440,173.56) -- (440,179) -- (434.56,179) -- cycle ;
%Shape: Rectangle [id:dp6971775190937775] 
\draw  [fill={rgb, 255:red, 0; green, 0; blue, 0 }  ,fill opacity=1 ] (396.9,173.56) -- (402.34,173.56) -- (402.34,179) -- (396.9,179) -- cycle ;
%Shape: Rectangle [id:dp6168137251469903] 
\draw  [fill={rgb, 255:red, 0; green, 0; blue, 0 }  ,fill opacity=1 ] (384.22,173.56) -- (389.66,173.56) -- (389.66,179) -- (384.22,179) -- cycle ;
%Shape: Rectangle [id:dp704046569533376] 
\draw  [fill={rgb, 255:red, 0; green, 0; blue, 0 }  ,fill opacity=1 ] (330.22,140.56) -- (335.66,140.56) -- (335.66,146) -- (330.22,146) -- cycle ;

% Text Node
\draw (147,96.4) node [anchor=north west][inner sep=0.75pt]    {$\mathcal{B}^{S}( f_{1})$};
% Text Node
\draw (367,96.4) node [anchor=north west][inner sep=0.75pt]    {$\mathcal{B}^{S}( f_{2})$};
% Text Node
\draw (84,222.4) node [anchor=north west][inner sep=0.75pt]    {$1000$};
% Text Node
\draw (275,222.4) node [anchor=north west][inner sep=0.75pt]    {$+$};
% Text Node
\draw (489,222.4) node [anchor=north west][inner sep=0.75pt]    {$=8000$};
% Text Node
\draw (129,221.2) node [anchor=north west][inner sep=0.75pt]    {$\cdot $};
% Text Node
\draw (154,203.4) node [anchor=north west][inner sep=0.75pt]    {$5000$};
% Text Node
\draw (154,243.4) node [anchor=north west][inner sep=0.75pt]    {$1000$};
% Text Node
\draw (199,221.2) node [anchor=north west][inner sep=0.75pt]    {$\cdot $};
% Text Node
\draw (224,203.4) node [anchor=north west][inner sep=0.75pt]    {$1000$};
% Text Node
\draw (224,243.4) node [anchor=north west][inner sep=0.75pt]    {$1000$};
% Text Node
\draw (304,222.4) node [anchor=north west][inner sep=0.75pt]    {$2000$};
% Text Node
\draw (349,221.2) node [anchor=north west][inner sep=0.75pt]    {$\cdot $};
% Text Node
\draw (374,203.4) node [anchor=north west][inner sep=0.75pt]    {$3000$};
% Text Node
\draw (374,243.4) node [anchor=north west][inner sep=0.75pt]    {$2000$};
% Text Node
\draw (419,221.2) node [anchor=north west][inner sep=0.75pt]    {$\cdot $};
% Text Node
\draw (444,203.4) node [anchor=north west][inner sep=0.75pt]    {$2000$};
% Text Node
\draw (444,243.4) node [anchor=north west][inner sep=0.75pt]    {$2000$};
% Text Node
\draw (114,123.4) node [anchor=north west][inner sep=0.75pt]  [font=\scriptsize]  {$\mathcal{B}_{s}$};
% Text Node
\draw (186,122.4) node [anchor=north west][inner sep=0.75pt]  [font=\scriptsize]  {$\Phi \left(\text{dbo:author}\right)$};
% Text Node
\draw (179,156.4) node [anchor=north west][inner sep=0.75pt]  [font=\scriptsize]  {$\Phi \left(\text{dbo:deathDate}\right)$};
% Text Node
\draw (77,154.4) node [anchor=north west][inner sep=0.75pt]  [font=\scriptsize]  {$\Phi \left(\text{dbo:nationality}\right)$};
% Text Node
\draw (334,123.4) node [anchor=north west][inner sep=0.75pt]  [font=\scriptsize]  {$\mathcal{B}_{s}$};
% Text Node
\draw (406,122.4) node [anchor=north west][inner sep=0.75pt]  [font=\scriptsize]  {$\Phi \left(\text{dbo:author}\right)$};
% Text Node
\draw (350,154.4) node [anchor=north west][inner sep=0.75pt]  [font=\scriptsize]  {$\Phi \left(\text{dbo:nationality}\right)$};;

\end{tikzpicture}

%% file: figures/cardex4.tex
\tikzset{every picture/.style={line width=0.75pt}} %set default line width to 0.75pt        

\begin{tikzpicture}[x=0.6pt,y=0.6pt,yscale=-1,xscale=1]
%uncomment if require: \path (0,887); %set diagram left start at 0, and has height of 887

%Rounded Rect [id:dp2190205672239841] 
\draw  [dash pattern={on 4.5pt off 4.5pt}] (87.78,69.03) .. controls (87.78,58.52) and (96.3,50) .. (106.82,50) -- (274.97,50) .. controls (285.48,50) and (294,58.52) .. (294,69.03) -- (294,126.14) .. controls (294,136.65) and (285.48,145.17) .. (274.97,145.17) -- (106.82,145.17) .. controls (96.3,145.17) and (87.78,136.65) .. (87.78,126.14) -- cycle ;
%Rounded Rect [id:dp9931506278268684] 
\draw  [dash pattern={on 4.5pt off 4.5pt}] (307.78,69.03) .. controls (307.78,58.52) and (316.3,50) .. (326.82,50) -- (494.97,50) .. controls (505.48,50) and (514,58.52) .. (514,69.03) -- (514,126.14) .. controls (514,136.65) and (505.48,145.17) .. (494.97,145.17) -- (326.82,145.17) .. controls (316.3,145.17) and (307.78,136.65) .. (307.78,126.14) -- cycle ;
%Rounded Rect [id:dp0380879084753456] 
\draw  [color={rgb, 255:red, 184; green, 233; blue, 134 }  ,draw opacity=1 ][line width=1.5]  (260,159.8) .. controls (260,156.6) and (262.6,154) .. (265.8,154) -- (301.2,154) .. controls (304.4,154) and (307,156.6) .. (307,159.8) -- (307,177.2) .. controls (307,180.4) and (304.4,183) .. (301.2,183) -- (265.8,183) .. controls (262.6,183) and (260,180.4) .. (260,177.2) -- cycle ;
%Rounded Rect [id:dp8187382752121503] 
\draw  [color={rgb, 255:red, 219; green, 126; blue, 136 }  ,draw opacity=1 ][line width=1.5]  (260,200.8) .. controls (260,197.6) and (262.6,195) .. (265.8,195) -- (301.2,195) .. controls (304.4,195) and (307,197.6) .. (307,200.8) -- (307,218.2) .. controls (307,221.4) and (304.4,224) .. (301.2,224) -- (265.8,224) .. controls (262.6,224) and (260,221.4) .. (260,218.2) -- cycle ;
%Straight Lines [id:da25661536692334963] 
\draw [line width=1.5]    (259,189) -- (306,189) ;
%Rounded Rect [id:dp7076134255402651] 
\draw  [color={rgb, 255:red, 155; green, 155; blue, 155 }  ,draw opacity=1 ][line width=1.5]  (94,111.25) .. controls (94,108.1) and (96.55,105.56) .. (99.69,105.56) -- (181.65,105.56) .. controls (184.79,105.56) and (187.34,108.1) .. (187.34,111.25) -- (187.34,128.31) .. controls (187.34,131.45) and (184.79,134) .. (181.65,134) -- (99.69,134) .. controls (96.55,134) and (94,131.45) .. (94,128.31) -- cycle ;
%Rounded Rect [id:dp4006216375756507] 
\draw  [color={rgb, 255:red, 155; green, 155; blue, 155 }  ,draw opacity=1 ][dash pattern={on 5.63pt off 4.5pt}][line width=1.5]  (314,80.25) .. controls (314,77.1) and (316.55,74.56) .. (319.69,74.56) -- (401.65,74.56) .. controls (404.79,74.56) and (407.34,77.1) .. (407.34,80.25) -- (407.34,97.31) .. controls (407.34,100.45) and (404.79,103) .. (401.65,103) -- (319.69,103) .. controls (316.55,103) and (314,100.45) .. (314,97.31) -- cycle ;
%Rounded Rect [id:dp06394315391465544] 
\draw  [color={rgb, 255:red, 184; green, 233; blue, 134 }  ,draw opacity=1 ][line width=1.5]  (194,79.25) .. controls (194,76.1) and (196.55,73.56) .. (199.69,73.56) -- (281.65,73.56) .. controls (284.79,73.56) and (287.34,76.1) .. (287.34,79.25) -- (287.34,96.31) .. controls (287.34,99.45) and (284.79,102) .. (281.65,102) -- (199.69,102) .. controls (196.55,102) and (194,99.45) .. (194,96.31) -- cycle ;
%Rounded Rect [id:dp8414431973606589] 
\draw  [color={rgb, 255:red, 219; green, 126; blue, 136 }  ,draw opacity=1 ][line width=1.5]  (94,79.25) .. controls (94,76.1) and (96.55,73.56) .. (99.69,73.56) -- (181.65,73.56) .. controls (184.79,73.56) and (187.34,76.1) .. (187.34,79.25) -- (187.34,96.31) .. controls (187.34,99.45) and (184.79,102) .. (181.65,102) -- (99.69,102) .. controls (96.55,102) and (94,99.45) .. (94,96.31) -- cycle ;
%Rounded Rect [id:dp9516626579948348] 
\draw  [color={rgb, 255:red, 219; green, 126; blue, 136 }  ,draw opacity=1 ][dash pattern={on 5.63pt off 4.5pt}][line width=1.5]  (322,159.8) .. controls (322,156.6) and (324.6,154) .. (327.8,154) -- (353.2,154) .. controls (356.4,154) and (359,156.6) .. (359,159.8) -- (359,177.2) .. controls (359,180.4) and (356.4,183) .. (353.2,183) -- (327.8,183) .. controls (324.6,183) and (322,180.4) .. (322,177.2) -- cycle ;
%Rounded Rect [id:dp3957394535972356] 
\draw  [color={rgb, 255:red, 155; green, 155; blue, 155 }  ,draw opacity=1 ][dash pattern={on 5.63pt off 4.5pt}][line width=1.5]  (322,199.8) .. controls (322,196.6) and (324.6,194) .. (327.8,194) -- (354.2,194) .. controls (357.4,194) and (360,196.6) .. (360,199.8) -- (360,217.2) .. controls (360,220.4) and (357.4,223) .. (354.2,223) -- (327.8,223) .. controls (324.6,223) and (322,220.4) .. (322,217.2) -- cycle ;
%Straight Lines [id:da4556970768625064] 
\draw [line width=1.5]    (321,189) -- (359,189) ;
%Rounded Rect [id:dp6634879433387079] 
\draw  [color={rgb, 255:red, 219; green, 126; blue, 136 }  ,draw opacity=1 ][dash pattern={on 5.63pt off 4.5pt}][line width=1.5]  (414,80.25) .. controls (414,77.1) and (416.55,74.56) .. (419.69,74.56) -- (501.65,74.56) .. controls (504.79,74.56) and (507.34,77.1) .. (507.34,80.25) -- (507.34,97.31) .. controls (507.34,100.45) and (504.79,103) .. (501.65,103) -- (419.69,103) .. controls (416.55,103) and (414,100.45) .. (414,97.31) -- cycle ;
%Rounded Rect [id:dp7544877638123171] 
\draw  [color={rgb, 255:red, 184; green, 233; blue, 134 }  ,draw opacity=1 ][dash pattern={on 5.63pt off 4.5pt}][line width=1.5]  (374,159.8) .. controls (374,156.6) and (376.6,154) .. (379.8,154) -- (405.2,154) .. controls (408.4,154) and (411,156.6) .. (411,159.8) -- (411,177.2) .. controls (411,180.4) and (408.4,183) .. (405.2,183) -- (379.8,183) .. controls (376.6,183) and (374,180.4) .. (374,177.2) -- cycle ;
%Rounded Rect [id:dp750321492093393] 
\draw  [color={rgb, 255:red, 155; green, 155; blue, 155 }  ,draw opacity=1 ][dash pattern={on 5.63pt off 4.5pt}][line width=1.5]  (374,199.8) .. controls (374,196.6) and (376.6,194) .. (379.8,194) -- (406.2,194) .. controls (409.4,194) and (412,196.6) .. (412,199.8) -- (412,217.2) .. controls (412,220.4) and (409.4,223) .. (406.2,223) -- (379.8,223) .. controls (376.6,223) and (374,220.4) .. (374,217.2) -- cycle ;
%Straight Lines [id:da31267436308160357] 
\draw [line width=1.5]    (373,189) -- (411,189) ;
%Rounded Rect [id:dp5405348663612818] 
\draw  [color={rgb, 255:red, 184; green, 233; blue, 134 }  ,draw opacity=1 ][dash pattern={on 5.63pt off 4.5pt}][line width=1.5]  (366,112.25) .. controls (366,109.1) and (368.55,106.56) .. (371.69,106.56) -- (453.65,106.56) .. controls (456.79,106.56) and (459.34,109.1) .. (459.34,112.25) -- (459.34,129.31) .. controls (459.34,132.45) and (456.79,135) .. (453.65,135) -- (371.69,135) .. controls (368.55,135) and (366,132.45) .. (366,129.31) -- cycle ;
%Straight Lines [id:da5552833395784121] 
\draw [color={rgb, 255:red, 125; green, 184; blue, 253 }  ,draw opacity=1 ][line width=1.5]    (159.25,134.4) -- (193.4,159.6) ;
%Straight Lines [id:da277684170090825] 
\draw [color={rgb, 255:red, 125; green, 184; blue, 253 }  ,draw opacity=1 ][line width=1.5]    (319.69,103) -- (231.09,158.4) ;
%Shape: Rectangle [id:dp809366852228819] 
\draw  [color={rgb, 255:red, 0; green, 0; blue, 0 }  ,draw opacity=1 ] (117.77,93.56) -- (139.55,93.56) -- (139.55,99) -- (117.77,99) -- cycle ;
%Shape: Rectangle [id:dp14565379689695024] 
\draw   (123.22,93.56) -- (128.66,93.56) -- (128.66,99) -- (123.22,99) -- cycle ;
%Shape: Rectangle [id:dp08438683653937318] 
\draw   (134.1,93.56) -- (139.55,93.56) -- (139.55,99) -- (134.1,99) -- cycle ;
%Shape: Rectangle [id:dp23436614644426168] 
\draw  [color={rgb, 255:red, 0; green, 0; blue, 0 }  ,draw opacity=1 ] (96,93.56) -- (117.77,93.56) -- (117.77,99) -- (96,99) -- cycle ;
%Shape: Rectangle [id:dp614643858977899] 
\draw   (101.44,93.56) -- (106.89,93.56) -- (106.89,99) -- (101.44,99) -- cycle ;
%Shape: Rectangle [id:dp6414431289735391] 
\draw   (112.33,93.56) -- (117.77,93.56) -- (117.77,99) -- (112.33,99) -- cycle ;
%Shape: Rectangle [id:dp8259723878929448] 
\draw  [fill={rgb, 255:red, 0; green, 0; blue, 0 }  ,fill opacity=1 ] (112.33,93.56) -- (117.77,93.56) -- (117.77,99) -- (112.33,99) -- cycle ;
%Shape: Rectangle [id:dp20518881145235013] 
\draw  [fill={rgb, 255:red, 0; green, 0; blue, 0 }  ,fill opacity=1 ] (106.89,93.56) -- (112.33,93.56) -- (112.33,99) -- (106.89,99) -- cycle ;
%Shape: Rectangle [id:dp27976547923798845] 
\draw  [color={rgb, 255:red, 0; green, 0; blue, 0 }  ,draw opacity=1 ] (163.23,93.56) -- (185,93.56) -- (185,99) -- (163.23,99) -- cycle ;
%Shape: Rectangle [id:dp5097497163252269] 
\draw   (168.67,93.56) -- (174.11,93.56) -- (174.11,99) -- (168.67,99) -- cycle ;
%Shape: Rectangle [id:dp29484322390314754] 
\draw   (179.56,93.56) -- (185,93.56) -- (185,99) -- (179.56,99) -- cycle ;
%Shape: Rectangle [id:dp30160446963561016] 
\draw  [color={rgb, 255:red, 0; green, 0; blue, 0 }  ,draw opacity=1 ] (141.45,93.56) -- (163.23,93.56) -- (163.23,99) -- (141.45,99) -- cycle ;
%Shape: Rectangle [id:dp6904143523898756] 
\draw   (146.9,93.56) -- (152.34,93.56) -- (152.34,99) -- (146.9,99) -- cycle ;
%Shape: Rectangle [id:dp7225532781862077] 
\draw   (157.78,93.56) -- (163.23,93.56) -- (163.23,99) -- (157.78,99) -- cycle ;
%Shape: Rectangle [id:dp23958469636329494] 
\draw  [fill={rgb, 255:red, 0; green, 0; blue, 0 }  ,fill opacity=1 ] (163.23,93.56) -- (168.67,93.56) -- (168.67,99) -- (163.23,99) -- cycle ;
%Shape: Rectangle [id:dp8561006286500586] 
\draw  [fill={rgb, 255:red, 0; green, 0; blue, 0 }  ,fill opacity=1 ] (179.56,93.56) -- (185,93.56) -- (185,99) -- (179.56,99) -- cycle ;
%Shape: Rectangle [id:dp7131056388820416] 
\draw  [fill={rgb, 255:red, 0; green, 0; blue, 0 }  ,fill opacity=1 ] (146.9,93.56) -- (152.34,93.56) -- (152.34,99) -- (146.9,99) -- cycle ;
%Shape: Rectangle [id:dp758553326378103] 
\draw  [fill={rgb, 255:red, 0; green, 0; blue, 0 }  ,fill opacity=1 ] (123.22,93.56) -- (128.66,93.56) -- (128.66,99) -- (123.22,99) -- cycle ;
%Shape: Rectangle [id:dp8961233103378866] 
\draw  [color={rgb, 255:red, 0; green, 0; blue, 0 }  ,draw opacity=1 ] (217.77,93.56) -- (239.55,93.56) -- (239.55,99) -- (217.77,99) -- cycle ;
%Shape: Rectangle [id:dp6433616885344674] 
\draw   (223.22,93.56) -- (228.66,93.56) -- (228.66,99) -- (223.22,99) -- cycle ;
%Shape: Rectangle [id:dp5537951117641917] 
\draw   (234.1,93.56) -- (239.55,93.56) -- (239.55,99) -- (234.1,99) -- cycle ;
%Shape: Rectangle [id:dp9121545008107032] 
\draw  [color={rgb, 255:red, 0; green, 0; blue, 0 }  ,draw opacity=1 ] (196,93.56) -- (217.77,93.56) -- (217.77,99) -- (196,99) -- cycle ;
%Shape: Rectangle [id:dp4573427357060552] 
\draw   (201.44,93.56) -- (206.89,93.56) -- (206.89,99) -- (201.44,99) -- cycle ;
%Shape: Rectangle [id:dp010359298186761734] 
\draw   (212.33,93.56) -- (217.77,93.56) -- (217.77,99) -- (212.33,99) -- cycle ;
%Shape: Rectangle [id:dp6672174222828738] 
\draw  [fill={rgb, 255:red, 0; green, 0; blue, 0 }  ,fill opacity=1 ] (206.89,93.56) -- (211.75,93.56) -- (211.75,99) -- (206.89,99) -- cycle ;
%Shape: Rectangle [id:dp7593500836109344] 
\draw  [fill={rgb, 255:red, 0; green, 0; blue, 0 }  ,fill opacity=1 ] (195.89,93.56) -- (201.33,93.56) -- (201.33,99) -- (195.89,99) -- cycle ;
%Shape: Rectangle [id:dp6837940755726547] 
\draw  [color={rgb, 255:red, 0; green, 0; blue, 0 }  ,draw opacity=1 ] (263.23,93.56) -- (285,93.56) -- (285,99) -- (263.23,99) -- cycle ;
%Shape: Rectangle [id:dp5392972732716229] 
\draw   (268.67,93.56) -- (274.11,93.56) -- (274.11,99) -- (268.67,99) -- cycle ;
%Shape: Rectangle [id:dp5359677965861114] 
\draw   (279.56,93.56) -- (285,93.56) -- (285,99) -- (279.56,99) -- cycle ;
%Shape: Rectangle [id:dp6465037619088907] 
\draw  [color={rgb, 255:red, 0; green, 0; blue, 0 }  ,draw opacity=1 ] (241.45,93.56) -- (263.23,93.56) -- (263.23,99) -- (241.45,99) -- cycle ;
%Shape: Rectangle [id:dp3513236099575058] 
\draw   (246.9,93.56) -- (252.34,93.56) -- (252.34,99) -- (246.9,99) -- cycle ;
%Shape: Rectangle [id:dp3330824968307433] 
\draw   (257.78,93.56) -- (263.23,93.56) -- (263.23,99) -- (257.78,99) -- cycle ;
%Shape: Rectangle [id:dp02985887119747832] 
\draw  [fill={rgb, 255:red, 0; green, 0; blue, 0 }  ,fill opacity=1 ] (263.23,93.56) -- (268.67,93.56) -- (268.67,99) -- (263.23,99) -- cycle ;
%Shape: Rectangle [id:dp20700381984983496] 
\draw  [fill={rgb, 255:red, 0; green, 0; blue, 0 }  ,fill opacity=1 ] (268.56,93.56) -- (274,93.56) -- (274,99) -- (268.56,99) -- cycle ;
%Shape: Rectangle [id:dp5473758615163772] 
\draw  [fill={rgb, 255:red, 0; green, 0; blue, 0 }  ,fill opacity=1 ] (257.9,93.56) -- (263.34,93.56) -- (263.34,99) -- (257.9,99) -- cycle ;
%Shape: Rectangle [id:dp5544917260686728] 
\draw  [fill={rgb, 255:red, 0; green, 0; blue, 0 }  ,fill opacity=1 ] (223.22,93.56) -- (228.66,93.56) -- (228.66,99) -- (223.22,99) -- cycle ;
%Shape: Rectangle [id:dp6074995035647714] 
\draw  [color={rgb, 255:red, 0; green, 0; blue, 0 }  ,draw opacity=1 ] (117.77,126.56) -- (139.55,126.56) -- (139.55,132) -- (117.77,132) -- cycle ;
%Shape: Rectangle [id:dp27938068840223174] 
\draw   (123.22,126.56) -- (128.66,126.56) -- (128.66,132) -- (123.22,132) -- cycle ;
%Shape: Rectangle [id:dp03181965529515718] 
\draw   (134.1,126.56) -- (139.55,126.56) -- (139.55,132) -- (134.1,132) -- cycle ;
%Shape: Rectangle [id:dp7420285155126686] 
\draw  [color={rgb, 255:red, 0; green, 0; blue, 0 }  ,draw opacity=1 ] (96,126.56) -- (117.77,126.56) -- (117.77,132) -- (96,132) -- cycle ;
%Shape: Rectangle [id:dp8163854094302424] 
\draw   (101.44,126.56) -- (106.89,126.56) -- (106.89,132) -- (101.44,132) -- cycle ;
%Shape: Rectangle [id:dp7354045942921992] 
\draw   (112.33,126.56) -- (117.77,126.56) -- (117.77,132) -- (112.33,132) -- cycle ;
%Shape: Rectangle [id:dp24365757152280865] 
\draw  [fill={rgb, 255:red, 0; green, 0; blue, 0 }  ,fill opacity=1 ] (107.33,126.56) -- (112.77,126.56) -- (112.77,132) -- (107.33,132) -- cycle ;
%Shape: Rectangle [id:dp1293829664272469] 
\draw  [fill={rgb, 255:red, 0; green, 0; blue, 0 }  ,fill opacity=1 ] (101.89,126.56) -- (107.33,126.56) -- (107.33,132) -- (101.89,132) -- cycle ;
%Shape: Rectangle [id:dp7830921190768055] 
\draw  [color={rgb, 255:red, 0; green, 0; blue, 0 }  ,draw opacity=1 ] (163.23,126.56) -- (185,126.56) -- (185,132) -- (163.23,132) -- cycle ;
%Shape: Rectangle [id:dp6193956797400524] 
\draw   (168.67,126.56) -- (174.11,126.56) -- (174.11,132) -- (168.67,132) -- cycle ;
%Shape: Rectangle [id:dp027820691294484168] 
\draw   (179.56,126.56) -- (185,126.56) -- (185,132) -- (179.56,132) -- cycle ;
%Shape: Rectangle [id:dp9629363350977311] 
\draw  [color={rgb, 255:red, 0; green, 0; blue, 0 }  ,draw opacity=1 ] (141.45,126.56) -- (163.23,126.56) -- (163.23,132) -- (141.45,132) -- cycle ;
%Shape: Rectangle [id:dp05279204502140444] 
\draw   (146.9,126.56) -- (152.34,126.56) -- (152.34,132) -- (146.9,132) -- cycle ;
%Shape: Rectangle [id:dp5396224362867755] 
\draw   (157.78,126.56) -- (163.23,126.56) -- (163.23,132) -- (157.78,132) -- cycle ;
%Shape: Rectangle [id:dp7168871573073952] 
\draw  [fill={rgb, 255:red, 0; green, 0; blue, 0 }  ,fill opacity=1 ] (157.5,126.56) -- (162.67,126.56) -- (162.67,132) -- (157.5,132) -- cycle ;
%Shape: Rectangle [id:dp22412830981770548] 
\draw  [fill={rgb, 255:red, 0; green, 0; blue, 0 }  ,fill opacity=1 ] (168.56,126.56) -- (174,126.56) -- (174,132) -- (168.56,132) -- cycle ;
%Shape: Rectangle [id:dp6693520872528866] 
\draw  [fill={rgb, 255:red, 0; green, 0; blue, 0 }  ,fill opacity=1 ] (146.9,126.56) -- (152.34,126.56) -- (152.34,132) -- (146.9,132) -- cycle ;
%Shape: Rectangle [id:dp029136444003242112] 
\draw  [fill={rgb, 255:red, 0; green, 0; blue, 0 }  ,fill opacity=1 ] (118.22,126.56) -- (123.66,126.56) -- (123.66,132) -- (118.22,132) -- cycle ;
%Shape: Rectangle [id:dp0979494509272012] 
\draw  [color={rgb, 255:red, 0; green, 0; blue, 0 }  ,draw opacity=1 ] (217.77,126.56) -- (239.55,126.56) -- (239.55,132) -- (217.77,132) -- cycle ;
%Shape: Rectangle [id:dp7458006857493811] 
\draw   (223.22,126.56) -- (228.66,126.56) -- (228.66,132) -- (223.22,132) -- cycle ;
%Shape: Rectangle [id:dp13315197109677912] 
\draw   (234.1,126.56) -- (239.55,126.56) -- (239.55,132) -- (234.1,132) -- cycle ;
%Shape: Rectangle [id:dp05728070719943412] 
\draw  [color={rgb, 255:red, 0; green, 0; blue, 0 }  ,draw opacity=1 ] (196,126.56) -- (217.77,126.56) -- (217.77,132) -- (196,132) -- cycle ;
%Shape: Rectangle [id:dp2380307339681228] 
\draw   (201.44,126.56) -- (206.89,126.56) -- (206.89,132) -- (201.44,132) -- cycle ;
%Shape: Rectangle [id:dp5952223595770433] 
\draw   (212.33,126.56) -- (217.77,126.56) -- (217.77,132) -- (212.33,132) -- cycle ;
%Shape: Rectangle [id:dp8933968734127776] 
\draw  [fill={rgb, 255:red, 0; green, 0; blue, 0 }  ,fill opacity=1 ] (212.33,126.56) -- (217.77,126.56) -- (217.77,132) -- (212.33,132) -- cycle ;
%Shape: Rectangle [id:dp3342458105605507] 
\draw  [fill={rgb, 255:red, 0; green, 0; blue, 0 }  ,fill opacity=1 ] (206.89,126.56) -- (212.33,126.56) -- (212.33,132) -- (206.89,132) -- cycle ;
%Shape: Rectangle [id:dp09584537027891848] 
\draw  [color={rgb, 255:red, 0; green, 0; blue, 0 }  ,draw opacity=1 ] (263.23,126.56) -- (285,126.56) -- (285,132) -- (263.23,132) -- cycle ;
%Shape: Rectangle [id:dp8352367057301456] 
\draw   (268.67,126.56) -- (274.11,126.56) -- (274.11,132) -- (268.67,132) -- cycle ;
%Shape: Rectangle [id:dp5033901255441389] 
\draw   (279.56,126.56) -- (285,126.56) -- (285,132) -- (279.56,132) -- cycle ;
%Shape: Rectangle [id:dp3480960888909408] 
\draw  [color={rgb, 255:red, 0; green, 0; blue, 0 }  ,draw opacity=1 ] (241.45,126.56) -- (263.23,126.56) -- (263.23,132) -- (241.45,132) -- cycle ;
%Shape: Rectangle [id:dp15057715187948384] 
\draw   (246.9,126.56) -- (252.34,126.56) -- (252.34,132) -- (246.9,132) -- cycle ;
%Shape: Rectangle [id:dp7377609085966137] 
\draw   (257.78,126.56) -- (263.23,126.56) -- (263.23,132) -- (257.78,132) -- cycle ;
%Shape: Rectangle [id:dp2828981034935061] 
\draw  [fill={rgb, 255:red, 0; green, 0; blue, 0 }  ,fill opacity=1 ] (246.23,126.56) -- (251.67,126.56) -- (251.67,132) -- (246.23,132) -- cycle ;
%Shape: Rectangle [id:dp3464351992499489] 
\draw  [fill={rgb, 255:red, 0; green, 0; blue, 0 }  ,fill opacity=1 ] (279.56,126.56) -- (285,126.56) -- (285,132) -- (279.56,132) -- cycle ;
%Shape: Rectangle [id:dp8050643994672504] 
\draw  [fill={rgb, 255:red, 0; green, 0; blue, 0 }  ,fill opacity=1 ] (241.9,126.56) -- (247.34,126.56) -- (247.34,132) -- (241.9,132) -- cycle ;
%Shape: Rectangle [id:dp5045560333448881] 
\draw  [fill={rgb, 255:red, 0; green, 0; blue, 0 }  ,fill opacity=1 ] (234.22,126.56) -- (239.66,126.56) -- (239.66,132) -- (234.22,132) -- cycle ;
%Shape: Rectangle [id:dp031679148939444346] 
\draw  [fill={rgb, 255:red, 0; green, 0; blue, 0 }  ,fill opacity=1 ] (234.22,93.56) -- (239.66,93.56) -- (239.66,99) -- (234.22,99) -- cycle ;
%Shape: Rectangle [id:dp684996410576178] 
\draw  [fill={rgb, 255:red, 0; green, 0; blue, 0 }  ,fill opacity=1 ] (123.22,126.56) -- (128.66,126.56) -- (128.66,132) -- (123.22,132) -- cycle ;
%Shape: Rectangle [id:dp7274827563061871] 
\draw  [color={rgb, 255:red, 0; green, 0; blue, 0 }  ,draw opacity=1 ] (337.77,93.56) -- (359.55,93.56) -- (359.55,99) -- (337.77,99) -- cycle ;
%Shape: Rectangle [id:dp08182146424072656] 
\draw   (343.22,93.56) -- (348.66,93.56) -- (348.66,99) -- (343.22,99) -- cycle ;
%Shape: Rectangle [id:dp08491476065865011] 
\draw   (354.1,93.56) -- (359.55,93.56) -- (359.55,99) -- (354.1,99) -- cycle ;
%Shape: Rectangle [id:dp19197746189894027] 
\draw  [color={rgb, 255:red, 0; green, 0; blue, 0 }  ,draw opacity=1 ] (316,93.56) -- (337.77,93.56) -- (337.77,99) -- (316,99) -- cycle ;
%Shape: Rectangle [id:dp47851453463035554] 
\draw   (321.44,93.56) -- (326.89,93.56) -- (326.89,99) -- (321.44,99) -- cycle ;
%Shape: Rectangle [id:dp6074569658584611] 
\draw   (332.33,93.56) -- (337.77,93.56) -- (337.77,99) -- (332.33,99) -- cycle ;
%Shape: Rectangle [id:dp984575994909423] 
\draw  [fill={rgb, 255:red, 0; green, 0; blue, 0 }  ,fill opacity=1 ] (326.33,93.56) -- (331.77,93.56) -- (331.77,99) -- (326.33,99) -- cycle ;
%Shape: Rectangle [id:dp9863229431580244] 
\draw  [fill={rgb, 255:red, 0; green, 0; blue, 0 }  ,fill opacity=1 ] (321.89,93.56) -- (327.33,93.56) -- (327.33,99) -- (321.89,99) -- cycle ;
%Shape: Rectangle [id:dp9293717160654238] 
\draw  [color={rgb, 255:red, 0; green, 0; blue, 0 }  ,draw opacity=1 ] (383.23,93.56) -- (405,93.56) -- (405,99) -- (383.23,99) -- cycle ;
%Shape: Rectangle [id:dp42308647361812635] 
\draw   (388.67,93.56) -- (394.11,93.56) -- (394.11,99) -- (388.67,99) -- cycle ;
%Shape: Rectangle [id:dp5074881824298463] 
\draw   (399.56,93.56) -- (405,93.56) -- (405,99) -- (399.56,99) -- cycle ;
%Shape: Rectangle [id:dp5052349435217836] 
\draw  [color={rgb, 255:red, 0; green, 0; blue, 0 }  ,draw opacity=1 ] (361.45,93.56) -- (383.23,93.56) -- (383.23,99) -- (361.45,99) -- cycle ;
%Shape: Rectangle [id:dp5597300643895734] 
\draw   (366.9,93.56) -- (372.34,93.56) -- (372.34,99) -- (366.9,99) -- cycle ;
%Shape: Rectangle [id:dp04724576786902368] 
\draw   (377.78,93.56) -- (383.23,93.56) -- (383.23,99) -- (377.78,99) -- cycle ;
%Shape: Rectangle [id:dp3357691376094324] 
\draw  [fill={rgb, 255:red, 0; green, 0; blue, 0 }  ,fill opacity=1 ] (383.23,93.56) -- (388.67,93.56) -- (388.67,99) -- (383.23,99) -- cycle ;
%Shape: Rectangle [id:dp5831067871199955] 
\draw  [fill={rgb, 255:red, 0; green, 0; blue, 0 }  ,fill opacity=1 ] (388.56,93.56) -- (394,93.56) -- (394,99) -- (388.56,99) -- cycle ;
%Shape: Rectangle [id:dp7373160255487945] 
\draw  [fill={rgb, 255:red, 0; green, 0; blue, 0 }  ,fill opacity=1 ] (366.9,93.56) -- (372.34,93.56) -- (372.34,99) -- (366.9,99) -- cycle ;
%Shape: Rectangle [id:dp9072245789958067] 
\draw  [fill={rgb, 255:red, 0; green, 0; blue, 0 }  ,fill opacity=1 ] (343.22,93.56) -- (348.66,93.56) -- (348.66,99) -- (343.22,99) -- cycle ;
%Shape: Rectangle [id:dp31950471767900335] 
\draw  [color={rgb, 255:red, 0; green, 0; blue, 0 }  ,draw opacity=1 ] (437.77,93.56) -- (459.55,93.56) -- (459.55,99) -- (437.77,99) -- cycle ;
%Shape: Rectangle [id:dp8075005818135271] 
\draw   (443.22,93.56) -- (448.66,93.56) -- (448.66,99) -- (443.22,99) -- cycle ;
%Shape: Rectangle [id:dp32299380001371636] 
\draw   (454.1,93.56) -- (459.55,93.56) -- (459.55,99) -- (454.1,99) -- cycle ;
%Shape: Rectangle [id:dp30737299652740013] 
\draw  [color={rgb, 255:red, 0; green, 0; blue, 0 }  ,draw opacity=1 ] (416,93.56) -- (437.77,93.56) -- (437.77,99) -- (416,99) -- cycle ;
%Shape: Rectangle [id:dp36163742991282655] 
\draw   (421.44,93.56) -- (426.89,93.56) -- (426.89,99) -- (421.44,99) -- cycle ;
%Shape: Rectangle [id:dp6983067421575124] 
\draw   (432.33,93.56) -- (437.77,93.56) -- (437.77,99) -- (432.33,99) -- cycle ;
%Shape: Rectangle [id:dp47841636033810797] 
\draw  [fill={rgb, 255:red, 0; green, 0; blue, 0 }  ,fill opacity=1 ] (432.33,93.56) -- (437.77,93.56) -- (437.77,99) -- (432.33,99) -- cycle ;
%Shape: Rectangle [id:dp24753829947737205] 
\draw  [fill={rgb, 255:red, 0; green, 0; blue, 0 }  ,fill opacity=1 ] (448.89,93.56) -- (454.33,93.56) -- (454.33,99) -- (448.89,99) -- cycle ;
%Shape: Rectangle [id:dp4461448596111536] 
\draw  [color={rgb, 255:red, 0; green, 0; blue, 0 }  ,draw opacity=1 ] (483.23,93.56) -- (505,93.56) -- (505,99) -- (483.23,99) -- cycle ;
%Shape: Rectangle [id:dp9423393954370953] 
\draw   (488.67,93.56) -- (494.11,93.56) -- (494.11,99) -- (488.67,99) -- cycle ;
%Shape: Rectangle [id:dp8255515750708937] 
\draw   (499.56,93.56) -- (505,93.56) -- (505,99) -- (499.56,99) -- cycle ;
%Shape: Rectangle [id:dp600384692594566] 
\draw  [color={rgb, 255:red, 0; green, 0; blue, 0 }  ,draw opacity=1 ] (461.45,93.56) -- (483.23,93.56) -- (483.23,99) -- (461.45,99) -- cycle ;
%Shape: Rectangle [id:dp4793051255313743] 
\draw   (466.9,93.56) -- (472.34,93.56) -- (472.34,99) -- (466.9,99) -- cycle ;
%Shape: Rectangle [id:dp9810100726159495] 
\draw   (477.78,93.56) -- (483.23,93.56) -- (483.23,99) -- (477.78,99) -- cycle ;
%Shape: Rectangle [id:dp6982066132806868] 
\draw  [fill={rgb, 255:red, 0; green, 0; blue, 0 }  ,fill opacity=1 ] (472.23,93.56) -- (477.67,93.56) -- (477.67,99) -- (472.23,99) -- cycle ;
%Shape: Rectangle [id:dp9030242909368408] 
\draw  [fill={rgb, 255:red, 0; green, 0; blue, 0 }  ,fill opacity=1 ] (494.56,93.56) -- (500,93.56) -- (500,99) -- (494.56,99) -- cycle ;
%Shape: Rectangle [id:dp6387480304711114] 
\draw  [fill={rgb, 255:red, 0; green, 0; blue, 0 }  ,fill opacity=1 ] (466.9,93.56) -- (472.34,93.56) -- (472.34,99) -- (466.9,99) -- cycle ;
%Shape: Rectangle [id:dp042607236727914444] 
\draw  [fill={rgb, 255:red, 0; green, 0; blue, 0 }  ,fill opacity=1 ] (443.22,93.56) -- (448.66,93.56) -- (448.66,99) -- (443.22,99) -- cycle ;
%Shape: Rectangle [id:dp6127158582040412] 
\draw  [color={rgb, 255:red, 0; green, 0; blue, 0 }  ,draw opacity=1 ] (390.77,126.56) -- (412.55,126.56) -- (412.55,132) -- (390.77,132) -- cycle ;
%Shape: Rectangle [id:dp22738416187931798] 
\draw   (396.22,126.56) -- (401.66,126.56) -- (401.66,132) -- (396.22,132) -- cycle ;
%Shape: Rectangle [id:dp26229615160724185] 
\draw   (407.1,126.56) -- (412.55,126.56) -- (412.55,132) -- (407.1,132) -- cycle ;
%Shape: Rectangle [id:dp9454223778354063] 
\draw  [color={rgb, 255:red, 0; green, 0; blue, 0 }  ,draw opacity=1 ] (369,126.56) -- (390.77,126.56) -- (390.77,132) -- (369,132) -- cycle ;
%Shape: Rectangle [id:dp0894022925803557] 
\draw   (374.44,126.56) -- (379.89,126.56) -- (379.89,132) -- (374.44,132) -- cycle ;
%Shape: Rectangle [id:dp3604223957998248] 
\draw   (385.33,126.56) -- (390.77,126.56) -- (390.77,132) -- (385.33,132) -- cycle ;
%Shape: Rectangle [id:dp9922505397974152] 
\draw  [fill={rgb, 255:red, 0; green, 0; blue, 0 }  ,fill opacity=1 ] (379.33,126.56) -- (384.77,126.56) -- (384.77,132) -- (379.33,132) -- cycle ;
%Shape: Rectangle [id:dp4460651345266271] 
\draw  [fill={rgb, 255:red, 0; green, 0; blue, 0 }  ,fill opacity=1 ] (374.89,126.56) -- (380.33,126.56) -- (380.33,132) -- (374.89,132) -- cycle ;
%Shape: Rectangle [id:dp7675757301425736] 
\draw  [color={rgb, 255:red, 0; green, 0; blue, 0 }  ,draw opacity=1 ] (436.23,126.56) -- (458,126.56) -- (458,132) -- (436.23,132) -- cycle ;
%Shape: Rectangle [id:dp7997841591995536] 
\draw   (441.67,126.56) -- (447.11,126.56) -- (447.11,132) -- (441.67,132) -- cycle ;
%Shape: Rectangle [id:dp06070955700650171] 
\draw   (452.56,126.56) -- (458,126.56) -- (458,132) -- (452.56,132) -- cycle ;
%Shape: Rectangle [id:dp5997552623895951] 
\draw  [color={rgb, 255:red, 0; green, 0; blue, 0 }  ,draw opacity=1 ] (414.45,126.56) -- (436.23,126.56) -- (436.23,132) -- (414.45,132) -- cycle ;
%Shape: Rectangle [id:dp11223549410199563] 
\draw   (419.9,126.56) -- (425.34,126.56) -- (425.34,132) -- (419.9,132) -- cycle ;
%Shape: Rectangle [id:dp6742902745925387] 
\draw   (430.78,126.56) -- (436.23,126.56) -- (436.23,132) -- (430.78,132) -- cycle ;
%Shape: Rectangle [id:dp12866846240528562] 
\draw  [fill={rgb, 255:red, 0; green, 0; blue, 0 }  ,fill opacity=1 ] (436.23,126.56) -- (441.67,126.56) -- (441.67,132) -- (436.23,132) -- cycle ;
%Shape: Rectangle [id:dp1564320426077903] 
\draw  [fill={rgb, 255:red, 0; green, 0; blue, 0 }  ,fill opacity=1 ] (441.56,126.56) -- (447,126.56) -- (447,132) -- (441.56,132) -- cycle ;
%Shape: Rectangle [id:dp23449485361209832] 
\draw  [fill={rgb, 255:red, 0; green, 0; blue, 0 }  ,fill opacity=1 ] (430.9,126.56) -- (436.34,126.56) -- (436.34,132) -- (430.9,132) -- cycle ;
%Shape: Rectangle [id:dp17514724430494988] 
\draw  [fill={rgb, 255:red, 0; green, 0; blue, 0 }  ,fill opacity=1 ] (396.22,126.56) -- (401.66,126.56) -- (401.66,132) -- (396.22,132) -- cycle ;
%Rounded Rect [id:dp8799880354946591] 
\draw  [color={rgb, 255:red, 219; green, 126; blue, 136 }  ,draw opacity=1 ][line width=1.5]  (127.3,180.05) .. controls (127.3,176.85) and (129.9,174.25) .. (133.1,174.25) -- (167.2,174.25) .. controls (170.4,174.25) and (173,176.85) .. (173,180.05) -- (173,197.45) .. controls (173,200.65) and (170.4,203.25) .. (167.2,203.25) -- (133.1,203.25) .. controls (129.9,203.25) and (127.3,200.65) .. (127.3,197.45) -- cycle ;
%Rounded Rect [id:dp27499242586656325] 
\draw  [color={rgb, 255:red, 155; green, 155; blue, 155 }  ,draw opacity=1 ][line width=1.5]  (192.3,200.05) .. controls (192.3,196.85) and (194.9,194.25) .. (198.1,194.25) -- (232.4,194.25) .. controls (235.6,194.25) and (238.2,196.85) .. (238.2,200.05) -- (238.2,217.45) .. controls (238.2,220.65) and (235.6,223.25) .. (232.4,223.25) -- (198.1,223.25) .. controls (194.9,223.25) and (192.3,220.65) .. (192.3,217.45) -- cycle ;
%Straight Lines [id:da1924710082374088] 
\draw [line width=1.5]    (185.3,189.25) -- (243.55,189.25) ;
%Rounded Rect [id:dp7589958592064084] 
\draw  [draw opacity=0][fill={rgb, 255:red, 125; green, 184; blue, 253 }  ,fill opacity=1 ][line width=1.5]  (186.3,160.05) .. controls (186.3,156.85) and (188.9,154.25) .. (192.1,154.25) -- (237,154.25) .. controls (240.2,154.25) and (242.8,156.85) .. (242.8,160.05) -- (242.8,177.45) .. controls (242.8,180.65) and (240.2,183.25) .. (237,183.25) -- (192.1,183.25) .. controls (188.9,183.25) and (186.3,180.65) .. (186.3,177.45) -- cycle ;
%Shape: Rectangle [id:dp9116542233287869] 
\draw  [color={rgb, 255:red, 0; green, 0; blue, 0 }  ,draw opacity=1 ] (201.97,174.33) -- (212.3,174.33) -- (212.3,176.92) -- (201.97,176.92) -- cycle ;
%Shape: Rectangle [id:dp4819609014357896] 
\draw   (204.55,174.33) -- (207.13,174.33) -- (207.13,176.92) -- (204.55,176.92) -- cycle ;
%Shape: Rectangle [id:dp47456781930064573] 
\draw   (209.72,174.33) -- (212.3,174.33) -- (212.3,176.92) -- (209.72,176.92) -- cycle ;
%Shape: Rectangle [id:dp07421718719060366] 
\draw  [color={rgb, 255:red, 0; green, 0; blue, 0 }  ,draw opacity=1 ] (191.63,174.33) -- (201.97,174.33) -- (201.97,176.92) -- (191.63,176.92) -- cycle ;
%Shape: Rectangle [id:dp3055724112633108] 
\draw   (194.22,174.33) -- (196.8,174.33) -- (196.8,176.92) -- (194.22,176.92) -- cycle ;
%Shape: Rectangle [id:dp9595631156137687] 
\draw   (199.38,174.33) -- (201.97,174.33) -- (201.97,176.92) -- (199.38,176.92) -- cycle ;
%Shape: Rectangle [id:dp8466031358576974] 
\draw  [fill={rgb, 255:red, 0; green, 0; blue, 0 }  ,fill opacity=1 ] (199.38,174.33) -- (201.97,174.33) -- (201.97,176.92) -- (199.38,176.92) -- cycle ;
%Shape: Rectangle [id:dp6844811022724133] 
\draw  [fill={rgb, 255:red, 0; green, 0; blue, 0 }  ,fill opacity=1 ] (196.8,174.33) -- (199.38,174.33) -- (199.38,176.92) -- (196.8,176.92) -- cycle ;
%Shape: Rectangle [id:dp15597682384988865] 
\draw  [fill={rgb, 255:red, 0; green, 0; blue, 0 }  ,fill opacity=1 ] (204.55,174.33) -- (207.13,174.33) -- (207.13,176.92) -- (204.55,176.92) -- cycle ;

% Text Node
\draw (165,52.4) node [anchor=north west][inner sep=0.75pt]    {$\mathcal{B}^{S}( f_{1})$};
% Text Node
\draw (385,52.4) node [anchor=north west][inner sep=0.75pt]    {$\mathcal{B}^{S}( f_{4})$};
% Text Node
\draw (244,179.2) node [anchor=north west][inner sep=0.75pt]    {$\cdot $};
% Text Node
\draw (266,161.4) node [anchor=north west][inner sep=0.75pt]    {$5000$};
% Text Node
\draw (266,201.4) node [anchor=north west][inner sep=0.75pt]    {$1000$};
% Text Node
\draw (306,179.2) node [anchor=north west][inner sep=0.75pt]    {$\cdot $};
% Text Node
\draw (328,161.4) node [anchor=north west][inner sep=0.75pt]    {$200$};
% Text Node
\draw (328,201.4) node [anchor=north west][inner sep=0.75pt]    {$200$};
% Text Node
\draw (358,179.2) node [anchor=north west][inner sep=0.75pt]    {$\cdot $};
% Text Node
\draw (380,161.4) node [anchor=north west][inner sep=0.75pt]    {$500$};
% Text Node
\draw (380,201.4) node [anchor=north west][inner sep=0.75pt]    {$200$};
% Text Node
\draw (418.8,178.2) node [anchor=north west][inner sep=0.75pt]    {$=625$};
% Text Node
\draw (132,76.4) node [anchor=north west][inner sep=0.75pt]  [font=\scriptsize]  {$\mathcal{B}_{s}$};
% Text Node
\draw (204,75.4) node [anchor=north west][inner sep=0.75pt]  [font=\scriptsize]  {$\Phi \left(\text{dbo:author}\right)$};
% Text Node
\draw (197,109.4) node [anchor=north west][inner sep=0.75pt]  [font=\scriptsize]  {$\Phi \left(\text{dbo:deathDate}\right)$};
% Text Node
\draw (95,107.4) node [anchor=north west][inner sep=0.75pt]  [font=\scriptsize]  {$\Phi \left(\text{dbo:nationality}\right)$};
% Text Node
\draw (352,76.4) node [anchor=north west][inner sep=0.75pt]  [font=\scriptsize]  {$\mathcal{B}_{s}$};
% Text Node
\draw (424,75.4) node [anchor=north west][inner sep=0.75pt]  [font=\scriptsize]  {$\Phi \left(\text{dbo:capital}\right)$};
% Text Node
\draw (368,107.4) node [anchor=north west][inner sep=0.75pt]  [font=\scriptsize]  {$\Phi \left(\text{dbo:currency}\right)$};
% Text Node
\draw (133.3,181.65) node [anchor=north west][inner sep=0.75pt]    {$1000$};
% Text Node
\draw (198.05,201.4) node [anchor=north west][inner sep=0.75pt]    {$1000$};
% Text Node
\draw (216.3,161.65) node [anchor=north west][inner sep=0.75pt]    {$50$};
% Text Node
\draw (172.3,179.45) node [anchor=north west][inner sep=0.75pt]    {$\cdot $};
% Text Node
\draw (193.97,159.65) node [anchor=north west][inner sep=0.75pt]    {$\cap $};

\end{tikzpicture}

%% file: figures/cardex5.tex
\tikzset{every picture/.style={line width=0.75pt}} %set default line width to 0.75pt        

\begin{tikzpicture}[x=0.6pt,y=0.6pt,yscale=-1,xscale=1]
%uncomment if require: \path (0,887); %set diagram left start at 0, and has height of 887

%Rounded Rect [id:dp966936906095217] 
\draw  [dash pattern={on 4.5pt off 4.5pt}] (107.78,72.03) .. controls (107.78,61.52) and (116.3,53) .. (126.82,53) -- (294.97,53) .. controls (305.48,53) and (314,61.52) .. (314,72.03) -- (314,129.14) .. controls (314,139.65) and (305.48,148.17) .. (294.97,148.17) -- (126.82,148.17) .. controls (116.3,148.17) and (107.78,139.65) .. (107.78,129.14) -- cycle ;
%Rounded Rect [id:dp7660610101797349] 
\draw  [dash pattern={on 4.5pt off 4.5pt}] (327.78,72.03) .. controls (327.78,61.52) and (336.3,53) .. (346.82,53) -- (514.97,53) .. controls (525.48,53) and (534,61.52) .. (534,72.03) -- (534,129.14) .. controls (534,139.65) and (525.48,148.17) .. (514.97,148.17) -- (346.82,148.17) .. controls (336.3,148.17) and (327.78,139.65) .. (327.78,129.14) -- cycle ;
%Rounded Rect [id:dp9677574419124895] 
\draw  [color={rgb, 255:red, 184; green, 233; blue, 134 }  ,draw opacity=1 ][line width=1.5]  (286,162.8) .. controls (286,159.6) and (288.6,157) .. (291.8,157) -- (327.2,157) .. controls (330.4,157) and (333,159.6) .. (333,162.8) -- (333,180.2) .. controls (333,183.4) and (330.4,186) .. (327.2,186) -- (291.8,186) .. controls (288.6,186) and (286,183.4) .. (286,180.2) -- cycle ;
%Rounded Rect [id:dp8099309515473015] 
\draw  [color={rgb, 255:red, 219; green, 126; blue, 136 }  ,draw opacity=1 ][line width=1.5]  (286,203.8) .. controls (286,200.6) and (288.6,198) .. (291.8,198) -- (327.2,198) .. controls (330.4,198) and (333,200.6) .. (333,203.8) -- (333,221.2) .. controls (333,224.4) and (330.4,227) .. (327.2,227) -- (291.8,227) .. controls (288.6,227) and (286,224.4) .. (286,221.2) -- cycle ;
%Straight Lines [id:da6457704907208008] 
\draw [line width=1.5]    (285,192) -- (332,192) ;
%Rounded Rect [id:dp09481359126316191] 
\draw  [color={rgb, 255:red, 155; green, 155; blue, 155 }  ,draw opacity=1 ][line width=1.5]  (167,114.25) .. controls (167,111.1) and (169.55,108.56) .. (172.69,108.56) -- (254.65,108.56) .. controls (257.79,108.56) and (260.34,111.1) .. (260.34,114.25) -- (260.34,131.31) .. controls (260.34,134.45) and (257.79,137) .. (254.65,137) -- (172.69,137) .. controls (169.55,137) and (167,134.45) .. (167,131.31) -- cycle ;
%Rounded Rect [id:dp9518362930197091] 
\draw  [color={rgb, 255:red, 155; green, 155; blue, 155 }  ,draw opacity=1 ][dash pattern={on 5.63pt off 4.5pt}][line width=1.5]  (334,82.75) .. controls (334,79.6) and (336.55,77.06) .. (339.69,77.06) -- (421.65,77.06) .. controls (424.79,77.06) and (427.34,79.6) .. (427.34,82.75) -- (427.34,99.81) .. controls (427.34,102.95) and (424.79,105.5) .. (421.65,105.5) -- (339.69,105.5) .. controls (336.55,105.5) and (334,102.95) .. (334,99.81) -- cycle ;
%Rounded Rect [id:dp3984632865591402] 
\draw  [color={rgb, 255:red, 184; green, 233; blue, 134 }  ,draw opacity=1 ][line width=1.5]  (214,82.25) .. controls (214,79.1) and (216.55,76.56) .. (219.69,76.56) -- (301.65,76.56) .. controls (304.79,76.56) and (307.34,79.1) .. (307.34,82.25) -- (307.34,99.31) .. controls (307.34,102.45) and (304.79,105) .. (301.65,105) -- (219.69,105) .. controls (216.55,105) and (214,102.45) .. (214,99.31) -- cycle ;
%Rounded Rect [id:dp6523890997176989] 
\draw  [color={rgb, 255:red, 219; green, 126; blue, 136 }  ,draw opacity=1 ][line width=1.5]  (114.25,82.25) .. controls (114.25,79.1) and (116.8,76.56) .. (119.94,76.56) -- (201.9,76.56) .. controls (205.04,76.56) and (207.59,79.1) .. (207.59,82.25) -- (207.59,99.31) .. controls (207.59,102.45) and (205.04,105) .. (201.9,105) -- (119.94,105) .. controls (116.8,105) and (114.25,102.45) .. (114.25,99.31) -- cycle ;
%Rounded Rect [id:dp6896525185884734] 
\draw  [color={rgb, 255:red, 219; green, 126; blue, 136 }  ,draw opacity=1 ][dash pattern={on 5.63pt off 4.5pt}][line width=1.5]  (348,162.8) .. controls (348,159.6) and (350.6,157) .. (353.8,157) -- (379.2,157) .. controls (382.4,157) and (385,159.6) .. (385,162.8) -- (385,180.2) .. controls (385,183.4) and (382.4,186) .. (379.2,186) -- (353.8,186) .. controls (350.6,186) and (348,183.4) .. (348,180.2) -- cycle ;
%Rounded Rect [id:dp4866042577199351] 
\draw  [color={rgb, 255:red, 155; green, 155; blue, 155 }  ,draw opacity=1 ][dash pattern={on 5.63pt off 4.5pt}][line width=1.5]  (348,202.8) .. controls (348,199.6) and (350.6,197) .. (353.8,197) -- (380.2,197) .. controls (383.4,197) and (386,199.6) .. (386,202.8) -- (386,220.2) .. controls (386,223.4) and (383.4,226) .. (380.2,226) -- (353.8,226) .. controls (350.6,226) and (348,223.4) .. (348,220.2) -- cycle ;
%Straight Lines [id:da1913488438425679] 
\draw [line width=1.5]    (347,192) -- (385,192) ;
%Rounded Rect [id:dp8275152382061634] 
\draw  [color={rgb, 255:red, 219; green, 126; blue, 136 }  ,draw opacity=1 ][dash pattern={on 5.63pt off 4.5pt}][line width=1.5]  (434,83.25) .. controls (434,80.1) and (436.55,77.56) .. (439.69,77.56) -- (521.65,77.56) .. controls (524.79,77.56) and (527.34,80.1) .. (527.34,83.25) -- (527.34,100.31) .. controls (527.34,103.45) and (524.79,106) .. (521.65,106) -- (439.69,106) .. controls (436.55,106) and (434,103.45) .. (434,100.31) -- cycle ;
%Rounded Rect [id:dp15880769336677258] 
\draw  [color={rgb, 255:red, 184; green, 233; blue, 134 }  ,draw opacity=1 ][dash pattern={on 5.63pt off 4.5pt}][line width=1.5]  (400,162.8) .. controls (400,159.6) and (402.6,157) .. (405.8,157) -- (431.2,157) .. controls (434.4,157) and (437,159.6) .. (437,162.8) -- (437,180.2) .. controls (437,183.4) and (434.4,186) .. (431.2,186) -- (405.8,186) .. controls (402.6,186) and (400,183.4) .. (400,180.2) -- cycle ;
%Rounded Rect [id:dp32502306086133137] 
\draw  [color={rgb, 255:red, 155; green, 155; blue, 155 }  ,draw opacity=1 ][dash pattern={on 5.63pt off 4.5pt}][line width=1.5]  (400,202.8) .. controls (400,199.6) and (402.6,197) .. (405.8,197) -- (432.2,197) .. controls (435.4,197) and (438,199.6) .. (438,202.8) -- (438,220.2) .. controls (438,223.4) and (435.4,226) .. (432.2,226) -- (405.8,226) .. controls (402.6,226) and (400,223.4) .. (400,220.2) -- cycle ;
%Straight Lines [id:da6841477099224821] 
\draw [line width=1.5]    (399,192) -- (437,192) ;
%Rounded Rect [id:dp40626226972839885] 
\draw  [color={rgb, 255:red, 184; green, 233; blue, 134 }  ,draw opacity=1 ][dash pattern={on 5.63pt off 4.5pt}][line width=1.5]  (334,115) .. controls (334,111.85) and (336.55,109.31) .. (339.69,109.31) -- (421.65,109.31) .. controls (424.79,109.31) and (427.34,111.85) .. (427.34,115) -- (427.34,132.06) .. controls (427.34,135.2) and (424.79,137.75) .. (421.65,137.75) -- (339.69,137.75) .. controls (336.55,137.75) and (334,135.2) .. (334,132.06) -- cycle ;
%Straight Lines [id:da914736403427958] 
\draw [color={rgb, 255:red, 248; green, 231; blue, 28 }  ,draw opacity=1 ][line width=1.5]    (186.6,137.2) -- (215.8,163) ;
%Straight Lines [id:da9578155562955049] 
\draw [color={rgb, 255:red, 248; green, 231; blue, 28 }  ,draw opacity=1 ][line width=1.5]    (339.69,105.5) -- (256.2,160.6) ;
%Shape: Rectangle [id:dp5033592828996503] 
\draw  [color={rgb, 255:red, 0; green, 0; blue, 0 }  ,draw opacity=1 ] (137.77,96.56) -- (159.55,96.56) -- (159.55,102) -- (137.77,102) -- cycle ;
%Shape: Rectangle [id:dp6553833836363036] 
\draw   (143.22,96.56) -- (148.66,96.56) -- (148.66,102) -- (143.22,102) -- cycle ;
%Shape: Rectangle [id:dp3982839769981218] 
\draw   (154.1,96.56) -- (159.55,96.56) -- (159.55,102) -- (154.1,102) -- cycle ;
%Shape: Rectangle [id:dp9144492883396299] 
\draw  [color={rgb, 255:red, 0; green, 0; blue, 0 }  ,draw opacity=1 ] (116,96.56) -- (137.77,96.56) -- (137.77,102) -- (116,102) -- cycle ;
%Shape: Rectangle [id:dp4157022946752308] 
\draw   (121.44,96.56) -- (126.89,96.56) -- (126.89,102) -- (121.44,102) -- cycle ;
%Shape: Rectangle [id:dp8116610445909572] 
\draw   (132.33,96.56) -- (137.77,96.56) -- (137.77,102) -- (132.33,102) -- cycle ;
%Shape: Rectangle [id:dp9316200870120502] 
\draw  [fill={rgb, 255:red, 0; green, 0; blue, 0 }  ,fill opacity=1 ] (121.33,96.56) -- (126.77,96.56) -- (126.77,102) -- (121.33,102) -- cycle ;
%Shape: Rectangle [id:dp6918959225556934] 
\draw  [fill={rgb, 255:red, 0; green, 0; blue, 0 }  ,fill opacity=1 ] (115.89,96.56) -- (121.33,96.56) -- (121.33,102) -- (115.89,102) -- cycle ;
%Shape: Rectangle [id:dp42369233483847635] 
\draw  [color={rgb, 255:red, 0; green, 0; blue, 0 }  ,draw opacity=1 ] (183.23,96.56) -- (205,96.56) -- (205,102) -- (183.23,102) -- cycle ;
%Shape: Rectangle [id:dp10141093701951398] 
\draw   (188.67,96.56) -- (194.11,96.56) -- (194.11,102) -- (188.67,102) -- cycle ;
%Shape: Rectangle [id:dp5913167909212861] 
\draw   (199.56,96.56) -- (205,96.56) -- (205,102) -- (199.56,102) -- cycle ;
%Shape: Rectangle [id:dp7762471337039698] 
\draw  [color={rgb, 255:red, 0; green, 0; blue, 0 }  ,draw opacity=1 ] (161.45,96.56) -- (183.23,96.56) -- (183.23,102) -- (161.45,102) -- cycle ;
%Shape: Rectangle [id:dp948550097899579] 
\draw   (166.9,96.56) -- (172.34,96.56) -- (172.34,102) -- (166.9,102) -- cycle ;
%Shape: Rectangle [id:dp4020811479441013] 
\draw   (177.78,96.56) -- (183.23,96.56) -- (183.23,102) -- (177.78,102) -- cycle ;
%Shape: Rectangle [id:dp273711545345393] 
\draw  [fill={rgb, 255:red, 0; green, 0; blue, 0 }  ,fill opacity=1 ] (177.23,96.56) -- (182.67,96.56) -- (182.67,102) -- (177.23,102) -- cycle ;
%Shape: Rectangle [id:dp26466019266299834] 
\draw  [fill={rgb, 255:red, 0; green, 0; blue, 0 }  ,fill opacity=1 ] (171.9,96.56) -- (177.34,96.56) -- (177.34,102) -- (171.9,102) -- cycle ;
%Shape: Rectangle [id:dp10744610918959663] 
\draw  [fill={rgb, 255:red, 0; green, 0; blue, 0 }  ,fill opacity=1 ] (126.22,96.56) -- (131.66,96.56) -- (131.66,102) -- (126.22,102) -- cycle ;
%Shape: Rectangle [id:dp9517810104354466] 
\draw  [color={rgb, 255:red, 0; green, 0; blue, 0 }  ,draw opacity=1 ] (237.77,96.56) -- (259.55,96.56) -- (259.55,102) -- (237.77,102) -- cycle ;
%Shape: Rectangle [id:dp5177455952240066] 
\draw   (243.22,96.56) -- (248.66,96.56) -- (248.66,102) -- (243.22,102) -- cycle ;
%Shape: Rectangle [id:dp6158930929706008] 
\draw   (254.1,96.56) -- (259.55,96.56) -- (259.55,102) -- (254.1,102) -- cycle ;
%Shape: Rectangle [id:dp0008549561288874186] 
\draw  [color={rgb, 255:red, 0; green, 0; blue, 0 }  ,draw opacity=1 ] (216,96.56) -- (237.77,96.56) -- (237.77,102) -- (216,102) -- cycle ;
%Shape: Rectangle [id:dp6534924826034666] 
\draw   (221.44,96.56) -- (226.89,96.56) -- (226.89,102) -- (221.44,102) -- cycle ;
%Shape: Rectangle [id:dp5179208529319574] 
\draw   (232.33,96.56) -- (237.77,96.56) -- (237.77,102) -- (232.33,102) -- cycle ;
%Shape: Rectangle [id:dp6370319617596009] 
\draw  [fill={rgb, 255:red, 0; green, 0; blue, 0 }  ,fill opacity=1 ] (232.33,96.56) -- (237.77,96.56) -- (237.77,102) -- (232.33,102) -- cycle ;
%Shape: Rectangle [id:dp1029005218524437] 
\draw  [fill={rgb, 255:red, 0; green, 0; blue, 0 }  ,fill opacity=1 ] (221.89,96.56) -- (227.33,96.56) -- (227.33,102) -- (221.89,102) -- cycle ;
%Shape: Rectangle [id:dp3664216631066237] 
\draw  [color={rgb, 255:red, 0; green, 0; blue, 0 }  ,draw opacity=1 ] (283.23,96.56) -- (305,96.56) -- (305,102) -- (283.23,102) -- cycle ;
%Shape: Rectangle [id:dp3598114339415095] 
\draw   (288.67,96.56) -- (294.11,96.56) -- (294.11,102) -- (288.67,102) -- cycle ;
%Shape: Rectangle [id:dp7589713181196447] 
\draw   (299.56,96.56) -- (305,96.56) -- (305,102) -- (299.56,102) -- cycle ;
%Shape: Rectangle [id:dp7339827782797104] 
\draw  [color={rgb, 255:red, 0; green, 0; blue, 0 }  ,draw opacity=1 ] (261.45,96.56) -- (283.23,96.56) -- (283.23,102) -- (261.45,102) -- cycle ;
%Shape: Rectangle [id:dp09645265406281245] 
\draw   (266.9,96.56) -- (272.34,96.56) -- (272.34,102) -- (266.9,102) -- cycle ;
%Shape: Rectangle [id:dp3029351948494421] 
\draw   (277.78,96.56) -- (283.23,96.56) -- (283.23,102) -- (277.78,102) -- cycle ;
%Shape: Rectangle [id:dp9808635469122418] 
\draw  [fill={rgb, 255:red, 0; green, 0; blue, 0 }  ,fill opacity=1 ] (272.23,96.56) -- (277.67,96.56) -- (277.67,102) -- (272.23,102) -- cycle ;
%Shape: Rectangle [id:dp07373327449708711] 
\draw  [fill={rgb, 255:red, 0; green, 0; blue, 0 }  ,fill opacity=1 ] (299.56,96.56) -- (305,96.56) -- (305,102) -- (299.56,102) -- cycle ;
%Shape: Rectangle [id:dp43274645973448167] 
\draw  [fill={rgb, 255:red, 0; green, 0; blue, 0 }  ,fill opacity=1 ] (261.9,96.56) -- (267.34,96.56) -- (267.34,102) -- (261.9,102) -- cycle ;
%Shape: Rectangle [id:dp768224287641539] 
\draw  [fill={rgb, 255:red, 0; green, 0; blue, 0 }  ,fill opacity=1 ] (249.22,96.56) -- (254.66,96.56) -- (254.66,102) -- (249.22,102) -- cycle ;
%Shape: Rectangle [id:dp9553721861682788] 
\draw  [color={rgb, 255:red, 0; green, 0; blue, 0 }  ,draw opacity=1 ] (190.77,129.56) -- (212.55,129.56) -- (212.55,135) -- (190.77,135) -- cycle ;
%Shape: Rectangle [id:dp9062454968628401] 
\draw   (196.22,129.56) -- (201.66,129.56) -- (201.66,135) -- (196.22,135) -- cycle ;
%Shape: Rectangle [id:dp6193648522647837] 
\draw   (207.1,129.56) -- (212.55,129.56) -- (212.55,135) -- (207.1,135) -- cycle ;
%Shape: Rectangle [id:dp4765122930147768] 
\draw  [color={rgb, 255:red, 0; green, 0; blue, 0 }  ,draw opacity=1 ] (169,129.56) -- (190.77,129.56) -- (190.77,135) -- (169,135) -- cycle ;
%Shape: Rectangle [id:dp9362289576706566] 
\draw   (174.44,129.56) -- (179.89,129.56) -- (179.89,135) -- (174.44,135) -- cycle ;
%Shape: Rectangle [id:dp7492361340228328] 
\draw   (185.33,129.56) -- (190.77,129.56) -- (190.77,135) -- (185.33,135) -- cycle ;
%Shape: Rectangle [id:dp581312992445255] 
\draw  [fill={rgb, 255:red, 0; green, 0; blue, 0 }  ,fill opacity=1 ] (185.33,129.56) -- (190.77,129.56) -- (190.77,135) -- (185.33,135) -- cycle ;
%Shape: Rectangle [id:dp6789459981923872] 
\draw  [fill={rgb, 255:red, 0; green, 0; blue, 0 }  ,fill opacity=1 ] (168.89,129.56) -- (174.33,129.56) -- (174.33,135) -- (168.89,135) -- cycle ;
%Shape: Rectangle [id:dp4007526214560658] 
\draw  [color={rgb, 255:red, 0; green, 0; blue, 0 }  ,draw opacity=1 ] (236.23,129.56) -- (258,129.56) -- (258,135) -- (236.23,135) -- cycle ;
%Shape: Rectangle [id:dp6356269763108078] 
\draw   (241.67,129.56) -- (247.11,129.56) -- (247.11,135) -- (241.67,135) -- cycle ;
%Shape: Rectangle [id:dp8735843021164262] 
\draw   (252.56,129.56) -- (258,129.56) -- (258,135) -- (252.56,135) -- cycle ;
%Shape: Rectangle [id:dp708702377659593] 
\draw  [color={rgb, 255:red, 0; green, 0; blue, 0 }  ,draw opacity=1 ] (214.45,129.56) -- (236.23,129.56) -- (236.23,135) -- (214.45,135) -- cycle ;
%Shape: Rectangle [id:dp15663738818746387] 
\draw   (219.9,129.56) -- (225.34,129.56) -- (225.34,135) -- (219.9,135) -- cycle ;
%Shape: Rectangle [id:dp7595969010152642] 
\draw   (230.78,129.56) -- (236.23,129.56) -- (236.23,135) -- (230.78,135) -- cycle ;
%Shape: Rectangle [id:dp9978326203270401] 
\draw  [fill={rgb, 255:red, 0; green, 0; blue, 0 }  ,fill opacity=1 ] (247.23,129.56) -- (252.67,129.56) -- (252.67,135) -- (247.23,135) -- cycle ;
%Shape: Rectangle [id:dp5918010892832809] 
\draw  [fill={rgb, 255:red, 0; green, 0; blue, 0 }  ,fill opacity=1 ] (252.56,129.56) -- (258,129.56) -- (258,135) -- (252.56,135) -- cycle ;
%Shape: Rectangle [id:dp1370164705963891] 
\draw  [fill={rgb, 255:red, 0; green, 0; blue, 0 }  ,fill opacity=1 ] (214.9,129.56) -- (220.34,129.56) -- (220.34,135) -- (214.9,135) -- cycle ;
%Shape: Rectangle [id:dp6458506209328039] 
\draw  [fill={rgb, 255:red, 0; green, 0; blue, 0 }  ,fill opacity=1 ] (202.22,129.56) -- (207.66,129.56) -- (207.66,135) -- (202.22,135) -- cycle ;
%Shape: Rectangle [id:dp9660146482595038] 
\draw  [fill={rgb, 255:red, 0; green, 0; blue, 0 }  ,fill opacity=1 ] (148.22,96.56) -- (153.66,96.56) -- (153.66,102) -- (148.22,102) -- cycle ;
%Shape: Rectangle [id:dp6706727097204953] 
\draw  [color={rgb, 255:red, 0; green, 0; blue, 0 }  ,draw opacity=1 ] (455.77,130.16) -- (477.55,130.16) -- (477.55,135.6) -- (455.77,135.6) -- cycle ;
%Shape: Rectangle [id:dp8370191941510071] 
\draw   (461.22,130.16) -- (466.66,130.16) -- (466.66,135.6) -- (461.22,135.6) -- cycle ;
%Shape: Rectangle [id:dp6930658619189607] 
\draw   (472.1,130.16) -- (477.55,130.16) -- (477.55,135.6) -- (472.1,135.6) -- cycle ;
%Shape: Rectangle [id:dp8367772425839086] 
\draw  [color={rgb, 255:red, 0; green, 0; blue, 0 }  ,draw opacity=1 ] (434,130.16) -- (455.77,130.16) -- (455.77,135.6) -- (434,135.6) -- cycle ;
%Shape: Rectangle [id:dp469363355299938] 
\draw   (439.44,130.16) -- (444.89,130.16) -- (444.89,135.6) -- (439.44,135.6) -- cycle ;
%Shape: Rectangle [id:dp37721783644624274] 
\draw   (450.33,130.16) -- (455.77,130.16) -- (455.77,135.6) -- (450.33,135.6) -- cycle ;
%Shape: Rectangle [id:dp9601085604236583] 
\draw  [fill={rgb, 255:red, 0; green, 0; blue, 0 }  ,fill opacity=1 ] (450.33,130.16) -- (455.77,130.16) -- (455.77,135.6) -- (450.33,135.6) -- cycle ;
%Shape: Rectangle [id:dp01927932715009517] 
\draw  [fill={rgb, 255:red, 0; green, 0; blue, 0 }  ,fill opacity=1 ] (444.89,130.16) -- (450.33,130.16) -- (450.33,135.6) -- (444.89,135.6) -- cycle ;
%Shape: Rectangle [id:dp4150106970351438] 
\draw  [color={rgb, 255:red, 0; green, 0; blue, 0 }  ,draw opacity=1 ] (501.23,130.16) -- (523,130.16) -- (523,135.6) -- (501.23,135.6) -- cycle ;
%Shape: Rectangle [id:dp6172673428971805] 
\draw   (506.67,130.16) -- (512.11,130.16) -- (512.11,135.6) -- (506.67,135.6) -- cycle ;
%Shape: Rectangle [id:dp1403114585707287] 
\draw   (517.56,130.16) -- (523,130.16) -- (523,135.6) -- (517.56,135.6) -- cycle ;
%Shape: Rectangle [id:dp7405688121908836] 
\draw  [color={rgb, 255:red, 0; green, 0; blue, 0 }  ,draw opacity=1 ] (479.45,130.16) -- (501.23,130.16) -- (501.23,135.6) -- (479.45,135.6) -- cycle ;
%Shape: Rectangle [id:dp7687904931759523] 
\draw   (484.9,130.16) -- (490.34,130.16) -- (490.34,135.6) -- (484.9,135.6) -- cycle ;
%Shape: Rectangle [id:dp49708085089448595] 
\draw   (495.78,130.16) -- (501.23,130.16) -- (501.23,135.6) -- (495.78,135.6) -- cycle ;
%Shape: Rectangle [id:dp6213890681250885] 
\draw  [fill={rgb, 255:red, 0; green, 0; blue, 0 }  ,fill opacity=1 ] (501.23,130.16) -- (506.67,130.16) -- (506.67,135.6) -- (501.23,135.6) -- cycle ;
%Shape: Rectangle [id:dp4698126604016818] 
\draw  [fill={rgb, 255:red, 0; green, 0; blue, 0 }  ,fill opacity=1 ] (512.56,130.16) -- (518,130.16) -- (518,135.6) -- (512.56,135.6) -- cycle ;
%Shape: Rectangle [id:dp5357102202581182] 
\draw  [fill={rgb, 255:red, 0; green, 0; blue, 0 }  ,fill opacity=1 ] (478.9,130.16) -- (484.34,130.16) -- (484.34,135.6) -- (478.9,135.6) -- cycle ;
%Shape: Rectangle [id:dp8112386787534698] 
\draw  [fill={rgb, 255:red, 0; green, 0; blue, 0 }  ,fill opacity=1 ] (440.22,130.16) -- (445.66,130.16) -- (445.66,135.6) -- (440.22,135.6) -- cycle ;
%Shape: Rectangle [id:dp22943811015177873] 
\draw  [color={rgb, 255:red, 0; green, 0; blue, 0 }  ,draw opacity=1 ] (356.77,96.56) -- (378.55,96.56) -- (378.55,102) -- (356.77,102) -- cycle ;
%Shape: Rectangle [id:dp6361270177370594] 
\draw   (362.22,96.56) -- (367.66,96.56) -- (367.66,102) -- (362.22,102) -- cycle ;
%Shape: Rectangle [id:dp7853212468270407] 
\draw   (373.1,96.56) -- (378.55,96.56) -- (378.55,102) -- (373.1,102) -- cycle ;
%Shape: Rectangle [id:dp08720934353424836] 
\draw  [color={rgb, 255:red, 0; green, 0; blue, 0 }  ,draw opacity=1 ] (335,96.56) -- (356.77,96.56) -- (356.77,102) -- (335,102) -- cycle ;
%Shape: Rectangle [id:dp3321333917575745] 
\draw   (340.44,96.56) -- (345.89,96.56) -- (345.89,102) -- (340.44,102) -- cycle ;
%Shape: Rectangle [id:dp24068770682179164] 
\draw   (351.33,96.56) -- (356.77,96.56) -- (356.77,102) -- (351.33,102) -- cycle ;
%Shape: Rectangle [id:dp5797492091021884] 
\draw  [fill={rgb, 255:red, 0; green, 0; blue, 0 }  ,fill opacity=1 ] (373.33,96.56) -- (378.77,96.56) -- (378.77,102) -- (373.33,102) -- cycle ;
%Shape: Rectangle [id:dp18847322091602303] 
\draw  [fill={rgb, 255:red, 0; green, 0; blue, 0 }  ,fill opacity=1 ] (334.89,96.56) -- (340.33,96.56) -- (340.33,102) -- (334.89,102) -- cycle ;
%Shape: Rectangle [id:dp21050149951965713] 
\draw  [color={rgb, 255:red, 0; green, 0; blue, 0 }  ,draw opacity=1 ] (402.23,96.56) -- (424,96.56) -- (424,102) -- (402.23,102) -- cycle ;
%Shape: Rectangle [id:dp6069674560930608] 
\draw   (407.67,96.56) -- (413.11,96.56) -- (413.11,102) -- (407.67,102) -- cycle ;
%Shape: Rectangle [id:dp819159797647965] 
\draw   (418.56,96.56) -- (424,96.56) -- (424,102) -- (418.56,102) -- cycle ;
%Shape: Rectangle [id:dp2692464850184928] 
\draw  [color={rgb, 255:red, 0; green, 0; blue, 0 }  ,draw opacity=1 ] (380.45,96.56) -- (402.23,96.56) -- (402.23,102) -- (380.45,102) -- cycle ;
%Shape: Rectangle [id:dp5064849269231425] 
\draw   (385.9,96.56) -- (391.34,96.56) -- (391.34,102) -- (385.9,102) -- cycle ;
%Shape: Rectangle [id:dp004334248459354484] 
\draw   (396.78,96.56) -- (402.23,96.56) -- (402.23,102) -- (396.78,102) -- cycle ;
%Shape: Rectangle [id:dp9858115206602445] 
\draw  [fill={rgb, 255:red, 0; green, 0; blue, 0 }  ,fill opacity=1 ] (402.23,96.56) -- (407.67,96.56) -- (407.67,102) -- (402.23,102) -- cycle ;
%Shape: Rectangle [id:dp6579666640967087] 
\draw  [fill={rgb, 255:red, 0; green, 0; blue, 0 }  ,fill opacity=1 ] (418.56,96.56) -- (424,96.56) -- (424,102) -- (418.56,102) -- cycle ;
%Shape: Rectangle [id:dp4073114928643373] 
\draw  [fill={rgb, 255:red, 0; green, 0; blue, 0 }  ,fill opacity=1 ] (380.9,96.56) -- (386.34,96.56) -- (386.34,102) -- (380.9,102) -- cycle ;
%Shape: Rectangle [id:dp0007217323904271655] 
\draw  [fill={rgb, 255:red, 0; green, 0; blue, 0 }  ,fill opacity=1 ] (368.22,96.56) -- (373.66,96.56) -- (373.66,102) -- (368.22,102) -- cycle ;
%Shape: Rectangle [id:dp14997884288616092] 
\draw  [color={rgb, 255:red, 0; green, 0; blue, 0 }  ,draw opacity=1 ] (456.77,96.56) -- (478.55,96.56) -- (478.55,102) -- (456.77,102) -- cycle ;
%Shape: Rectangle [id:dp5188069347602424] 
\draw   (462.22,96.56) -- (467.66,96.56) -- (467.66,102) -- (462.22,102) -- cycle ;
%Shape: Rectangle [id:dp6392285876718438] 
\draw   (473.1,96.56) -- (478.55,96.56) -- (478.55,102) -- (473.1,102) -- cycle ;
%Shape: Rectangle [id:dp11459934654110004] 
\draw  [color={rgb, 255:red, 0; green, 0; blue, 0 }  ,draw opacity=1 ] (435,96.56) -- (456.77,96.56) -- (456.77,102) -- (435,102) -- cycle ;
%Shape: Rectangle [id:dp6812832269092226] 
\draw   (440.44,96.56) -- (445.89,96.56) -- (445.89,102) -- (440.44,102) -- cycle ;
%Shape: Rectangle [id:dp5624518012105442] 
\draw   (451.33,96.56) -- (456.77,96.56) -- (456.77,102) -- (451.33,102) -- cycle ;
%Shape: Rectangle [id:dp08376723156782828] 
\draw  [fill={rgb, 255:red, 0; green, 0; blue, 0 }  ,fill opacity=1 ] (456.33,96.56) -- (461.77,96.56) -- (461.77,102) -- (456.33,102) -- cycle ;
%Shape: Rectangle [id:dp9443188152927212] 
\draw  [fill={rgb, 255:red, 0; green, 0; blue, 0 }  ,fill opacity=1 ] (434.89,96.56) -- (440.33,96.56) -- (440.33,102) -- (434.89,102) -- cycle ;
%Shape: Rectangle [id:dp3869962032070253] 
\draw  [color={rgb, 255:red, 0; green, 0; blue, 0 }  ,draw opacity=1 ] (502.23,96.56) -- (524,96.56) -- (524,102) -- (502.23,102) -- cycle ;
%Shape: Rectangle [id:dp8311600279861484] 
\draw   (507.67,96.56) -- (513.11,96.56) -- (513.11,102) -- (507.67,102) -- cycle ;
%Shape: Rectangle [id:dp8942104551175005] 
\draw   (518.56,96.56) -- (524,96.56) -- (524,102) -- (518.56,102) -- cycle ;
%Shape: Rectangle [id:dp157220850559468] 
\draw  [color={rgb, 255:red, 0; green, 0; blue, 0 }  ,draw opacity=1 ] (480.45,96.56) -- (502.23,96.56) -- (502.23,102) -- (480.45,102) -- cycle ;
%Shape: Rectangle [id:dp2420980444347961] 
\draw   (485.9,96.56) -- (491.34,96.56) -- (491.34,102) -- (485.9,102) -- cycle ;
%Shape: Rectangle [id:dp03169838334227337] 
\draw   (496.78,96.56) -- (502.23,96.56) -- (502.23,102) -- (496.78,102) -- cycle ;
%Shape: Rectangle [id:dp14322883191259572] 
\draw  [fill={rgb, 255:red, 0; green, 0; blue, 0 }  ,fill opacity=1 ] (497.23,96.56) -- (502.67,96.56) -- (502.67,102) -- (497.23,102) -- cycle ;
%Shape: Rectangle [id:dp557491502912643] 
\draw  [fill={rgb, 255:red, 0; green, 0; blue, 0 }  ,fill opacity=1 ] (518.56,96.56) -- (524,96.56) -- (524,102) -- (518.56,102) -- cycle ;
%Shape: Rectangle [id:dp9236036706669019] 
\draw  [fill={rgb, 255:red, 0; green, 0; blue, 0 }  ,fill opacity=1 ] (491.9,96.56) -- (497.34,96.56) -- (497.34,102) -- (491.9,102) -- cycle ;
%Shape: Rectangle [id:dp03810594561912173] 
\draw  [fill={rgb, 255:red, 0; green, 0; blue, 0 }  ,fill opacity=1 ] (462.22,96.56) -- (467.66,96.56) -- (467.66,102) -- (462.22,102) -- cycle ;
%Shape: Rectangle [id:dp009801702885619767] 
\draw  [color={rgb, 255:red, 0; green, 0; blue, 0 }  ,draw opacity=1 ] (357.77,129.56) -- (379.55,129.56) -- (379.55,135) -- (357.77,135) -- cycle ;
%Shape: Rectangle [id:dp12819581187815232] 
\draw   (363.22,129.56) -- (368.66,129.56) -- (368.66,135) -- (363.22,135) -- cycle ;
%Shape: Rectangle [id:dp4720455255371949] 
\draw   (374.1,129.56) -- (379.55,129.56) -- (379.55,135) -- (374.1,135) -- cycle ;
%Shape: Rectangle [id:dp7462260979089286] 
\draw  [color={rgb, 255:red, 0; green, 0; blue, 0 }  ,draw opacity=1 ] (336,129.56) -- (357.77,129.56) -- (357.77,135) -- (336,135) -- cycle ;
%Shape: Rectangle [id:dp40177025872055194] 
\draw   (341.44,129.56) -- (346.89,129.56) -- (346.89,135) -- (341.44,135) -- cycle ;
%Shape: Rectangle [id:dp629788628901606] 
\draw   (352.33,129.56) -- (357.77,129.56) -- (357.77,135) -- (352.33,135) -- cycle ;
%Shape: Rectangle [id:dp7715994681018318] 
\draw  [fill={rgb, 255:red, 0; green, 0; blue, 0 }  ,fill opacity=1 ] (352.33,129.56) -- (357.77,129.56) -- (357.77,135) -- (352.33,135) -- cycle ;
%Shape: Rectangle [id:dp20826654895246788] 
\draw  [fill={rgb, 255:red, 0; green, 0; blue, 0 }  ,fill opacity=1 ] (340.89,129.56) -- (346.33,129.56) -- (346.33,135) -- (340.89,135) -- cycle ;
%Shape: Rectangle [id:dp9327222097833238] 
\draw  [color={rgb, 255:red, 0; green, 0; blue, 0 }  ,draw opacity=1 ] (403.23,129.56) -- (425,129.56) -- (425,135) -- (403.23,135) -- cycle ;
%Shape: Rectangle [id:dp16027622712367695] 
\draw   (408.67,129.56) -- (414.11,129.56) -- (414.11,135) -- (408.67,135) -- cycle ;
%Shape: Rectangle [id:dp18456953746654658] 
\draw   (419.56,129.56) -- (425,129.56) -- (425,135) -- (419.56,135) -- cycle ;
%Shape: Rectangle [id:dp46731153990051155] 
\draw  [color={rgb, 255:red, 0; green, 0; blue, 0 }  ,draw opacity=1 ] (381.45,129.56) -- (403.23,129.56) -- (403.23,135) -- (381.45,135) -- cycle ;
%Shape: Rectangle [id:dp786624669100937] 
\draw   (386.9,129.56) -- (392.34,129.56) -- (392.34,135) -- (386.9,135) -- cycle ;
%Shape: Rectangle [id:dp491334523307034] 
\draw   (397.78,129.56) -- (403.23,129.56) -- (403.23,135) -- (397.78,135) -- cycle ;
%Shape: Rectangle [id:dp6990096772418418] 
\draw  [fill={rgb, 255:red, 0; green, 0; blue, 0 }  ,fill opacity=1 ] (403.23,129.56) -- (408.67,129.56) -- (408.67,135) -- (403.23,135) -- cycle ;
%Shape: Rectangle [id:dp05942783495456516] 
\draw  [fill={rgb, 255:red, 0; green, 0; blue, 0 }  ,fill opacity=1 ] (419.56,129.56) -- (425,129.56) -- (425,135) -- (419.56,135) -- cycle ;
%Shape: Rectangle [id:dp19611475546243518] 
\draw  [fill={rgb, 255:red, 0; green, 0; blue, 0 }  ,fill opacity=1 ] (407.9,129.56) -- (413.34,129.56) -- (413.34,135) -- (407.9,135) -- cycle ;
%Shape: Rectangle [id:dp44094289564908695] 
\draw  [fill={rgb, 255:red, 0; green, 0; blue, 0 }  ,fill opacity=1 ] (363.22,129.56) -- (368.66,129.56) -- (368.66,135) -- (363.22,135) -- cycle ;
%Rounded Rect [id:dp11400070986650868] 
\draw  [color={rgb, 255:red, 219; green, 126; blue, 136 }  ,draw opacity=1 ][line width=1.5]  (146.3,182.05) .. controls (146.3,178.85) and (148.9,176.25) .. (152.1,176.25) -- (186.2,176.25) .. controls (189.4,176.25) and (192,178.85) .. (192,182.05) -- (192,199.45) .. controls (192,202.65) and (189.4,205.25) .. (186.2,205.25) -- (152.1,205.25) .. controls (148.9,205.25) and (146.3,202.65) .. (146.3,199.45) -- cycle ;
%Rounded Rect [id:dp03113644602908827] 
\draw  [color={rgb, 255:red, 155; green, 155; blue, 155 }  ,draw opacity=1 ][line width=1.5]  (212.3,202.05) .. controls (212.3,198.85) and (214.9,196.25) .. (218.1,196.25) -- (252.4,196.25) .. controls (255.6,196.25) and (258.2,198.85) .. (258.2,202.05) -- (258.2,219.45) .. controls (258.2,222.65) and (255.6,225.25) .. (252.4,225.25) -- (218.1,225.25) .. controls (214.9,225.25) and (212.3,222.65) .. (212.3,219.45) -- cycle ;
%Straight Lines [id:da6540614729615439] 
\draw [line width=1.5]    (204.3,191.25) -- (265.6,191.25) ;
%Rounded Rect [id:dp05496390135064089] 
\draw  [draw opacity=0][fill={rgb, 255:red, 248; green, 231; blue, 28 }  ,fill opacity=1 ][line width=1.5]  (205.3,162.05) .. controls (205.3,158.85) and (207.9,156.25) .. (211.1,156.25) -- (259.8,156.25) .. controls (263,156.25) and (265.6,158.85) .. (265.6,162.05) -- (265.6,179.45) .. controls (265.6,182.65) and (263,185.25) .. (259.8,185.25) -- (211.1,185.25) .. controls (207.9,185.25) and (205.3,182.65) .. (205.3,179.45) -- cycle ;
%Shape: Rectangle [id:dp29506669754240156] 
\draw  [color={rgb, 255:red, 0; green, 0; blue, 0 }  ,draw opacity=1 ] (220.97,176.33) -- (231.3,176.33) -- (231.3,178.92) -- (220.97,178.92) -- cycle ;
%Shape: Rectangle [id:dp9708100224496945] 
\draw   (223.55,176.33) -- (226.13,176.33) -- (226.13,178.92) -- (223.55,178.92) -- cycle ;
%Shape: Rectangle [id:dp8144538916957773] 
\draw   (228.72,176.33) -- (231.3,176.33) -- (231.3,178.92) -- (228.72,178.92) -- cycle ;
%Shape: Rectangle [id:dp6269036810527563] 
\draw  [color={rgb, 255:red, 0; green, 0; blue, 0 }  ,draw opacity=1 ] (210.63,176.33) -- (220.97,176.33) -- (220.97,178.92) -- (210.63,178.92) -- cycle ;
%Shape: Rectangle [id:dp6279952279458694] 
\draw   (213.22,176.33) -- (215.8,176.33) -- (215.8,178.92) -- (213.22,178.92) -- cycle ;
%Shape: Rectangle [id:dp7622311277782168] 
\draw   (218.38,176.33) -- (220.97,176.33) -- (220.97,178.92) -- (218.38,178.92) -- cycle ;
%Shape: Rectangle [id:dp3069752369808537] 
\draw  [fill={rgb, 255:red, 0; green, 0; blue, 0 }  ,fill opacity=1 ] (218.38,176.33) -- (220.97,176.33) -- (220.97,178.92) -- (218.38,178.92) -- cycle ;
%Shape: Rectangle [id:dp8957417200359413] 
\draw  [fill={rgb, 255:red, 0; green, 0; blue, 0 }  ,fill opacity=1 ] (215.8,176.33) -- (218.38,176.33) -- (218.38,178.92) -- (215.8,178.92) -- cycle ;
%Shape: Rectangle [id:dp5471789044826444] 
\draw  [fill={rgb, 255:red, 0; green, 0; blue, 0 }  ,fill opacity=1 ] (223.55,176.33) -- (226.13,176.33) -- (226.13,178.92) -- (223.55,178.92) -- cycle ;

% Text Node
\draw (185,55.4) node [anchor=north west][inner sep=0.75pt]    {$\mathcal{B}^{S}( f_{2})$};
% Text Node
\draw (405,55.4) node [anchor=north west][inner sep=0.75pt]    {$\mathcal{B}^{S}( f_{3})$};
% Text Node
\draw (270,182.2) node [anchor=north west][inner sep=0.75pt]    {$\cdot $};
% Text Node
\draw (292,164.4) node [anchor=north west][inner sep=0.75pt]    {$3000$};
% Text Node
\draw (292,204.4) node [anchor=north west][inner sep=0.75pt]    {$2000$};
% Text Node
\draw (332,182.2) node [anchor=north west][inner sep=0.75pt]    {$\cdot $};
% Text Node
\draw (354,164.4) node [anchor=north west][inner sep=0.75pt]    {$100$};
% Text Node
\draw (354,204.4) node [anchor=north west][inner sep=0.75pt]    {$100$};
% Text Node
\draw (384,182.2) node [anchor=north west][inner sep=0.75pt]    {$\cdot $};
% Text Node
\draw (406,164.4) node [anchor=north west][inner sep=0.75pt]    {$150$};
% Text Node
\draw (406,204.4) node [anchor=north west][inner sep=0.75pt]    {$100$};
% Text Node
\draw (444.8,181.2) node [anchor=north west][inner sep=0.75pt]    {$=225$};
% Text Node
\draw (152,79.4) node [anchor=north west][inner sep=0.75pt]  [font=\scriptsize]  {$\mathcal{B}_{s}$};
% Text Node
\draw (224,78.4) node [anchor=north west][inner sep=0.75pt]  [font=\scriptsize]  {$\Phi \left(\text{dbo:author}\right)$};
% Text Node
\draw (168,110.15) node [anchor=north west][inner sep=0.75pt]  [font=\scriptsize]  {$\Phi \left(\text{dbo:nationality}\right)$};
% Text Node
\draw (435,113) node [anchor=north west][inner sep=0.75pt]  [font=\scriptsize]  {$\Phi \left(\text{dbo:population}\right)$};
% Text Node
\draw (371,79.4) node [anchor=north west][inner sep=0.75pt]  [font=\scriptsize]  {$\mathcal{B}_{s}$};
% Text Node
\draw (443,78.4) node [anchor=north west][inner sep=0.75pt]  [font=\scriptsize]  {$\Phi \left(\text{dbo:capital}\right)$};
% Text Node
\draw (335,110.4) node [anchor=north west][inner sep=0.75pt]  [font=\scriptsize]  {$\Phi \left(\text{dbo:currency}\right)$};
% Text Node
\draw (152.3,183.65) node [anchor=north west][inner sep=0.75pt]    {$2000$};
% Text Node
\draw (218.05,203.4) node [anchor=north west][inner sep=0.75pt]    {$2000$};
% Text Node
\draw (235.3,163.65) node [anchor=north west][inner sep=0.75pt]    {$100$};
% Text Node
\draw (191.3,181.45) node [anchor=north west][inner sep=0.75pt]    {$\cdot $};
% Text Node
\draw (212.97,161.65) node [anchor=north west][inner sep=0.75pt]    {$\cap $};

\end{tikzpicture}

%% file: figures/cardex7.tex
\tikzset{every picture/.style={line width=0.75pt}} %set default line width to 0.75pt        

\begin{tikzpicture}[x=0.6pt,y=0.6pt,yscale=-1,xscale=1]
%uncomment if require: \path (0,233); %set diagram left start at 0, and has height of 233

%Rounded Rect [id:dp4944366072053469] 
\draw  [color={rgb, 255:red, 0; green, 0; blue, 0 }  ,draw opacity=1 ][dash pattern={on 4.5pt off 4.5pt}] (252.89,58.76) .. controls (252.89,48.25) and (261.41,39.73) .. (271.93,39.73) -- (440.07,39.73) .. controls (450.59,39.73) and (459.11,48.25) .. (459.11,58.76) -- (459.11,115.86) .. controls (459.11,126.37) and (450.59,134.9) .. (440.07,134.9) -- (271.93,134.9) .. controls (261.41,134.9) and (252.89,126.37) .. (252.89,115.86) -- cycle ;
%Rounded Rect [id:dp6057302861490522] 
\draw  [color={rgb, 255:red, 0; green, 0; blue, 0 }  ,draw opacity=1 ][dash pattern={on 4.5pt off 4.5pt}] (37.89,58.76) .. controls (37.89,48.25) and (46.41,39.73) .. (56.93,39.73) -- (225.07,39.73) .. controls (235.59,39.73) and (244.11,48.25) .. (244.11,58.76) -- (244.11,115.86) .. controls (244.11,126.37) and (235.59,134.9) .. (225.07,134.9) -- (56.93,134.9) .. controls (46.41,134.9) and (37.89,126.37) .. (37.89,115.86) -- cycle ;
%Rounded Rect [id:dp9859190253312992] 
\draw  [color={rgb, 255:red, 125; green, 184; blue, 253 }  ,draw opacity=1 ][line width=1.5]  (40.75,102.5) .. controls (40.75,99.35) and (43.3,96.81) .. (46.44,96.81) -- (128.4,96.81) .. controls (131.54,96.81) and (134.09,99.35) .. (134.09,102.5) -- (134.09,119.56) .. controls (134.09,122.7) and (131.54,125.25) .. (128.4,125.25) -- (46.44,125.25) .. controls (43.3,125.25) and (40.75,122.7) .. (40.75,119.56) -- cycle ;
%Rounded Rect [id:dp42779392841749186] 
\draw  [color={rgb, 255:red, 155; green, 155; blue, 155 }  ,draw opacity=1 ][line width=1.5]  (130.5,164.8) .. controls (130.5,161.6) and (133.1,159) .. (136.3,159) -- (162.7,159) .. controls (165.9,159) and (168.5,161.6) .. (168.5,164.8) -- (168.5,182.2) .. controls (168.5,185.4) and (165.9,188) .. (162.7,188) -- (136.3,188) .. controls (133.1,188) and (130.5,185.4) .. (130.5,182.2) -- cycle ;
%Rounded Rect [id:dp342457580670087] 
\draw  [color={rgb, 255:red, 125; green, 184; blue, 253 }  ,draw opacity=1 ][dash pattern={on 5.63pt off 4.5pt}][line width=1.5]  (193.5,184.8) .. controls (193.5,181.6) and (196.1,179) .. (199.3,179) -- (225.7,179) .. controls (228.9,179) and (231.5,181.6) .. (231.5,184.8) -- (231.5,202.2) .. controls (231.5,205.4) and (228.9,208) .. (225.7,208) -- (199.3,208) .. controls (196.1,208) and (193.5,205.4) .. (193.5,202.2) -- cycle ;
%Straight Lines [id:da360661908610052] 
\draw [line width=1.5]    (183.5,174) -- (241.75,174) ;
%Rounded Rect [id:dp28413375701819643] 
\draw  [draw opacity=0][fill={rgb, 255:red, 125; green, 184; blue, 253 }  ,fill opacity=1 ][line width=1.5]  (184.5,144.8) .. controls (184.5,141.6) and (187.1,139) .. (190.3,139) -- (235.2,139) .. controls (238.4,139) and (241,141.6) .. (241,144.8) -- (241,162.2) .. controls (241,165.4) and (238.4,168) .. (235.2,168) -- (190.3,168) .. controls (187.1,168) and (184.5,165.4) .. (184.5,162.2) -- cycle ;
%Shape: Rectangle [id:dp17625667653484345] 
\draw  [color={rgb, 255:red, 0; green, 0; blue, 0 }  ,draw opacity=1 ] (200.17,159.08) -- (210.5,159.08) -- (210.5,161.67) -- (200.17,161.67) -- cycle ;
%Shape: Rectangle [id:dp10489283230654078] 
\draw   (202.75,159.08) -- (205.33,159.08) -- (205.33,161.67) -- (202.75,161.67) -- cycle ;
%Shape: Rectangle [id:dp5257698067867771] 
\draw   (207.92,159.08) -- (210.5,159.08) -- (210.5,161.67) -- (207.92,161.67) -- cycle ;
%Shape: Rectangle [id:dp5100475150694872] 
\draw  [color={rgb, 255:red, 0; green, 0; blue, 0 }  ,draw opacity=1 ] (189.83,159.08) -- (200.17,159.08) -- (200.17,161.67) -- (189.83,161.67) -- cycle ;
%Shape: Rectangle [id:dp17271991681253063] 
\draw   (192.42,159.08) -- (195,159.08) -- (195,161.67) -- (192.42,161.67) -- cycle ;
%Shape: Rectangle [id:dp8294110408631215] 
\draw   (197.58,159.08) -- (200.17,159.08) -- (200.17,161.67) -- (197.58,161.67) -- cycle ;
%Shape: Rectangle [id:dp5073446993860504] 
\draw  [fill={rgb, 255:red, 0; green, 0; blue, 0 }  ,fill opacity=1 ] (197.58,159.08) -- (200.17,159.08) -- (200.17,161.67) -- (197.58,161.67) -- cycle ;
%Shape: Rectangle [id:dp5941885199194303] 
\draw  [fill={rgb, 255:red, 0; green, 0; blue, 0 }  ,fill opacity=1 ] (195,159.08) -- (197.58,159.08) -- (197.58,161.67) -- (195,161.67) -- cycle ;
%Shape: Rectangle [id:dp021482975385067826] 
\draw  [fill={rgb, 255:red, 0; green, 0; blue, 0 }  ,fill opacity=1 ] (202.75,159.08) -- (205.33,159.08) -- (205.33,161.67) -- (202.75,161.67) -- cycle ;

%Shape: Rectangle [id:dp759310445132021] 
\draw  [color={rgb, 255:red, 0; green, 0; blue, 0 }  ,draw opacity=1 ] (64.77,83.56) -- (86.55,83.56) -- (86.55,89) -- (64.77,89) -- cycle ;
%Shape: Rectangle [id:dp975015237767249] 
\draw   (70.22,83.56) -- (75.66,83.56) -- (75.66,89) -- (70.22,89) -- cycle ;
%Shape: Rectangle [id:dp6157250338722847] 
\draw   (81.1,83.56) -- (86.55,83.56) -- (86.55,89) -- (81.1,89) -- cycle ;
%Shape: Rectangle [id:dp2808616241648826] 
\draw  [color={rgb, 255:red, 0; green, 0; blue, 0 }  ,draw opacity=1 ] (43,83.56) -- (64.77,83.56) -- (64.77,89) -- (43,89) -- cycle ;
%Shape: Rectangle [id:dp6133787016066401] 
\draw   (48.44,83.56) -- (53.89,83.56) -- (53.89,89) -- (48.44,89) -- cycle ;
%Shape: Rectangle [id:dp028923780433613433] 
\draw   (59.33,83.56) -- (64.77,83.56) -- (64.77,89) -- (59.33,89) -- cycle ;
%Shape: Rectangle [id:dp879806085952972] 
\draw  [fill={rgb, 255:red, 0; green, 0; blue, 0 }  ,fill opacity=1 ] (59.33,83.56) -- (64.77,83.56) -- (64.77,89) -- (59.33,89) -- cycle ;
%Shape: Rectangle [id:dp5838024664927027] 
\draw  [fill={rgb, 255:red, 0; green, 0; blue, 0 }  ,fill opacity=1 ] (53.89,83.56) -- (59.33,83.56) -- (59.33,89) -- (53.89,89) -- cycle ;
%Shape: Rectangle [id:dp07830779693672596] 
\draw  [color={rgb, 255:red, 0; green, 0; blue, 0 }  ,draw opacity=1 ] (110.23,83.56) -- (132,83.56) -- (132,89) -- (110.23,89) -- cycle ;
%Shape: Rectangle [id:dp4125478363477736] 
\draw   (115.67,83.56) -- (121.11,83.56) -- (121.11,89) -- (115.67,89) -- cycle ;
%Shape: Rectangle [id:dp0017953389666957031] 
\draw   (126.56,83.56) -- (132,83.56) -- (132,89) -- (126.56,89) -- cycle ;
%Shape: Rectangle [id:dp42019925213501985] 
\draw  [color={rgb, 255:red, 0; green, 0; blue, 0 }  ,draw opacity=1 ] (88.45,83.56) -- (110.23,83.56) -- (110.23,89) -- (88.45,89) -- cycle ;
%Shape: Rectangle [id:dp8715578025659171] 
\draw   (93.9,83.56) -- (99.34,83.56) -- (99.34,89) -- (93.9,89) -- cycle ;
%Shape: Rectangle [id:dp3868532531017791] 
\draw   (104.78,83.56) -- (110.23,83.56) -- (110.23,89) -- (104.78,89) -- cycle ;
%Shape: Rectangle [id:dp7660775409077611] 
\draw  [fill={rgb, 255:red, 0; green, 0; blue, 0 }  ,fill opacity=1 ] (110.23,83.56) -- (115.67,83.56) -- (115.67,89) -- (110.23,89) -- cycle ;
%Shape: Rectangle [id:dp5977334071669691] 
\draw  [fill={rgb, 255:red, 0; green, 0; blue, 0 }  ,fill opacity=1 ] (126.56,83.56) -- (132,83.56) -- (132,89) -- (126.56,89) -- cycle ;
%Shape: Rectangle [id:dp5193025828364821] 
\draw  [fill={rgb, 255:red, 0; green, 0; blue, 0 }  ,fill opacity=1 ] (93.9,83.56) -- (99.34,83.56) -- (99.34,89) -- (93.9,89) -- cycle ;
%Shape: Rectangle [id:dp6599558668953142] 
\draw  [fill={rgb, 255:red, 0; green, 0; blue, 0 }  ,fill opacity=1 ] (70.22,83.56) -- (75.66,83.56) -- (75.66,89) -- (70.22,89) -- cycle ;
%Shape: Rectangle [id:dp9934260474523886] 
\draw  [color={rgb, 255:red, 0; green, 0; blue, 0 }  ,draw opacity=1 ] (164.77,83.56) -- (186.55,83.56) -- (186.55,89) -- (164.77,89) -- cycle ;
%Shape: Rectangle [id:dp26967492156752293] 
\draw   (170.22,83.56) -- (175.66,83.56) -- (175.66,89) -- (170.22,89) -- cycle ;
%Shape: Rectangle [id:dp07824006367469294] 
\draw   (181.1,83.56) -- (186.55,83.56) -- (186.55,89) -- (181.1,89) -- cycle ;
%Shape: Rectangle [id:dp506616467061946] 
\draw  [color={rgb, 255:red, 0; green, 0; blue, 0 }  ,draw opacity=1 ] (143,83.56) -- (164.77,83.56) -- (164.77,89) -- (143,89) -- cycle ;
%Shape: Rectangle [id:dp22521518217839231] 
\draw   (148.44,83.56) -- (153.89,83.56) -- (153.89,89) -- (148.44,89) -- cycle ;
%Shape: Rectangle [id:dp7962180443867344] 
\draw   (159.33,83.56) -- (164.77,83.56) -- (164.77,89) -- (159.33,89) -- cycle ;
%Shape: Rectangle [id:dp18881730759854787] 
\draw  [fill={rgb, 255:red, 0; green, 0; blue, 0 }  ,fill opacity=1 ] (153.89,83.56) -- (158.75,83.56) -- (158.75,89) -- (153.89,89) -- cycle ;
%Shape: Rectangle [id:dp46565796851345576] 
\draw  [fill={rgb, 255:red, 0; green, 0; blue, 0 }  ,fill opacity=1 ] (142.89,83.56) -- (148.33,83.56) -- (148.33,89) -- (142.89,89) -- cycle ;
%Shape: Rectangle [id:dp6477067797008845] 
\draw  [color={rgb, 255:red, 0; green, 0; blue, 0 }  ,draw opacity=1 ] (210.23,83.56) -- (232,83.56) -- (232,89) -- (210.23,89) -- cycle ;
%Shape: Rectangle [id:dp32051066497240543] 
\draw   (215.67,83.56) -- (221.11,83.56) -- (221.11,89) -- (215.67,89) -- cycle ;
%Shape: Rectangle [id:dp45638971169502274] 
\draw   (226.56,83.56) -- (232,83.56) -- (232,89) -- (226.56,89) -- cycle ;
%Shape: Rectangle [id:dp7136307831671067] 
\draw  [color={rgb, 255:red, 0; green, 0; blue, 0 }  ,draw opacity=1 ] (188.45,83.56) -- (210.23,83.56) -- (210.23,89) -- (188.45,89) -- cycle ;
%Shape: Rectangle [id:dp3928645308804092] 
\draw   (193.9,83.56) -- (199.34,83.56) -- (199.34,89) -- (193.9,89) -- cycle ;
%Shape: Rectangle [id:dp08348846284738098] 
\draw   (204.78,83.56) -- (210.23,83.56) -- (210.23,89) -- (204.78,89) -- cycle ;
%Shape: Rectangle [id:dp21070043343475164] 
\draw  [fill={rgb, 255:red, 0; green, 0; blue, 0 }  ,fill opacity=1 ] (210.23,83.56) -- (215.67,83.56) -- (215.67,89) -- (210.23,89) -- cycle ;
%Shape: Rectangle [id:dp07916099167750523] 
\draw  [fill={rgb, 255:red, 0; green, 0; blue, 0 }  ,fill opacity=1 ] (215.56,83.56) -- (221,83.56) -- (221,89) -- (215.56,89) -- cycle ;
%Shape: Rectangle [id:dp7010739436281225] 
\draw  [fill={rgb, 255:red, 0; green, 0; blue, 0 }  ,fill opacity=1 ] (204.9,83.56) -- (210.34,83.56) -- (210.34,89) -- (204.9,89) -- cycle ;
%Shape: Rectangle [id:dp15826847287772328] 
\draw  [fill={rgb, 255:red, 0; green, 0; blue, 0 }  ,fill opacity=1 ] (170.22,83.56) -- (175.66,83.56) -- (175.66,89) -- (170.22,89) -- cycle ;
%Shape: Rectangle [id:dp9648861040865324] 
\draw  [color={rgb, 255:red, 0; green, 0; blue, 0 }  ,draw opacity=1 ] (64.77,116.56) -- (86.55,116.56) -- (86.55,122) -- (64.77,122) -- cycle ;
%Shape: Rectangle [id:dp7341425329513045] 
\draw   (70.22,116.56) -- (75.66,116.56) -- (75.66,122) -- (70.22,122) -- cycle ;
%Shape: Rectangle [id:dp06902897481523151] 
\draw   (81.1,116.56) -- (86.55,116.56) -- (86.55,122) -- (81.1,122) -- cycle ;
%Shape: Rectangle [id:dp45325492343584495] 
\draw  [color={rgb, 255:red, 0; green, 0; blue, 0 }  ,draw opacity=1 ] (43,116.56) -- (64.77,116.56) -- (64.77,122) -- (43,122) -- cycle ;
%Shape: Rectangle [id:dp37230385141820543] 
\draw   (48.44,116.56) -- (53.89,116.56) -- (53.89,122) -- (48.44,122) -- cycle ;
%Shape: Rectangle [id:dp7976646346305731] 
\draw   (59.33,116.56) -- (64.77,116.56) -- (64.77,122) -- (59.33,122) -- cycle ;
%Shape: Rectangle [id:dp5045601086583449] 
\draw  [fill={rgb, 255:red, 0; green, 0; blue, 0 }  ,fill opacity=1 ] (54.33,116.56) -- (59.77,116.56) -- (59.77,122) -- (54.33,122) -- cycle ;
%Shape: Rectangle [id:dp7697522265290504] 
\draw  [fill={rgb, 255:red, 0; green, 0; blue, 0 }  ,fill opacity=1 ] (48.89,116.56) -- (54.33,116.56) -- (54.33,122) -- (48.89,122) -- cycle ;
%Shape: Rectangle [id:dp2932848872427918] 
\draw  [color={rgb, 255:red, 0; green, 0; blue, 0 }  ,draw opacity=1 ] (110.23,116.56) -- (132,116.56) -- (132,122) -- (110.23,122) -- cycle ;
%Shape: Rectangle [id:dp2164806334093532] 
\draw   (115.67,116.56) -- (121.11,116.56) -- (121.11,122) -- (115.67,122) -- cycle ;
%Shape: Rectangle [id:dp9983122650480425] 
\draw   (126.56,116.56) -- (132,116.56) -- (132,122) -- (126.56,122) -- cycle ;
%Shape: Rectangle [id:dp6226783798940934] 
\draw  [color={rgb, 255:red, 0; green, 0; blue, 0 }  ,draw opacity=1 ] (88.45,116.56) -- (110.23,116.56) -- (110.23,122) -- (88.45,122) -- cycle ;
%Shape: Rectangle [id:dp3800794047473234] 
\draw   (93.9,116.56) -- (99.34,116.56) -- (99.34,122) -- (93.9,122) -- cycle ;
%Shape: Rectangle [id:dp6628949846758166] 
\draw   (104.78,116.56) -- (110.23,116.56) -- (110.23,122) -- (104.78,122) -- cycle ;
%Shape: Rectangle [id:dp057342954047830075] 
\draw  [fill={rgb, 255:red, 0; green, 0; blue, 0 }  ,fill opacity=1 ] (104.5,116.56) -- (109.67,116.56) -- (109.67,122) -- (104.5,122) -- cycle ;
%Shape: Rectangle [id:dp098240575200087] 
\draw  [fill={rgb, 255:red, 0; green, 0; blue, 0 }  ,fill opacity=1 ] (115.56,116.56) -- (121,116.56) -- (121,122) -- (115.56,122) -- cycle ;
%Shape: Rectangle [id:dp7294894641396354] 
\draw  [fill={rgb, 255:red, 0; green, 0; blue, 0 }  ,fill opacity=1 ] (93.9,116.56) -- (99.34,116.56) -- (99.34,122) -- (93.9,122) -- cycle ;
%Shape: Rectangle [id:dp44013813675823665] 
\draw  [fill={rgb, 255:red, 0; green, 0; blue, 0 }  ,fill opacity=1 ] (65.22,116.56) -- (70.66,116.56) -- (70.66,122) -- (65.22,122) -- cycle ;
%Shape: Rectangle [id:dp8055884755260532] 
\draw  [color={rgb, 255:red, 0; green, 0; blue, 0 }  ,draw opacity=1 ] (164.77,116.56) -- (186.55,116.56) -- (186.55,122) -- (164.77,122) -- cycle ;
%Shape: Rectangle [id:dp5728861106243691] 
\draw   (170.22,116.56) -- (175.66,116.56) -- (175.66,122) -- (170.22,122) -- cycle ;
%Shape: Rectangle [id:dp8192476218284442] 
\draw   (181.1,116.56) -- (186.55,116.56) -- (186.55,122) -- (181.1,122) -- cycle ;
%Shape: Rectangle [id:dp8794576123412207] 
\draw  [color={rgb, 255:red, 0; green, 0; blue, 0 }  ,draw opacity=1 ] (143,116.56) -- (164.77,116.56) -- (164.77,122) -- (143,122) -- cycle ;
%Shape: Rectangle [id:dp7136464192942481] 
\draw   (148.44,116.56) -- (153.89,116.56) -- (153.89,122) -- (148.44,122) -- cycle ;
%Shape: Rectangle [id:dp1105896510615324] 
\draw   (159.33,116.56) -- (164.77,116.56) -- (164.77,122) -- (159.33,122) -- cycle ;
%Shape: Rectangle [id:dp07329595727617078] 
\draw  [fill={rgb, 255:red, 0; green, 0; blue, 0 }  ,fill opacity=1 ] (159.33,116.56) -- (164.77,116.56) -- (164.77,122) -- (159.33,122) -- cycle ;
%Shape: Rectangle [id:dp6306775252313667] 
\draw  [fill={rgb, 255:red, 0; green, 0; blue, 0 }  ,fill opacity=1 ] (153.89,116.56) -- (159.33,116.56) -- (159.33,122) -- (153.89,122) -- cycle ;
%Shape: Rectangle [id:dp5424132194263426] 
\draw  [color={rgb, 255:red, 0; green, 0; blue, 0 }  ,draw opacity=1 ] (210.23,116.56) -- (232,116.56) -- (232,122) -- (210.23,122) -- cycle ;
%Shape: Rectangle [id:dp4306049105681876] 
\draw   (215.67,116.56) -- (221.11,116.56) -- (221.11,122) -- (215.67,122) -- cycle ;
%Shape: Rectangle [id:dp010941713837690492] 
\draw   (226.56,116.56) -- (232,116.56) -- (232,122) -- (226.56,122) -- cycle ;
%Shape: Rectangle [id:dp7539350644925137] 
\draw  [color={rgb, 255:red, 0; green, 0; blue, 0 }  ,draw opacity=1 ] (188.45,116.56) -- (210.23,116.56) -- (210.23,122) -- (188.45,122) -- cycle ;
%Shape: Rectangle [id:dp4847653205418142] 
\draw   (193.9,116.56) -- (199.34,116.56) -- (199.34,122) -- (193.9,122) -- cycle ;
%Shape: Rectangle [id:dp08272979770206035] 
\draw   (204.78,116.56) -- (210.23,116.56) -- (210.23,122) -- (204.78,122) -- cycle ;
%Shape: Rectangle [id:dp3557494453282398] 
\draw  [fill={rgb, 255:red, 0; green, 0; blue, 0 }  ,fill opacity=1 ] (193.23,116.56) -- (198.67,116.56) -- (198.67,122) -- (193.23,122) -- cycle ;
%Shape: Rectangle [id:dp8719612229773848] 
\draw  [fill={rgb, 255:red, 0; green, 0; blue, 0 }  ,fill opacity=1 ] (226.56,116.56) -- (232,116.56) -- (232,122) -- (226.56,122) -- cycle ;
%Shape: Rectangle [id:dp15697937335443146] 
\draw  [fill={rgb, 255:red, 0; green, 0; blue, 0 }  ,fill opacity=1 ] (188.9,116.56) -- (194.34,116.56) -- (194.34,122) -- (188.9,122) -- cycle ;
%Shape: Rectangle [id:dp8942307747464165] 
\draw  [fill={rgb, 255:red, 0; green, 0; blue, 0 }  ,fill opacity=1 ] (181.22,116.56) -- (186.66,116.56) -- (186.66,122) -- (181.22,122) -- cycle ;
%Shape: Rectangle [id:dp1290704865914869] 
\draw  [fill={rgb, 255:red, 0; green, 0; blue, 0 }  ,fill opacity=1 ] (181.22,83.56) -- (186.66,83.56) -- (186.66,89) -- (181.22,89) -- cycle ;
%Shape: Rectangle [id:dp6487935135331143] 
\draw  [fill={rgb, 255:red, 0; green, 0; blue, 0 }  ,fill opacity=1 ] (70.22,116.56) -- (75.66,116.56) -- (75.66,122) -- (70.22,122) -- cycle ;
%Rounded Rect [id:dp23709154552920975] 
\draw  [color={rgb, 255:red, 125; green, 184; blue, 253 }  ,draw opacity=1 ][dash pattern={on 5.63pt off 4.5pt}][line width=1.5]  (259.5,74.25) .. controls (259.5,71.1) and (262.05,68.56) .. (265.19,68.56) -- (347.15,68.56) .. controls (350.29,68.56) and (352.84,71.1) .. (352.84,74.25) -- (352.84,91.31) .. controls (352.84,94.45) and (350.29,97) .. (347.15,97) -- (265.19,97) .. controls (262.05,97) and (259.5,94.45) .. (259.5,91.31) -- cycle ;
%Shape: Rectangle [id:dp0519177450068693] 
\draw  [color={rgb, 255:red, 0; green, 0; blue, 0 }  ,draw opacity=1 ] (282.77,88.56) -- (304.55,88.56) -- (304.55,94) -- (282.77,94) -- cycle ;
%Shape: Rectangle [id:dp6052939021434991] 
\draw   (288.22,88.56) -- (293.66,88.56) -- (293.66,94) -- (288.22,94) -- cycle ;
%Shape: Rectangle [id:dp16670608020063915] 
\draw   (299.1,88.56) -- (304.55,88.56) -- (304.55,94) -- (299.1,94) -- cycle ;
%Shape: Rectangle [id:dp8314380758135606] 
\draw  [color={rgb, 255:red, 0; green, 0; blue, 0 }  ,draw opacity=1 ] (261,88.56) -- (282.77,88.56) -- (282.77,94) -- (261,94) -- cycle ;
%Shape: Rectangle [id:dp6135744346406096] 
\draw   (266.44,88.56) -- (271.89,88.56) -- (271.89,94) -- (266.44,94) -- cycle ;
%Shape: Rectangle [id:dp8709200640091259] 
\draw   (277.33,88.56) -- (282.77,88.56) -- (282.77,94) -- (277.33,94) -- cycle ;
%Shape: Rectangle [id:dp5864089062509791] 
\draw  [fill={rgb, 255:red, 0; green, 0; blue, 0 }  ,fill opacity=1 ] (271.33,88.56) -- (276.77,88.56) -- (276.77,94) -- (271.33,94) -- cycle ;
%Shape: Rectangle [id:dp7604364398872725] 
\draw  [fill={rgb, 255:red, 0; green, 0; blue, 0 }  ,fill opacity=1 ] (266.89,88.56) -- (272.33,88.56) -- (272.33,94) -- (266.89,94) -- cycle ;
%Shape: Rectangle [id:dp7290824873305203] 
\draw  [color={rgb, 255:red, 0; green, 0; blue, 0 }  ,draw opacity=1 ] (328.23,88.56) -- (350,88.56) -- (350,94) -- (328.23,94) -- cycle ;
%Shape: Rectangle [id:dp17541154693222638] 
\draw   (333.67,88.56) -- (339.11,88.56) -- (339.11,94) -- (333.67,94) -- cycle ;
%Shape: Rectangle [id:dp7622589297254321] 
\draw   (344.56,88.56) -- (350,88.56) -- (350,94) -- (344.56,94) -- cycle ;
%Shape: Rectangle [id:dp37296760971643617] 
\draw  [color={rgb, 255:red, 0; green, 0; blue, 0 }  ,draw opacity=1 ] (306.45,88.56) -- (328.23,88.56) -- (328.23,94) -- (306.45,94) -- cycle ;
%Shape: Rectangle [id:dp0017845178648092652] 
\draw   (311.9,88.56) -- (317.34,88.56) -- (317.34,94) -- (311.9,94) -- cycle ;
%Shape: Rectangle [id:dp5214294809310611] 
\draw   (322.78,88.56) -- (328.23,88.56) -- (328.23,94) -- (322.78,94) -- cycle ;
%Shape: Rectangle [id:dp9098404144438652] 
\draw  [fill={rgb, 255:red, 0; green, 0; blue, 0 }  ,fill opacity=1 ] (328.23,88.56) -- (333.67,88.56) -- (333.67,94) -- (328.23,94) -- cycle ;
%Shape: Rectangle [id:dp8774563238342413] 
\draw  [fill={rgb, 255:red, 0; green, 0; blue, 0 }  ,fill opacity=1 ] (333.56,88.56) -- (339,88.56) -- (339,94) -- (333.56,94) -- cycle ;
%Shape: Rectangle [id:dp635658944512437] 
\draw  [fill={rgb, 255:red, 0; green, 0; blue, 0 }  ,fill opacity=1 ] (311.9,88.56) -- (317.34,88.56) -- (317.34,94) -- (311.9,94) -- cycle ;
%Shape: Rectangle [id:dp9105212927774133] 
\draw  [fill={rgb, 255:red, 0; green, 0; blue, 0 }  ,fill opacity=1 ] (288.22,88.56) -- (293.66,88.56) -- (293.66,94) -- (288.22,94) -- cycle ;
%Shape: Rectangle [id:dp8232549220129979] 
\draw  [color={rgb, 255:red, 0; green, 0; blue, 0 }  ,draw opacity=1 ] (382.77,88.56) -- (404.55,88.56) -- (404.55,94) -- (382.77,94) -- cycle ;
%Shape: Rectangle [id:dp09095697393465041] 
\draw   (388.22,88.56) -- (393.66,88.56) -- (393.66,94) -- (388.22,94) -- cycle ;
%Shape: Rectangle [id:dp4788170945747149] 
\draw   (399.1,88.56) -- (404.55,88.56) -- (404.55,94) -- (399.1,94) -- cycle ;
%Shape: Rectangle [id:dp9186664064940906] 
\draw  [color={rgb, 255:red, 0; green, 0; blue, 0 }  ,draw opacity=1 ] (361,88.56) -- (382.77,88.56) -- (382.77,94) -- (361,94) -- cycle ;
%Shape: Rectangle [id:dp6218823545423343] 
\draw   (366.44,88.56) -- (371.89,88.56) -- (371.89,94) -- (366.44,94) -- cycle ;
%Shape: Rectangle [id:dp0822334552888967] 
\draw   (377.33,88.56) -- (382.77,88.56) -- (382.77,94) -- (377.33,94) -- cycle ;
%Shape: Rectangle [id:dp9967150649763646] 
\draw  [fill={rgb, 255:red, 0; green, 0; blue, 0 }  ,fill opacity=1 ] (377.33,88.56) -- (382.77,88.56) -- (382.77,94) -- (377.33,94) -- cycle ;
%Shape: Rectangle [id:dp6919579603384958] 
\draw  [fill={rgb, 255:red, 0; green, 0; blue, 0 }  ,fill opacity=1 ] (393.89,88.56) -- (399.33,88.56) -- (399.33,94) -- (393.89,94) -- cycle ;
%Shape: Rectangle [id:dp7224231116068004] 
\draw  [color={rgb, 255:red, 0; green, 0; blue, 0 }  ,draw opacity=1 ] (428.23,88.56) -- (450,88.56) -- (450,94) -- (428.23,94) -- cycle ;
%Shape: Rectangle [id:dp35649549887185306] 
\draw   (433.67,88.56) -- (439.11,88.56) -- (439.11,94) -- (433.67,94) -- cycle ;
%Shape: Rectangle [id:dp9143159359478957] 
\draw   (444.56,88.56) -- (450,88.56) -- (450,94) -- (444.56,94) -- cycle ;
%Shape: Rectangle [id:dp5721397357435282] 
\draw  [color={rgb, 255:red, 0; green, 0; blue, 0 }  ,draw opacity=1 ] (406.45,88.56) -- (428.23,88.56) -- (428.23,94) -- (406.45,94) -- cycle ;
%Shape: Rectangle [id:dp6959748679530873] 
\draw   (411.9,88.56) -- (417.34,88.56) -- (417.34,94) -- (411.9,94) -- cycle ;
%Shape: Rectangle [id:dp15810087126016492] 
\draw   (422.78,88.56) -- (428.23,88.56) -- (428.23,94) -- (422.78,94) -- cycle ;
%Shape: Rectangle [id:dp8575230926750776] 
\draw  [fill={rgb, 255:red, 0; green, 0; blue, 0 }  ,fill opacity=1 ] (417.23,88.56) -- (422.67,88.56) -- (422.67,94) -- (417.23,94) -- cycle ;
%Shape: Rectangle [id:dp43830285537240143] 
\draw  [fill={rgb, 255:red, 0; green, 0; blue, 0 }  ,fill opacity=1 ] (439.56,88.56) -- (445,88.56) -- (445,94) -- (439.56,94) -- cycle ;
%Shape: Rectangle [id:dp386838864790245] 
\draw  [fill={rgb, 255:red, 0; green, 0; blue, 0 }  ,fill opacity=1 ] (411.9,88.56) -- (417.34,88.56) -- (417.34,94) -- (411.9,94) -- cycle ;
%Shape: Rectangle [id:dp03328427046632809] 
\draw  [fill={rgb, 255:red, 0; green, 0; blue, 0 }  ,fill opacity=1 ] (388.22,88.56) -- (393.66,88.56) -- (393.66,94) -- (388.22,94) -- cycle ;
%Shape: Rectangle [id:dp2452443458150808] 
\draw  [color={rgb, 255:red, 0; green, 0; blue, 0 }  ,draw opacity=1 ] (335.77,121.56) -- (357.55,121.56) -- (357.55,127) -- (335.77,127) -- cycle ;
%Shape: Rectangle [id:dp9192009027930365] 
\draw   (341.22,121.56) -- (346.66,121.56) -- (346.66,127) -- (341.22,127) -- cycle ;
%Shape: Rectangle [id:dp36996555009665155] 
\draw   (352.1,121.56) -- (357.55,121.56) -- (357.55,127) -- (352.1,127) -- cycle ;
%Shape: Rectangle [id:dp43976797219344266] 
\draw   (319.44,121.56) -- (324.89,121.56) -- (324.89,127) -- (319.44,127) -- cycle ;
%Shape: Rectangle [id:dp23405294385636655] 
\draw   (330.33,121.56) -- (335.77,121.56) -- (335.77,127) -- (330.33,127) -- cycle ;
%Shape: Rectangle [id:dp17522710369275885] 
\draw  [fill={rgb, 255:red, 0; green, 0; blue, 0 }  ,fill opacity=1 ] (324.33,121.56) -- (329.77,121.56) -- (329.77,127) -- (324.33,127) -- cycle ;
%Shape: Rectangle [id:dp3342272591122667] 
\draw  [fill={rgb, 255:red, 0; green, 0; blue, 0 }  ,fill opacity=1 ] (319.89,121.56) -- (325.33,121.56) -- (325.33,127) -- (319.89,127) -- cycle ;
%Shape: Rectangle [id:dp836934855547566] 
\draw  [color={rgb, 255:red, 0; green, 0; blue, 0 }  ,draw opacity=1 ] (381.23,121.56) -- (403,121.56) -- (403,127) -- (381.23,127) -- cycle ;
%Shape: Rectangle [id:dp15987746321848662] 
\draw   (386.67,121.56) -- (392.11,121.56) -- (392.11,127) -- (386.67,127) -- cycle ;
%Shape: Rectangle [id:dp7639472984379454] 
\draw   (397.56,121.56) -- (403,121.56) -- (403,127) -- (397.56,127) -- cycle ;
%Shape: Rectangle [id:dp23152896925316946] 
\draw  [color={rgb, 255:red, 0; green, 0; blue, 0 }  ,draw opacity=1 ] (359.45,121.56) -- (381.23,121.56) -- (381.23,127) -- (359.45,127) -- cycle ;
%Shape: Rectangle [id:dp006350902922161228] 
\draw   (364.9,121.56) -- (370.34,121.56) -- (370.34,127) -- (364.9,127) -- cycle ;
%Shape: Rectangle [id:dp9196496410849807] 
\draw   (375.78,121.56) -- (381.23,121.56) -- (381.23,127) -- (375.78,127) -- cycle ;
%Shape: Rectangle [id:dp650129974980868] 
\draw  [fill={rgb, 255:red, 0; green, 0; blue, 0 }  ,fill opacity=1 ] (381.23,121.56) -- (386.67,121.56) -- (386.67,127) -- (381.23,127) -- cycle ;
%Shape: Rectangle [id:dp06858559067969527] 
\draw  [fill={rgb, 255:red, 0; green, 0; blue, 0 }  ,fill opacity=1 ] (386.56,121.56) -- (392,121.56) -- (392,127) -- (386.56,127) -- cycle ;
%Shape: Rectangle [id:dp9111528268058434] 
\draw  [fill={rgb, 255:red, 0; green, 0; blue, 0 }  ,fill opacity=1 ] (375.9,121.56) -- (381.34,121.56) -- (381.34,127) -- (375.9,127) -- cycle ;
%Shape: Rectangle [id:dp4612804266286159] 
\draw  [fill={rgb, 255:red, 0; green, 0; blue, 0 }  ,fill opacity=1 ] (341.22,121.56) -- (346.66,121.56) -- (346.66,127) -- (341.22,127) -- cycle ;
%Straight Lines [id:da704009738201635] 
\draw [color={rgb, 255:red, 125; green, 184; blue, 253 }  ,draw opacity=1 ][line width=1.5]    (278.75,97.5) -- (231,144) ;
%Straight Lines [id:da022335544047055733] 
\draw [color={rgb, 255:red, 125; green, 184; blue, 253 }  ,draw opacity=1 ][line width=1.5]    (128.4,125.25) -- (189.5,146.5) ;
%Rounded Rect [id:dp5588681606157239] 
\draw  [color={rgb, 255:red, 219; green, 126; blue, 136 }  ,draw opacity=1 ][line width=1.5]  (255.5,184.8) .. controls (255.5,181.6) and (258.1,179) .. (261.3,179) -- (295.2,179) .. controls (298.4,179) and (301,181.6) .. (301,184.8) -- (301,202.2) .. controls (301,205.4) and (298.4,208) .. (295.2,208) -- (261.3,208) .. controls (258.1,208) and (255.5,205.4) .. (255.5,202.2) -- cycle ;
%Straight Lines [id:da8640529752630756] 
\draw [line width=1.5]    (253.5,174) -- (302.5,174) ;
%Rounded Rect [id:dp6028177760885459] 
\draw  [color={rgb, 255:red, 184; green, 233; blue, 134 }  ,draw opacity=1 ][line width=1.5]  (140.75,69.5) .. controls (140.75,66.35) and (143.3,63.81) .. (146.44,63.81) -- (228.4,63.81) .. controls (231.54,63.81) and (234.09,66.35) .. (234.09,69.5) -- (234.09,86.56) .. controls (234.09,89.7) and (231.54,92.25) .. (228.4,92.25) -- (146.44,92.25) .. controls (143.3,92.25) and (140.75,89.7) .. (140.75,86.56) -- cycle ;
%Rounded Rect [id:dp39414956519032895] 
\draw  [color={rgb, 255:red, 219; green, 126; blue, 136 }  ,draw opacity=1 ][line width=1.5]  (40.75,69.5) .. controls (40.75,66.35) and (43.3,63.81) .. (46.44,63.81) -- (128.4,63.81) .. controls (131.54,63.81) and (134.09,66.35) .. (134.09,69.5) -- (134.09,86.56) .. controls (134.09,89.7) and (131.54,92.25) .. (128.4,92.25) -- (46.44,92.25) .. controls (43.3,92.25) and (40.75,89.7) .. (40.75,86.56) -- cycle ;
%Rounded Rect [id:dp44564360191282293] 
\draw  [color={rgb, 255:red, 184; green, 233; blue, 134 }  ,draw opacity=1 ][line width=1.5]  (255.5,144.8) .. controls (255.5,141.6) and (258.1,139) .. (261.3,139) -- (295.2,139) .. controls (298.4,139) and (301,141.6) .. (301,144.8) -- (301,162.2) .. controls (301,165.4) and (298.4,168) .. (295.2,168) -- (261.3,168) .. controls (258.1,168) and (255.5,165.4) .. (255.5,162.2) -- cycle ;

% Text Node
\draw (327,43.4) node [anchor=north west][inner sep=0.75pt]    {$\mathcal{B}^{S}( f_{4})$};
% Text Node
\draw (302.63,165.2) node [anchor=north west][inner sep=0.75pt]    {$=625$};
% Text Node
\draw (112,43.4) node [anchor=north west][inner sep=0.75pt]    {$\mathcal{B}^{S}( f_{1})$};
% Text Node
\draw (136.5,166.4) node [anchor=north west][inner sep=0.75pt]    {$500$};
% Text Node
\draw (199.25,186.15) node [anchor=north west][inner sep=0.75pt]    {$200$};
% Text Node
\draw (214.5,146.4) node [anchor=north west][inner sep=0.75pt]    {$50$};
% Text Node
\draw (170.5,164.2) node [anchor=north west][inner sep=0.75pt]    {$\cdot $};
% Text Node
\draw (192.17,144.4) node [anchor=north west][inner sep=0.75pt]    {$\cap $};
% Text Node
\draw (79,66.4) node [anchor=north west][inner sep=0.75pt]  [font=\scriptsize]  {$\mathcal{B}_{s}$};
% Text Node
\draw (151,65.4) node [anchor=north west][inner sep=0.75pt]  [font=\scriptsize]  {$\Phi \left(\text{dbo:author}\right)$};
% Text Node
\draw (144,99.4) node [anchor=north west][inner sep=0.75pt]  [font=\scriptsize]  {$\Phi \left(\text{dbo:deathDate}\right)$};
% Text Node
\draw (42,97.4) node [anchor=north west][inner sep=0.75pt]  [font=\scriptsize]  {$\Phi \left(\text{dbo:nationality}\right)$};
% Text Node
\draw (297,71.4) node [anchor=north west][inner sep=0.75pt]  [font=\scriptsize]  {$\mathcal{B}_{s}$};
% Text Node
\draw (369,70.4) node [anchor=north west][inner sep=0.75pt]  [font=\scriptsize]  {$\Phi \left(\text{dbo:capital}\right)$};
% Text Node
\draw (313,102.4) node [anchor=north west][inner sep=0.75pt]  [font=\scriptsize]  {$\Phi \left(\text{dbo:currency}\right)$};
% Text Node
\draw (261.25,186.15) node [anchor=north west][inner sep=0.75pt]    {$1000$};
% Text Node
\draw (240.5,164.2) node [anchor=north west][inner sep=0.75pt]    {$\cdot $};
% Text Node
\draw (261.25,146.15) node [anchor=north west][inner sep=0.75pt]    {$5000$};

\end{tikzpicture}

%% file: figures/cardex8.tex
\tikzset{every picture/.style={line width=0.75pt}} %set default line width to 0.75pt        

\begin{tikzpicture}[x=0.6pt,y=0.6pt,yscale=-1,xscale=1]

%Rounded Rect [id:dp5337362719505643] 
\draw  [color={rgb, 255:red, 0; green, 0; blue, 0 }  ,draw opacity=1 ][dash pattern={on 4.5pt off 4.5pt}] (252.89,36.76) .. controls (252.89,26.25) and (261.41,17.73) .. (271.93,17.73) -- (440.07,17.73) .. controls (450.59,17.73) and (459.11,26.25) .. (459.11,36.76) -- (459.11,93.86) .. controls (459.11,104.37) and (450.59,112.9) .. (440.07,112.9) -- (271.93,112.9) .. controls (261.41,112.9) and (252.89,104.37) .. (252.89,93.86) -- cycle ;
%Rounded Rect [id:dp1991640965840593] 
\draw  [color={rgb, 255:red, 0; green, 0; blue, 0 }  ,draw opacity=1 ][dash pattern={on 4.5pt off 4.5pt}] (37.89,36.76) .. controls (37.89,26.25) and (46.41,17.73) .. (56.93,17.73) -- (225.07,17.73) .. controls (235.59,17.73) and (244.11,26.25) .. (244.11,36.76) -- (244.11,93.86) .. controls (244.11,104.37) and (235.59,112.9) .. (225.07,112.9) -- (56.93,112.9) .. controls (46.41,112.9) and (37.89,104.37) .. (37.89,93.86) -- cycle ;
%Shape: Rectangle [id:dp027043687427108387] 
\draw  [color={rgb, 255:red, 0; green, 0; blue, 0 }  ,draw opacity=1 ] (383.77,99.41) -- (405.55,99.41) -- (405.55,104.85) -- (383.77,104.85) -- cycle ;
%Shape: Rectangle [id:dp35637978026126793] 
\draw   (389.22,99.41) -- (394.66,99.41) -- (394.66,104.85) -- (389.22,104.85) -- cycle ;
%Shape: Rectangle [id:dp7240756054635308] 
\draw   (400.1,99.41) -- (405.55,99.41) -- (405.55,104.85) -- (400.1,104.85) -- cycle ;
%Shape: Rectangle [id:dp6453365833687256] 
\draw  [color={rgb, 255:red, 0; green, 0; blue, 0 }  ,draw opacity=1 ] (362,99.41) -- (383.77,99.41) -- (383.77,104.85) -- (362,104.85) -- cycle ;
%Shape: Rectangle [id:dp33836770589303156] 
\draw   (367.44,99.41) -- (372.89,99.41) -- (372.89,104.85) -- (367.44,104.85) -- cycle ;
%Shape: Rectangle [id:dp5675819180851581] 
\draw   (378.33,99.41) -- (383.77,99.41) -- (383.77,104.85) -- (378.33,104.85) -- cycle ;
%Shape: Rectangle [id:dp3164841278987345] 
\draw  [fill={rgb, 255:red, 0; green, 0; blue, 0 }  ,fill opacity=1 ] (378.33,99.41) -- (383.77,99.41) -- (383.77,104.85) -- (378.33,104.85) -- cycle ;
%Shape: Rectangle [id:dp03284967801813754] 
\draw  [fill={rgb, 255:red, 0; green, 0; blue, 0 }  ,fill opacity=1 ] (372.89,99.41) -- (378.33,99.41) -- (378.33,104.85) -- (372.89,104.85) -- cycle ;
%Shape: Rectangle [id:dp4140020567081759] 
\draw  [color={rgb, 255:red, 0; green, 0; blue, 0 }  ,draw opacity=1 ] (429.23,99.41) -- (451,99.41) -- (451,104.85) -- (429.23,104.85) -- cycle ;
%Shape: Rectangle [id:dp6144131535144788] 
\draw   (434.67,99.41) -- (440.11,99.41) -- (440.11,104.85) -- (434.67,104.85) -- cycle ;
%Shape: Rectangle [id:dp17285050764246712] 
\draw   (445.56,99.41) -- (451,99.41) -- (451,104.85) -- (445.56,104.85) -- cycle ;
%Shape: Rectangle [id:dp7716672821850609] 
\draw  [color={rgb, 255:red, 0; green, 0; blue, 0 }  ,draw opacity=1 ] (407.45,99.41) -- (429.23,99.41) -- (429.23,104.85) -- (407.45,104.85) -- cycle ;
%Shape: Rectangle [id:dp5032146343607005] 
\draw   (412.9,99.41) -- (418.34,99.41) -- (418.34,104.85) -- (412.9,104.85) -- cycle ;
%Shape: Rectangle [id:dp44749674714698506] 
\draw   (423.78,99.41) -- (429.23,99.41) -- (429.23,104.85) -- (423.78,104.85) -- cycle ;
%Shape: Rectangle [id:dp7503518374006767] 
\draw  [fill={rgb, 255:red, 0; green, 0; blue, 0 }  ,fill opacity=1 ] (429.23,99.41) -- (434.67,99.41) -- (434.67,104.85) -- (429.23,104.85) -- cycle ;
%Shape: Rectangle [id:dp9915754700605127] 
\draw  [fill={rgb, 255:red, 0; green, 0; blue, 0 }  ,fill opacity=1 ] (440.56,99.41) -- (446,99.41) -- (446,104.85) -- (440.56,104.85) -- cycle ;
%Shape: Rectangle [id:dp8277111960569304] 
\draw  [fill={rgb, 255:red, 0; green, 0; blue, 0 }  ,fill opacity=1 ] (406.9,99.41) -- (412.34,99.41) -- (412.34,104.85) -- (406.9,104.85) -- cycle ;
%Shape: Rectangle [id:dp7571572308262134] 
\draw  [fill={rgb, 255:red, 0; green, 0; blue, 0 }  ,fill opacity=1 ] (368.22,99.41) -- (373.66,99.41) -- (373.66,104.85) -- (368.22,104.85) -- cycle ;
%Shape: Rectangle [id:dp6526781753559012] 
\draw  [color={rgb, 255:red, 0; green, 0; blue, 0 }  ,draw opacity=1 ] (284.77,65.81) -- (306.55,65.81) -- (306.55,71.25) -- (284.77,71.25) -- cycle ;
%Shape: Rectangle [id:dp06693408194024109] 
\draw   (290.22,65.81) -- (295.66,65.81) -- (295.66,71.25) -- (290.22,71.25) -- cycle ;
%Shape: Rectangle [id:dp7787337841982618] 
\draw   (301.1,65.81) -- (306.55,65.81) -- (306.55,71.25) -- (301.1,71.25) -- cycle ;
%Shape: Rectangle [id:dp938796016548658] 
\draw  [color={rgb, 255:red, 0; green, 0; blue, 0 }  ,draw opacity=1 ] (263,65.81) -- (284.77,65.81) -- (284.77,71.25) -- (263,71.25) -- cycle ;
%Shape: Rectangle [id:dp5698873118731258] 
\draw   (268.44,65.81) -- (273.89,65.81) -- (273.89,71.25) -- (268.44,71.25) -- cycle ;
%Shape: Rectangle [id:dp4134331727597642] 
\draw   (279.33,65.81) -- (284.77,65.81) -- (284.77,71.25) -- (279.33,71.25) -- cycle ;
%Shape: Rectangle [id:dp4166651172229019] 
\draw  [fill={rgb, 255:red, 0; green, 0; blue, 0 }  ,fill opacity=1 ] (301.33,65.81) -- (306.77,65.81) -- (306.77,71.25) -- (301.33,71.25) -- cycle ;
%Shape: Rectangle [id:dp1410352392516373] 
\draw  [fill={rgb, 255:red, 0; green, 0; blue, 0 }  ,fill opacity=1 ] (262.89,65.81) -- (268.33,65.81) -- (268.33,71.25) -- (262.89,71.25) -- cycle ;
%Shape: Rectangle [id:dp0591222497903543] 
\draw  [color={rgb, 255:red, 0; green, 0; blue, 0 }  ,draw opacity=1 ] (330.23,65.81) -- (352,65.81) -- (352,71.25) -- (330.23,71.25) -- cycle ;
%Shape: Rectangle [id:dp5762271031567084] 
\draw   (335.67,65.81) -- (341.11,65.81) -- (341.11,71.25) -- (335.67,71.25) -- cycle ;
%Shape: Rectangle [id:dp137311189040503] 
\draw   (346.56,65.81) -- (352,65.81) -- (352,71.25) -- (346.56,71.25) -- cycle ;
%Shape: Rectangle [id:dp27794044137178997] 
\draw  [color={rgb, 255:red, 0; green, 0; blue, 0 }  ,draw opacity=1 ] (308.45,65.81) -- (330.23,65.81) -- (330.23,71.25) -- (308.45,71.25) -- cycle ;
%Shape: Rectangle [id:dp9018191734445775] 
\draw   (313.9,65.81) -- (319.34,65.81) -- (319.34,71.25) -- (313.9,71.25) -- cycle ;
%Shape: Rectangle [id:dp027568296039877538] 
\draw   (324.78,65.81) -- (330.23,65.81) -- (330.23,71.25) -- (324.78,71.25) -- cycle ;
%Shape: Rectangle [id:dp663984995668203] 
\draw  [fill={rgb, 255:red, 0; green, 0; blue, 0 }  ,fill opacity=1 ] (330.23,65.81) -- (335.67,65.81) -- (335.67,71.25) -- (330.23,71.25) -- cycle ;
%Shape: Rectangle [id:dp005985020653322737] 
\draw  [fill={rgb, 255:red, 0; green, 0; blue, 0 }  ,fill opacity=1 ] (346.56,65.81) -- (352,65.81) -- (352,71.25) -- (346.56,71.25) -- cycle ;
%Shape: Rectangle [id:dp5069250241375403] 
\draw  [fill={rgb, 255:red, 0; green, 0; blue, 0 }  ,fill opacity=1 ] (308.9,65.81) -- (314.34,65.81) -- (314.34,71.25) -- (308.9,71.25) -- cycle ;
%Shape: Rectangle [id:dp5232770958899274] 
\draw  [fill={rgb, 255:red, 0; green, 0; blue, 0 }  ,fill opacity=1 ] (296.22,65.81) -- (301.66,65.81) -- (301.66,71.25) -- (296.22,71.25) -- cycle ;
%Shape: Rectangle [id:dp7112096870415118] 
\draw  [color={rgb, 255:red, 0; green, 0; blue, 0 }  ,draw opacity=1 ] (384.77,65.81) -- (406.55,65.81) -- (406.55,71.25) -- (384.77,71.25) -- cycle ;
%Shape: Rectangle [id:dp6359414664856482] 
\draw   (390.22,65.81) -- (395.66,65.81) -- (395.66,71.25) -- (390.22,71.25) -- cycle ;
%Shape: Rectangle [id:dp41553413043866805] 
\draw   (401.1,65.81) -- (406.55,65.81) -- (406.55,71.25) -- (401.1,71.25) -- cycle ;
%Shape: Rectangle [id:dp11865282533834454] 
\draw  [color={rgb, 255:red, 0; green, 0; blue, 0 }  ,draw opacity=1 ] (363,65.81) -- (384.77,65.81) -- (384.77,71.25) -- (363,71.25) -- cycle ;
%Shape: Rectangle [id:dp6433197518846232] 
\draw   (368.44,65.81) -- (373.89,65.81) -- (373.89,71.25) -- (368.44,71.25) -- cycle ;
%Shape: Rectangle [id:dp2659283596423684] 
\draw   (379.33,65.81) -- (384.77,65.81) -- (384.77,71.25) -- (379.33,71.25) -- cycle ;
%Shape: Rectangle [id:dp9118231799222272] 
\draw  [fill={rgb, 255:red, 0; green, 0; blue, 0 }  ,fill opacity=1 ] (384.33,65.81) -- (389.77,65.81) -- (389.77,71.25) -- (384.33,71.25) -- cycle ;
%Shape: Rectangle [id:dp8135547949167822] 
\draw  [fill={rgb, 255:red, 0; green, 0; blue, 0 }  ,fill opacity=1 ] (362.89,65.81) -- (368.33,65.81) -- (368.33,71.25) -- (362.89,71.25) -- cycle ;
%Shape: Rectangle [id:dp6800653542677554] 
\draw  [color={rgb, 255:red, 0; green, 0; blue, 0 }  ,draw opacity=1 ] (430.23,65.81) -- (452,65.81) -- (452,71.25) -- (430.23,71.25) -- cycle ;
%Shape: Rectangle [id:dp12290052227336723] 
\draw   (435.67,65.81) -- (441.11,65.81) -- (441.11,71.25) -- (435.67,71.25) -- cycle ;
%Shape: Rectangle [id:dp5220589832438873] 
\draw   (446.56,65.81) -- (452,65.81) -- (452,71.25) -- (446.56,71.25) -- cycle ;
%Shape: Rectangle [id:dp07922882010913457] 
\draw  [color={rgb, 255:red, 0; green, 0; blue, 0 }  ,draw opacity=1 ] (408.45,65.81) -- (430.23,65.81) -- (430.23,71.25) -- (408.45,71.25) -- cycle ;
%Shape: Rectangle [id:dp6747551094684505] 
\draw   (413.9,65.81) -- (419.34,65.81) -- (419.34,71.25) -- (413.9,71.25) -- cycle ;
%Shape: Rectangle [id:dp10985358859576044] 
\draw   (424.78,65.81) -- (430.23,65.81) -- (430.23,71.25) -- (424.78,71.25) -- cycle ;
%Shape: Rectangle [id:dp03971546729084152] 
\draw  [fill={rgb, 255:red, 0; green, 0; blue, 0 }  ,fill opacity=1 ] (425.23,65.81) -- (430.67,65.81) -- (430.67,71.25) -- (425.23,71.25) -- cycle ;
%Shape: Rectangle [id:dp12959825193023333] 
\draw  [fill={rgb, 255:red, 0; green, 0; blue, 0 }  ,fill opacity=1 ] (446.56,65.81) -- (452,65.81) -- (452,71.25) -- (446.56,71.25) -- cycle ;
%Shape: Rectangle [id:dp5906151640503468] 
\draw  [fill={rgb, 255:red, 0; green, 0; blue, 0 }  ,fill opacity=1 ] (419.9,65.81) -- (425.34,65.81) -- (425.34,71.25) -- (419.9,71.25) -- cycle ;
%Shape: Rectangle [id:dp39448431271544915] 
\draw  [fill={rgb, 255:red, 0; green, 0; blue, 0 }  ,fill opacity=1 ] (390.22,65.81) -- (395.66,65.81) -- (395.66,71.25) -- (390.22,71.25) -- cycle ;
%Shape: Rectangle [id:dp7603653772448877] 
\draw  [color={rgb, 255:red, 0; green, 0; blue, 0 }  ,draw opacity=1 ] (285.77,98.81) -- (307.55,98.81) -- (307.55,104.25) -- (285.77,104.25) -- cycle ;
%Shape: Rectangle [id:dp6393709617634158] 
\draw   (291.22,98.81) -- (296.66,98.81) -- (296.66,104.25) -- (291.22,104.25) -- cycle ;
%Shape: Rectangle [id:dp5879468469147441] 
\draw   (302.1,98.81) -- (307.55,98.81) -- (307.55,104.25) -- (302.1,104.25) -- cycle ;
%Shape: Rectangle [id:dp5508783868060436] 
\draw  [color={rgb, 255:red, 0; green, 0; blue, 0 }  ,draw opacity=1 ] (264,98.81) -- (285.77,98.81) -- (285.77,104.25) -- (264,104.25) -- cycle ;
%Shape: Rectangle [id:dp3363952957826243] 
\draw   (269.44,98.81) -- (274.89,98.81) -- (274.89,104.25) -- (269.44,104.25) -- cycle ;
%Shape: Rectangle [id:dp765771268648129] 
\draw   (280.33,98.81) -- (285.77,98.81) -- (285.77,104.25) -- (280.33,104.25) -- cycle ;
%Shape: Rectangle [id:dp6326777924851636] 
\draw  [fill={rgb, 255:red, 0; green, 0; blue, 0 }  ,fill opacity=1 ] (280.33,98.81) -- (285.77,98.81) -- (285.77,104.25) -- (280.33,104.25) -- cycle ;
%Shape: Rectangle [id:dp8757960938849737] 
\draw  [fill={rgb, 255:red, 0; green, 0; blue, 0 }  ,fill opacity=1 ] (268.89,98.81) -- (274.33,98.81) -- (274.33,104.25) -- (268.89,104.25) -- cycle ;
%Shape: Rectangle [id:dp25960864053827126] 
\draw  [color={rgb, 255:red, 0; green, 0; blue, 0 }  ,draw opacity=1 ] (331.23,98.81) -- (353,98.81) -- (353,104.25) -- (331.23,104.25) -- cycle ;
%Shape: Rectangle [id:dp24834459632736727] 
\draw   (336.67,98.81) -- (342.11,98.81) -- (342.11,104.25) -- (336.67,104.25) -- cycle ;
%Shape: Rectangle [id:dp8847465664824542] 
\draw   (347.56,98.81) -- (353,98.81) -- (353,104.25) -- (347.56,104.25) -- cycle ;
%Shape: Rectangle [id:dp40912916587022297] 
\draw  [color={rgb, 255:red, 0; green, 0; blue, 0 }  ,draw opacity=1 ] (309.45,98.81) -- (331.23,98.81) -- (331.23,104.25) -- (309.45,104.25) -- cycle ;
%Shape: Rectangle [id:dp39952228010123225] 
\draw   (314.9,98.81) -- (320.34,98.81) -- (320.34,104.25) -- (314.9,104.25) -- cycle ;
%Shape: Rectangle [id:dp6835959479050414] 
\draw   (325.78,98.81) -- (331.23,98.81) -- (331.23,104.25) -- (325.78,104.25) -- cycle ;
%Shape: Rectangle [id:dp9882161356238017] 
\draw  [fill={rgb, 255:red, 0; green, 0; blue, 0 }  ,fill opacity=1 ] (331.23,98.81) -- (336.67,98.81) -- (336.67,104.25) -- (331.23,104.25) -- cycle ;
%Shape: Rectangle [id:dp8709359540754332] 
\draw  [fill={rgb, 255:red, 0; green, 0; blue, 0 }  ,fill opacity=1 ] (347.56,98.81) -- (353,98.81) -- (353,104.25) -- (347.56,104.25) -- cycle ;
%Shape: Rectangle [id:dp1037800941547451] 
\draw  [fill={rgb, 255:red, 0; green, 0; blue, 0 }  ,fill opacity=1 ] (335.9,98.81) -- (341.34,98.81) -- (341.34,104.25) -- (335.9,104.25) -- cycle ;
%Shape: Rectangle [id:dp5524691509672242] 
\draw  [fill={rgb, 255:red, 0; green, 0; blue, 0 }  ,fill opacity=1 ] (291.22,98.81) -- (296.66,98.81) -- (296.66,104.25) -- (291.22,104.25) -- cycle ;
%Rounded Rect [id:dp979345087360473] 
\draw  [color={rgb, 255:red, 248; green, 231; blue, 28 }  ,draw opacity=1 ][line width=1.5]  (97,83.5) .. controls (97,80.35) and (99.55,77.81) .. (102.69,77.81) -- (184.65,77.81) .. controls (187.79,77.81) and (190.34,80.35) .. (190.34,83.5) -- (190.34,100.56) .. controls (190.34,103.7) and (187.79,106.25) .. (184.65,106.25) -- (102.69,106.25) .. controls (99.55,106.25) and (97,103.7) .. (97,100.56) -- cycle ;
%Rounded Rect [id:dp5750317077913616] 
\draw  [color={rgb, 255:red, 248; green, 231; blue, 28 }  ,draw opacity=1 ][dash pattern={on 5.63pt off 4.5pt}][line width=1.5]  (261,52.5) .. controls (261,49.35) and (263.55,46.81) .. (266.69,46.81) -- (348.65,46.81) .. controls (351.79,46.81) and (354.34,49.35) .. (354.34,52.5) -- (354.34,69.56) .. controls (354.34,72.7) and (351.79,75.25) .. (348.65,75.25) -- (266.69,75.25) .. controls (263.55,75.25) and (261,72.7) .. (261,69.56) -- cycle ;
%Shape: Rectangle [id:dp6837977406303503] 
\draw  [color={rgb, 255:red, 0; green, 0; blue, 0 }  ,draw opacity=1 ] (67.77,65.81) -- (89.55,65.81) -- (89.55,71.25) -- (67.77,71.25) -- cycle ;
%Shape: Rectangle [id:dp7701844152689241] 
\draw   (73.22,65.81) -- (78.66,65.81) -- (78.66,71.25) -- (73.22,71.25) -- cycle ;
%Shape: Rectangle [id:dp4097784298570798] 
\draw   (84.1,65.81) -- (89.55,65.81) -- (89.55,71.25) -- (84.1,71.25) -- cycle ;
%Shape: Rectangle [id:dp7374067645866113] 
\draw  [color={rgb, 255:red, 0; green, 0; blue, 0 }  ,draw opacity=1 ] (46,65.81) -- (67.77,65.81) -- (67.77,71.25) -- (46,71.25) -- cycle ;
%Shape: Rectangle [id:dp24108045324071192] 
\draw   (51.44,65.81) -- (56.89,65.81) -- (56.89,71.25) -- (51.44,71.25) -- cycle ;
%Shape: Rectangle [id:dp23216314173959496] 
\draw   (62.33,65.81) -- (67.77,65.81) -- (67.77,71.25) -- (62.33,71.25) -- cycle ;
%Shape: Rectangle [id:dp4096936474250271] 
\draw  [fill={rgb, 255:red, 0; green, 0; blue, 0 }  ,fill opacity=1 ] (51.33,65.81) -- (56.77,65.81) -- (56.77,71.25) -- (51.33,71.25) -- cycle ;
%Shape: Rectangle [id:dp8932996797944499] 
\draw  [fill={rgb, 255:red, 0; green, 0; blue, 0 }  ,fill opacity=1 ] (45.89,65.81) -- (51.33,65.81) -- (51.33,71.25) -- (45.89,71.25) -- cycle ;
%Shape: Rectangle [id:dp8965542849889673] 
\draw  [color={rgb, 255:red, 0; green, 0; blue, 0 }  ,draw opacity=1 ] (113.23,65.81) -- (135,65.81) -- (135,71.25) -- (113.23,71.25) -- cycle ;
%Shape: Rectangle [id:dp7152071388595123] 
\draw   (118.67,65.81) -- (124.11,65.81) -- (124.11,71.25) -- (118.67,71.25) -- cycle ;
%Shape: Rectangle [id:dp4902970034517551] 
\draw   (129.56,65.81) -- (135,65.81) -- (135,71.25) -- (129.56,71.25) -- cycle ;
%Shape: Rectangle [id:dp44616104650710464] 
\draw  [color={rgb, 255:red, 0; green, 0; blue, 0 }  ,draw opacity=1 ] (91.45,65.81) -- (113.23,65.81) -- (113.23,71.25) -- (91.45,71.25) -- cycle ;
%Shape: Rectangle [id:dp5800394335970014] 
\draw   (96.9,65.81) -- (102.34,65.81) -- (102.34,71.25) -- (96.9,71.25) -- cycle ;
%Shape: Rectangle [id:dp26410715853784117] 
\draw   (107.78,65.81) -- (113.23,65.81) -- (113.23,71.25) -- (107.78,71.25) -- cycle ;
%Shape: Rectangle [id:dp7508156618619276] 
\draw  [fill={rgb, 255:red, 0; green, 0; blue, 0 }  ,fill opacity=1 ] (107.23,65.81) -- (112.67,65.81) -- (112.67,71.25) -- (107.23,71.25) -- cycle ;
%Shape: Rectangle [id:dp5695326996775272] 
\draw  [fill={rgb, 255:red, 0; green, 0; blue, 0 }  ,fill opacity=1 ] (101.9,65.81) -- (107.34,65.81) -- (107.34,71.25) -- (101.9,71.25) -- cycle ;
%Shape: Rectangle [id:dp09410830520351832] 
\draw  [fill={rgb, 255:red, 0; green, 0; blue, 0 }  ,fill opacity=1 ] (56.22,65.81) -- (61.66,65.81) -- (61.66,71.25) -- (56.22,71.25) -- cycle ;
%Shape: Rectangle [id:dp1549918675054498] 
\draw  [color={rgb, 255:red, 0; green, 0; blue, 0 }  ,draw opacity=1 ] (167.77,65.81) -- (189.55,65.81) -- (189.55,71.25) -- (167.77,71.25) -- cycle ;
%Shape: Rectangle [id:dp6983738205655214] 
\draw   (173.22,65.81) -- (178.66,65.81) -- (178.66,71.25) -- (173.22,71.25) -- cycle ;
%Shape: Rectangle [id:dp5662588998351693] 
\draw   (184.1,65.81) -- (189.55,65.81) -- (189.55,71.25) -- (184.1,71.25) -- cycle ;
%Shape: Rectangle [id:dp5276658491018045] 
\draw  [color={rgb, 255:red, 0; green, 0; blue, 0 }  ,draw opacity=1 ] (146,65.81) -- (167.77,65.81) -- (167.77,71.25) -- (146,71.25) -- cycle ;
%Shape: Rectangle [id:dp023920111165606994] 
\draw   (151.44,65.81) -- (156.89,65.81) -- (156.89,71.25) -- (151.44,71.25) -- cycle ;
%Shape: Rectangle [id:dp52645810601853] 
\draw   (162.33,65.81) -- (167.77,65.81) -- (167.77,71.25) -- (162.33,71.25) -- cycle ;
%Shape: Rectangle [id:dp07987417316674583] 
\draw  [fill={rgb, 255:red, 0; green, 0; blue, 0 }  ,fill opacity=1 ] (162.33,65.81) -- (167.77,65.81) -- (167.77,71.25) -- (162.33,71.25) -- cycle ;
%Shape: Rectangle [id:dp9718672142817159] 
\draw  [fill={rgb, 255:red, 0; green, 0; blue, 0 }  ,fill opacity=1 ] (151.89,65.81) -- (157.33,65.81) -- (157.33,71.25) -- (151.89,71.25) -- cycle ;
%Shape: Rectangle [id:dp009235460991394273] 
\draw  [color={rgb, 255:red, 0; green, 0; blue, 0 }  ,draw opacity=1 ] (213.23,65.81) -- (235,65.81) -- (235,71.25) -- (213.23,71.25) -- cycle ;
%Shape: Rectangle [id:dp95387330219417] 
\draw   (218.67,65.81) -- (224.11,65.81) -- (224.11,71.25) -- (218.67,71.25) -- cycle ;
%Shape: Rectangle [id:dp4283681757864932] 
\draw   (229.56,65.81) -- (235,65.81) -- (235,71.25) -- (229.56,71.25) -- cycle ;
%Shape: Rectangle [id:dp6858864029096099] 
\draw  [color={rgb, 255:red, 0; green, 0; blue, 0 }  ,draw opacity=1 ] (191.45,65.81) -- (213.23,65.81) -- (213.23,71.25) -- (191.45,71.25) -- cycle ;
%Shape: Rectangle [id:dp5496436070296301] 
\draw   (196.9,65.81) -- (202.34,65.81) -- (202.34,71.25) -- (196.9,71.25) -- cycle ;
%Shape: Rectangle [id:dp8388631955865972] 
\draw   (207.78,65.81) -- (213.23,65.81) -- (213.23,71.25) -- (207.78,71.25) -- cycle ;
%Shape: Rectangle [id:dp029936534740579557] 
\draw  [fill={rgb, 255:red, 0; green, 0; blue, 0 }  ,fill opacity=1 ] (202.23,65.81) -- (207.67,65.81) -- (207.67,71.25) -- (202.23,71.25) -- cycle ;
%Shape: Rectangle [id:dp727271271275483] 
\draw  [fill={rgb, 255:red, 0; green, 0; blue, 0 }  ,fill opacity=1 ] (229.56,65.81) -- (235,65.81) -- (235,71.25) -- (229.56,71.25) -- cycle ;
%Shape: Rectangle [id:dp8089585809256357] 
\draw  [fill={rgb, 255:red, 0; green, 0; blue, 0 }  ,fill opacity=1 ] (191.9,65.81) -- (197.34,65.81) -- (197.34,71.25) -- (191.9,71.25) -- cycle ;
%Shape: Rectangle [id:dp7216003673999061] 
\draw  [fill={rgb, 255:red, 0; green, 0; blue, 0 }  ,fill opacity=1 ] (179.22,65.81) -- (184.66,65.81) -- (184.66,71.25) -- (179.22,71.25) -- cycle ;
%Shape: Rectangle [id:dp48495271890854097] 
\draw  [color={rgb, 255:red, 0; green, 0; blue, 0 }  ,draw opacity=1 ] (120.77,98.81) -- (142.55,98.81) -- (142.55,104.25) -- (120.77,104.25) -- cycle ;
%Shape: Rectangle [id:dp11538919082734678] 
\draw   (126.22,98.81) -- (131.66,98.81) -- (131.66,104.25) -- (126.22,104.25) -- cycle ;
%Shape: Rectangle [id:dp3685346169860526] 
\draw   (137.1,98.81) -- (142.55,98.81) -- (142.55,104.25) -- (137.1,104.25) -- cycle ;
%Shape: Rectangle [id:dp8101731429825046] 
\draw  [color={rgb, 255:red, 0; green, 0; blue, 0 }  ,draw opacity=1 ] (99,98.81) -- (120.77,98.81) -- (120.77,104.25) -- (99,104.25) -- cycle ;
%Shape: Rectangle [id:dp889994322478662] 
\draw   (104.44,98.81) -- (109.89,98.81) -- (109.89,104.25) -- (104.44,104.25) -- cycle ;
%Shape: Rectangle [id:dp855643224118878] 
\draw   (115.33,98.81) -- (120.77,98.81) -- (120.77,104.25) -- (115.33,104.25) -- cycle ;
%Shape: Rectangle [id:dp4588966325795628] 
\draw  [fill={rgb, 255:red, 0; green, 0; blue, 0 }  ,fill opacity=1 ] (115.33,98.81) -- (120.77,98.81) -- (120.77,104.25) -- (115.33,104.25) -- cycle ;
%Shape: Rectangle [id:dp07955502403268144] 
\draw  [fill={rgb, 255:red, 0; green, 0; blue, 0 }  ,fill opacity=1 ] (98.89,98.81) -- (104.33,98.81) -- (104.33,104.25) -- (98.89,104.25) -- cycle ;
%Shape: Rectangle [id:dp5099564470149034] 
\draw  [color={rgb, 255:red, 0; green, 0; blue, 0 }  ,draw opacity=1 ] (166.23,98.81) -- (188,98.81) -- (188,104.25) -- (166.23,104.25) -- cycle ;
%Shape: Rectangle [id:dp5254233394758896] 
\draw   (171.67,98.81) -- (177.11,98.81) -- (177.11,104.25) -- (171.67,104.25) -- cycle ;
%Shape: Rectangle [id:dp7479522750563787] 
\draw   (182.56,98.81) -- (188,98.81) -- (188,104.25) -- (182.56,104.25) -- cycle ;
%Shape: Rectangle [id:dp29174866068114347] 
\draw  [color={rgb, 255:red, 0; green, 0; blue, 0 }  ,draw opacity=1 ] (144.45,98.81) -- (166.23,98.81) -- (166.23,104.25) -- (144.45,104.25) -- cycle ;
%Shape: Rectangle [id:dp005718383367462221] 
\draw   (149.9,98.81) -- (155.34,98.81) -- (155.34,104.25) -- (149.9,104.25) -- cycle ;
%Shape: Rectangle [id:dp5430989924736439] 
\draw   (160.78,98.81) -- (166.23,98.81) -- (166.23,104.25) -- (160.78,104.25) -- cycle ;
%Shape: Rectangle [id:dp12633547666812173] 
\draw  [fill={rgb, 255:red, 0; green, 0; blue, 0 }  ,fill opacity=1 ] (177.23,98.81) -- (182.67,98.81) -- (182.67,104.25) -- (177.23,104.25) -- cycle ;
%Shape: Rectangle [id:dp2707568869111783] 
\draw  [fill={rgb, 255:red, 0; green, 0; blue, 0 }  ,fill opacity=1 ] (182.56,98.81) -- (188,98.81) -- (188,104.25) -- (182.56,104.25) -- cycle ;
%Shape: Rectangle [id:dp9043742951036082] 
\draw  [fill={rgb, 255:red, 0; green, 0; blue, 0 }  ,fill opacity=1 ] (144.9,98.81) -- (150.34,98.81) -- (150.34,104.25) -- (144.9,104.25) -- cycle ;
%Shape: Rectangle [id:dp9889378794256738] 
\draw  [fill={rgb, 255:red, 0; green, 0; blue, 0 }  ,fill opacity=1 ] (132.22,98.81) -- (137.66,98.81) -- (137.66,104.25) -- (132.22,104.25) -- cycle ;
%Shape: Rectangle [id:dp6544196875496333] 
\draw  [fill={rgb, 255:red, 0; green, 0; blue, 0 }  ,fill opacity=1 ] (78.22,65.81) -- (83.66,65.81) -- (83.66,71.25) -- (78.22,71.25) -- cycle ;
%Straight Lines [id:da8925403000335385] 
\draw [color={rgb, 255:red, 248; green, 231; blue, 28 }  ,draw opacity=1 ][line width=1.5]    (164,106.5) -- (196,121.5) ;
%Straight Lines [id:da9762517701149744] 
\draw [color={rgb, 255:red, 248; green, 231; blue, 28 }  ,draw opacity=1 ][line width=1.5]    (266.69,75.25) -- (235,122) ;
%Rounded Rect [id:dp11355893688486751] 
\draw  [color={rgb, 255:red, 155; green, 155; blue, 155 }  ,draw opacity=1 ][dash pattern={on 5.63pt off 4.5pt}][line width=1.5]  (124.5,142.8) .. controls (124.5,139.6) and (127.1,137) .. (130.3,137) -- (156.7,137) .. controls (159.9,137) and (162.5,139.6) .. (162.5,142.8) -- (162.5,160.2) .. controls (162.5,163.4) and (159.9,166) .. (156.7,166) -- (130.3,166) .. controls (127.1,166) and (124.5,163.4) .. (124.5,160.2) -- cycle ;
%Rounded Rect [id:dp04922230912056358] 
\draw  [color={rgb, 255:red, 248; green, 231; blue, 28 }  ,draw opacity=1 ][dash pattern={on 5.63pt off 4.5pt}][line width=1.5]  (193.5,162.8) .. controls (193.5,159.6) and (196.1,157) .. (199.3,157) -- (225.7,157) .. controls (228.9,157) and (231.5,159.6) .. (231.5,162.8) -- (231.5,180.2) .. controls (231.5,183.4) and (228.9,186) .. (225.7,186) -- (199.3,186) .. controls (196.1,186) and (193.5,183.4) .. (193.5,180.2) -- cycle ;
%Straight Lines [id:da4058098587847835] 
\draw [line width=1.5]    (177.5,152) -- (240,152) ;
%Rounded Rect [id:dp46742396135527353] 
\draw  [draw opacity=0][fill={rgb, 255:red, 248; green, 231; blue, 28 }  ,fill opacity=1 ][line width=1.5]  (178.5,122) .. controls (178.5,118.8) and (181.1,116.2) .. (184.3,116.2) -- (234.2,116.2) .. controls (237.4,116.2) and (240,118.8) .. (240,122) -- (240,139.4) .. controls (240,142.6) and (237.4,145.2) .. (234.2,145.2) -- (184.3,145.2) .. controls (181.1,145.2) and (178.5,142.6) .. (178.5,139.4) -- cycle ;
%Shape: Rectangle [id:dp7828802491912566] 
\draw  [color={rgb, 255:red, 0; green, 0; blue, 0 }  ,draw opacity=1 ] (194.17,137.08) -- (204.5,137.08) -- (204.5,139.67) -- (194.17,139.67) -- cycle ;
%Shape: Rectangle [id:dp2728430883877362] 
\draw   (196.75,137.08) -- (199.33,137.08) -- (199.33,139.67) -- (196.75,139.67) -- cycle ;
%Shape: Rectangle [id:dp5408681729059397] 
\draw   (201.92,137.08) -- (204.5,137.08) -- (204.5,139.67) -- (201.92,139.67) -- cycle ;
%Shape: Rectangle [id:dp993717660654201] 
\draw  [color={rgb, 255:red, 0; green, 0; blue, 0 }  ,draw opacity=1 ] (183.83,137.08) -- (194.17,137.08) -- (194.17,139.67) -- (183.83,139.67) -- cycle ;
%Shape: Rectangle [id:dp2876814609901691] 
\draw   (186.42,137.08) -- (189,137.08) -- (189,139.67) -- (186.42,139.67) -- cycle ;
%Shape: Rectangle [id:dp24797893933063586] 
\draw   (191.58,137.08) -- (194.17,137.08) -- (194.17,139.67) -- (191.58,139.67) -- cycle ;
%Shape: Rectangle [id:dp9722531136258419] 
\draw  [fill={rgb, 255:red, 0; green, 0; blue, 0 }  ,fill opacity=1 ] (191.58,137.08) -- (194.17,137.08) -- (194.17,139.67) -- (191.58,139.67) -- cycle ;
%Shape: Rectangle [id:dp5501982398752098] 
\draw  [fill={rgb, 255:red, 0; green, 0; blue, 0 }  ,fill opacity=1 ] (189,137.08) -- (191.58,137.08) -- (191.58,139.67) -- (189,139.67) -- cycle ;
%Shape: Rectangle [id:dp8215761735287193] 
\draw  [fill={rgb, 255:red, 0; green, 0; blue, 0 }  ,fill opacity=1 ] (196.75,137.08) -- (199.33,137.08) -- (199.33,139.67) -- (196.75,139.67) -- cycle ;

%Rounded Rect [id:dp6282620304647393] 
\draw  [color={rgb, 255:red, 219; green, 126; blue, 136 }  ,draw opacity=1 ][dash pattern={on 5.63pt off 4.5pt}][line width=1.5]  (256.5,162.8) .. controls (256.5,159.6) and (259.1,157) .. (262.3,157) -- (296.2,157) .. controls (299.4,157) and (302,159.6) .. (302,162.8) -- (302,180.2) .. controls (302,183.4) and (299.4,186) .. (296.2,186) -- (262.3,186) .. controls (259.1,186) and (256.5,183.4) .. (256.5,180.2) -- cycle ;
%Straight Lines [id:da2806466942493768] 
\draw [line width=1.5]    (254.5,152) -- (303.5,152) ;
%Rounded Rect [id:dp2984569211994006] 
\draw  [color={rgb, 255:red, 184; green, 233; blue, 134 }  ,draw opacity=1 ][dash pattern={on 5.63pt off 4.5pt}][line width=1.5]  (256.5,122.8) .. controls (256.5,119.6) and (259.1,117) .. (262.3,117) -- (296.2,117) .. controls (299.4,117) and (302,119.6) .. (302,122.8) -- (302,140.2) .. controls (302,143.4) and (299.4,146) .. (296.2,146) -- (262.3,146) .. controls (259.1,146) and (256.5,143.4) .. (256.5,140.2) -- cycle ;
%Rounded Rect [id:dp023042655557532044] 
\draw  [color={rgb, 255:red, 184; green, 233; blue, 134 }  ,draw opacity=1 ][dash pattern={on 5.63pt off 4.5pt}][line width=1.5]  (143.75,51.5) .. controls (143.75,48.35) and (146.3,45.81) .. (149.44,45.81) -- (231.4,45.81) .. controls (234.54,45.81) and (237.09,48.35) .. (237.09,51.5) -- (237.09,68.56) .. controls (237.09,71.7) and (234.54,74.25) .. (231.4,74.25) -- (149.44,74.25) .. controls (146.3,74.25) and (143.75,71.7) .. (143.75,68.56) -- cycle ;
%Rounded Rect [id:dp5657410487599877] 
\draw  [color={rgb, 255:red, 219; green, 126; blue, 136 }  ,draw opacity=1 ][dash pattern={on 5.63pt off 4.5pt}][line width=1.5]  (43.75,51.5) .. controls (43.75,48.35) and (46.3,45.81) .. (49.44,45.81) -- (131.4,45.81) .. controls (134.54,45.81) and (137.09,48.35) .. (137.09,51.5) -- (137.09,68.56) .. controls (137.09,71.7) and (134.54,74.25) .. (131.4,74.25) -- (49.44,74.25) .. controls (46.3,74.25) and (43.75,71.7) .. (43.75,68.56) -- cycle ;

% Text Node
\draw (327,21.4) node [anchor=north west][inner sep=0.75pt]    {$\mathcal{B}^{S}( f_{3})$};
% Text Node
\draw (112,21.4) node [anchor=north west][inner sep=0.75pt]    {$\mathcal{B}^{S}( f_{2})$};
% Text Node
\draw (363,82.25) node [anchor=north west][inner sep=0.75pt]  [font=\scriptsize]  {$\Phi \left(\text{dbo:population}\right)$};
% Text Node
\draw (299,48.65) node [anchor=north west][inner sep=0.75pt]  [font=\scriptsize]  {$\mathcal{B}_{s}$};
% Text Node
\draw (371,47.65) node [anchor=north west][inner sep=0.75pt]  [font=\scriptsize]  {$\Phi \left(\text{dbo:capital}\right)$};
% Text Node
\draw (263,79.65) node [anchor=north west][inner sep=0.75pt]  [font=\scriptsize]  {$\Phi \left(\text{dbo:currency}\right)$};
% Text Node
\draw (82,48.65) node [anchor=north west][inner sep=0.75pt]  [font=\scriptsize]  {$\mathcal{B}_{s}$};
% Text Node
\draw (154,47.65) node [anchor=north west][inner sep=0.75pt]  [font=\scriptsize]  {$\Phi \left(\text{dbo:author}\right)$};
% Text Node
\draw (98,79.65) node [anchor=north west][inner sep=0.75pt]  [font=\scriptsize]  {$\Phi \left(\text{dbo:nationality}\right)$};
% Text Node
\draw (303.63,143.2) node [anchor=north west][inner sep=0.75pt]    {$=225$};
% Text Node
\draw (130.5,144.4) node [anchor=north west][inner sep=0.75pt]    {$150$};
% Text Node
\draw (199.25,164.15) node [anchor=north west][inner sep=0.75pt]    {$100$};
% Text Node
\draw (208.5,124.4) node [anchor=north west][inner sep=0.75pt]    {$100$};
% Text Node
\draw (164.5,142.2) node [anchor=north west][inner sep=0.75pt]    {$\cdot $};
% Text Node
\draw (187,122.4) node [anchor=north west][inner sep=0.75pt]    {$\cap $};
% Text Node
\draw (262.25,163.65) node [anchor=north west][inner sep=0.75pt]    {$2000$};
% Text Node
\draw (241.5,142.2) node [anchor=north west][inner sep=0.75pt]    {$\cdot $};
% Text Node
\draw (262.25,124.15) node [anchor=north west][inner sep=0.75pt]    {$3000$};

\end{tikzpicture}

%% file: figures/pex1.tex
\tikzset{every picture/.style={line width=0.75pt}} %set default line width to 0.75pt        

\begin{tikzpicture}[x=0.5pt,y=0.5pt,yscale=-1,xscale=1]

%Straight Lines [id:da27545794922948463] 
\draw    (355,137) -- (387,158) ;
%Rounded Rect [id:dp5392002078936551] 
\draw  [draw opacity=0][fill={rgb, 255:red, 125; green, 184; blue, 253 }  ,fill opacity=1 ] (363,163.2) .. controls (363,159.22) and (366.22,156) .. (370.2,156) -- (422.8,156) .. controls (426.78,156) and (430,159.22) .. (430,163.2) -- (430,184.8) .. controls (430,188.78) and (426.78,192) .. (422.8,192) -- (370.2,192) .. controls (366.22,192) and (363,188.78) .. (363,184.8) -- cycle ;
%Straight Lines [id:da794463235166469] 
\draw    (342,135) -- (298,169) ;
%Rounded Rect [id:dp33675717106656167] 
\draw  [draw opacity=0][fill={rgb, 255:red, 125; green, 184; blue, 253 }  ,fill opacity=1 ] (267,176.2) .. controls (267,172.22) and (270.22,169) .. (274.2,169) -- (326.8,169) .. controls (330.78,169) and (334,172.22) .. (334,176.2) -- (334,197.8) .. controls (334,201.78) and (330.78,205) .. (326.8,205) -- (274.2,205) .. controls (270.22,205) and (267,201.78) .. (267,197.8) -- cycle ;

% Text Node
%\draw (360,121.4) node [anchor=north west][inner sep=0.75pt]    {$n_1$};
% Text Node
\draw (338,121.4) node [anchor=north west][inner sep=0.75pt]    {$\bigcup $};
% Text Node
\draw (366,162.4) node [anchor=north west][inner sep=0.75pt]    {$[[ P_{1}]]_{f_{2}}^{n_3}$};
% Text Node
\draw (270,175.4) node [anchor=north west][inner sep=0.75pt]    {$[[ P_{1}]]_{f_{1}}^{n_2}$};

\end{tikzpicture}

%% file: figures/pex2.tex
\tikzset{every picture/.style={line width=0.75pt}} %set default line width to 0.75pt        

\begin{tikzpicture}[x=0.5pt,y=0.5pt,yscale=-1,xscale=1]

%Straight Lines [id:da27545794922948463] 
\draw    (355,137) -- (387,158) ;
%Rounded Rect [id:dp5392002078936551] 
\draw  [draw opacity=0][fill={rgb, 255:red, 248; green, 231; blue, 28 }  ,fill opacity=1 ] (363,163.2) .. controls (363,159.22) and (366.22,156) .. (370.2,156) -- (422.8,156) .. controls (426.78,156) and (430,159.22) .. (430,163.2) -- (430,184.8) .. controls (430,188.78) and (426.78,192) .. (422.8,192) -- (370.2,192) .. controls (366.22,192) and (363,188.78) .. (363,184.8) -- cycle ;
%Straight Lines [id:da794463235166469] 
\draw    (342,135) -- (298,169) ;
%Rounded Rect [id:dp33675717106656167] 
\draw  [draw opacity=0][fill={rgb, 255:red, 248; green, 231; blue, 28 }  ,fill opacity=1 ] (267,176.2) .. controls (267,172.22) and (270.22,169) .. (274.2,169) -- (326.8,169) .. controls (330.78,169) and (334,172.22) .. (334,176.2) -- (334,197.8) .. controls (334,201.78) and (330.78,205) .. (326.8,205) -- (274.2,205) .. controls (270.22,205) and (267,201.78) .. (267,197.8) -- cycle ;

% Text Node
%\draw (360,121.4) node [anchor=north west][inner sep=0.75pt]    {$n_1$};
% Text Node
\draw (338,121.4) node [anchor=north west][inner sep=0.75pt]    {$\bigcup $};
% Text Node
\draw (366,162.4) node [anchor=north west][inner sep=0.75pt]    {$[[ P_{2}]]_{f_{4}}^{n_2}$};
% Text Node
\draw (270,175.4) node [anchor=north west][inner sep=0.75pt]    {$[[ P_{2}]]_{f_{3}}^{n_3}$};

\end{tikzpicture}

%% file: figures/pex3.tex
\tikzset{every picture/.style={line width=0.75pt}} %set default line width to 0.75pt        

\begin{tikzpicture}[x=0.5pt,y=0.5pt,yscale=-1,xscale=1]

%Rounded Rect [id:dp33675717106656167] 
\draw  [draw opacity=0][fill={rgb, 255:red, 184; green, 233; blue, 134 }  ,fill opacity=1 ] (267,176.2) .. controls (267,172.22) and (270.22,169) .. (274.2,169) -- (326.8,169) .. controls (330.78,169) and (334,172.22) .. (334,176.2) -- (334,197.8) .. controls (334,201.78) and (330.78,205) .. (326.8,205) -- (274.2,205) .. controls (270.22,205) and (267,201.78) .. (267,197.8) -- cycle ;

% Text Node
\draw (270,175.4) node [anchor=north west][inner sep=0.75pt]    {$[[ P_{3}]]_{f_{5}}^{n_1}$};

\end{tikzpicture}

%% file: figures/pex4.tex
\tikzset{every picture/.style={line width=0.75pt}} %set default line width to 0.75pt        

\begin{tikzpicture}[x=0.5pt,y=0.5pt,yscale=-1,xscale=1]

%Straight Lines [id:da3752946459185349] 
\draw    (355,136) -- (416,181) ;
%Shape: Path Data [id:dp5219564963280906] 
\draw  [fill={rgb, 255:red, 117; green, 174; blue, 240 }  ,fill opacity=1 ] (416,191) -- (416,175) -- (426.5,183) -- (416,191) -- cycle (437,175) -- (437,191) -- (426.5,183) -- (437,175) -- cycle ;
%Straight Lines [id:da4803394967495148] 
\draw    (437,183) -- (458,209) ;
%Rounded Rect [id:dp39534133932958815] 
\draw  [draw opacity=0][fill={rgb, 255:red, 125; green, 184; blue, 253 }  ,fill opacity=1 ] (430,215.2) .. controls (430,211.22) and (433.22,208) .. (437.2,208) -- (489.8,208) .. controls (493.78,208) and (497,211.22) .. (497,215.2) -- (497,236.8) .. controls (497,240.78) and (493.78,244) .. (489.8,244) -- (437.2,244) .. controls (433.22,244) and (430,240.78) .. (430,236.8) -- cycle ;
%Rounded Rect [id:dp16614415676250782] 
\draw  [draw opacity=0][fill={rgb, 255:red, 248; green, 231; blue, 28 }  ,fill opacity=1 ] (355,224.2) .. controls (355,220.22) and (358.22,217) .. (362.2,217) -- (414.8,217) .. controls (418.78,217) and (422,220.22) .. (422,224.2) -- (422,245.8) .. controls (422,249.78) and (418.78,253) .. (414.8,253) -- (362.2,253) .. controls (358.22,253) and (355,249.78) .. (355,245.8) -- cycle ;
%Straight Lines [id:da2966351312571295] 
\draw    (416,185) -- (395.5,217) ;
%Straight Lines [id:da33666794339638273] 
\draw    (343,141) -- (287,185) ;
%Shape: Path Data [id:dp8388109606125012] 
\draw  [fill={rgb, 255:red, 117; green, 174; blue, 240 }  ,fill opacity=1 ] (266,191) -- (266,175) -- (276.5,183) -- (266,191) -- cycle (287,175) -- (287,191) -- (276.5,183) -- (287,175) -- cycle ;
%Straight Lines [id:da8953237556687728] 
\draw    (287,186) -- (308,209) ;
%Rounded Rect [id:dp2695898301949107] 
\draw  [draw opacity=0][fill={rgb, 255:red, 125; green, 184; blue, 253 }  ,fill opacity=1 ] (280,215.2) .. controls (280,211.22) and (283.22,208) .. (287.2,208) -- (339.8,208) .. controls (343.78,208) and (347,211.22) .. (347,215.2) -- (347,236.8) .. controls (347,240.78) and (343.78,244) .. (339.8,244) -- (287.2,244) .. controls (283.22,244) and (280,240.78) .. (280,236.8) -- cycle ;
%Rounded Rect [id:dp39993870397322084] 
\draw  [draw opacity=0][fill={rgb, 255:red, 248; green, 231; blue, 28 }  ,fill opacity=1 ] (205,224.2) .. controls (205,220.22) and (208.22,217) .. (212.2,217) -- (264.8,217) .. controls (268.78,217) and (272,220.22) .. (272,224.2) -- (272,245.8) .. controls (272,249.78) and (268.78,253) .. (264.8,253) -- (212.2,253) .. controls (208.22,253) and (205,249.78) .. (205,245.8) -- cycle ;
%Straight Lines [id:da036147633593597184] 
\draw    (266,185) -- (245.5,217) ;

% Text Node
%\draw (360,121.4) node [anchor=north west][inner sep=0.75pt]    {$n_1$};
% Text Node
\draw (338,121.4) node [anchor=north west][inner sep=0.75pt]    {$\bigcup $};
% Text Node
\draw (433,214.4) node [anchor=north west][inner sep=0.75pt]    {$[[ P_{1}]]_{f_{2}}^{n_3}$};
% Text Node
\draw (438,163.4) node [anchor=north west][inner sep=0.75pt]    {$n_{3}$};
% Text Node
\draw (358,223.4) node [anchor=north west][inner sep=0.75pt]    {$[[ P_{2}]]_{f_{3}}^{n_3}$};
% Text Node
\draw (283,214.4) node [anchor=north west][inner sep=0.75pt]    {$[[ P_{1}]]_{f_{1}}^{n_2}$};
% Text Node
\draw (279,158.4) node [anchor=north west][inner sep=0.75pt]    {$n_{2}$};
% Text Node
\draw (208,223.4) node [anchor=north west][inner sep=0.75pt]    {$[[ P_{2}]]_{f_{4}}^{n_2}$};

\end{tikzpicture}

%% file: figures/pex5.tex
\tikzset{every picture/.style={line width=0.75pt}} %set default line width to 0.75pt        

\begin{tikzpicture}[x=0.5pt,y=0.5pt,yscale=-1,xscale=1]

%Straight Lines [id:da4594681119369949] 
\draw    (375,156) -- (407,191) ;
%Rounded Rect [id:dp5905130256643163] 
\draw  [draw opacity=0][fill={rgb, 255:red, 248; green, 231; blue, 28 }  ,fill opacity=1 ] (383,194.2) .. controls (383,190.22) and (386.22,187) .. (390.2,187) -- (442.8,187) .. controls (446.78,187) and (450,190.22) .. (450,194.2) -- (450,215.8) .. controls (450,219.78) and (446.78,223) .. (442.8,223) -- (390.2,223) .. controls (386.22,223) and (383,219.78) .. (383,215.8) -- cycle ;
%Straight Lines [id:da1999178830191941] 
\draw    (363,161) -- (328,192) ;
%Rounded Rect [id:dp7963881155353636] 
\draw  [draw opacity=0][fill={rgb, 255:red, 248; green, 231; blue, 28 }  ,fill opacity=1 ] (291,197.2) .. controls (291,193.22) and (294.22,190) .. (298.2,190) -- (350.8,190) .. controls (354.78,190) and (358,193.22) .. (358,197.2) -- (358,218.8) .. controls (358,222.78) and (354.78,226) .. (350.8,226) -- (298.2,226) .. controls (294.22,226) and (291,222.78) .. (291,218.8) -- cycle ;
%Straight Lines [id:da9405540978921029] 
\draw    (422,114) -- (443,140) ;
%Straight Lines [id:da6518242455801789] 
\draw    (400.5,113.5) -- (375,152) ;
%Rounded Rect [id:dp8348235367784094] 
\draw  [draw opacity=0][fill={rgb, 255:red, 184; green, 233; blue, 134 }  ,fill opacity=1 ] (425,148.2) .. controls (425,144.22) and (428.22,141) .. (432.2,141) -- (484.8,141) .. controls (488.78,141) and (492,144.22) .. (492,148.2) -- (492,169.8) .. controls (492,173.78) and (488.78,177) .. (484.8,177) -- (432.2,177) .. controls (428.22,177) and (425,173.78) .. (425,169.8) -- cycle ;

% Text Node
%\draw (363,127) node [anchor=north west][inner sep=0.75pt]    {$n_1$};
% Text Node
\draw (358,141.4) node [anchor=north west][inner sep=0.75pt]    {$\bigcup $};
% Text Node
\draw (386,193.4) node [anchor=north west][inner sep=0.75pt]    {$[[ P_{2}]]_{f_{4}}^{n_2}$};
% Text Node
\draw (294,196.4) node [anchor=north west][inner sep=0.75pt]    {$[[ P_{2}]]_{f_{3}}^{n_3}$};
% Text Node
\draw (428,147.4) node [anchor=north west][inner sep=0.75pt]    {$[[ P_{3}]]_{f_{5}}^{n_1}$};
% Text Node
\draw (428,98.4) node [anchor=north west][inner sep=0.75pt]    {$n_{1}$};
% Text Node
\draw (398,100.4) node [anchor=north west][inner sep=0.75pt]    {$\bigtimes $};

\end{tikzpicture}

%% file: figures/pex6.tex
\tikzset{every picture/.style={line width=0.75pt}} %set default line width to 0.75pt        

\begin{tikzpicture}[x=0.5pt,y=0.5pt,yscale=-1,xscale=1]

%Straight Lines [id:da4594681119369949] 
\draw    (375,156) -- (407,191) ;
%Rounded Rect [id:dp5905130256643163] 
\draw  [draw opacity=0][fill={rgb, 255:red, 125; green, 184; blue, 253 }  ,fill opacity=1 ] (383,194.2) .. controls (383,190.22) and (386.22,187) .. (390.2,187) -- (442.8,187) .. controls (446.78,187) and (450,190.22) .. (450,194.2) -- (450,215.8) .. controls (450,219.78) and (446.78,223) .. (442.8,223) -- (390.2,223) .. controls (386.22,223) and (383,219.78) .. (383,215.8) -- cycle ;
%Straight Lines [id:da1999178830191941] 
\draw    (363,161) -- (328,192) ;
%Rounded Rect [id:dp7963881155353636] 
\draw  [draw opacity=0][fill={rgb, 255:red, 125; green, 184; blue, 253 }  ,fill opacity=1 ] (291,197.2) .. controls (291,193.22) and (294.22,190) .. (298.2,190) -- (350.8,190) .. controls (354.78,190) and (358,193.22) .. (358,197.2) -- (358,218.8) .. controls (358,222.78) and (354.78,226) .. (350.8,226) -- (298.2,226) .. controls (294.22,226) and (291,222.78) .. (291,218.8) -- cycle ;
%Straight Lines [id:da9405540978921029] 
\draw    (422,114) -- (443,140) ;
%Straight Lines [id:da6518242455801789] 
\draw    (400.5,113.5) -- (375,152) ;
%Rounded Rect [id:dp8348235367784094] 
\draw  [draw opacity=0][fill={rgb, 255:red, 184; green, 233; blue, 134 }  ,fill opacity=1 ] (425,148.2) .. controls (425,144.22) and (428.22,141) .. (432.2,141) -- (484.8,141) .. controls (488.78,141) and (492,144.22) .. (492,148.2) -- (492,169.8) .. controls (492,173.78) and (488.78,177) .. (484.8,177) -- (432.2,177) .. controls (428.22,177) and (425,173.78) .. (425,169.8) -- cycle ;
%Shape: Path Data [id:dp16764352462852905] 
\draw  [fill={rgb, 255:red, 117; green, 174; blue, 240 }  ,fill opacity=1 ] (402,123) -- (402,107) -- (412.5,115) -- (402,123) -- cycle (423,107) -- (423,123) -- (412.5,115) -- (423,107) -- cycle ;

% Text Node
%\draw (363,127) node [anchor=north west][inner sep=0.75pt]    {$n_1$};
% Text Node
\draw (358,141.4) node [anchor=north west][inner sep=0.75pt]    {$\bigcup $};
% Text Node
\draw (386,193.4) node [anchor=north west][inner sep=0.75pt]    {$[[ P_{1}]]_{f_{2}}^{n_3}$};
% Text Node
\draw (294,196.4) node [anchor=north west][inner sep=0.75pt]    {$[[ P_{1}]]_{f_{1}}^{n_2}$};
% Text Node
\draw (428,147.4) node [anchor=north west][inner sep=0.75pt]    {$[[ P_{3}]]_{f_{5}}^{n_1}$};
% Text Node
\draw (428,98.4) node [anchor=north west][inner sep=0.75pt]    {$n_{1}$};

\end{tikzpicture}

%% file: figures/plan.tex
\tikzset{every picture/.style={line width=0.75pt}} %set default line width to 0.75pt        

\begin{tikzpicture}[x=0.5pt,y=0.5pt,yscale=-1,xscale=1]
%uncomment if require: \path (0,357); %set diagram left start at 0, and has height of 357

%Shape: Path Data [id:dp9878579699825297] 
\draw  [fill={rgb, 255:red, 117; green, 174; blue, 240 }  ,fill opacity=1 ] (361,82) -- (361,66) -- (371.5,74) -- (361,82) -- cycle (382,66) -- (382,82) -- (371.5,74) -- (382,66) -- cycle ;
%Straight Lines [id:da9641184210280794] 
\draw    (382,74) -- (403,100) ;
%Straight Lines [id:da4079373688457316] 
\draw    (360.5,73.5) -- (335,112) ;
%Rounded Rect [id:dp16209261154995547] 
\draw  [draw opacity=0][fill={rgb, 255:red, 184; green, 233; blue, 134 }  ,fill opacity=1 ] (385,108.2) .. controls (385,104.22) and (388.22,101) .. (392.2,101) -- (444.8,101) .. controls (448.78,101) and (452,104.22) .. (452,108.2) -- (452,129.8) .. controls (452,133.78) and (448.78,137) .. (444.8,137) -- (392.2,137) .. controls (388.22,137) and (385,133.78) .. (385,129.8) -- cycle ;
%Straight Lines [id:da10759439648769964] 
\draw    (335,116) -- (396,161) ;
%Shape: Path Data [id:dp5377764604823965] 
\draw  [fill={rgb, 255:red, 117; green, 174; blue, 240 }  ,fill opacity=1 ] (396,171) -- (396,155) -- (406.5,163) -- (396,171) -- cycle (417,155) -- (417,171) -- (406.5,163) -- (417,155) -- cycle ;
%Straight Lines [id:da9644949002198261] 
\draw    (417,163) -- (438,189) ;
%Rounded Rect [id:dp9946112722684539] 
\draw  [draw opacity=0][fill={rgb, 255:red, 125; green, 184; blue, 253 }  ,fill opacity=1 ] (410,195.2) .. controls (410,191.22) and (413.22,188) .. (417.2,188) -- (469.8,188) .. controls (473.78,188) and (477,191.22) .. (477,195.2) -- (477,216.8) .. controls (477,220.78) and (473.78,224) .. (469.8,224) -- (417.2,224) .. controls (413.22,224) and (410,220.78) .. (410,216.8) -- cycle ;
%Rounded Rect [id:dp8975832188035411] 
\draw  [draw opacity=0][fill={rgb, 255:red, 248; green, 231; blue, 28 }  ,fill opacity=1 ] (335,204.2) .. controls (335,200.22) and (338.22,197) .. (342.2,197) -- (394.8,197) .. controls (398.78,197) and (402,200.22) .. (402,204.2) -- (402,225.8) .. controls (402,229.78) and (398.78,233) .. (394.8,233) -- (342.2,233) .. controls (338.22,233) and (335,229.78) .. (335,225.8) -- cycle ;
%Straight Lines [id:da8341529811243695] 
\draw    (396,165) -- (375.5,197) ;
%Straight Lines [id:da6616176490084112] 
\draw    (323,121) -- (267,165) ;
%Shape: Path Data [id:dp3566490902416758] 
\draw  [fill={rgb, 255:red, 117; green, 174; blue, 240 }  ,fill opacity=1 ] (246,171) -- (246,155) -- (256.5,163) -- (246,171) -- cycle (267,155) -- (267,171) -- (256.5,163) -- (267,155) -- cycle ;
%Straight Lines [id:da7937898753551095] 
\draw    (267,166) -- (288,189) ;
%Rounded Rect [id:dp1007361945790608] 
\draw  [draw opacity=0][fill={rgb, 255:red, 125; green, 184; blue, 253 }  ,fill opacity=1 ] (260,195.2) .. controls (260,191.22) and (263.22,188) .. (267.2,188) -- (319.8,188) .. controls (323.78,188) and (327,191.22) .. (327,195.2) -- (327,216.8) .. controls (327,220.78) and (323.78,224) .. (319.8,224) -- (267.2,224) .. controls (263.22,224) and (260,220.78) .. (260,216.8) -- cycle ;
%Rounded Rect [id:dp15453529843862168] 
\draw  [draw opacity=0][fill={rgb, 255:red, 248; green, 231; blue, 28 }  ,fill opacity=1 ] (185,204.2) .. controls (185,200.22) and (188.22,197) .. (192.2,197) -- (244.8,197) .. controls (248.78,197) and (252,200.22) .. (252,204.2) -- (252,225.8) .. controls (252,229.78) and (248.78,233) .. (244.8,233) -- (192.2,233) .. controls (188.22,233) and (185,229.78) .. (185,225.8) -- cycle ;
%Straight Lines [id:da6162016201598419] 
\draw    (246,165) -- (225.5,197) ;

% Text Node
\draw (388,107.4) node [anchor=north west][inner sep=0.75pt]    {$[[ P_{3}]]_{f_{5}}^{n_1}$};
% Text Node
\draw (388,58.4) node [anchor=north west][inner sep=0.75pt]    {$n_{1}$};
% Text Node
%\draw (323,87) node [anchor=north west][inner sep=0.75pt]    {$n_1$};
% Text Node
\draw (318,101.4) node [anchor=north west][inner sep=0.75pt]    {$\bigcup $};
% Text Node
\draw (413,194.4) node [anchor=north west][inner sep=0.75pt]    {$[[ P_{1}]]_{f_{2}}^{n_3}$};
% Text Node
\draw (418,143.4) node [anchor=north west][inner sep=0.75pt]    {$n_{3}$};
% Text Node
\draw (338,203.4) node [anchor=north west][inner sep=0.75pt]    {$[[ P_{2}]]_{f_{3}}^{n_3}$};
% Text Node
\draw (263,194.4) node [anchor=north west][inner sep=0.75pt]    {$[[ P_{1}]]_{f_{1}}^{n_2}$};
% Text Node
\draw (259,138.4) node [anchor=north west][inner sep=0.75pt]    {$n_{2}$};
% Text Node
\draw (188,203.4) node [anchor=north west][inner sep=0.75pt]    {$[[ P_{2}]]_{f_{4}}^{n_2}$};

\end{tikzpicture}

%% file: queryprocessing.tex
Until now, we have described in Section~\ref{sec:queryoptimization} how \system{} obtains a query execution plan using compatibility graphs and locality information provided by the SPBF indexes.
In this section, we detail how \system{} evaluates a query given a query execution plan.

Given a BGP $P$, a compatibility graph $G^C=G^C(P,I^S)$, and a query execution plan $\Pi$ over $P$ and $G^C$, \system{} processes $P$ by processing the operations specified in $\Pi$ and, in doing so, delegating joins and Cartesian products to the nodes specified in $\Pi$.
The intermediate results from previous steps are used as input to subqueries at a later stage in the query execution plan.
In case of a distributed join, the intermediate results are transferred along with the partial query to use local bind joins similar to~\cite{DBLP:journals/corr/abs-2002-09172,brtpf}.
%At each step, solution mappings from previously evaluated subqueries are used as bindings.
%This is similar to how SPF~\cite{DBLP:journals/corr/abs-2002-09172} and brTPF~\cite{brtpf} process queries.
To formalize how star patterns in the query execution plan are processed over the fragments, we define a so-called \emph{selector} function in line with related work~\cite{DBLP:conf/www/AebeloeMH21,DBLP:journals/corr/abs-2002-09172,brtpf}. 
The selector function returns the results of processing the star pattern over a fragment given a set of solution mappings, i.e., the set of stars in the fragment that constitute the answer to the star pattern, as follows:
%The selector function is formalized in line with related work~\cite{DBLP:conf/www/AebeloeMH21,DBLP:journals/corr/abs-2002-09172,brtpf} as follows:
%To achieve this, in line with~\cite{DBLP:conf/www/AebeloeMH21,DBLP:journals/corr/abs-2002-09172,brtpf}, each \system{} node contains a \emph{selector} function that formalizes how star pattern requests are processed over a fragment in a node $n$'s local datastore.
%Furthermore, since partitions in \system{} are based on characteristic sets, the selector function is in line with the star pattern-based selector function presented in~\cite{DBLP:journals/corr/abs-2002-09172}.
%Let $[[P]]_f$ denote the answer to a star pattern $P$ over a fragment $f$ and let $[[P]]_f$ be a set of solution mappings.
%Then, a set of triples $T$ are matching triples to $P$ if there exists a solution mapping $\mu\in[[P]]_f$ such that $T=\mu[P]$.
%Formally, the selector function is defined as follows.

\begin{definition}[Selector Function~\cite{DBLP:conf/www/AebeloeMH21,DBLP:journals/corr/abs-2002-09172,brtpf}]\label{def:selector}
Given a node $n$, a star pattern $P$, and a finite set of distinct solution mappings $\Omega$, the star pattern-based selector function for $P$ and $\Omega$, denoted $s_{(P,\Omega)}$ is for every fragment $f$ in $n$'s local datastore defined as follows.

$$
	 s_{(P,\Omega)}(f) =
	\begin{cases}
		\{t\in T\mid T\subseteq f \land T[P]\} & \text{if } \Omega=\emptyset  \\
		\{t\in T\mid T\subseteq f \land \exists \mu\in [[P]]_f,\mu'\in\Omega:\mu[P]=T\land\mu'\subseteq\mu \} & \text{ otherwise.}
	\end{cases} $$
\end{definition}

In line with~\cite{DBLP:conf/www/AebeloeMH21,DBLP:journals/corr/abs-2002-09172,brtpf}, and to avoid long-running requests on each node, we apply pagination to the results of star pattern requests, i.e., we group the results into reasonably sized pages to avoid excessive data transfer.
The page size used in our experimental evaluation (Section~\ref{sec:experiments}) is the page size recommended by related work~\cite{DBLP:conf/www/AebeloeMH21,DBLP:journals/corr/abs-2002-09172,brtpf}, i.e., $100$.
However, for ease of presentation, we assume that all results can fit into one page when presenting the approach to query processing.
Furthermore, to avoid underestimating costs caused by the selector function returning some duplicate values (e.g., when the same subject has multiple object values for a specific predicate), our implementation always uses $card_S$ (Equation~\ref{eq:cardnondistinctsingle}) and $card^{\bowtie}_S$ (Equation~\ref{eq:cardjoinnondistinct}) for cardinality estimations, regardless of whether or not the \texttt{DISTINCT} keyword is given.
Last, given a star pattern $P$, a node $n$, a fragment $f_i$, and a finite set of solution mappings $\Omega$, $sel_n(f_i,P,\Omega)$ denotes the result of invoking $s_{(P,\Omega)}(f_i)$ on $n$.

\begin{algorithm*}[htb]
\caption{Evaluate a join plan}
\label{algo:process}
\begin{algorithmic}[1]
\Statex \textbf{Input:} A join plan $\Pi$; a node $n$; a set of solution mappings $\Omega$
\Statex \textbf{Output:} A set of solution mappings $\Omega$
\Function{evaluatePlan}{$\Pi$,$n$,$\Omega=\{\emptyset\}$}\label{ln:evalfunc}
%\IfThen{$\Pi=\Pi_{\emptyset}$}{\Return $\Omega$;}
%\If{$\Pi=\Pi_{\emptyset}$}
%\State \Return $\emptyset$;\label{ln:retempty2}
\If{$\Pi=\Pi_1\times^{n_i}\Pi_2$}\label{ln:cartstart}
\State $\Omega_1\leftarrow$ \texttt{evaluatePlan}$(\Pi_1,n_i,\Omega)$;\label{ln:cartmid1}
\State $\Omega_2\leftarrow$ \texttt{evaluatePlan}$(\Pi_2,n_i,\Omega)$;\label{ln:cartmid2}
\State $\Omega\leftarrow \Omega_1\times\Omega_2$;\label{ln:cartend}
\ElsIf{$\Pi=\Pi_1\bowtie^{n_i}\Pi_2$}\label{ln:joinstart}
\State $\Omega\leftarrow$ \texttt{evaluatePlan}$(\Pi_1,n_i,\Omega)$;\label{ln:joinmid}
\State $\Omega\leftarrow$ \texttt{evaluatePlan}$(\Pi_2,n_i,\Omega)$;\label{ln:joinend1}
%\State \Return \texttt{delegateJoin}$(\Pi',P,f,n_i,\Omega)$;
\ElsIf{$\Pi=\Pi_1\cup\Pi_2$}\label{ln:unionstart}
\State $\Omega_1\leftarrow$ \texttt{evaluatePlan}$(\Pi_1,n,\Omega)$;\label{ln:unionmid1}
\State $\Omega_2\leftarrow$ \texttt{evaluatePlan}$(\Pi_2,n,\Omega)$;\label{ln:unionmid2}
\State $\Omega\leftarrow \Omega_1\cup\Omega_2$;\label{ln:unionend}
\ElsIf{$\Pi=[[P]]_f$}\label{ln:spstart}
\State $N\leftarrow I^S_n.\eta(f)$;
\IfThen{$n\in N$}{$n_i\leftarrow n$;}\label{ln:selif}
\ElseThen{$n_i\leftarrow$ \texttt{takeOne}$(N)$;}\label{ln:selelse}
\State $\phi\leftarrow sel_{n_i}(f,P,\Omega)$;\label{ln:selector}
\State $\Omega\leftarrow\Omega\bowtie\{\mu\mid dom(\mu)=vars(P)$ and $\mu[P]\in\phi\}$;\label{ln:spend}
\EndIf
%\ElseThen{\Return $\Omega$;}
\State \Return $\Omega$;\label{ln:retevalfunc}
\EndFunction\label{ln:evalfuncend}
%\Function{delegateJoin}{$\Pi$,$P$,$f$,$n$,$\Omega$}
%\State $\Pi\leftarrow\Pi_1\bowtie^{n_i}\Pi_2$;
%\State $\Pi_1,\Pi_2\leftarrow\Pi_1,\Pi_2$ where $\Pi=\Pi_1\bowtie^{n}\Pi_2$;
%\State $\Pi_1\leftarrow\Pi_1$ where $\Pi=\Pi_1\bowtie^{n}\Pi_2$;
%\State $\Pi_2\leftarrow\Pi_2$ where $\Pi=\Pi_1\bowtie^{n}\Pi_2$;
%\State $n_i\leftarrow n_i$ where $\Pi=\Pi_1\bowtie^{n}\Pi_2$;
%\State $\Omega\leftarrow$ \texttt{evaluatePlan}$(\Pi,n,\Omega)$;
%\State \Return \texttt{processPattern}$(P,f,n,\Omega)$;
%\EndFunction
%\Function{processUnion}{$\Pi_1$,$\Pi_2$,$n$,$\Omega$}
%\State $\Omega_1\leftarrow$ \texttt{evaluatePlan}$(\Pi_1,n,\Omega)$;
%\State $\Omega_2\leftarrow$ \texttt{evaluatePlan}$(\Pi_2,n,\Omega)$;
%\State \Return $\Omega_1\cup\Omega_2$;
%\EndFunction
%\Function{processPattern}{$P$,$f$,$n$,$\Omega$}
%\State $N\leftarrow I^S_n.\eta(f)$;
%\IfThen{$n\in N$}{$n_i\leftarrow n$;}
%\ElseThen{$n_i\leftarrow$ \texttt{takeOne}$(N)$;}
%\State $\phi\leftarrow sel_{n_i}(f,P,\Omega)$;
%\State \Return $\Omega\bowtie\{\mu\mid dom(\mu)=vars(P)$ and $\mu[P]\in\phi\}$;
%\EndFunction
\end{algorithmic}
\end{algorithm*}

Let $I^S_n$ denote a node $n$'s SPBF index.
%The empty join plan (i.e., the plan containing no star patterns) is denoted $\Pi_{\emptyset}$.
%Furthermore, the notation $|\Pi|$ denotes the number of star patterns in $\Pi$.
The \texttt{evaluatePlan} function in Algorithm~\ref{algo:process} defines a recursive function that processes a query execution plan on a node $n$ by using the selector function defined in Definition~\ref{def:selector} for selections in the plan and making recursive calls to the nodes specified in the plan. 

\begin{figure*}[b!]
\centering
\begin{subfigure}[b]{.49\textwidth}
  \centering
  \input{figures/procex1.tex}
  \caption{Delegating subqueries to $n_2$ and $n_3$}\label{subfig:procex1}
\end{subfigure}
\begin{subfigure}[b]{.49\textwidth}
  \centering
  \input{figures/procex2.tex}
  \caption{Processing $[[P_3]]_{f_5}$ locally on node $n_1$}\label{subfig:procex2}
\end{subfigure}
\caption{Processing $\Pi$ in Figure~\ref{subfig:planex} on $n_1$ by (a) delegating $[[P_2]]_{f_4}\bowtie^{n_2} [[P_1]]_{f_1}$ to $n_2$ and $[[P_2]]_{f_3}\bowtie^{n_3} [[P_1]]_{f_2}$ to $n_3$ concurrently and (b) processing the join between these 850 results and $[[P_3]]_{f_5}$ locally on $n_1$ to achieve the $154$ results (solid arrows denote neighbors, dotted arrows subquery delegation, and dashed arrows transferring of intermediate results).
$n_1$ can send intermediate results to $n_3$ since it is within its horizon.}
\label{fig:procex}
\end{figure*}
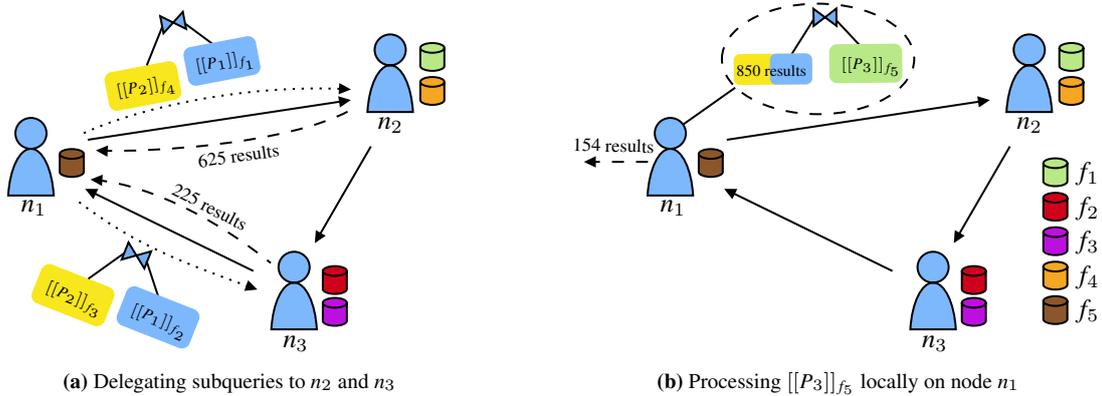

Consider, for instance, the query execution plan $\Pi$ shown in Figure~\ref{subfig:planex} for query $Q$ in Figure~\ref{subfig:running_q} processed by node $n_1$ in the running example.
Figure~\ref{fig:procex} shows an overview of which parts of the query are sent to which node during query processing.
Since $\Pi$ is of type join, the function enters the if statement in line~\ref{ln:joinstart}.
Here, the function first makes a recursive call (since the join was delegated to node $n_1$) with the left-most subplan, i.e., $\Pi_1=([[P_2]]_{f_4}\bowtie^{n_2} [[P_1]]_{f_1})\cup ([[P_2]]_{f_3}\bowtie^{n_3} [[P_1]]_{f_2})$ (visualized in Figure~\ref{subfig:pex4}), in line~\ref{ln:joinmid}.

Since $\Pi_1$ is of type union, Algorithm~\ref{algo:process} in lines~\ref{ln:unionmid1}-\ref{ln:unionmid2} makes two recursive calls for the two subplans $\Pi_1=[[P_2]]_{f_4}\bowtie^{n_2} [[P_1]]_{f_1}$ and $\Pi_2=[[P_2]]_{f_3}\bowtie^{n_3} [[P_1]]_{f_2}$.
Note that these two recursive calls can be processed concurrently and indeed is done so in the implementation of \system{}.
This step is shown in Figure~\ref{subfig:procex1} where $\Pi_1$ is sent to node $n_2$ and $\Pi_2$ is sent to node $n_3$.
Since both subplans follow the same structure, and thus the same evaluation process, we will only explain what happens when processing $\Pi_1$.

When processing the plan $\Pi_1$ from above, Algorithm~\ref{algo:process} first calls the \texttt{evaluatePlan} on node $n_2$ for the subplan $[[P_2]]_{f_4}$, i.e., the selection for $P_2$ over $f_4$, in line~\ref{ln:joinmid}.
%The if-else statement on lines~\ref{ln:selif}-\ref{ln:selelse} selects the node on which to process that star pattern.
The \texttt{takeOne} function in line~\ref{ln:selelse} selects a random node with the fragment in its local datastore if the node that processes the subquery does not store the fragment locally.
In this case, since $n_2$ stores $f_4$, it calls the selector function for $P_2$ over $f_4$ locally in line~\ref{ln:selector}.
The $500$ results of processing $P_2$ over $f_4$ (cf. Table~\ref{tab:cardinalities}) are then joined with the singleton set of bindings $\Omega$ that includes the empty mapping (i.e., a mapping compatible with any mapping) in line~\ref{ln:spend} and returned in line~\ref{ln:retevalfunc}.

Upon receiving the $500$ results in line~\ref{ln:joinmid}, Algorithm~\ref{algo:process} makes another recursive call in line~\ref{ln:joinend1} to \texttt{evaluatePlan} on node $n_2$ for the subplan $[[P_1]]_{f_1}$, i.e., the selection for $P_1$ over $f_1$ with the $500$ intermediate results in $\Omega$.
Again, $n_2$ calls the local selector for $P_1$ over $f_1$ using the intermediate results in $\Omega$ as bindings.
This results in $625$ intermediate results in $\Omega$ that are the result of processing $P_1\bowtie P_2$ over $f_1$ and $f_4$, which are returned by the function in line~\ref{ln:retevalfunc}.

While $n_2$ found the $625$ results from processing $[[P_2]]_{f_4}\bowtie^{n_2} [[P_1]]_{f_1}$ in the recursive call in line~\ref{ln:unionmid1}, $n_3$ found the additional $225$ results of processing $[[P_2]]_{f_3}\bowtie^{n_3} [[P_1]]_{f_2}$ in the recursive call in line~\ref{ln:unionmid2} following the same steps as described above for $n_2$.
In line~\ref{ln:unionend}, these results are combined and 850 bindings are returned in line~\ref{ln:retevalfunc}, which is visualized on Figure~\ref{subfig:procex1} as $n_2$ returning $625$ results to $n_1$ and $n_3$ returning $225$ results to $n_1$.

The $850$ intermediate results in $\Omega$ found by processing $([[P_2]]_{f_4}\bowtie^{n_2} [[P_1]]_{f_1})\cup ([[P_2]]_{f_3}\bowtie^{n_3} [[P_1]]_{f_2})$ in line~\ref{ln:joinmid} are used as bindings for the recursive call made in line~\ref{ln:joinend1} for the subplan $[[P_3]]_{f_5}$.
This is visualized in Figure~\ref{subfig:procex2}. % where $[[P_3]]_{f_5}$ is joined with the $850$ earlier results lovally on node $n_1$.
Since $n_1$ stores $f_5$ locally, it calls the local selector for $P_3$ over $f_5$ and $\Omega$ in line~\ref{ln:selector}.
The $154$ results of processing $P_3$ over $f_5$ are joined with $\Omega$ in line~\ref{ln:spend} and returned as the final results in line~\ref{ln:retevalfunc}.

%Notice that the recursive calls on lines~\ref{ln:cartmid1}-\ref{ln:cartmid2} and lines~\ref{ln:unionmid1}-\ref{ln:unionmid2}, i.e., for the Cartesian product and union operators, can be processed concurrently and indeed is done so in the implementation of \system{}, while the recursive calls on lines~\ref{ln:joinmid}-\ref{ln:joinend1}, i.e., for the join operator, have to be processed sequentially since the call on line~\ref{ln:joinend1} uses the intermediate bindings found from the call on line~\ref{ln:joinmid}.
As mentioned above, our implementation uses pagination of the results meaning, for instance, when processing the subplan $[[P_2]]_{f_4}$ in line~\ref{ln:joinmid}, the $500$ results would be split into multiple pages.
In the implementation of \system{}, nodes at subsequent steps in the pipeline start processing joins as soon as they receive some intermediate bindings.
For instance, in the running example, $n_1$ starts processing the join between $P_2\bowtie P_1\bowtie P_3$ locally as soon as it receives results for $P_2\bowtie P_1$ from either $n_2$ or $n_3$.

%% file: figures/procex1.tex
\tikzset{every picture/.style={line width=0.75pt}} %set default line width to 0.75pt        

\begin{tikzpicture}[x=0.5pt,y=0.5pt,yscale=-1,xscale=1]
%uncomment if require: \path (0,344); %set diagram left start at 0, and has height of 344

%Flowchart: Magnetic Disk [id:dp7247643729600729] 
\draw  [fill={rgb, 255:red, 208; green, 2; blue, 27 }  ,fill opacity=1 ] (344,234) -- (344,247.65) .. controls (344,249.68) and (339.97,251.33) .. (335,251.33) .. controls (330.03,251.33) and (326,249.68) .. (326,247.65) -- (326,234)(344,234) .. controls (344,236.03) and (339.97,237.68) .. (335,237.68) .. controls (330.03,237.68) and (326,236.03) .. (326,234) .. controls (326,231.97) and (330.03,230.33) .. (335,230.33) .. controls (339.97,230.33) and (344,231.97) .. (344,234) -- cycle ;
%Straight Lines [id:da5568957457791845] 
\draw    (150,135) -- (344.03,105.45) ;
\draw [shift={(347,105)}, rotate = 171.34] [fill={rgb, 255:red, 0; green, 0; blue, 0 }  ][line width=0.08]  [draw opacity=0] (8.93,-4.29) -- (0,0) -- (8.93,4.29) -- cycle    ;
%Straight Lines [id:da07882088901935402] 
\draw    (362,140) -- (321.5,210.4) ;
\draw [shift={(320,213)}, rotate = 299.91] [fill={rgb, 255:red, 0; green, 0; blue, 0 }  ][line width=0.08]  [draw opacity=0] (8.93,-4.29) -- (0,0) -- (8.93,4.29) -- cycle    ;
%Straight Lines [id:da1587072735599474] 
\draw    (274,234) -- (150.69,173.32) ;
\draw [shift={(148,172)}, rotate = 26.2] [fill={rgb, 255:red, 0; green, 0; blue, 0 }  ][line width=0.08]  [draw opacity=0] (8.93,-4.29) -- (0,0) -- (8.93,4.29) -- cycle    ;
%Flowchart: Magnetic Disk [id:dp8135312481761536] 
\draw  [fill={rgb, 255:red, 139; green, 87; blue, 42 }  ,fill opacity=1 ] (147,145.68) -- (147,159.32) .. controls (147,161.35) and (142.97,163) .. (138,163) .. controls (133.03,163) and (129,161.35) .. (129,159.32) -- (129,145.68)(147,145.68) .. controls (147,147.7) and (142.97,149.35) .. (138,149.35) .. controls (133.03,149.35) and (129,147.7) .. (129,145.68) .. controls (129,143.65) and (133.03,142) .. (138,142) .. controls (142.97,142) and (147,143.65) .. (147,145.68) -- cycle ;
%Flowchart: Magnetic Disk [id:dp32085119665350426] 
\draw  [fill={rgb, 255:red, 184; green, 233; blue, 134 }  ,fill opacity=1 ] (417,65.68) -- (417,79.33) .. controls (417,81.35) and (412.97,83) .. (408,83) .. controls (403.03,83) and (399,81.35) .. (399,79.33) -- (399,65.68)(417,65.68) .. controls (417,67.7) and (412.97,69.35) .. (408,69.35) .. controls (403.03,69.35) and (399,67.7) .. (399,65.68) .. controls (399,63.65) and (403.03,62) .. (408,62) .. controls (412.97,62) and (417,63.65) .. (417,65.68) -- cycle ;
%Flowchart: Magnetic Disk [id:dp7693312823758679] 
\draw  [fill={rgb, 255:red, 245; green, 166; blue, 35 }  ,fill opacity=1 ] (417,91.33) -- (417,104.98) .. controls (417,107) and (412.97,108.65) .. (408,108.65) .. controls (403.03,108.65) and (399,107) .. (399,104.98) -- (399,91.33)(417,91.33) .. controls (417,93.35) and (412.97,95) .. (408,95) .. controls (403.03,95) and (399,93.35) .. (399,91.33) .. controls (399,89.3) and (403.03,87.65) .. (408,87.65) .. controls (412.97,87.65) and (417,89.3) .. (417,91.33) -- cycle ;
%Flowchart: Magnetic Disk [id:dp05248014479362606] 
\draw  [fill={rgb, 255:red, 189; green, 16; blue, 224 }  ,fill opacity=1 ] (344,258) -- (344,271.65) .. controls (344,273.68) and (339.97,275.33) .. (335,275.33) .. controls (330.03,275.33) and (326,273.68) .. (326,271.65) -- (326,258)(344,258) .. controls (344,260.03) and (339.97,261.68) .. (335,261.68) .. controls (330.03,261.68) and (326,260.03) .. (326,258) .. controls (326,255.97) and (330.03,254.33) .. (335,254.33) .. controls (339.97,254.33) and (344,255.97) .. (344,258) -- cycle ;
%Shape: Path Data [id:dp617733302986398] 
\draw  [fill={rgb, 255:red, 125; green, 184; blue, 253 }  ,fill opacity=1 ] (107.47,141.81) .. controls (116.18,141.81) and (123.39,157.16) .. (124.5,177.04) -- (90.43,177.04) .. controls (91.54,157.16) and (98.75,141.81) .. (107.47,141.81) -- cycle (107.47,118.5) .. controls (113.9,118.5) and (119.12,123.72) .. (119.12,130.16) .. controls (119.12,136.59) and (113.9,141.81) .. (107.47,141.81) .. controls (101.03,141.81) and (95.81,136.59) .. (95.81,130.16) .. controls (95.81,123.72) and (101.03,118.5) .. (107.47,118.5) -- cycle ;
%Shape: Path Data [id:dp6568901603528909] 
\draw  [fill={rgb, 255:red, 125; green, 184; blue, 253 }  ,fill opacity=1 ] (304.47,242.81) .. controls (313.18,242.81) and (320.39,258.16) .. (321.5,278.04) -- (287.43,278.04) .. controls (288.54,258.16) and (295.75,242.81) .. (304.47,242.81) -- cycle (304.47,219.5) .. controls (310.9,219.5) and (316.12,224.72) .. (316.12,231.16) .. controls (316.12,237.59) and (310.9,242.81) .. (304.47,242.81) .. controls (298.03,242.81) and (292.81,237.59) .. (292.81,231.16) .. controls (292.81,224.72) and (298.03,219.5) .. (304.47,219.5) -- cycle ;
%Shape: Path Data [id:dp5771478236615337] 
\draw  [fill={rgb, 255:red, 125; green, 184; blue, 253 }  ,fill opacity=1 ] (376.47,78.81) .. controls (385.18,78.81) and (392.39,94.16) .. (393.5,114.04) -- (359.43,114.04) .. controls (360.54,94.16) and (367.75,78.81) .. (376.47,78.81) -- cycle (376.47,55.5) .. controls (382.9,55.5) and (388.12,60.72) .. (388.12,67.16) .. controls (388.12,73.59) and (382.9,78.81) .. (376.47,78.81) .. controls (370.03,78.81) and (364.81,73.59) .. (364.81,67.16) .. controls (364.81,60.72) and (370.03,55.5) .. (376.47,55.5) -- cycle ;
%Curve Lines [id:da7740646856163652] 
\draw  [dash pattern={on 0.84pt off 2.51pt}]  (147,130) .. controls (182.64,113.17) and (263.36,92.42) .. (344.54,96.86) ;
\draw [shift={(347,97)}, rotate = 183.49] [fill={rgb, 255:red, 0; green, 0; blue, 0 }  ][line width=0.08]  [draw opacity=0] (8.93,-4.29) -- (0,0) -- (8.93,4.29) -- cycle    ;
%Shape: Path Data [id:dp06724058068156558] 
\draw  [fill={rgb, 255:red, 117; green, 174; blue, 240 }  ,fill opacity=1 ] (206.72,54.64) -- (204.24,41.72) -- (213.96,46.55) -- (206.72,54.64) -- cycle (221.19,38.47) -- (223.67,51.38) -- (213.96,46.55) -- (221.19,38.47) -- cycle ;
%Straight Lines [id:da1900578872058143] 
\draw    (222.9,47.35) -- (243.42,62.66) ;
%Rounded Rect [id:dp9999661890302846] 
\draw  [draw opacity=0][fill={rgb, 255:red, 125; green, 184; blue, 253 }  ,fill opacity=1 ] (221.77,72.01) .. controls (221.16,68.8) and (223.26,65.69) .. (226.47,65.08) -- (268.94,56.92) .. controls (272.15,56.31) and (275.25,58.41) .. (275.87,61.62) -- (279.22,79.06) .. controls (279.83,82.27) and (277.73,85.37) .. (274.52,85.99) -- (232.05,94.14) .. controls (228.84,94.76) and (225.74,92.66) .. (225.12,89.45) -- cycle ;
%Rounded Rect [id:dp7148296446402947] 
\draw  [draw opacity=0][fill={rgb, 255:red, 248; green, 231; blue, 28 }  ,fill opacity=1 ] (162.62,90.9) .. controls (162,87.69) and (164.1,84.59) .. (167.31,83.97) -- (209.78,75.82) .. controls (212.99,75.2) and (216.09,77.3) .. (216.71,80.51) -- (220.06,97.95) .. controls (220.68,101.16) and (218.57,104.27) .. (215.36,104.88) -- (172.89,113.04) .. controls (169.68,113.65) and (166.58,111.55) .. (165.96,108.34) -- cycle ;
%Straight Lines [id:da8331873765539162] 
\draw    (205.79,49.8) -- (194.2,78.81) ;
%Shape: Path Data [id:dp9538974737214193] 
\draw  [fill={rgb, 255:red, 117; green, 174; blue, 240 }  ,fill opacity=1 ] (175.35,225.39) -- (180.48,212.42) -- (186.42,222.27) -- (175.35,225.39) -- cycle (197.5,219.15) -- (192.37,232.12) -- (186.42,222.27) -- (197.5,219.15) -- cycle ;
%Straight Lines [id:da7503177552825675] 
\draw    (194.05,226.56) -- (203.06,254.12) ;
%Rounded Rect [id:dp8112288274980909] 
\draw  [draw opacity=0][fill={rgb, 255:red, 125; green, 184; blue, 253 }  ,fill opacity=1 ] (178.94,249.5) .. controls (180.21,246.28) and (183.86,244.7) .. (187.08,245.97) -- (229.73,262.84) .. controls (232.95,264.11) and (234.53,267.76) .. (233.26,270.98) -- (226.33,288.5) .. controls (225.05,291.72) and (221.41,293.3) .. (218.18,292.02) -- (175.54,275.16) .. controls (172.31,273.88) and (170.73,270.24) .. (172.01,267.01) -- cycle ;
%Rounded Rect [id:dp5483697856534778] 
\draw  [draw opacity=0][fill={rgb, 255:red, 248; green, 231; blue, 28 }  ,fill opacity=1 ] (115.24,232.75) .. controls (116.52,229.52) and (120.17,227.94) .. (123.39,229.22) -- (166.04,246.08) .. controls (169.26,247.36) and (170.84,251.01) .. (169.56,254.23) -- (162.64,271.74) .. controls (161.36,274.97) and (157.71,276.55) .. (154.49,275.27) -- (111.85,258.4) .. controls (108.62,257.13) and (107.04,253.48) .. (108.32,250.26) -- cycle ;
%Straight Lines [id:da10103808794553759] 
\draw    (177.27,220.53) -- (150.39,239.9) ;
%Curve Lines [id:da3372699848397547] 
\draw  [dash pattern={on 0.84pt off 2.51pt}]  (143.5,182) .. controls (185.86,214.51) and (220.45,227.61) .. (274.52,247.1) ;
\draw [shift={(277,248)}, rotate = 199.82] [fill={rgb, 255:red, 0; green, 0; blue, 0 }  ][line width=0.08]  [draw opacity=0] (8.93,-4.29) -- (0,0) -- (8.93,4.29) -- cycle    ;
%Curve Lines [id:da2805777594926425] 
\draw  [dash pattern={on 4.5pt off 4.5pt}]  (155.54,165.28) .. controls (201.69,169.6) and (253.85,200.7) .. (287,228) ;
\draw [shift={(152,165)}, rotate = 3.65] [fill={rgb, 255:red, 0; green, 0; blue, 0 }  ][line width=0.08]  [draw opacity=0] (8.93,-4.29) -- (0,0) -- (8.93,4.29) -- cycle    ;
%Curve Lines [id:da9930808201881932] 
\draw  [dash pattern={on 4.5pt off 4.5pt}]  (159.26,143.16) .. controls (196.43,144.62) and (299.44,138.22) .. (346,113) ;
\draw [shift={(156,143)}, rotate = 3.37] [fill={rgb, 255:red, 0; green, 0; blue, 0 }  ][line width=0.08]  [draw opacity=0] (8.93,-4.29) -- (0,0) -- (8.93,4.29) -- cycle    ;

% Text Node
\draw (98,178.4) node [anchor=north west][inner sep=0.75pt]    {$n_{1}$};
% Text Node
\draw (365,115.4) node [anchor=north west][inner sep=0.75pt]    {$n_{2}$};
% Text Node
\draw (294,280.4) node [anchor=north west][inner sep=0.75pt]    {$n_{3}$};
% Text Node
\draw (227.04,72.74) node [anchor=north west][inner sep=0.75pt]  [font=\tiny,rotate=-349.13]  {$[[ P_{1}]]_{f_{1}}$};
% Text Node
\draw (167.88,91.63) node [anchor=north west][inner sep=0.75pt]  [font=\tiny,rotate=-349.13]  {$[[ P_{2}]]_{f_{4}}$};
% Text Node
\draw (183.87,254.41) node [anchor=north west][inner sep=0.75pt]  [font=\tiny,rotate=-21.58]  {$[[ P_{1}]]_{f_{2}}$};
% Text Node
\draw (120.18,237.66) node [anchor=north west][inner sep=0.75pt]  [font=\tiny,rotate=-21.58]  {$[[ P_{2}]]_{f_{3}}$};
% Text Node
\draw (215.74,164.03) node [anchor=north west][inner sep=0.75pt]  [font=\scriptsize,rotate=-27.82] [align=left] {225 results};
% Text Node
\draw (229.6,144.58) node [anchor=north west][inner sep=0.75pt]  [font=\scriptsize,rotate=-351.85] [align=left] {625 results};

\end{tikzpicture}

%% file: figures/procex2.tex
\tikzset{every picture/.style={line width=0.75pt}} %set default line width to 0.75pt        

\begin{tikzpicture}[x=0.5pt,y=0.5pt,yscale=-1,xscale=1]
%uncomment if require: \path (0,316); %set diagram left start at 0, and has height of 316

%Straight Lines [id:da37704984932826724] 
\draw    (234.97,29.3) -- (218.49,55.56) ;
%Flowchart: Magnetic Disk [id:dp033147064678631155] 
\draw  [fill={rgb, 255:red, 208; green, 2; blue, 27 }  ,fill opacity=1 ] (364,219) -- (364,232.65) .. controls (364,234.68) and (359.97,236.33) .. (355,236.33) .. controls (350.03,236.33) and (346,234.68) .. (346,232.65) -- (346,219)(364,219) .. controls (364,221.03) and (359.97,222.68) .. (355,222.68) .. controls (350.03,222.68) and (346,221.03) .. (346,219) .. controls (346,216.97) and (350.03,215.33) .. (355,215.33) .. controls (359.97,215.33) and (364,216.97) .. (364,219) -- cycle ;
%Straight Lines [id:da49610200616268807] 
\draw    (170,120) -- (364.03,90.45) ;
\draw [shift={(367,90)}, rotate = 171.34] [fill={rgb, 255:red, 0; green, 0; blue, 0 }  ][line width=0.08]  [draw opacity=0] (8.93,-4.29) -- (0,0) -- (8.93,4.29) -- cycle    ;
%Straight Lines [id:da7464309454454302] 
\draw    (382,125) -- (341.5,195.4) ;
\draw [shift={(340,198)}, rotate = 299.91] [fill={rgb, 255:red, 0; green, 0; blue, 0 }  ][line width=0.08]  [draw opacity=0] (8.93,-4.29) -- (0,0) -- (8.93,4.29) -- cycle    ;
%Straight Lines [id:da5027382613891735] 
\draw    (294,219) -- (170.69,158.32) ;
\draw [shift={(168,157)}, rotate = 26.2] [fill={rgb, 255:red, 0; green, 0; blue, 0 }  ][line width=0.08]  [draw opacity=0] (8.93,-4.29) -- (0,0) -- (8.93,4.29) -- cycle    ;
%Flowchart: Magnetic Disk [id:dp15170467018485534] 
\draw  [fill={rgb, 255:red, 139; green, 87; blue, 42 }  ,fill opacity=1 ] (167,130.68) -- (167,144.32) .. controls (167,146.35) and (162.97,148) .. (158,148) .. controls (153.03,148) and (149,146.35) .. (149,144.32) -- (149,130.68)(167,130.68) .. controls (167,132.7) and (162.97,134.35) .. (158,134.35) .. controls (153.03,134.35) and (149,132.7) .. (149,130.68) .. controls (149,128.65) and (153.03,127) .. (158,127) .. controls (162.97,127) and (167,128.65) .. (167,130.68) -- cycle ;
%Flowchart: Magnetic Disk [id:dp06442421923580233] 
\draw  [fill={rgb, 255:red, 184; green, 233; blue, 134 }  ,fill opacity=1 ] (437,50.68) -- (437,64.33) .. controls (437,66.35) and (432.97,68) .. (428,68) .. controls (423.03,68) and (419,66.35) .. (419,64.33) -- (419,50.68)(437,50.68) .. controls (437,52.7) and (432.97,54.35) .. (428,54.35) .. controls (423.03,54.35) and (419,52.7) .. (419,50.68) .. controls (419,48.65) and (423.03,47) .. (428,47) .. controls (432.97,47) and (437,48.65) .. (437,50.68) -- cycle ;
%Flowchart: Magnetic Disk [id:dp9287473688285852] 
\draw  [fill={rgb, 255:red, 245; green, 166; blue, 35 }  ,fill opacity=1 ] (437,76.33) -- (437,89.98) .. controls (437,92) and (432.97,93.65) .. (428,93.65) .. controls (423.03,93.65) and (419,92) .. (419,89.98) -- (419,76.33)(437,76.33) .. controls (437,78.35) and (432.97,80) .. (428,80) .. controls (423.03,80) and (419,78.35) .. (419,76.33) .. controls (419,74.3) and (423.03,72.65) .. (428,72.65) .. controls (432.97,72.65) and (437,74.3) .. (437,76.33) -- cycle ;
%Flowchart: Magnetic Disk [id:dp3488463391684711] 
\draw  [fill={rgb, 255:red, 189; green, 16; blue, 224 }  ,fill opacity=1 ] (364,243) -- (364,256.65) .. controls (364,258.68) and (359.97,260.33) .. (355,260.33) .. controls (350.03,260.33) and (346,258.68) .. (346,256.65) -- (346,243)(364,243) .. controls (364,245.03) and (359.97,246.68) .. (355,246.68) .. controls (350.03,246.68) and (346,245.03) .. (346,243) .. controls (346,240.97) and (350.03,239.33) .. (355,239.33) .. controls (359.97,239.33) and (364,240.97) .. (364,243) -- cycle ;
%Shape: Path Data [id:dp24217160671675186] 
\draw  [fill={rgb, 255:red, 125; green, 184; blue, 253 }  ,fill opacity=1 ] (127.72,126.81) .. controls (136.43,126.81) and (143.64,142.16) .. (144.75,162.04) -- (110.68,162.04) .. controls (111.79,142.16) and (119,126.81) .. (127.72,126.81) -- cycle (127.72,103.5) .. controls (134.15,103.5) and (139.37,108.72) .. (139.37,115.16) .. controls (139.37,121.59) and (134.15,126.81) .. (127.72,126.81) .. controls (121.28,126.81) and (116.06,121.59) .. (116.06,115.16) .. controls (116.06,108.72) and (121.28,103.5) .. (127.72,103.5) -- cycle ;
%Shape: Path Data [id:dp2239835795713876] 
\draw  [fill={rgb, 255:red, 125; green, 184; blue, 253 }  ,fill opacity=1 ] (324.47,227.81) .. controls (333.18,227.81) and (340.39,243.16) .. (341.5,263.04) -- (307.43,263.04) .. controls (308.54,243.16) and (315.75,227.81) .. (324.47,227.81) -- cycle (324.47,204.5) .. controls (330.9,204.5) and (336.12,209.72) .. (336.12,216.16) .. controls (336.12,222.59) and (330.9,227.81) .. (324.47,227.81) .. controls (318.03,227.81) and (312.81,222.59) .. (312.81,216.16) .. controls (312.81,209.72) and (318.03,204.5) .. (324.47,204.5) -- cycle ;
%Shape: Path Data [id:dp800037048658433] 
\draw  [fill={rgb, 255:red, 125; green, 184; blue, 253 }  ,fill opacity=1 ] (396.47,63.81) .. controls (405.18,63.81) and (412.39,79.16) .. (413.5,99.04) -- (379.43,99.04) .. controls (380.54,79.16) and (387.75,63.81) .. (396.47,63.81) -- cycle (396.47,40.5) .. controls (402.9,40.5) and (408.12,45.72) .. (408.12,52.16) .. controls (408.12,58.59) and (402.9,63.81) .. (396.47,63.81) .. controls (390.03,63.81) and (384.81,58.59) .. (384.81,52.16) .. controls (384.81,45.72) and (390.03,40.5) .. (396.47,40.5) -- cycle ;
%Shape: Path Data [id:dp17325202374598914] 
\draw  [fill={rgb, 255:red, 117; green, 174; blue, 240 }  ,fill opacity=1 ] (235.02,34.19) -- (234.9,21.14) -- (243.52,27.59) -- (235.02,34.19) -- cycle (252.03,20.98) -- (252.15,34.04) -- (243.52,27.59) -- (252.03,20.98) -- cycle ;
%Straight Lines [id:da09246515829863444] 
\draw    (252.11,29.96) -- (269.42,48.56) ;
%Rounded Rect [id:dp7370595321234379] 
\draw  [draw opacity=0][fill={rgb, 255:red, 184; green, 233; blue, 134 }  ,fill opacity=1 ] (246.62,53.83) .. controls (246.59,50.59) and (249.2,47.93) .. (252.44,47.9) -- (295.36,47.51) .. controls (298.61,47.48) and (301.26,50.08) .. (301.29,53.33) -- (301.46,70.95) .. controls (301.49,74.2) and (298.88,76.85) .. (295.63,76.88) -- (252.72,77.28) .. controls (249.47,77.31) and (246.82,74.7) .. (246.79,71.46) -- cycle ;
%Rounded Rect [id:dp9746043167003058] 
\draw  [draw opacity=0][fill={rgb, 255:red, 248; green, 231; blue, 28 }  ,fill opacity=1 ] (175.13,60.02) .. controls (175.11,57.45) and (177.18,55.34) .. (179.75,55.32) -- (229.15,54.86) .. controls (231.73,54.84) and (233.83,56.91) .. (233.86,59.48) -- (233.99,73.46) .. controls (234.01,76.04) and (231.94,78.14) .. (229.37,78.17) -- (179.97,78.62) .. controls (177.39,78.65) and (175.29,76.58) .. (175.26,74.01) -- cycle ;
%Rounded Rect [id:dp002959824358406915] 
\draw  [draw opacity=0][fill={rgb, 255:red, 125; green, 184; blue, 253 }  ,fill opacity=1 ] (202.28,59.77) .. controls (202.25,57.2) and (204.32,55.09) .. (206.89,55.07) -- (229.15,54.86) .. controls (231.73,54.84) and (233.83,56.91) .. (233.86,59.48) -- (233.99,73.46) .. controls (234.01,76.04) and (231.94,78.14) .. (229.37,78.17) -- (207.11,78.37) .. controls (204.54,78.4) and (202.43,76.33) .. (202.41,73.75) -- cycle ;

%Flowchart: Magnetic Disk [id:dp8547137372111436] 
\draw  [fill={rgb, 255:red, 184; green, 233; blue, 134 }  ,fill opacity=1 ] (425,137.68) -- (425,151.32) .. controls (425,153.35) and (420.97,155) .. (416,155) .. controls (411.03,155) and (407,153.35) .. (407,151.32) -- (407,137.68)(425,137.68) .. controls (425,139.7) and (420.97,141.35) .. (416,141.35) .. controls (411.03,141.35) and (407,139.7) .. (407,137.68) .. controls (407,135.65) and (411.03,134) .. (416,134) .. controls (420.97,134) and (425,135.65) .. (425,137.68) -- cycle ;
%Flowchart: Magnetic Disk [id:dp5104317597541549] 
\draw  [fill={rgb, 255:red, 208; green, 2; blue, 27 }  ,fill opacity=1 ] (425,163.68) -- (425,177.33) .. controls (425,179.35) and (420.97,181) .. (416,181) .. controls (411.03,181) and (407,179.35) .. (407,177.33) -- (407,163.68)(425,163.68) .. controls (425,165.7) and (420.97,167.35) .. (416,167.35) .. controls (411.03,167.35) and (407,165.7) .. (407,163.68) .. controls (407,161.65) and (411.03,160) .. (416,160) .. controls (420.97,160) and (425,161.65) .. (425,163.68) -- cycle ;
%Flowchart: Magnetic Disk [id:dp13006308699885505] 
\draw  [fill={rgb, 255:red, 189; green, 16; blue, 224 }  ,fill opacity=1 ] (425,189.68) -- (425,203.33) .. controls (425,205.35) and (420.97,207) .. (416,207) .. controls (411.03,207) and (407,205.35) .. (407,203.33) -- (407,189.68)(425,189.68) .. controls (425,191.7) and (420.97,193.35) .. (416,193.35) .. controls (411.03,193.35) and (407,191.7) .. (407,189.68) .. controls (407,187.65) and (411.03,186) .. (416,186) .. controls (420.97,186) and (425,187.65) .. (425,189.68) -- cycle ;
%Flowchart: Magnetic Disk [id:dp5592896367142556] 
\draw  [fill={rgb, 255:red, 245; green, 166; blue, 35 }  ,fill opacity=1 ] (425,215.68) -- (425,229.33) .. controls (425,231.35) and (420.97,233) .. (416,233) .. controls (411.03,233) and (407,231.35) .. (407,229.33) -- (407,215.68)(425,215.68) .. controls (425,217.7) and (420.97,219.35) .. (416,219.35) .. controls (411.03,219.35) and (407,217.7) .. (407,215.68) .. controls (407,213.65) and (411.03,212) .. (416,212) .. controls (420.97,212) and (425,213.65) .. (425,215.68) -- cycle ;
%Flowchart: Magnetic Disk [id:dp6717726752034842] 
\draw  [fill={rgb, 255:red, 139; green, 87; blue, 42 }  ,fill opacity=1 ] (425,241.68) -- (425,255.33) .. controls (425,257.35) and (420.97,259) .. (416,259) .. controls (411.03,259) and (407,257.35) .. (407,255.33) -- (407,241.68)(425,241.68) .. controls (425,243.7) and (420.97,245.35) .. (416,245.35) .. controls (411.03,245.35) and (407,243.7) .. (407,241.68) .. controls (407,239.65) and (411.03,238) .. (416,238) .. controls (420.97,238) and (425,239.65) .. (425,241.68) -- cycle ;
%Shape: Ellipse [id:dp6042473335526607] 
\draw  [dash pattern={on 4.5pt off 4.5pt}] (166.33,59) .. controls (166.33,35.99) and (198.57,17.33) .. (238.33,17.33) .. controls (278.1,17.33) and (310.33,35.99) .. (310.33,59) .. controls (310.33,82.01) and (278.1,100.67) .. (238.33,100.67) .. controls (198.57,100.67) and (166.33,82.01) .. (166.33,59) -- cycle ;
%Straight Lines [id:da36261477220911353] 
\draw    (176.33,80.67) -- (137,108.67) ;
%Straight Lines [id:da34215768303909233] 
\draw  [dash pattern={on 4.5pt off 4.5pt}]  (111,136.67) -- (66,136.67) ;
\draw [shift={(63,136.67)}, rotate = 360] [fill={rgb, 255:red, 0; green, 0; blue, 0 }  ][line width=0.08]  [draw opacity=0] (8.93,-4.29) -- (0,0) -- (8.93,4.29) -- cycle    ;

% Text Node
\draw (118,163.4) node [anchor=north west][inner sep=0.75pt]    {$n_{1}$};
% Text Node
\draw (385,100.4) node [anchor=north west][inner sep=0.75pt]    {$n_{2}$};
% Text Node
\draw (314,265.4) node [anchor=north west][inner sep=0.75pt]    {$n_{3}$};
% Text Node
\draw (253.23,55.33) node [anchor=north west][inner sep=0.75pt]  [font=\tiny,rotate=-359.47]  {$[[ P_{3}]]_{f_{5}}$};
% Text Node
\draw (429,134.4) node [anchor=north west][inner sep=0.75pt]    {$f_{1}$};
% Text Node
\draw (429,160.4) node [anchor=north west][inner sep=0.75pt]    {$f_{2}$};
% Text Node
\draw (429,186.4) node [anchor=north west][inner sep=0.75pt]    {$f_{3}$};
% Text Node
\draw (429,212.4) node [anchor=north west][inner sep=0.75pt]    {$f_{4}$};
% Text Node
\draw (429,238.4) node [anchor=north west][inner sep=0.75pt]    {$f_{5}$};
% Text Node
\draw (175,61.51) node [anchor=north west][inner sep=0.75pt]  [font=\tiny] [align=left] {850 results};
% Text Node
\draw (52.56,116.56) node [anchor=north west][inner sep=0.75pt]  [font=\scriptsize,rotate=-1.17] [align=left] {154 results};

\end{tikzpicture}

%% file: experiments.tex
The experimental evaluation compares \system{} with two state-of-the-art approaches building on P2P systems: \piqnic~\cite{DBLP:conf/esws/AebeloeMH19} and \colchain~\cite{DBLP:conf/www/AebeloeMH21} with the query optimization approach outlined in~\cite{ppbfs}.
To do this, we implemented the fragmentation, indexing, and cardinality estimation approach as a separate package in Java 8 and modified \piqnic's and \colchain's query processors to use it.
Like \colchain{} and \piqnic, \system's query processor is implemented as an extension to Apache Jena\footnote{\url{https://jena.apache.org}}.
Fragments in our implementation are stored as HDT files~\cite{DBLP:journals/ws/FernandezMGPA13}, allowing for efficient processing of the star patterns.
%In this section, we go into detail with the experimental setup and experimental results.
We provide all source code, experimental setup (queries, datasets, etc.), and the full experimental results on our website\footnote{\texttt{https://relweb.cs.aau.dk/lothbrok}}.

%To assess the efficiency of using \system{} to optimize decentralized SPARQL queries, we compare the performance and efficiency of state-of-the-art decentralized architectures both with and without the use of \system{}.
%Specifically, to test \system{} in different setups, we extended two separate architectures with \system; \colchain~\cite{DBLP:conf/www/AebeloeMH21} and \piqnic~\cite{DBLP:conf/esws/AebeloeMH19}.
%To assess \system{}'s performance and scalability in different scenarios, we use both synthetic and real-world datasets and queries.
%We compare \colchain{} and \piqnic{} with and without \system.
%
%Moreover, we discuss the implementation details.

%\subsection{Implementation Details}\label{subsec:impl}

\subsection{Experimental Setup}
In this section, we detail the experimental setup, including a characterization of the used datasets and queries, the hardware and software setup, experimental configuration, as well as the evaluation metrics.

\textit{Datasets and Queries.} 
To test the scalability of the approaches when the network is under heavy load, and to assess the impact of the query pattern on performance and network usage, we ran experiments with the synthetic WatDiv~\cite{watdiv} benchmark using different dataset sizes: 10 million triples to 1 billion triples.
Furthermore, to test \system{} in a realistic setting where users would upload several interlinked datasets to a network, and ask queries with varying complexity, we ran experiments using a well-known benchmark suite for federated RDF engines called LargeRDFBench~\cite{largerdfbench}.
LargeRDFBench comprises 13 different, interlinked datasets with over a billion triples in total.
To provide a fair comparison between the systems with and without \system, we created an equal number of fragments for both fragmentations: characteristic sets (Section~\ref{subsec:partitioning}) and predicate-based. %fragmentation approach used by the original version of \piqnic{} and \colchain.
To do this, we iteratively merged the characteristic set fragments with the fewest number of subjects into larger fragments following the approach outlined in Section~\ref{subsec:partitioning} until the number of fragments equalled the number of predicate-based fragments.
The characteristics of the datasets are shown in Table~\ref{tbl:datasets}.
Furthermore, to assess the impact of reducing the number of characteristic sets on query completeness, we ran similar experiments where we did not create an equal number of fragments for \system{}, i.e., where we created one fragment for each characteristic set that describes at least 50 subjects and provide the results on our website\footnotemark[3];
since these results are quite similar to the ones presented in this section, we will not report on them further.

\begin{table*}[htb!]
\centering
\caption{Characteristics of the used datasets}\label{tbl:datasets}
\begin{tabular}{lrrrrr}
\hline
\textbf{Dataset}     & \textbf{\#triples}    & \textbf{\#subjects}\hspace{1ex} & \textbf{\#predicates}\hspace{1ex} & \textbf{\#objects}  \\ \hline
\texttt{LargeRDFBench}     & 1,003,960,176 & 165,785,212  & 2,160  & 326,209,517 \\
\hspace{2ex}\textit{LinkedTCGA-M}  & 415,030,327 & 83,006,609  & 6 & 166,106,744   \\ 
\hspace{2ex}\textit{LinkedTCGA-E}  & 344,576,146 & 57,429,904  & 7 & 84,403,402  \\
\hspace{2ex}\textit{LinkedTCGA-A}  & 35,329,868 & 5,782,962  & 383 & 8,329,393 \\
\hspace{2ex}\textit{ChEBI}  & 4,772,706 & 50,477  & 28 & 772,138 \\  
\hspace{2ex}\textit{DBPedia-Subset}  & 42,849,609 & 9,495,865  & 1,063 & 13,620,028 \\  
\hspace{2ex}\textit{DrugBank}  & 517,023 & 19,693  & 119 & 276,142   \\  
\hspace{2ex}\textit{GeoNames}  & 107,950,085 & 7,479,714  & 26 & 35,799,392   \\
\hspace{2ex}\textit{Jamendo}  & 1,049,647 & 335,925  & 26 & 440,686  \\  
\hspace{2ex}\textit{KEGG}  & 1,090,830 & 34,260  & 21 & 939,258  \\  
\hspace{2ex}\textit{LinkedMDB}  & 6,147,996 & 694,400  & 222 & 2,052,959 \\ 
\hspace{2ex}\textit{NYT}  & 335,198 & 21,666  & 36 & 191,538  \\  
\hspace{2ex}\textit{SWDF}  & 103,595 & 11,974  & 118 & 37,547  \\ 
\hspace{2ex}\textit{Affymetrix}  & 44,207,146 & 1,421,763  & 105 & 13,240,270  \\  
\hline
\texttt{watdiv10M}   & 10,916,457    & 521,585     & 86            & 1,005,832 \\
\texttt{watdiv100M}  & 108,997,714   & 5,212,385   & 86            & 9,753,266  \\
\texttt{watdiv1000M} & 1,092,155,948 & 52,120,385  & 86  & 92,220,397 \\
%\texttt{watdiv10B} & 10,920,048,634 & 521,200,385  & 86  & 837,127,565  & \todo{-} & 86\\\hline
\end{tabular}
\end{table*}

LargeRDFBench includes 40 different queries~\cite{largerdfbench} that are divided into five different categories of varying complexity and result set sizes: Simple (S), Complex (C), Large Data (L), and Complex and High Data Sources (CH).
%Furthermore, to study scalability and the impact of the number of star-shaped subqueries on the query processor, as well as to stress-test \system{}, 

\begin{figure*}[tb!]
\centering
\begin{subfigure}[b]{0.48\textwidth}
  \centering
  \includegraphics[width=\textwidth]{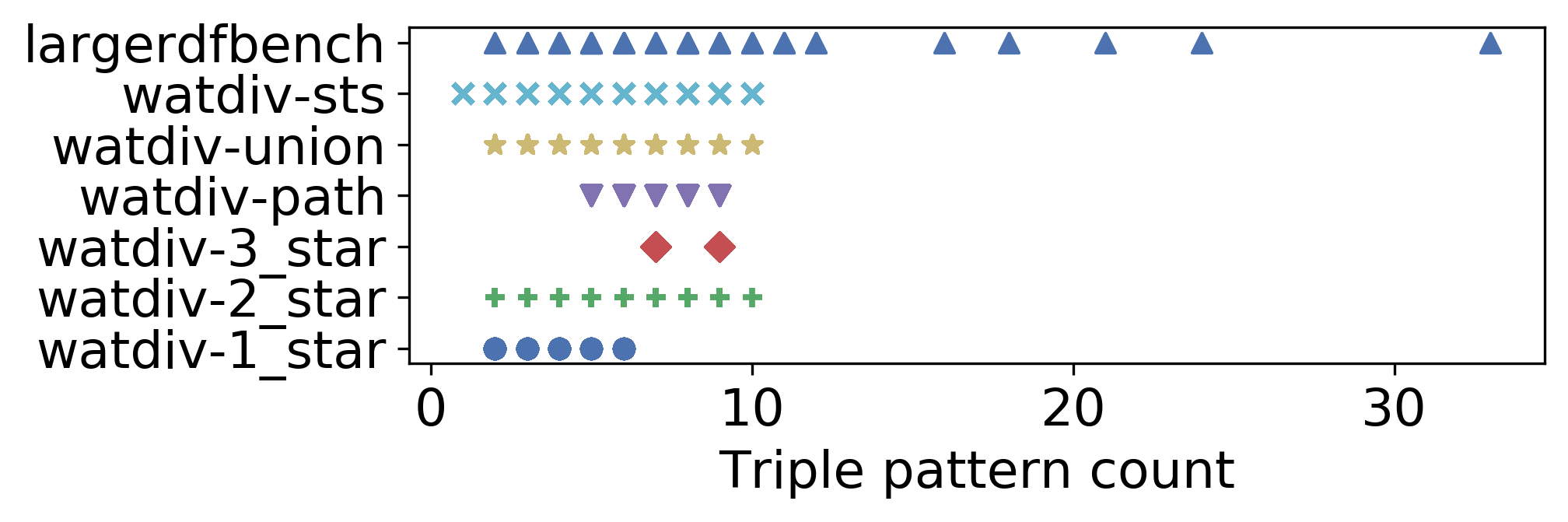}
  \caption{Triple pattern count (\#TP)}\label{subfig:tpcount}
\end{subfigure}
\begin{subfigure}[b]{0.48\textwidth}
  \centering
  \includegraphics[width=\textwidth]{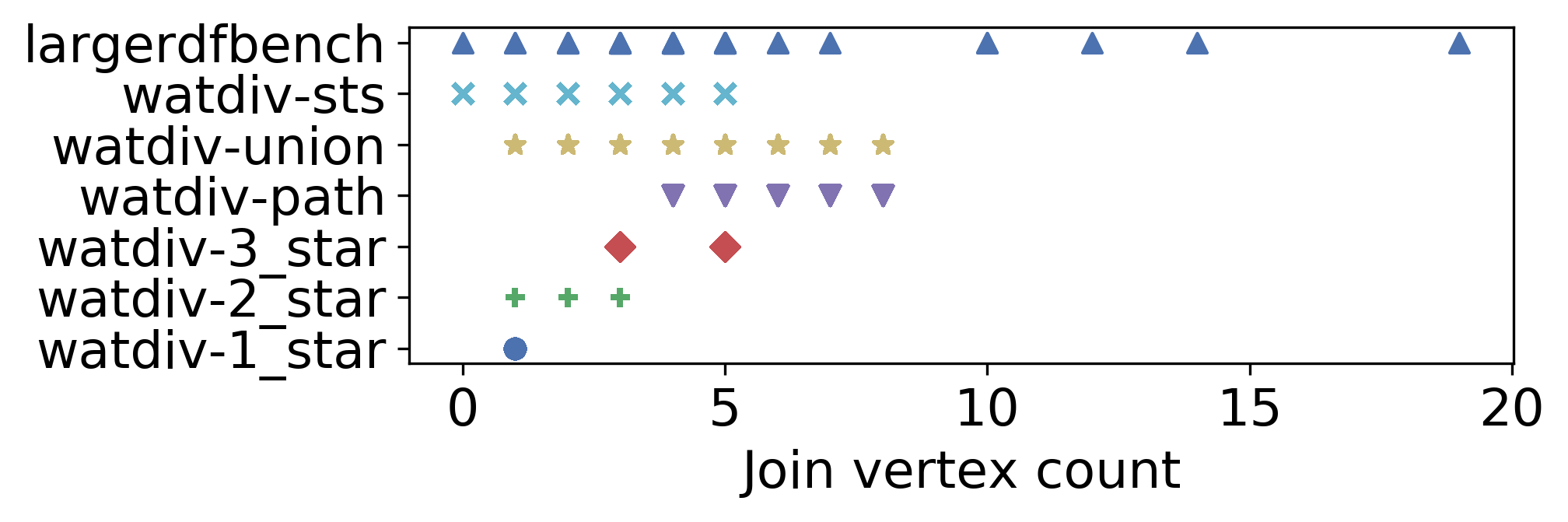}
  \caption{Join vertex count (\#JV)}\label{subfig:jvcount}
\end{subfigure}
\begin{subfigure}[b]{0.48\textwidth}
  \centering
  \includegraphics[width=\textwidth]{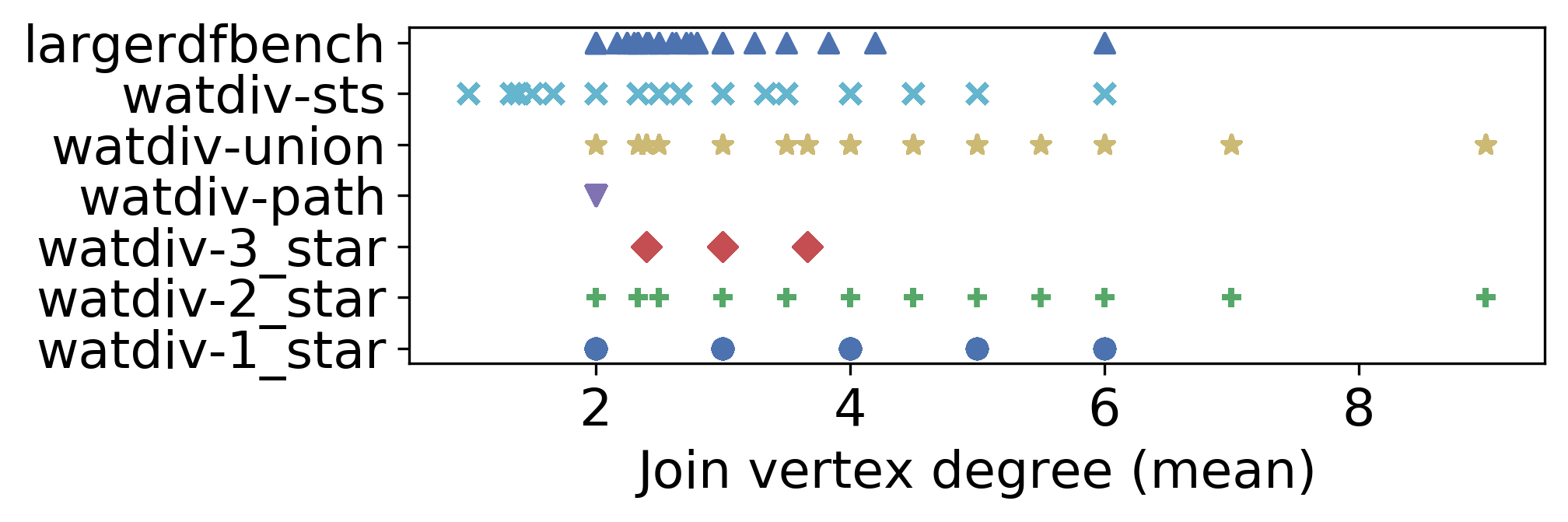}
  \caption{Join vertex degree (\textsc{deg})}\label{subfig:jvdegree}
\end{subfigure}
\begin{subfigure}[b]{0.48\textwidth}
  \centering
  \includegraphics[width=\textwidth]{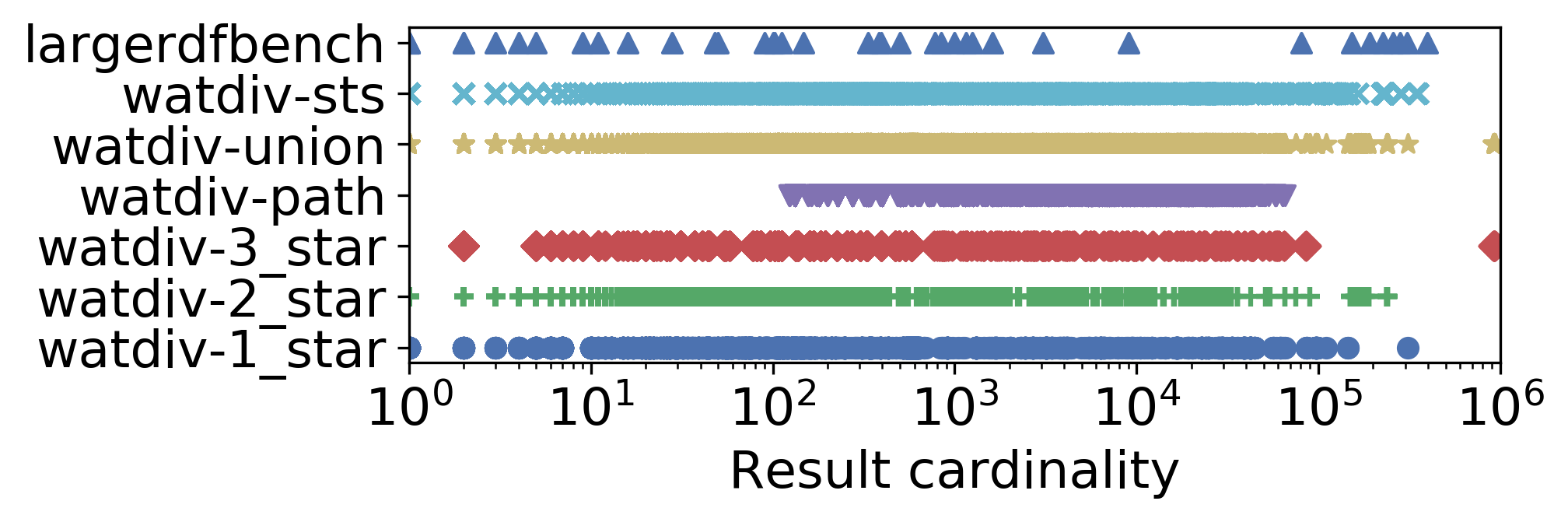}
  \caption{Result cardinality (\#Results)}\label{subfig:card}
\end{subfigure}
\begin{subfigure}[b]{0.48\textwidth}
  \centering
  \includegraphics[width=\textwidth]{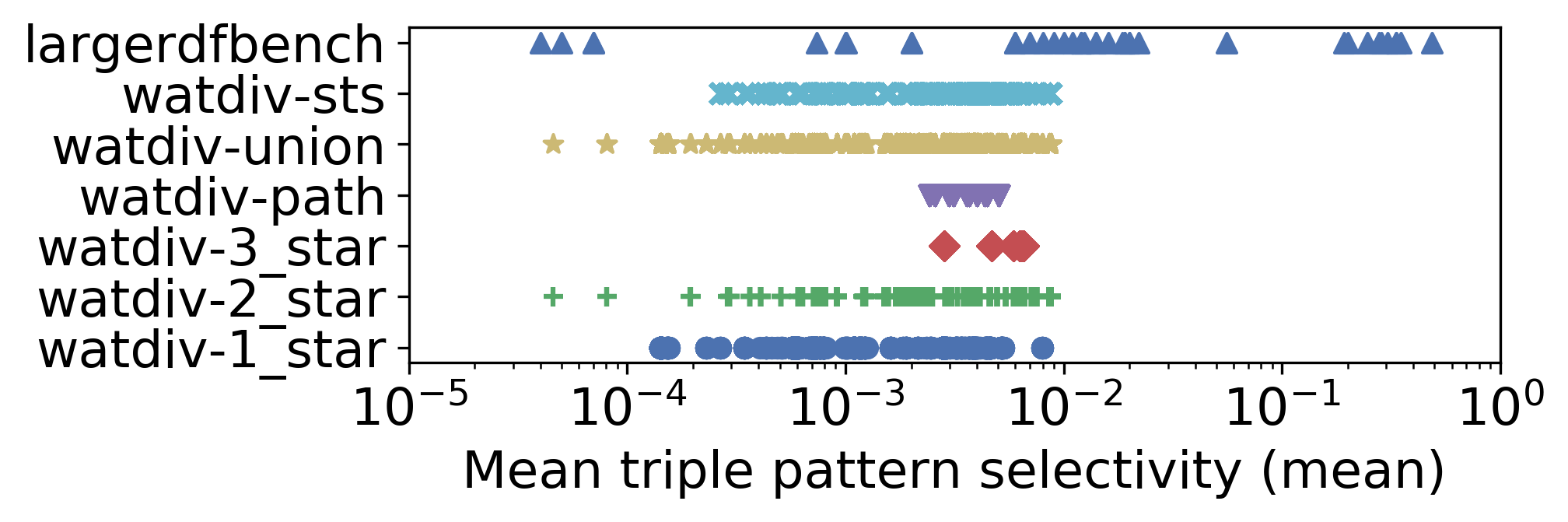}
  \caption{Mean TP selectivity (\textsc{sel}$_{\mathcal{G}}(tp)$)}\label{subfig:sel}
\end{subfigure}
\begin{subfigure}[b]{0.48\textwidth}
  \centering
  \scalebox{0.8}{
  \begin{tabular}{llll}
\hline
\textbf{Query load} & \textbf{SS} & \textbf{SO} & \textbf{OO} \\ \hline
watdiv-1\_star      & 100\%       & 0\%         & 0\%         \\
watdiv-2\_stars     & 55.38\%     & 34.31\%     & 10.31\%     \\
watdiv-3\_stars     & 57.67\%     & 30.77\%     & 11.53\%     \\
watdiv-paths        & 0\%         & 100\%       & 0\%         \\
watdiv-union        & 53.27\%     & 41.27\%     & 5.46\%      \\
watdiv-sts          & 55.57\%     & 35.83\%     & 8.6\%       \\
largerdfbench       & 41.41\%     & 45.81\%     & 12.78\%     \\ \hline
\end{tabular}
}
  \caption{Join vertex type over each query load}\label{subfig:jvtype}
\end{subfigure}
\caption{Characteristics of all query loads (WatDiv query loads over \texttt{watdiv100M}; statistics over the \texttt{watdiv10M} and \texttt{watdiv100M} datasets can be found on our website$^3$).}
\label{fig:qchar}
\end{figure*}

For WatDiv, we used WatDiv \emph{star query loads} from~\cite{DBLP:journals/corr/abs-2002-09172} consisting of 1-3 star patterns, called the \texttt{watdiv-1\_star}, \texttt{watdiv-2\_star}, and \texttt{watdiv-3\_star} query loads, as well as a query load consisting of path queries, i.e., queries where each star pattern only has one triple pattern, called the \texttt{watdiv\_path} query load.
Each of these query loads consists of 6,400 different queries.
Furthermore, we combine the aforementioned query loads into a single query load called \texttt{watdiv-union}.
Last, we created a query load with 19,968 queries from the WatDiv stress testing query templates (156 per node) called \texttt{watdiv-sts}.
The complete set of queries is available on our website\footnotemark[3].
Figure~\ref{fig:qchar} shows an overview of the following characteristics of each load~\cite{DBLP:journals/corr/abs-2002-09172,DBLP:conf/semweb/AlucHOD14}:
Triple pattern count \#TP (Figure~\ref{subfig:tpcount}),
join vertex count \#JV (Figure~\ref{subfig:jvcount}),
join vertex degree \textsc{deg} (Figure~\ref{subfig:jvdegree}),
result cardinality \#Results (Figure~\ref{subfig:card}),
mean triple pattern selectivity \textsc{sel}$_{\mathcal{G}}(tp)$ (Figure~\ref{subfig:sel}), and
join vertex type (Figure~\ref{subfig:jvtype}).

\textit{Experimental Configuration.} 
We compare the following systems:
%(i) \piqnic~\cite{DBLP:conf/esws/AebeloeMH19} using flooding to process queries (\piqnic),
(1) \piqnic~\cite{DBLP:conf/esws/AebeloeMH19} using PPBF indexes~\cite{ppbfs} (\piqnic),
(2) \system{} on top of \piqnic{} (\system$_{\mathtt{\piqnic}}$),
(3) \colchain~\cite{DBLP:conf/www/AebeloeMH21} using PPBF indexes (\colchain), and
(4) \system{} on top of \colchain{} (\system$_{\mathtt{\colchain}}$).
All configurations were run on networks with 128 nodes.
To assess the scalability of \system{} under load, we ran 156 \texttt{watdiv-sts} queries concurrently on each node over 8 different configurations where $2^i$ nodes issue queries concurrently such that $0\leq i\leq 7$ (i.e., up to all 128 nodes).
Furthermore, to analyze the impact of the query pattern on performance, we ran the WatDiv star query loads over each WatDiv dataset size such that for each star query load, each node issued 50 queries.
Lastly, we tested the performance of \system{} over each individual query in LargeRDFBench by running the queries sequentially in random order on three randomly selected nodes and report the average result.

\textit{Hardware Configuration.} For all configurations and P2P systems, we ran 128 nodes concurrently on a virtual machine (VM) with 128 vCPU cores with a clock speed of 2.5GHz, 64KB L1 cache, 512KB L2 cache, 8192KB L3, and a total of 2TB main memory.
To spread out resources evenly across nodes, all nodes were restricted to use 1 vCPU core and 15GB memory, enforced using the \texttt{-Xmx} and \texttt{-XX:ActiveProcessorCount} options for the JVM.
Furthermore, to simulate a more realistic scenario, where nodes are not run on the same machine, we simulated a connection speed of 20 MB/s.

\textit{Evaluation Metrics.}
We used measured the following metrics:
\begin{itemize}
\item \textit{Workload Time (WT):} The amount of time (in milliseconds) it takes to complete an entire workload including queries that time out.
\item \textit{Throughput (TP):} The number of completed queries in the workload divided by the total workload time (i.e., number of queries per minute).
\item \textit{Number of Timeouts (NTO):} The number of queries that timed out (timeout being 1200 seconds).
\item \textit{Query Execution Time (QET):} The amount of time (in milliseconds) elapsed between when a query is issued and when its processing has finished.
\item \textit{Query Response Time (QRT):} The amount of time (in milliseconds) elapsed between when a query is issued and when the first result is computed.
\item \textit{Query Optimization Time (QOT):} The amount of time (in miliseconds) elapsed between when a query is issued and when the optimizer has finished (i.e., when query execution starts).
\item \textit{Number of Requests (REQ):} The number of requests made between nodes when processing a query (including requests made from nodes that have been delegated subqueries).
\item \textit{Number of Transferred Bytes (NTB):} The amount of data (in bytes) transferred between nodes when processing a query (including data transferred to and from nodes that have been delegated subqueries).
\item \textit{Number of Relevant Nodes (NRN):} The number of distinct nodes that replicate fragments containing relevant data to a query.
\item \textit{Number of Relevant Fragments (NRF):} The number of distinct fragments containing relevant data to a query.
\end{itemize}

\textit{Software Configuration.}
Unless otherwise specified, we used the following parameters when running the systems.
For \colchain, we used the following parameters recommended in~\cite{DBLP:conf/www/AebeloeMH21}: Community Size: 20, Number of Communities: 200.
For \piqnic, we use the following parameters recommended in~\cite{DBLP:conf/esws/AebeloeMH19}: Time-to-Live (number of hops): 5, Number of Neighbors: 5.
The replication factor for \piqnic{} (i.e., the percentage of nodes replicating each fragment) was matched with the size of the communities in \colchain{} to provide a better comparison.
Nodes were randomly assigned neighbors throughout the network.
The page size (i.e., how many results can be returned with each request, was set to $100$.
Furthermore, to limit the size of HTTP requests, the number of results that each system was allowed to attach to each request (i.e., $|\Omega|$ in Section~\ref{sec:queryprocessing}) was set to $|\Omega|=30$
The timeout for all systems and queries was set to $20$ minutes (1,200 seconds).

%In the remainder of this section, we present the experimental results.
%However, we present only the most interesting results in this section; Appendix~\ref{app:largerdfbench} contains additional experimental results.
%Furthermore, we provide all the results on our website\footnotemark[4].

\begin{figure*}[tb!]
\centering
\includegraphics[width=.6\textwidth]{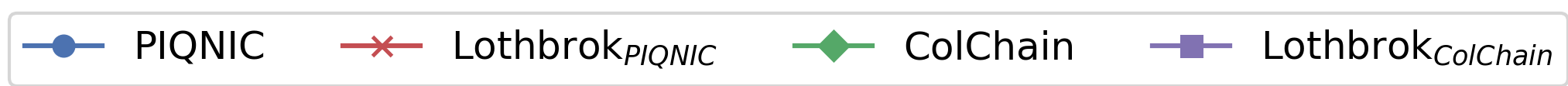}
\begin{subfigure}[b]{0.48\textwidth}
  \includegraphics[width=\textwidth]{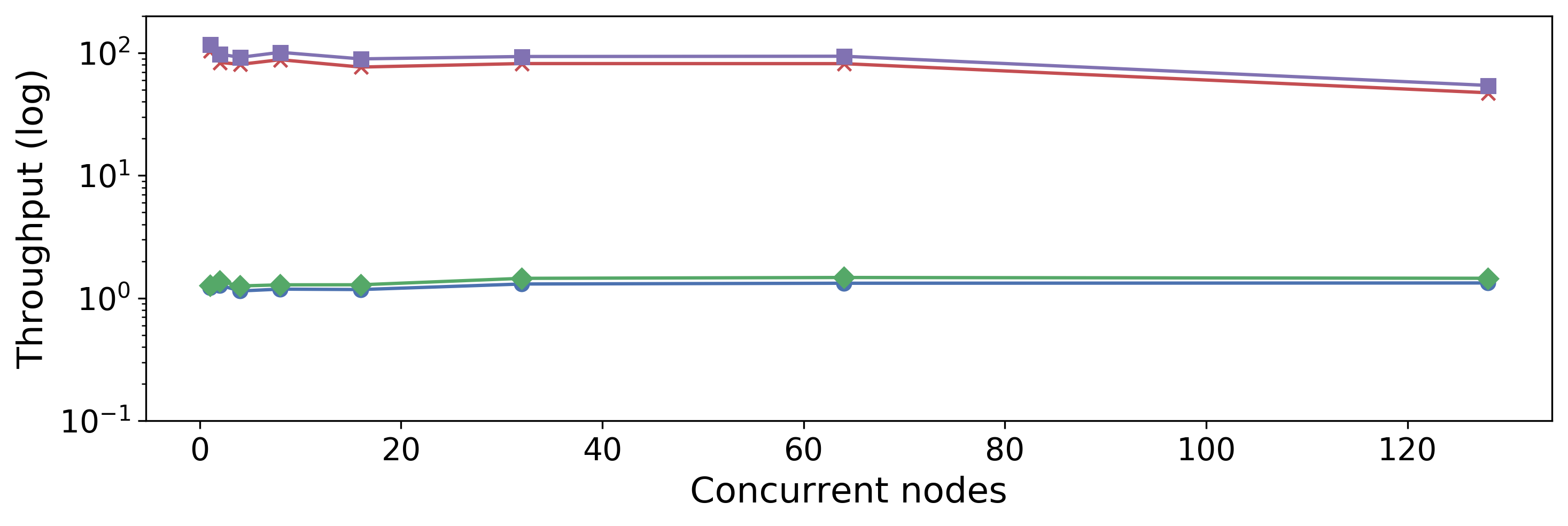}
  \caption{Throughput (TP) over \texttt{watdiv10M}}\label{subfig:scal_10M}
\end{subfigure}
\begin{subfigure}[b]{0.48\textwidth}
  \includegraphics[width=\textwidth]{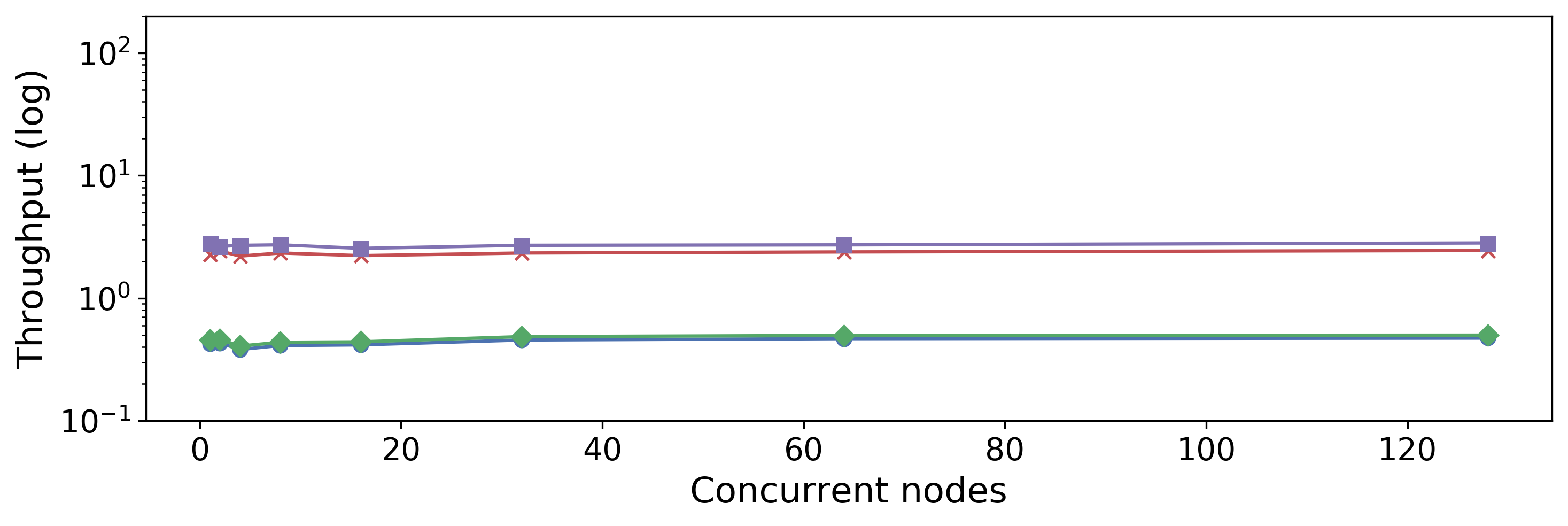}
  \caption{Throughput (TP) over \texttt{watdiv100M}}\label{subfig:scal_100M}
\end{subfigure}
\begin{subfigure}[b]{0.48\textwidth}
  \includegraphics[width=\textwidth]{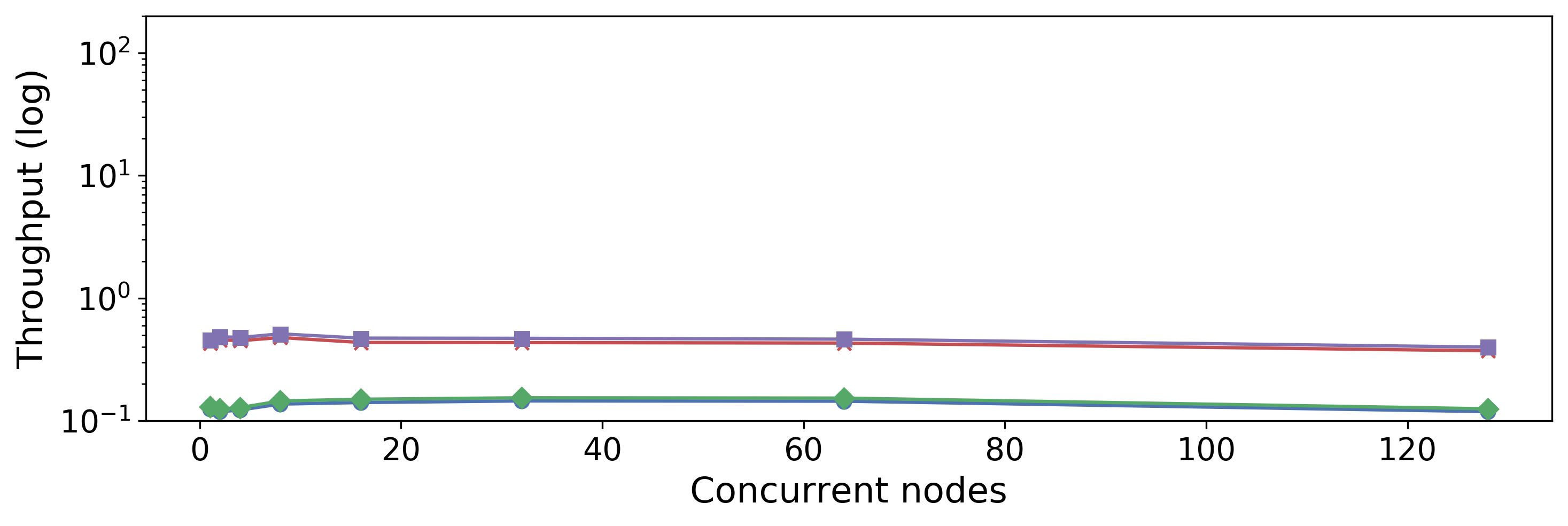}
  \caption{Throughput (TP) over \texttt{watdiv1000M}}\label{subfig:scal_1000M}
\end{subfigure}
\begin{subfigure}[b]{0.48\textwidth}
  \includegraphics[width=\textwidth]{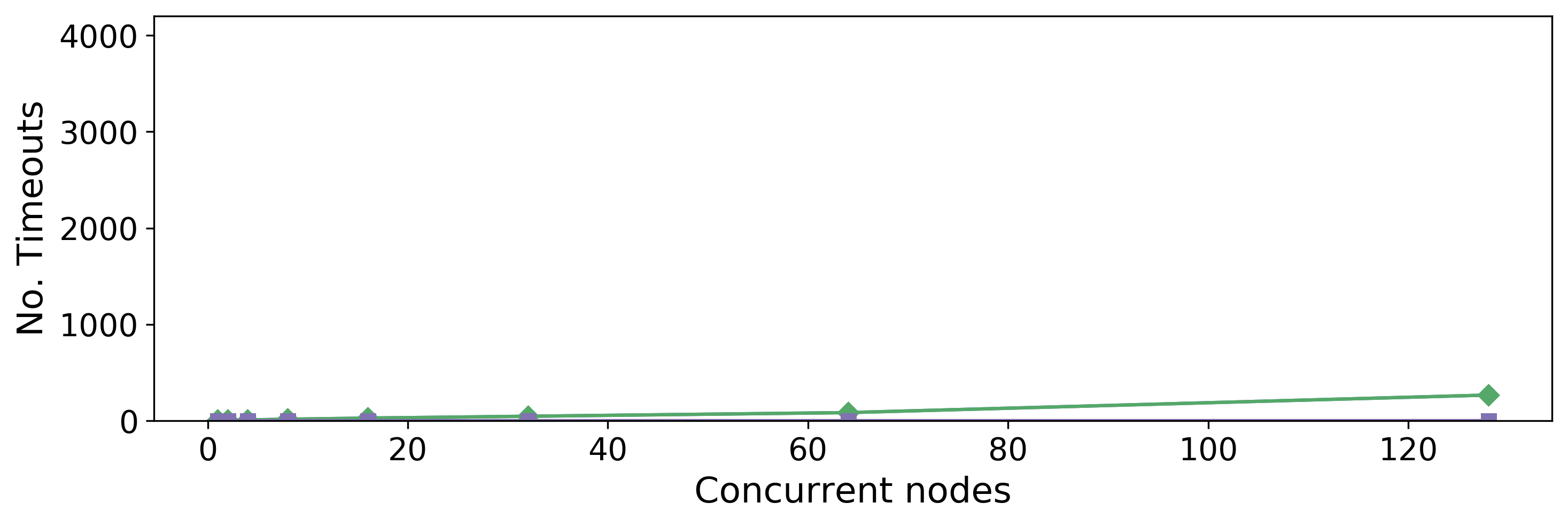}
  \caption{Number of timeouts (NTO) over \texttt{watdiv10M}}\label{subfig:to_10M}
\end{subfigure}
\begin{subfigure}[b]{0.48\textwidth}
  \includegraphics[width=\textwidth]{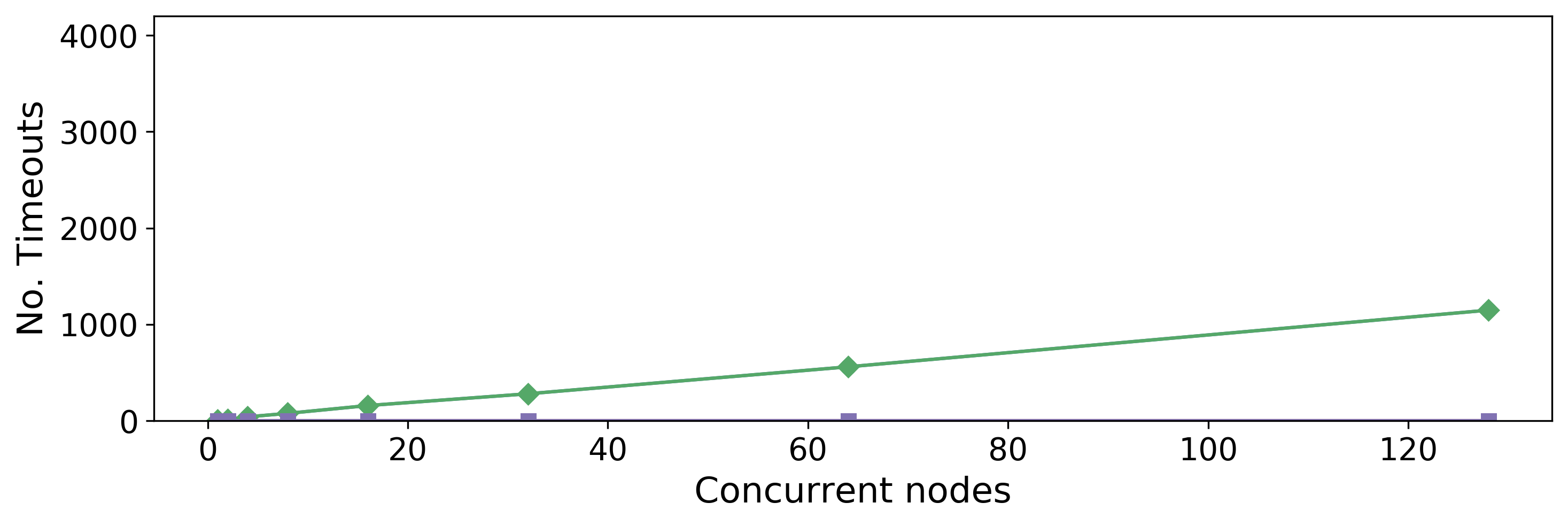}
  \caption{Number of timeouts (NTO) over \texttt{watdiv100M}}\label{subfig:to_100M}
\end{subfigure}
\begin{subfigure}[b]{0.48\textwidth}
  \includegraphics[width=\textwidth]{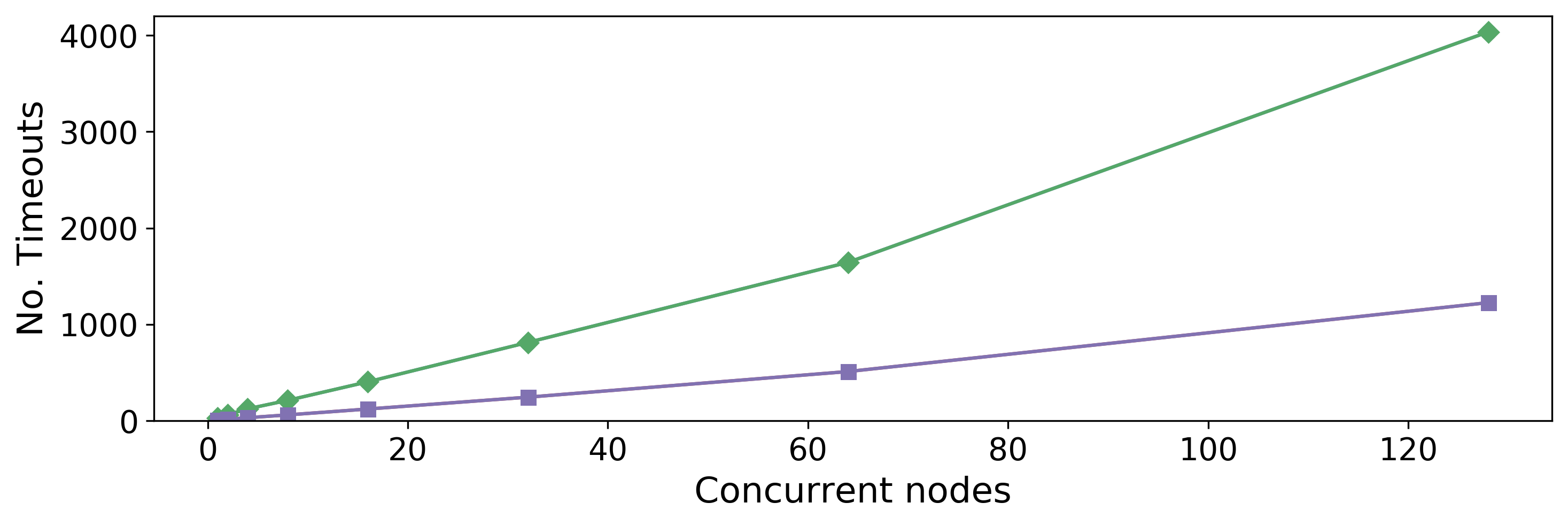}
  \caption{Number of timeouts (NTO) over \texttt{watdiv1000M}}\label{subfig:to_1000M}
\end{subfigure}
\begin{subfigure}[b]{0.48\textwidth}
  \includegraphics[width=\textwidth]{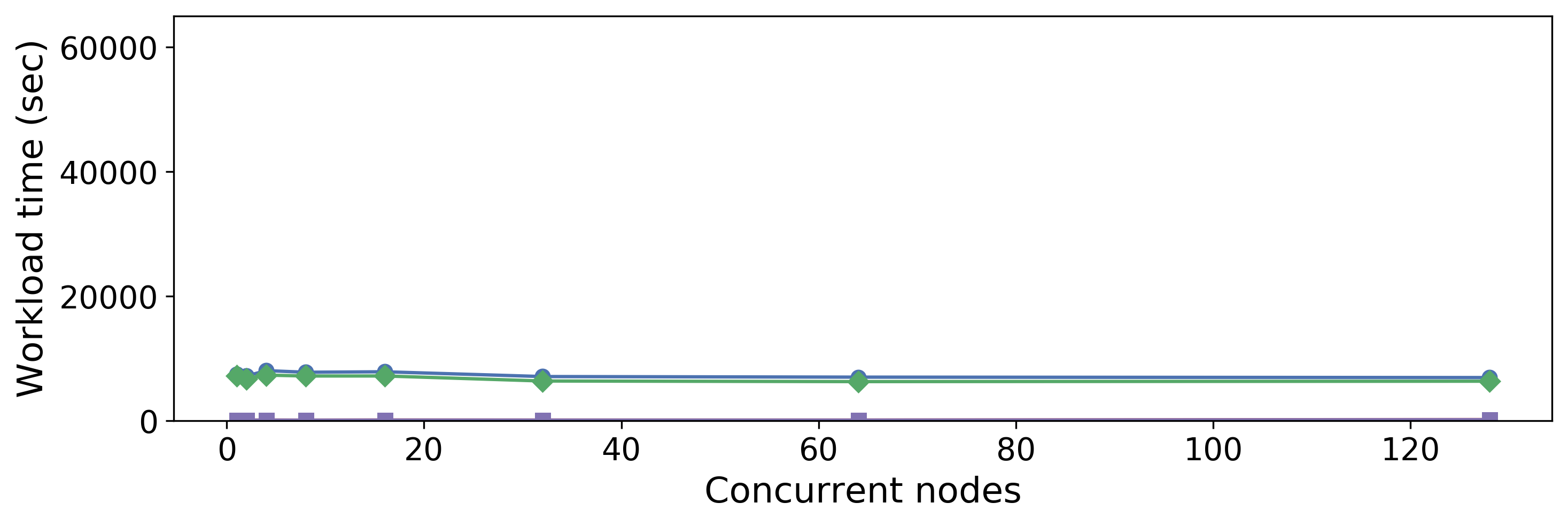}
  \caption{Workload time (WT) over \texttt{watdiv10M}}\label{subfig:wt_10M}
\end{subfigure}
\begin{subfigure}[b]{0.48\textwidth}
  \includegraphics[width=\textwidth]{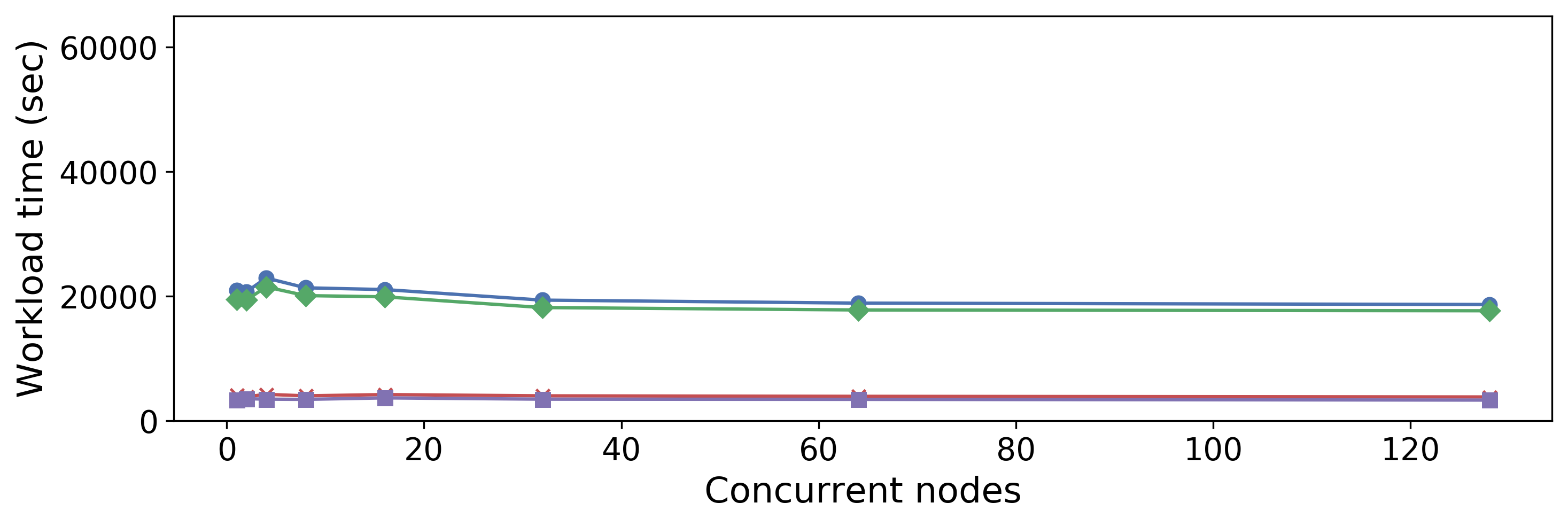}
  \caption{Workload time (WT) over \texttt{watdiv100M}}\label{subfig:wt_100M}
\end{subfigure}
\begin{subfigure}[b]{0.48\textwidth}
  \includegraphics[width=\textwidth]{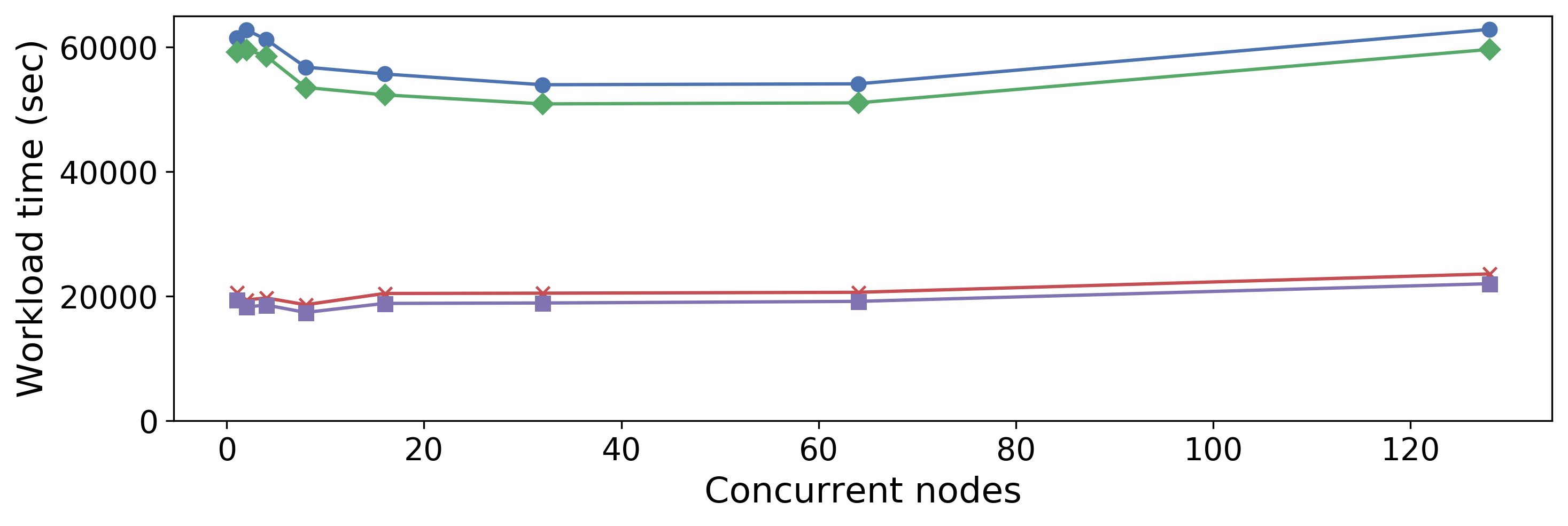}
  \caption{Workload time (WT) over \texttt{watdiv1000M}}\label{subfig:wt_1000M}
\end{subfigure}
\caption{Throughput (TP), number of timeouts (NTO), and workload time (WT) for \texttt{watdiv-sts} over the \texttt{watdiv10M}, \texttt{watdiv100M}, and \texttt{watdiv1000M} datasets.}
\label{fig:scalability}
\end{figure*}

\subsection{Scalability under Load}\label{subsec:scalability}
In these experiments, we ran the \texttt{watdiv-sts} queries over each WatDiv dataset in configurations where $2^i$ nodes issued 156 queries from the \texttt{watdiv-sts} query load concurrently such that $0\leq i\leq 7$.
Figures~\ref{subfig:scal_10M}-\ref{subfig:scal_1000M} show the throughput (TP) of the \texttt{watdiv-sts} query load over each configuration in the scalability tests for the \texttt{watdiv10M} (Figure~\ref{subfig:scal_10M}), \texttt{watdiv100M} (Figure~\ref{subfig:scal_100M}), and \texttt{watdiv1000M} (Figure~\ref{subfig:scal_1000M}) datasets in logarithmic scale.
Clearly, \system{} has a significantly higher throughput across all datasets and configurations compared to the approaches that do not include \system{} (i.e., \piqnic{} and \colchain).
In fact, for \texttt{watdiv10M}, this increase in throughput is close to two orders of magnitude.
While the increase in throughput that \system{} provides is smaller for both \texttt{watdiv100M} and \texttt{watdiv1000M}, \system{} still increases the throughput by close to an order of magnitude for these datasets.
Furthermore, while some results show that \colchain{} has a slightly higher throughput than \piqnic, both with and without \system{} on top, this difference is relatively negligible. % and is most likely due to the specific network topology in the neighborhood of each node.
Last, the results show that the throughput of \system{} is relatively stable when increasing numbers of nodes issue queries concurrently.
In fact, even when every node in the network issue queries concurrently, the throughput is relatively close to the highest throughput throughout the configurations.
%The slight increase in throughput for a small number of nodes issuing queries concurrently can be explained by the network being able to the network being able to utilize the resources across the nodes more efficiently, thus being able to complete more queries in the same amount of time.

Figures~\ref{subfig:to_10M}-\ref{subfig:to_1000M} show the number of queries that timed out (TO) of the \texttt{watdiv-sts} query load over each configuration for each WatDiv dataset.
As expected, the number of timeouts increases relatively linearly with the number of nodes issuing queries concurrently.
This is due to the fact that when more nodes issue queries, more queries in total are executed, meaning the total number of the queries that time out increases.
Generally, the queries that time out correspond to query templates that result in a large number of intermediate results, e.g., by using the \texttt{owl:sameAs} predicate.
Furthermore, \piqnic{} and \colchain{} incur significantly more timeouts without \system{} compared to with \system.
In fact, for both \texttt{watdiv10M} and \texttt{watdiv100M}, \system{} experiences no timeouts while \piqnic{} and \colchain{} experience 267 timeouts for \texttt{watdiv10M} and 1,148 timeouts for \texttt{watdiv1000M}.
Even for \texttt{watdiv1000M}, the number of timeouts experienced by \system{} is just 1,151 while \piqnic{} and \colchain{} both experience 4,036 timeouts.
Furthermore, \piqnic{} and \colchain{} incur the exact same number of timeouts.

Figures~\ref{subfig:wt_10M}-\ref{subfig:wt_1000M} show the workload time (WT) for each configuration.
In line with the throughput and number of timeouts, \system{} incurs a significantly lower average workload time than \piqnic{} and \colchain{} across all experiments and datasets.
The slight decrease in the workload time for fewer nodes can be attributed to the network being able to process more queries concurrently when the overall load is relatively low.
Nevertheless, the average workload time only increases slightly even when all nodes issue queries concurrently.

Overall, our experimental results show that, even when the network is under heavy query processing load, \system{} increases the query throughput and decreases the average workload time significantly compared to state-of-the-art decentralized systems.
In fact, the increase in performance is up to two orders of magnitude.
As a result, \system{} is also able to finish more queries without timing out.
%As such, they show that \system{} increases performance significantly compared to the state of the art decentralized systems while being able to handle increasin query loads quite well.

\subsection{Impact of Query Pattern}\label{subsec:querypattern}
To test the impact of the query pattern on the performance of \system{}, we ran the \texttt{watdiv-1\_star}, \texttt{watdiv-2\_star}, \texttt{watdiv-3\_star}, \texttt{watdiv-path}, \texttt{watdiv-union}, and \texttt{watdiv-sts} query loads on each system; the \texttt{watdiv-sts} queries consist of, on average, more selective star patterns compared to the other WatDiv query loads (Figure~\ref{fig:qchar}).
%Furthermore, we include the \texttt{watdiv-sts} queries in this comparison as well

\begin{figure*}[tb!]
\centering
\includegraphics[width=.6\textwidth]{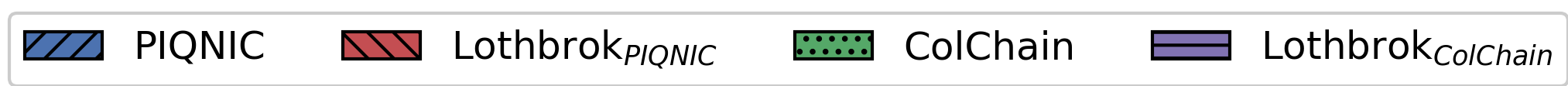}
\begin{subfigure}[b]{0.48\textwidth}
  \includegraphics[width=\textwidth]{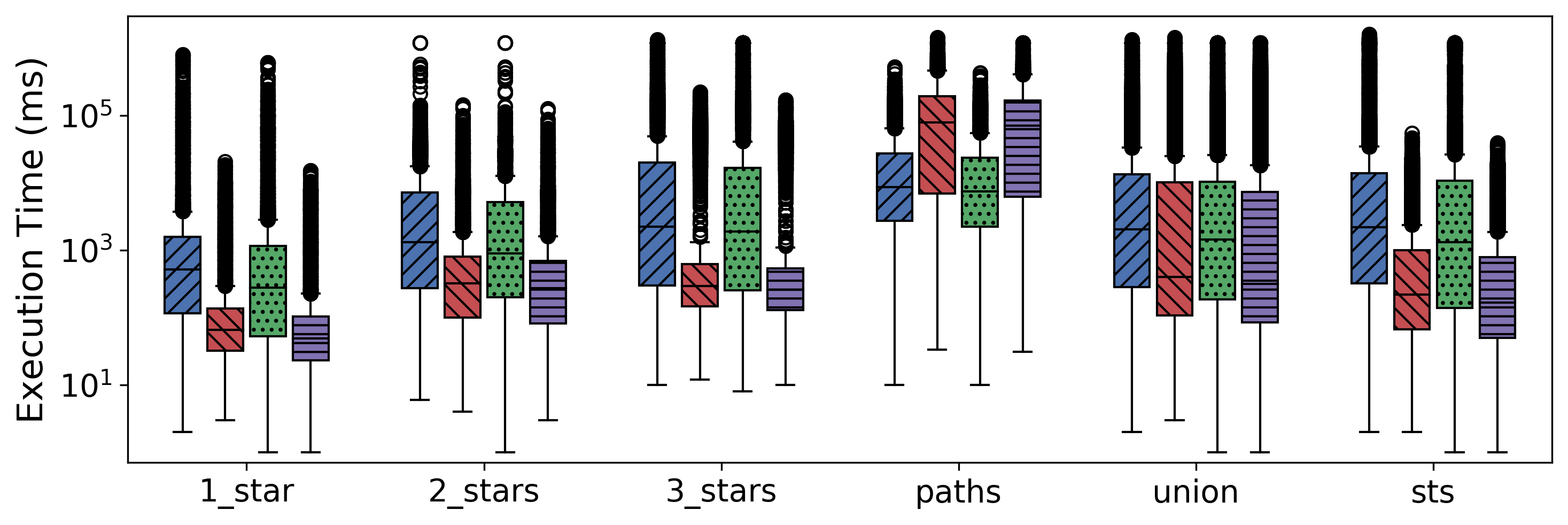}
  \caption{Query execution time (QET) over \texttt{watdiv10M}}\label{subfig:qet_10M}
\end{subfigure}
\begin{subfigure}[b]{0.48\textwidth}
  \includegraphics[width=\textwidth]{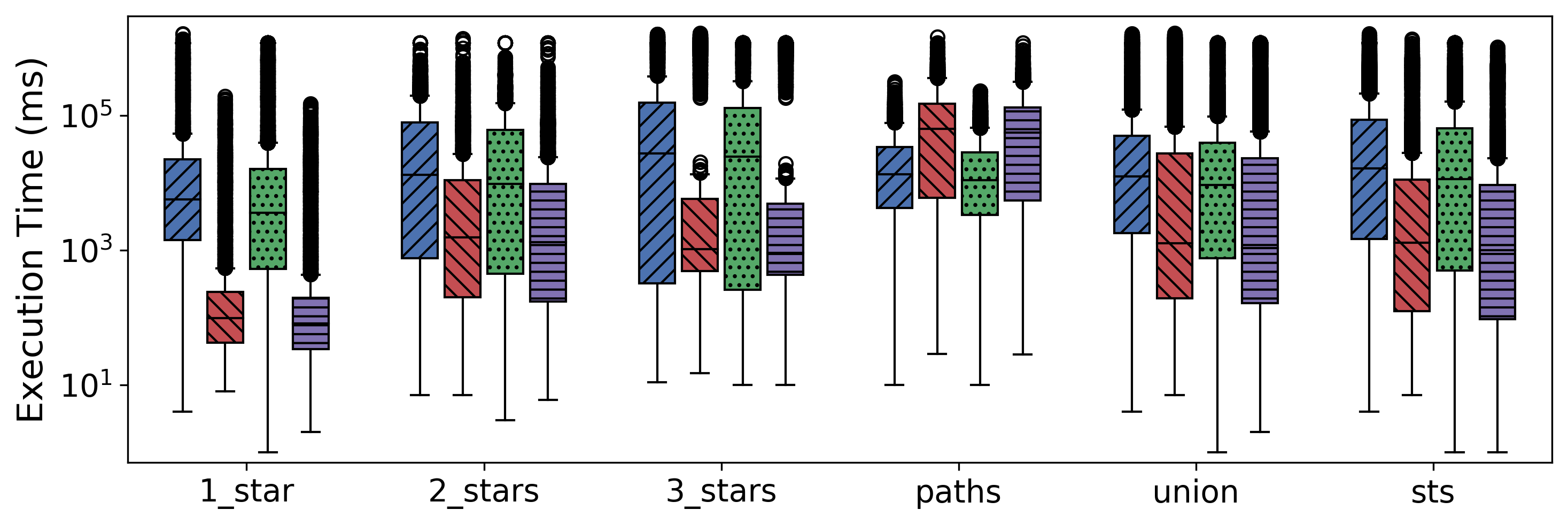}
  \caption{Query execution time (QET) over \texttt{watdiv100M}}\label{subfig:qet_100M}
\end{subfigure}
\begin{subfigure}[b]{0.48\textwidth}
  \includegraphics[width=\textwidth]{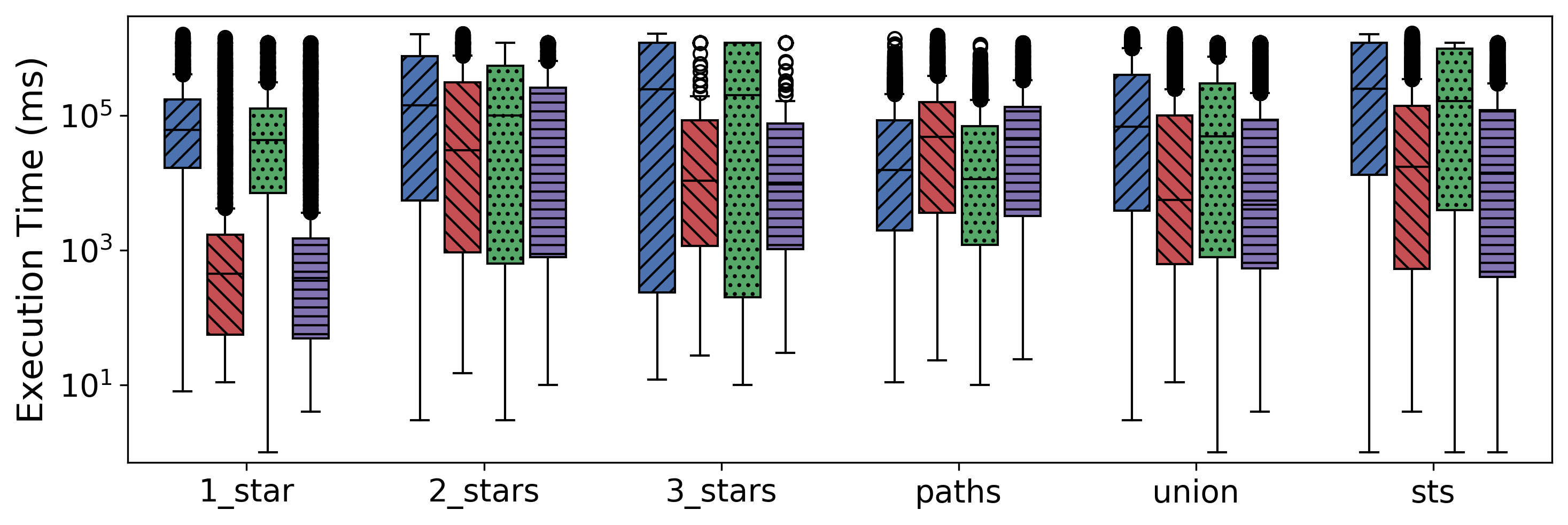}
  \caption{Query execution time (QET) over \texttt{watdiv1000M}}\label{subfig:qet_1000M}
\end{subfigure}
\begin{subfigure}[b]{0.48\textwidth}
  \includegraphics[width=\textwidth]{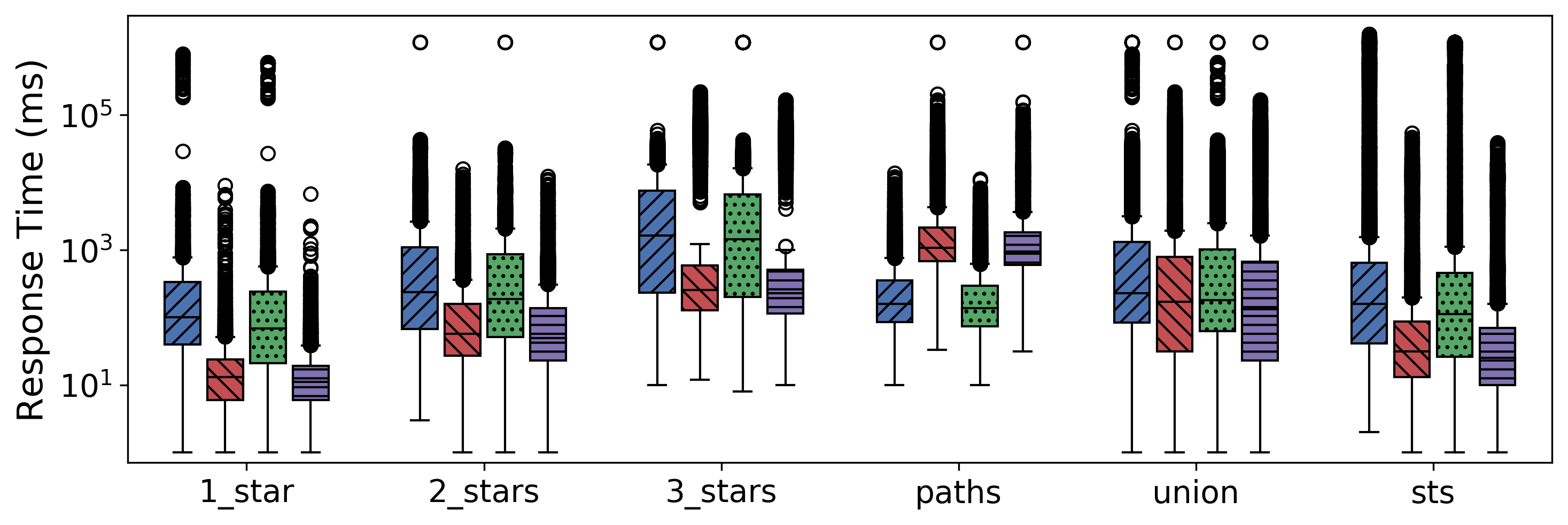}
  \caption{Query response time (QRT) over \texttt{watdiv10M}}\label{subfig:qrt_10M}
\end{subfigure}
\begin{subfigure}[b]{0.48\textwidth}
  \includegraphics[width=\textwidth]{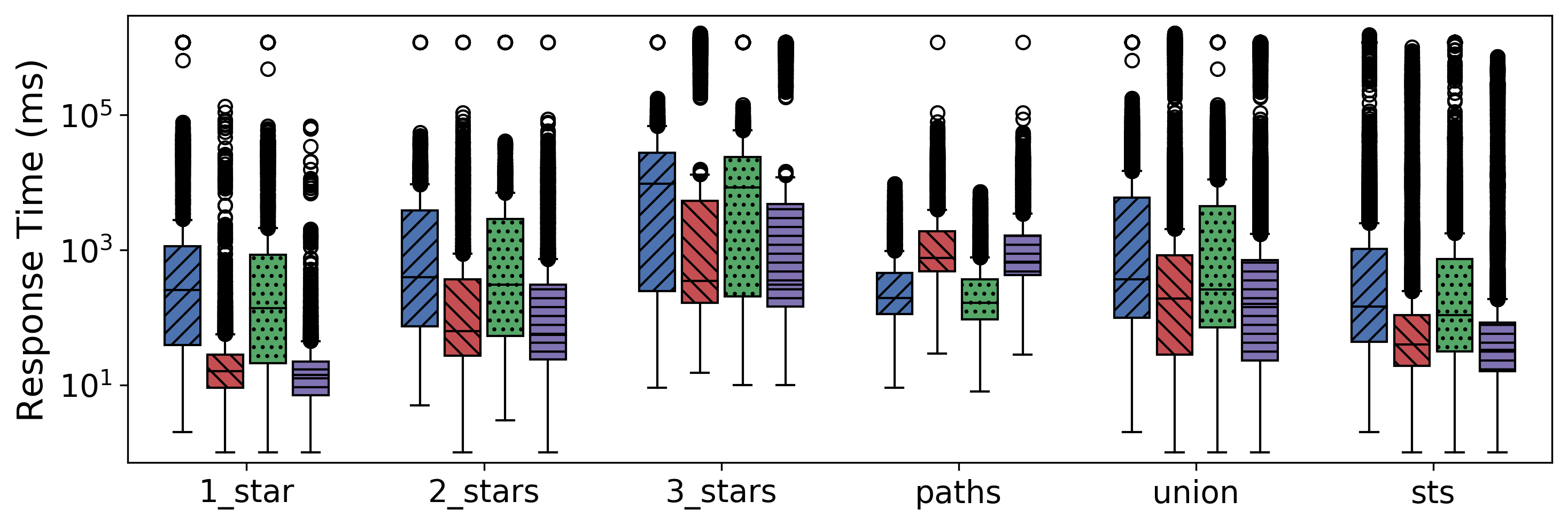}
  \caption{Query response time (QRT) over \texttt{watdiv100M}}\label{subfig:qrt_100M}
\end{subfigure}
\begin{subfigure}[b]{0.48\textwidth}
  \includegraphics[width=\textwidth]{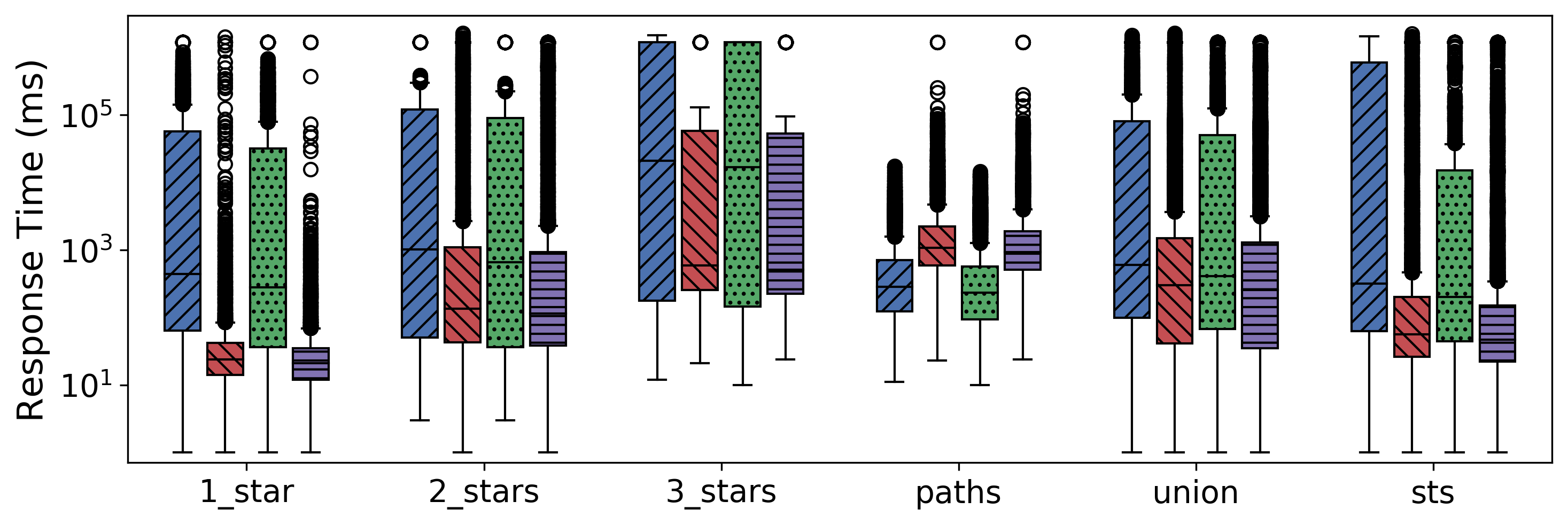}
  \caption{Query response time (QRT) over \texttt{watdiv1000M}}\label{subfig:qrt_1000M}
\end{subfigure}
\caption{Query execution time (QET) and query response time (QRT) for the WatDiv datasets and star queries.}
\label{fig:shape_perf}
\end{figure*}

Figures~\ref{subfig:qet_10M}-\ref{subfig:qet_1000M} show the execution time (QET) for each WatDiv query load over each WatDiv dataset, and Figures~\ref{subfig:qrt_10M}-\ref{subfig:qrt_1000M} show the response time (QRT) for each WatDiv query load in logarithmic scale.
Our results show that \system{} has significantly better performance across all datasets for almost every query load.
As expected, the improvement in performance is more significant for the query loads with a lower number of star patterns.
This is due to the fact that since the star patterns within these queries represent a large part of the query, \system{} has to issue fewer requests overall, lowering the network overhead.
For instance, the queries in the \texttt{watdiv-1\_star} query load can by \system{} be answered by issuing 0.89 requests per 90 results\footnote{Even though one request can fetch up to 90 results, the average number of requests is lower than 1 since the nodes store some data locally.}, whereas \piqnic{} and \colchain{} have to issue 9.27 requests per 90 results on average, for \texttt{watdiv1000M} in our experiments.
In the \texttt{watdiv-3\_star} query load, the improvement in performance is more modest across the datasets since each star pattern is a relatively small part of the query resulting in a higher number of requests; however, on average, we still see a performance increase of up to an order of magnitude.

\begin{figure*}[tb!]
\centering
\includegraphics[width=.6\textwidth]{figures/results/shape/legend3.png}
\begin{subfigure}[b]{0.48\textwidth}
  \includegraphics[width=\textwidth]{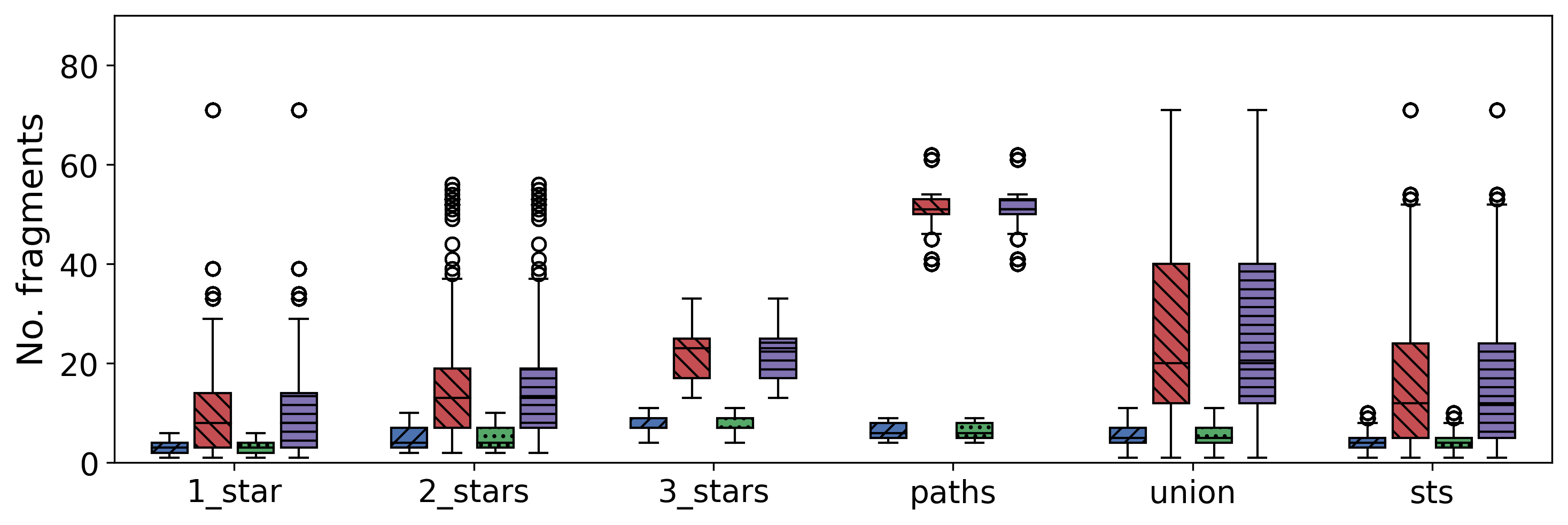}
  \caption{Number of relevant fragments (NRF) over \texttt{watdiv10M}}\label{subfig:nrf_10M}
\end{subfigure}
\begin{subfigure}[b]{0.48\textwidth}
  \includegraphics[width=\textwidth]{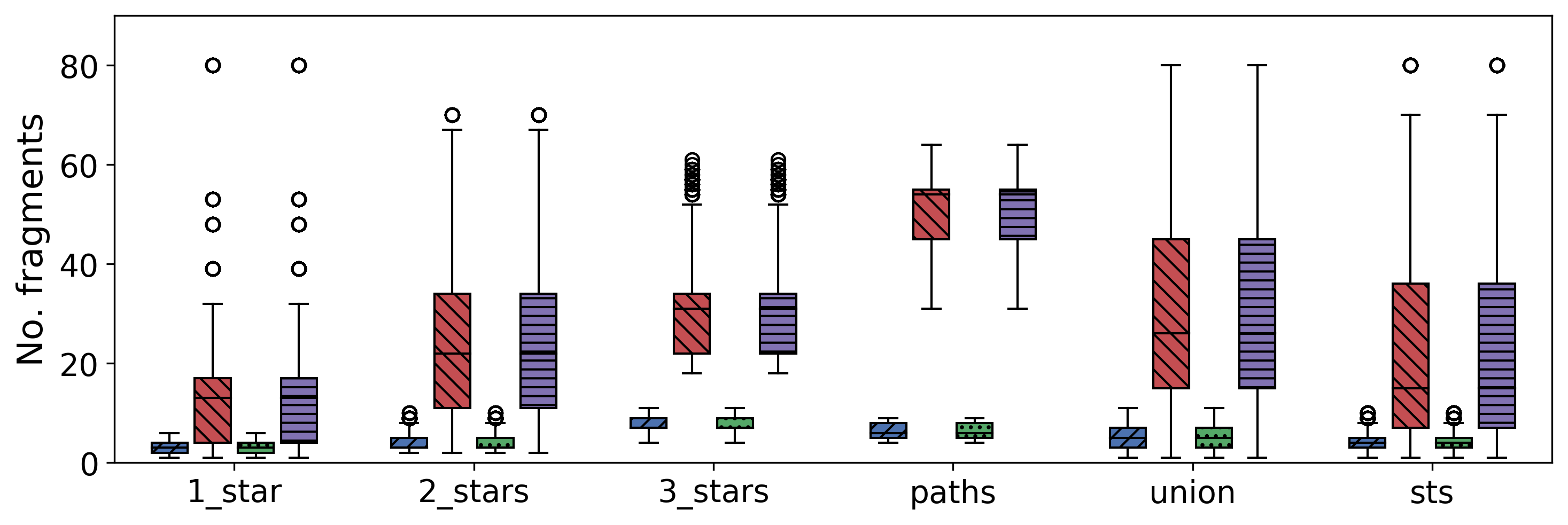}
  \caption{Number of relevant fragments (NRF) over \texttt{watdiv100M}}\label{subfig:nrf_100M}
\end{subfigure}
\begin{subfigure}[b]{0.48\textwidth}
  \includegraphics[width=\textwidth]{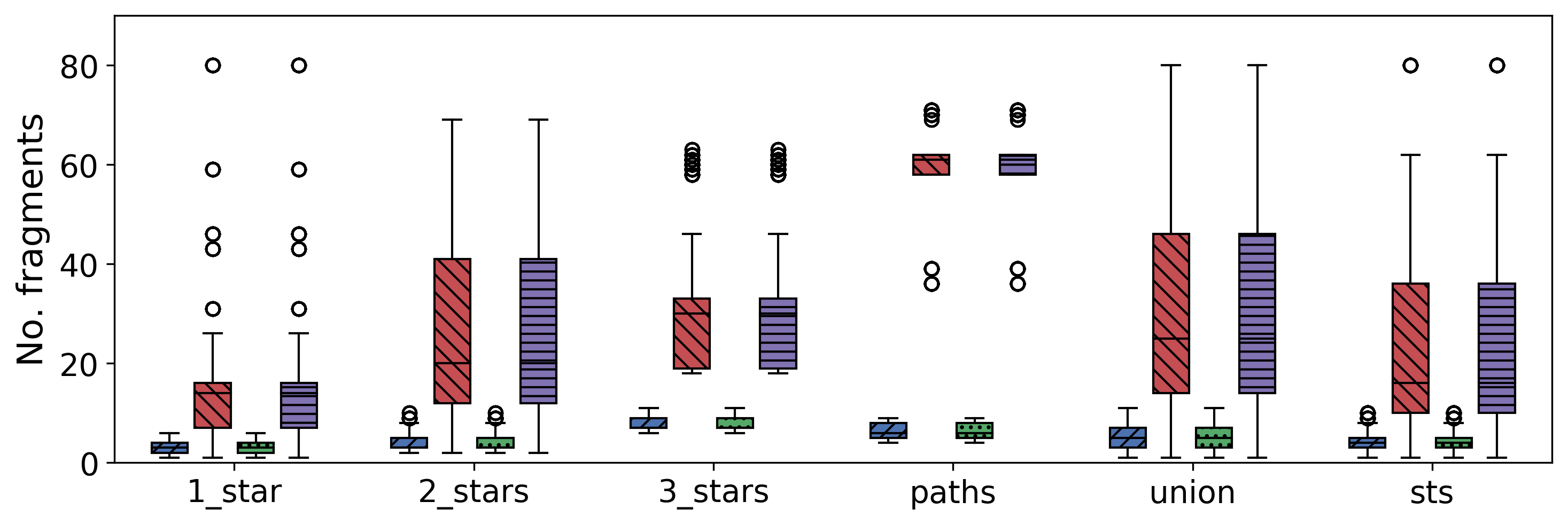}
  \caption{Number of relevant fragments (NRF) over \texttt{watdiv1000M}}\label{subfig:nrf_1000M}
\end{subfigure}
\begin{subfigure}[b]{0.48\textwidth}
  \includegraphics[width=\textwidth]{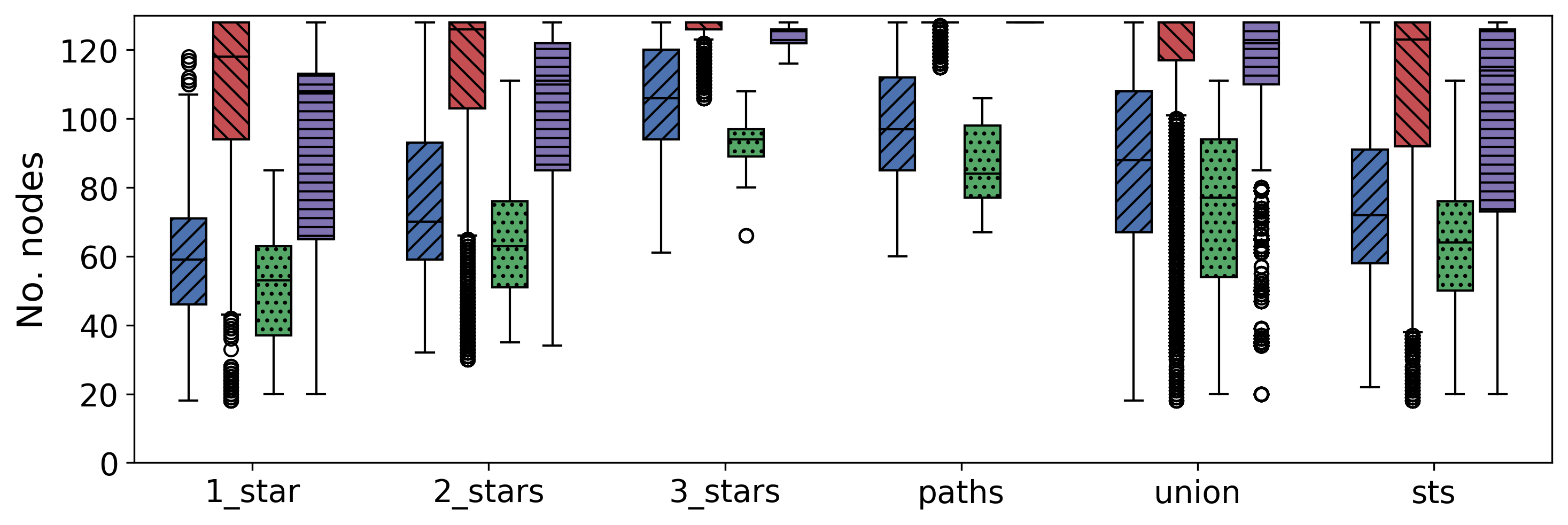}
  \caption{Number of relevant nodes (NRN) over \texttt{watdiv10M}}\label{subfig:nrn_10M}
\end{subfigure}
\begin{subfigure}[b]{0.48\textwidth}
  \includegraphics[width=\textwidth]{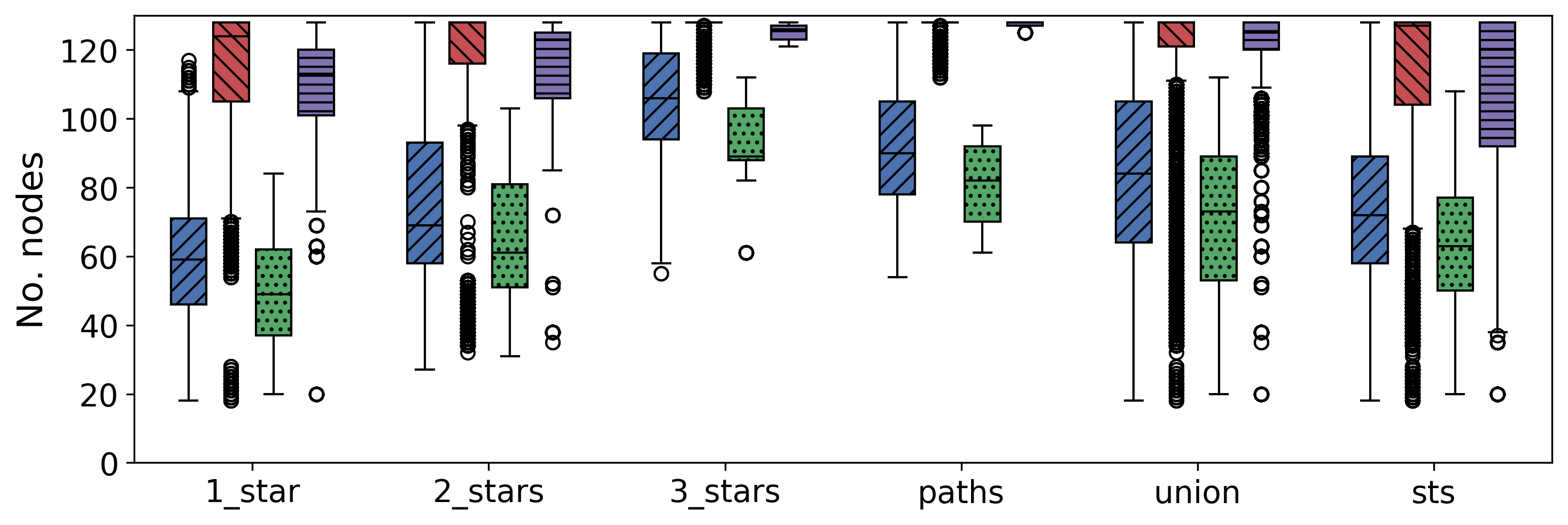}
  \caption{Number of relevant nodes (NRN) over \texttt{watdiv100M}}\label{subfig:nrn_100M}
\end{subfigure}
\begin{subfigure}[b]{0.48\textwidth}
  \includegraphics[width=\textwidth]{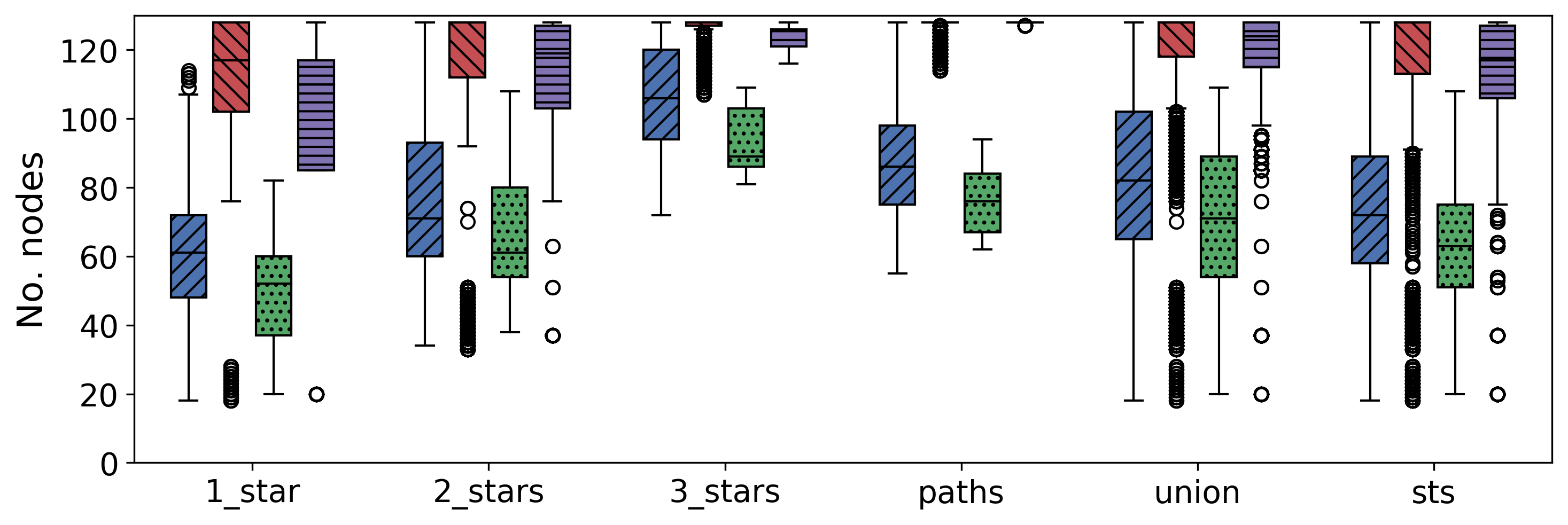}
  \caption{Number of relevant nodes (NRN) over \texttt{watdiv1000M}}\label{subfig:nrn_1000M}
\end{subfigure}
\caption{Number of relevant fragments (NRF) and number of relevant nodes (NRN) for the WatDiv datasets and star queries.}
\label{fig:shape_resource}
\end{figure*}

We notice that for the \texttt{watdiv-path} query load, \system{} actually has a slightly worse performance both in terms of QET and QRT compared to \piqnic{} and \colchain{} due to higher network usage.
Figure~\ref{fig:shape_resource} shows the number of relevant fragments (NRF) and the number of relevant nodes (NRN) for each query load over each dataset after optimization (similar figures are provided for NRF and NRN before optimization on our website\footnotemark[3]).
Analyzing these results, we see that the decreased performance for \texttt{watdiv-path} is caused by \system{} having a significantly larger number of relevant fragments and by extension a larger number of relevant nodes compared to \piqnic{} and \colchain.
In fact, this is the case for all the WatDiv query loads (9 times larger for \texttt{watdiv-path} while up to 5 times larger for the other query loads); however, for the other query loads, this is compensated by the increased performance that the query optimization approach provides.
This analysis is corroborated by the number of fragments pruned during optimization for each query load (figures provided on our website\footnotemark[3]); the \texttt{watdiv-path} query load has significantly less pruned fragments compared to the other query loads except \texttt{watdiv-1\_star}. %, suggesting that the optimizer 
For \piqnic{} and \colchain{}, the number of relevant fragments will always equal the number of unique predicates in the query since one fragment is created per predicate; however, due to fragmenting the data based on characteristic sets, \system{} can encounter multiple fragments for each unique predicate in the query. 
Furthermore, the number of relevant fragments is, on average, more than twice as high for \system{} over the \texttt{watdiv-path} query load than over the other query loads.
This is because most of the path queries use common predicates like \texttt{owl:sameAs}.

Nevertheless, the slightly worse performance for \system{} over \texttt{watdiv-path} is compensated by the significantly improved performance over the other query loads, so we still see a performance increase for the \texttt{watdiv-union} query load.
As such, our experimental results show that \system{} is generally able to increase performance over queries with star-shaped subqueries (i.e., all other queries than path queries) significantly and that the increase in performance depends on the shape of the query; queries with fewer but larger star patterns (cf. Figure~\ref{subfig:jvdegree}) show a bigger performance increase than queries with many but small star patterns.

\begin{figure*}[tb!]
\centering
\includegraphics[width=.6\textwidth]{figures/results/shape/legend3.png}
\begin{subfigure}[b]{0.48\textwidth}
  \includegraphics[width=\textwidth]{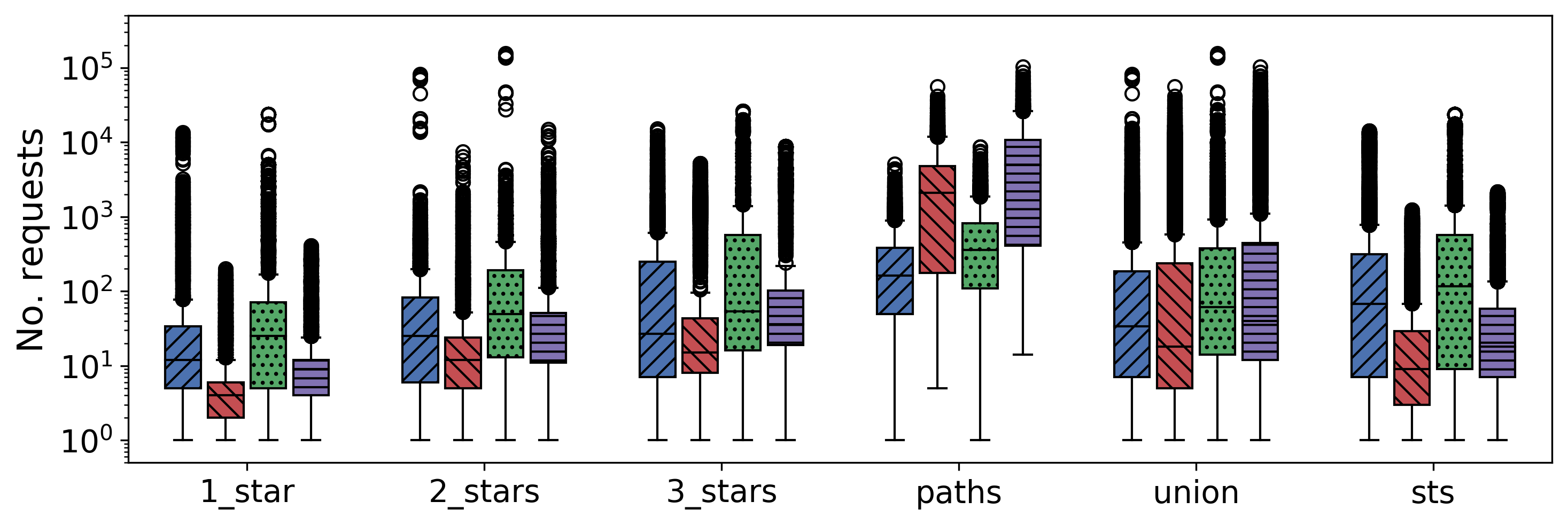}
  \caption{Number of requests (REQ) over \texttt{watdiv10M}}\label{subfig:req_10M}
\end{subfigure}
\begin{subfigure}[b]{0.48\textwidth}
  \includegraphics[width=\textwidth]{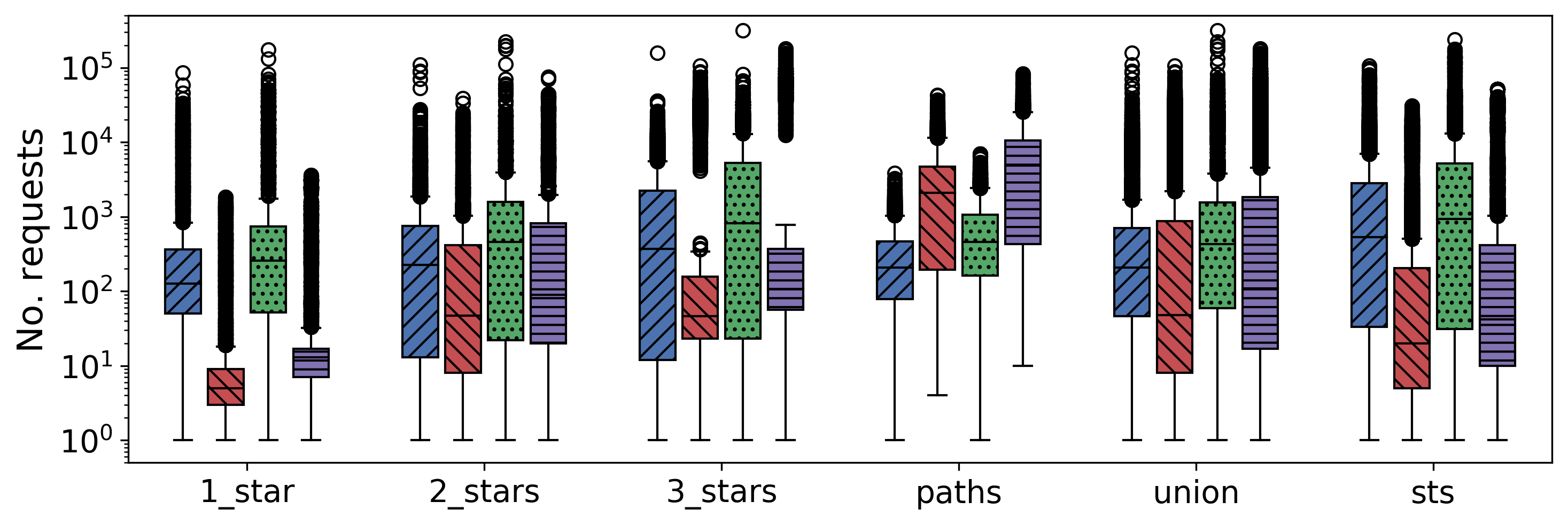}
  \caption{Number of requests (REQ) over \texttt{watdiv100M}}\label{subfig:req_100M}
\end{subfigure}
\begin{subfigure}[b]{0.48\textwidth}
  \includegraphics[width=\textwidth]{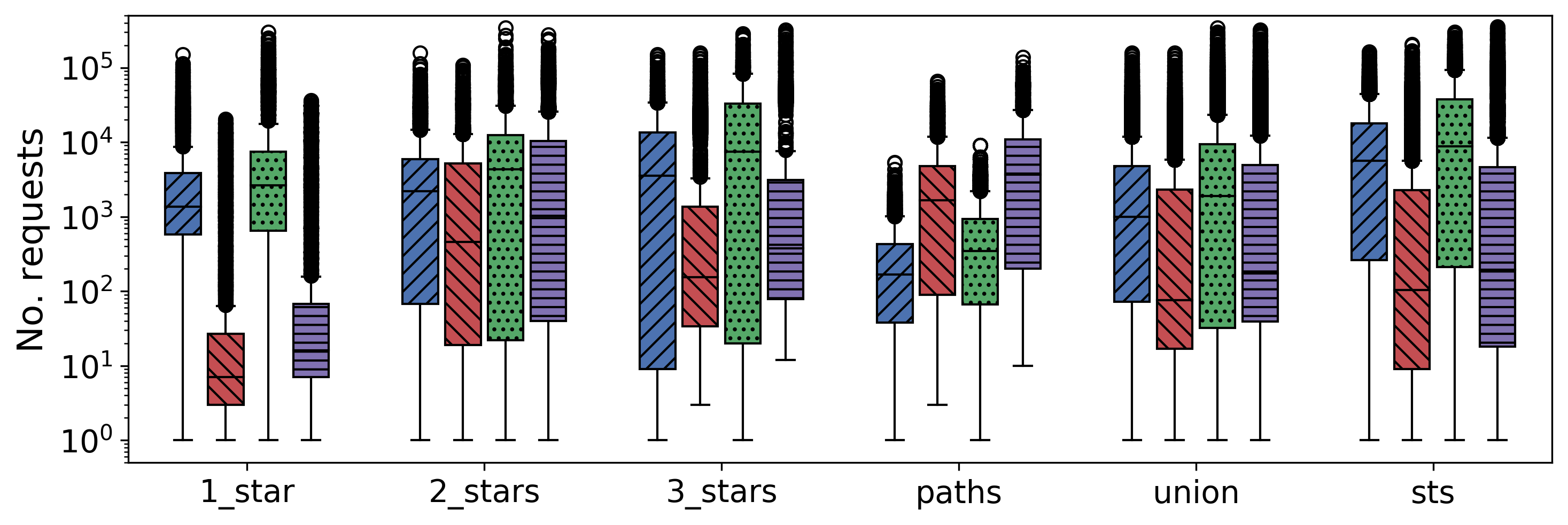}
  \caption{Number of requests (REQ) over \texttt{watdiv1000M}}\label{subfig:req_1000M}
\end{subfigure}
\begin{subfigure}[b]{0.48\textwidth}
  \includegraphics[width=\textwidth]{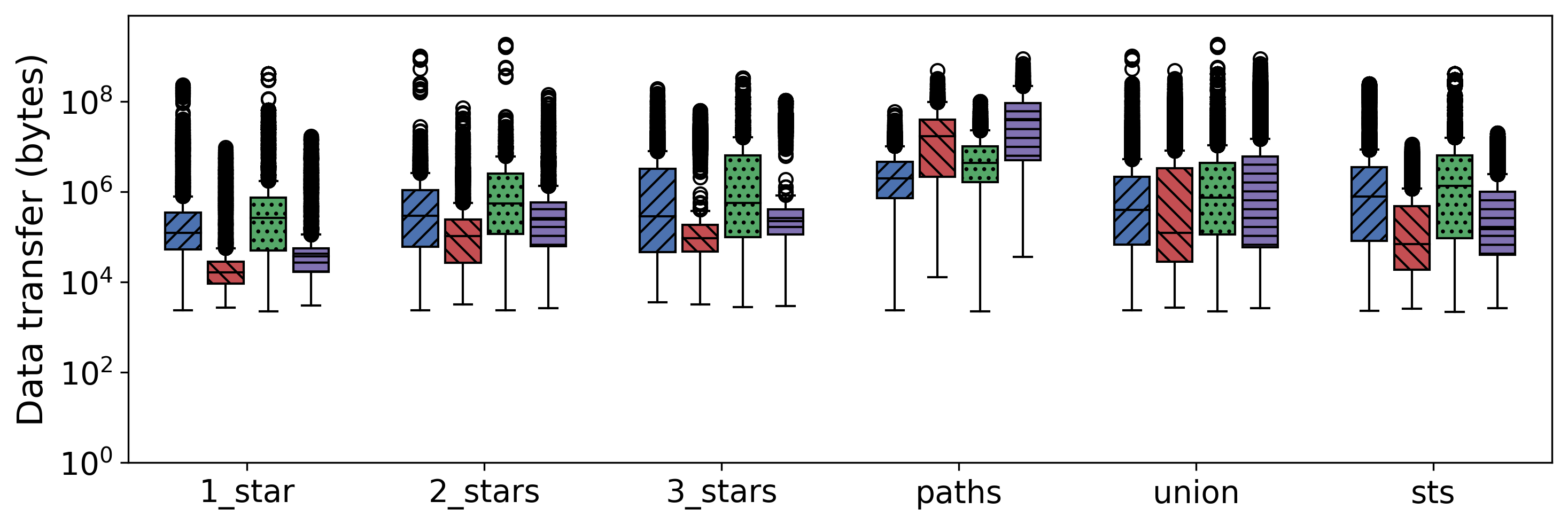}
  \caption{Number of transferred bytes (NTB) over \texttt{watdiv10M}}\label{subfig:ntb_10M}
\end{subfigure}
\begin{subfigure}[b]{0.48\textwidth}
  \includegraphics[width=\textwidth]{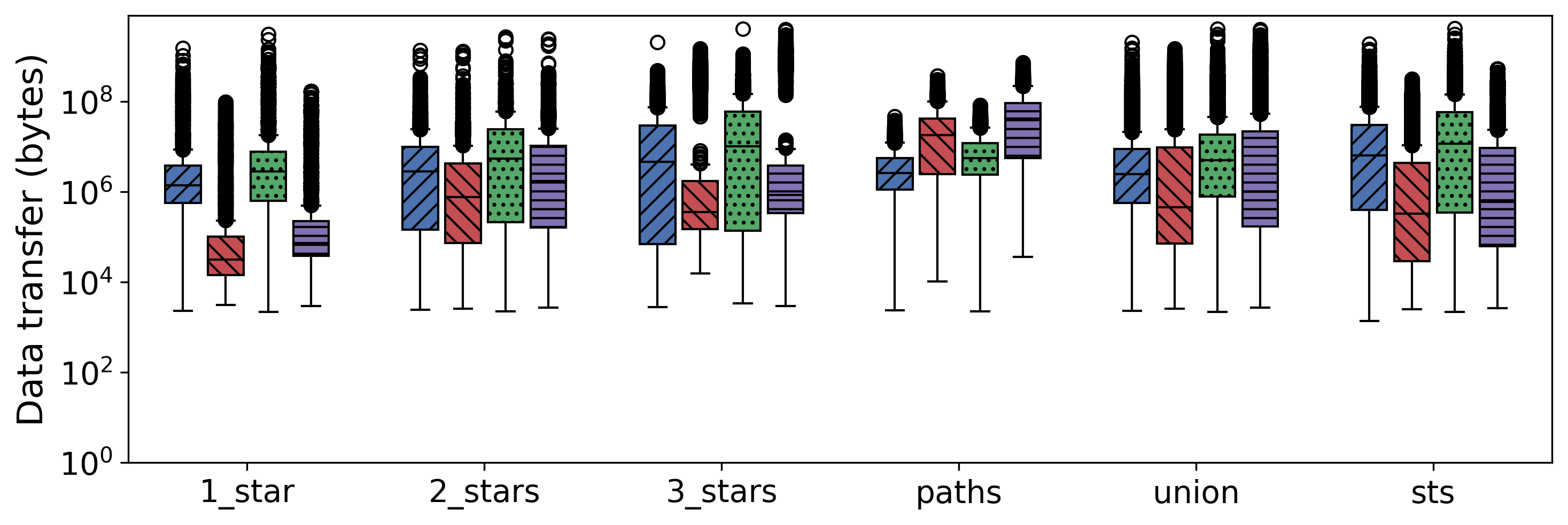}
  \caption{Number of transferred bytes (NTB) over \texttt{watdiv100M}}\label{subfig:ntb_100M}
\end{subfigure}
\begin{subfigure}[b]{0.48\textwidth}
  \includegraphics[width=\textwidth]{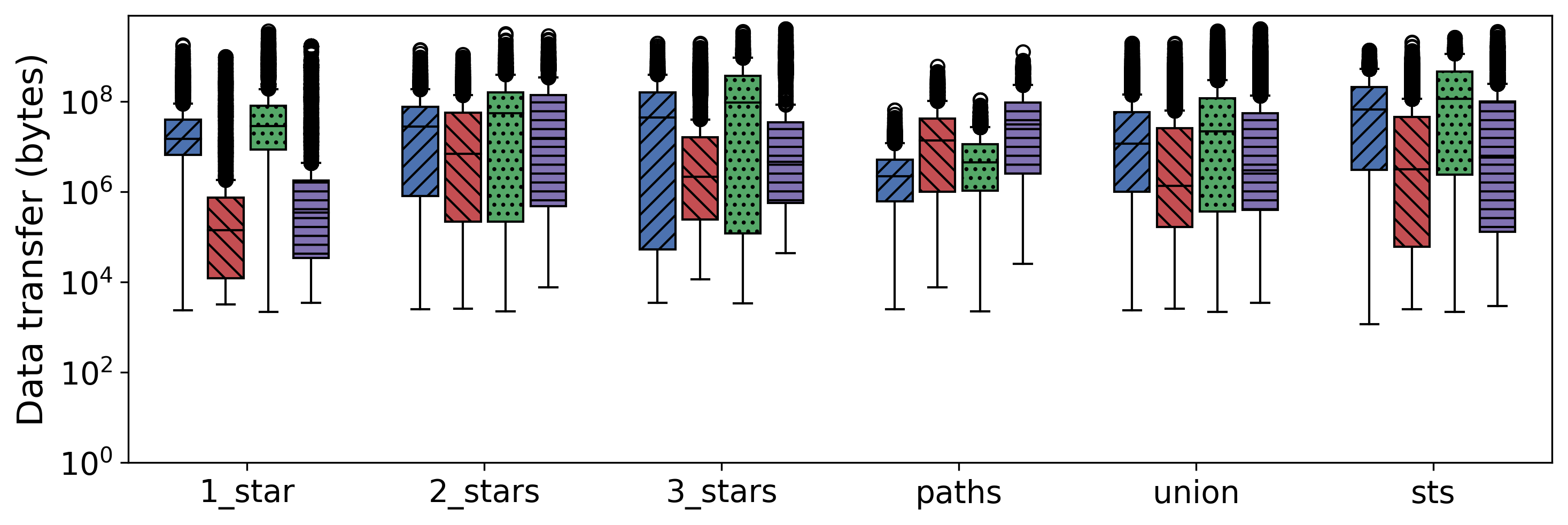}
  \caption{Number of transferred bytes (NTB) over \texttt{watdiv1000M}}\label{subfig:ntb_1000M}
\end{subfigure}
\caption{Number of requests (REQ) and number of transferred bytes (NTB) for the WatDiv datasets and star queries.}
\label{fig:net_us}
\end{figure*}

\subsection{Network Usage}\label{subsec:network}
Figure~\ref{fig:net_us} shows the network usage when processing WatDiv queries over each WatDiv dataset in terms of the number of requests (Figures~\ref{subfig:req_10M}-\ref{subfig:req_1000M}) and the number of transferred bytes (Figures~\ref{subfig:ntb_10M}-\ref{subfig:ntb_1000M}) in logarithmic scale.
\system{} incurs a significant lower network overhead for all query loads except \texttt{watdiv-path} despite the larger number of relevant fragments as discussed in Section~\ref{subsec:querypattern}.
This is caused by \system{} having to send significantly fewer requests for each star pattern since a star pattern can be processed entirely over the relevant fragments, even if there are more fragments (and thus nodes) to send the requests to.
Again, the query loads with a smaller number of star patterns see a larger decrease in network usage since larger parts of the queries can be processed by individual nodes.
Since the queries in the \texttt{watdiv-path} query load do not benefit from the star pattern-based query processing, the network usage is slightly higher; however, even still, the \texttt{watdiv-union} shows an improvement in the network usage for \system.
These results are in line with the experiments shown in Sections~\ref{subsec:scalability} and~\ref{subsec:querypattern} and support the hypothesis that \system{} increases performance by lowering the network overhead when processing queries, compared to state-of-the-art systems such as \piqnic{} and \colchain{}.

\subsection{Performance of Individual Queries}\label{subsec:individual}
\begin{figure*}[tb!]
\centering
\includegraphics[width=.6\textwidth]{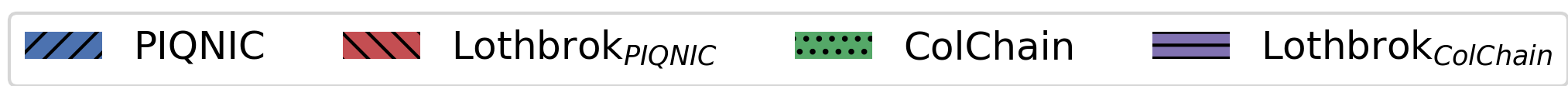}
%\vspace{-5ex}
\begin{subfigure}[b]{\textwidth}
  \includegraphics[width=\textwidth]{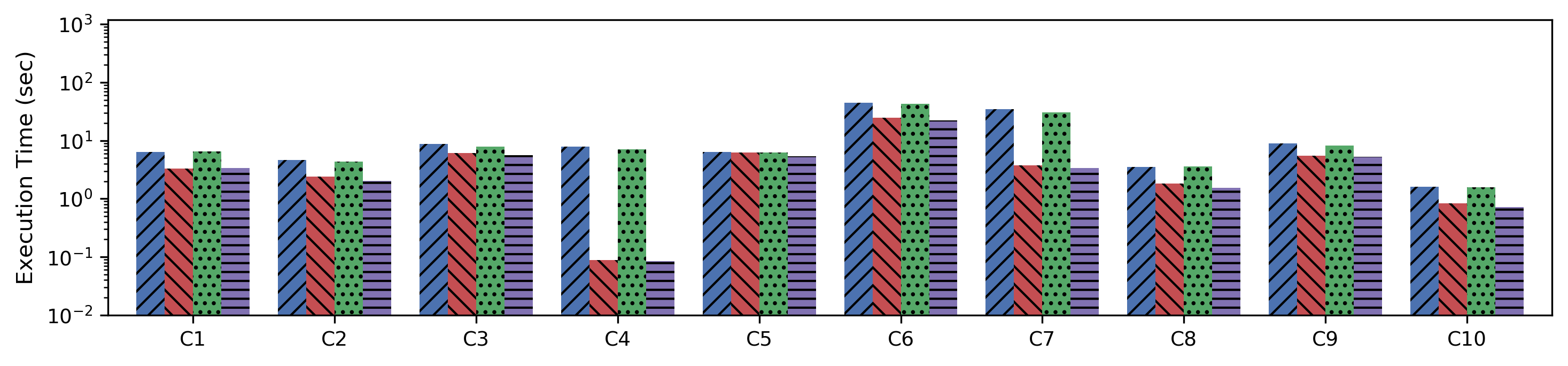}
  \caption{Execution time (QET)}\label{subfig:qet_c}
\end{subfigure}
\begin{subfigure}[b]{\textwidth}
  \includegraphics[width=\textwidth]{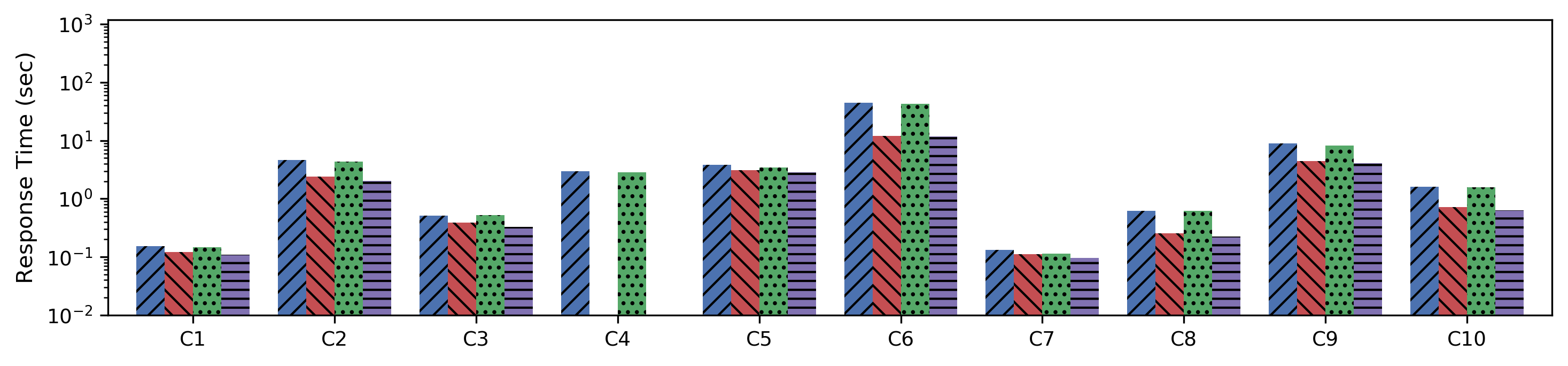}
  \caption{Response time (QRT)}\label{subfig:qrt_c}
\end{subfigure}
\begin{subfigure}[b]{\textwidth}
  \includegraphics[width=\textwidth]{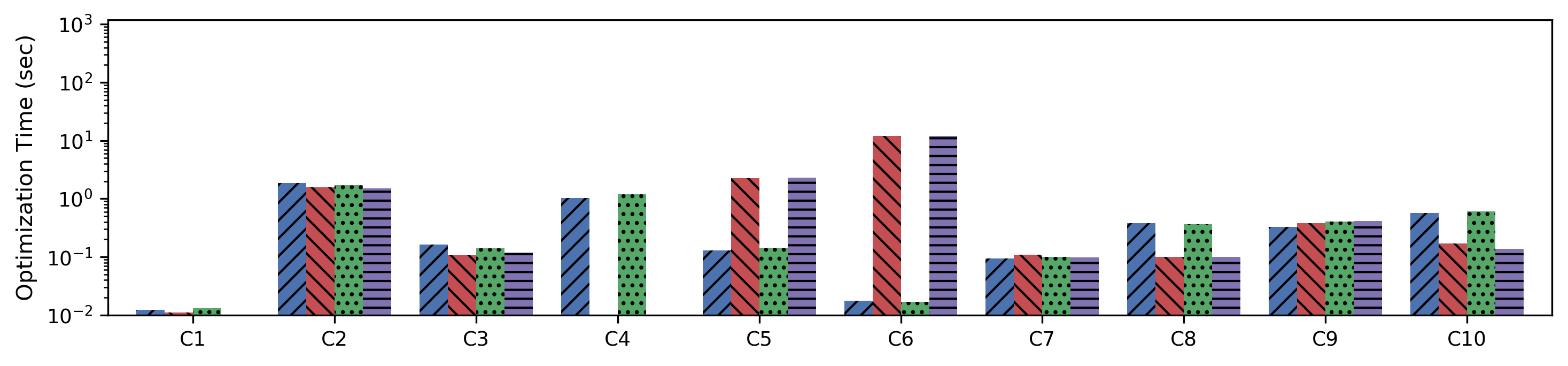}
  \caption{Optimization time (QOT)}\label{subfig:qot_c}
\end{subfigure}
\caption{Query execution time (a), response time (b), and optimization time (c) for the C query load over LargeRDFBench.}
\label{fig:perf_c}
\end{figure*}

In these experiments, we ran the LargeRDFBench queries three times on each system sequentially to test the performance of those individual queries and report the average results.
Figure~\ref{fig:perf_c} shows the execution time (Figure~\ref{subfig:qet_c}), response time (Figure~\ref{subfig:qrt_c}), and optimization time (Figure~\ref{subfig:qot_c}) for the C query load over LargeRDFBench in logarithmic scale.
Similar figures for the other LargeRDFBench query loads are provided on our website\footnotemark[3].
The results in Figure~\ref{fig:perf_c} are similar to the remaining query loads; we show the C query load since this query load had the most diversity in the performance across the queries.

While, in our experiments, \system{} provides an improvement for the execution time (Figure~\ref{subfig:qet_c}) across all the queries in LargeRDFBench, the improvement varies based on the query shape in line with the findings of~\cite{DBLP:journals/corr/abs-2002-09172,DBLP:conf/www/AzzamAMKPH21} and the query shape experiments shown in Section~\ref{subsec:querypattern}. %; since the LargeRDFBench data is less structured than the WatDiv data, the increase in performance is relatively smaller as well.
For instance, query C4 consists of one highly selective star pattern with 6 unique predicates.
\system{} is thus able to answer C4 with one request to the only fragment with that predicate combination, while \piqnic{} and \colchain{} have to send at least one request per triple pattern.
Hence, \system{} has around two orders of magnitude better performance for this particular query.
On the other hand, query C5 consists of four star patterns, two of which contain only one triple pattern with one of them being the very common \texttt{rdfs:label} predicate.
As a result, \system{} has more than twice the number of relevant fragments for C5 compared to both \piqnic{} and \colchain{}.
Nevertheless, \system{} still has slightly improved performance for C5 compared to \piqnic{} and \colchain{} since the query still contains two star patterns with three triple patterns each, meaning the increased optimization and communication overhead that the additional relevant fragments entail is offset by the benefits of processing the star patterns over the individual fragments.
The response times (Figure~\ref{subfig:qrt_c}) show a similar comparison between the systems as the execution times (Figure~\ref{subfig:qet_c}) with the exception of query C4.
Again, the reason being that \system{} can process this query with a single request, and therefore the first result is obtained immediately after receiving the response to the request.

However, the optimization times (Figure~\ref{subfig:qot_c}) differ quite significantly depending on the number of relevant fragments to the query.
For instance, queries like C5 and C6 (that contain a star pattern consisting of a single triple pattern with a very common predicate) incur a significant number of relevant fragments for \system{} (286 for C5 and 144 for C6) and thus a higher optimization time.
This is the case, since a higher number of relevant fragments means a higher number of SPBFs have to be intersected which represents an overhead.
In all of these cases, however, the benefits of processing entire star patterns over the fragments, in terms of decreased network overhead mean that the overall execution time is still lower for \system.
This is especially the case for C6, which contains a star pattern with 6 triple patterns that in \piqnic{} and \colchain{} have to be processed individually.
On the other hand, queries like C4 that contain few very selective star patterns have a low optimization time for \system{}, since each star pattern have very few relevant fragments.
In the case of C4, \piqnic{} and \colchain{} have a relatively high number of relevant fragments due to one of the predicates being the common \texttt{owl:sameAs} predicate that occurs in multiple datasets.
As a result, \piqnic{} and \colchain{} have a significantly higher optimization time for this query compared to \system{}.

\begin{figure*}[tb!]
\centering
\includegraphics[width=.6\textwidth]{figures/results/performance/legend1.png}\\
\begin{subfigure}[b]{0.38\textwidth}
  \centering
  \includegraphics[width=\textwidth]{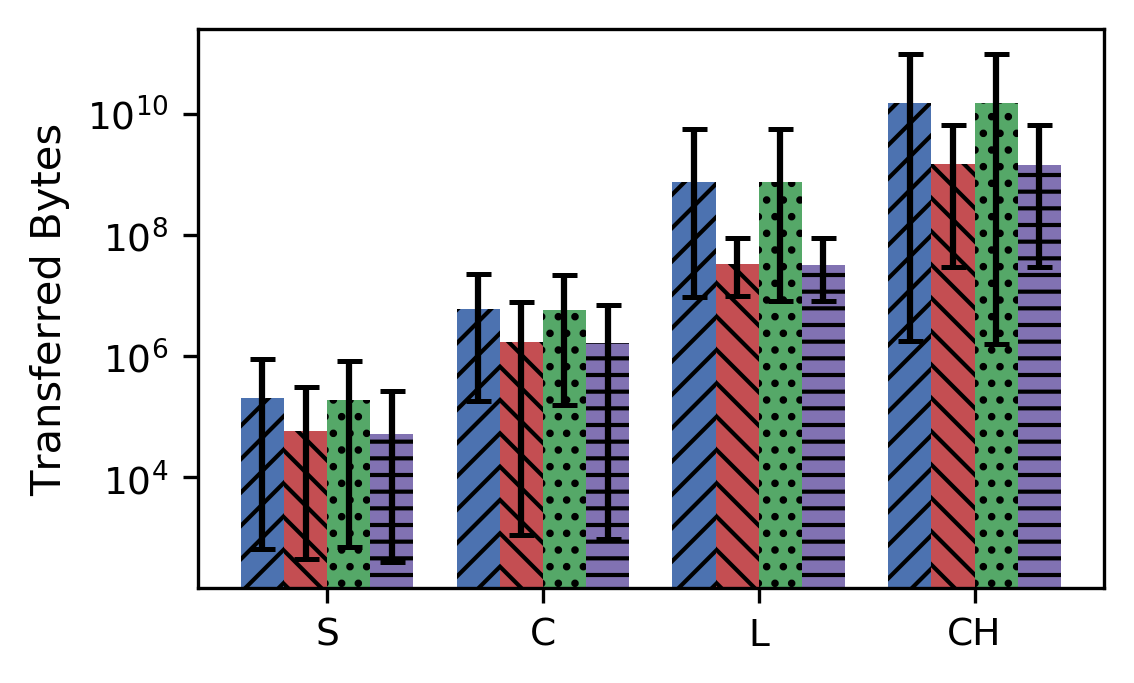}
  \caption{Number of Transferred Bytes (NTB)}\label{subfig:ntb}
\end{subfigure}
\begin{subfigure}[b]{0.38\textwidth}
  \centering
  \includegraphics[width=\textwidth]{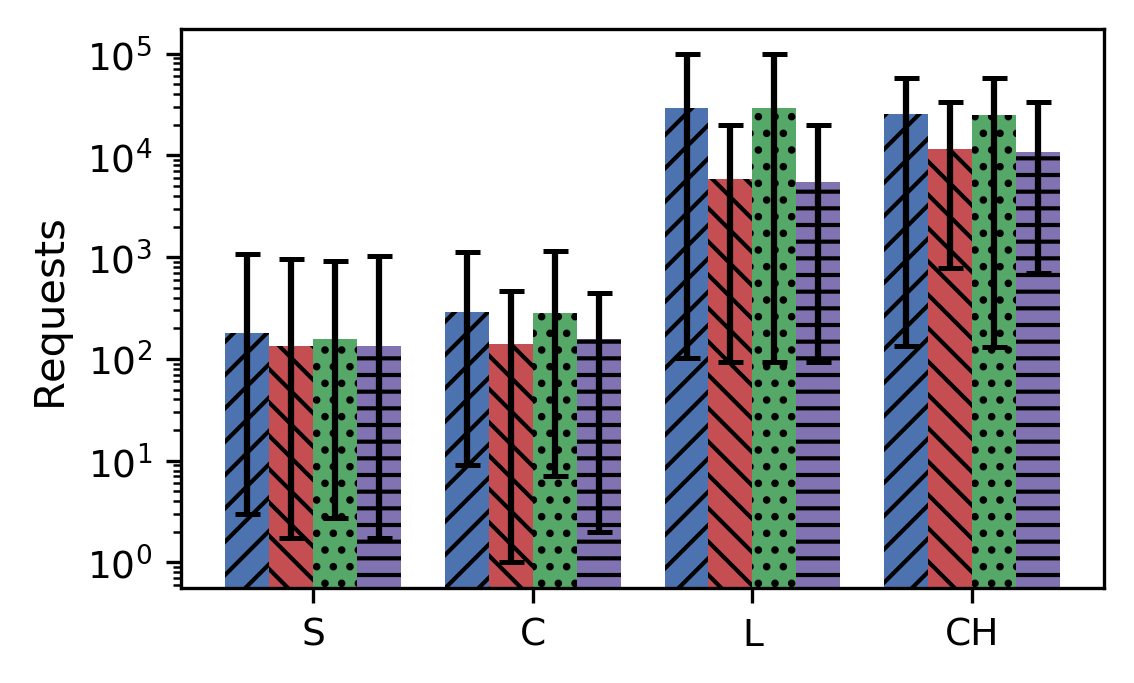}
  \caption{Number of Requests (REQ)}\label{subfig:req}
\end{subfigure}
\begin{subfigure}[b]{0.38\textwidth}
  \centering
  \includegraphics[width=\textwidth]{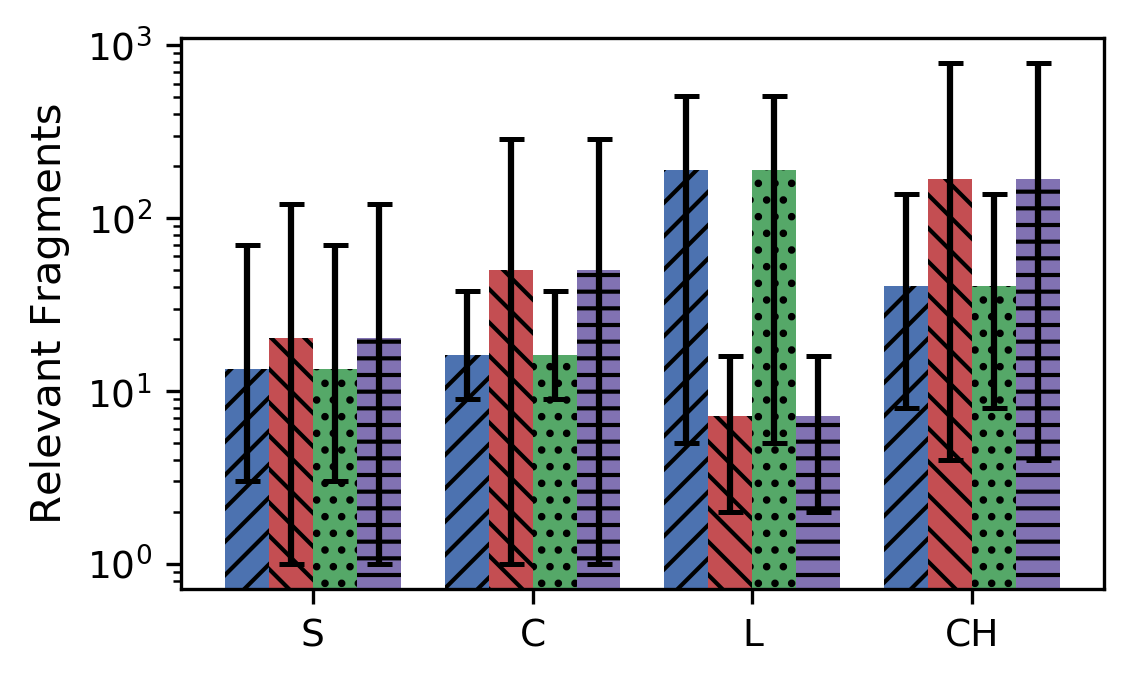}
  \caption{Number of Relevant Fragments (NRF)}\label{subfig:nrf}
\end{subfigure}
\begin{subfigure}[b]{0.38\textwidth}
  \centering
  \includegraphics[width=\textwidth]{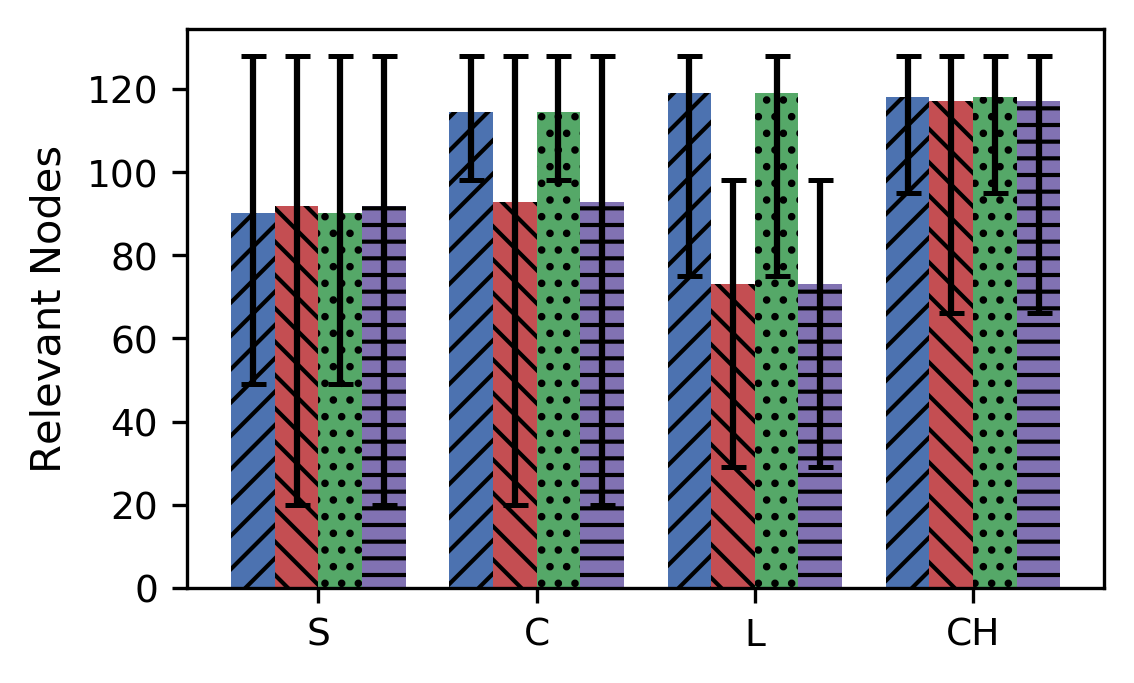}
  \caption{Number of Relevant Nodes (NRN)}\label{subfig:nrn}
\end{subfigure}
\caption{Number of Transferred Bytes (NTB) (a), Number of Requests (REQ) (b), Number of Relevant Fragments (NRF) (c), and Number of Relevant Nodes (NRN) (d) for each LArgeRDFBench query load.}
\label{fig:res_perf_ntb_req}
\end{figure*}

Figure~\ref{fig:res_perf_ntb_req} shows the number of transferred bytes (Figure~\ref{subfig:ntb}), the number of requests (Figure~\ref{subfig:req}), the number of relevant fragments (Figure~\ref{subfig:nrf}), and the number of relevant nodes (Figure~\ref{subfig:nrn}) for each LargeRDFBench query load in logarithmic scale.
We provide figures displaying each measure in Figure~\ref{fig:res_perf_ntb_req} for each individual LargeRDFBench query on our website\footnotemark[3].
As with the experiments shown in Section~\ref{subsec:network}, \system{} clearly incurs a lower network usage than both \piqnic{} and \colchain{}, both in terms of data transfer (Figure~\ref{subfig:ntb}) and the number of requests made (Figure~\ref{subfig:req}).
This, together with the performance experiments, shows that \system{} is able to reduce the network overhead significantly across all query loads and, in doing so, increase the performance overall.

Interestingly, while for most query loads, \system{} has a higher number of relevant fragments (Figure~\ref{subfig:nrf}) in line with the experiments presented in Section~\ref{subsec:querypattern}, for the L query load, \system{} has a lower number of relevant fragments in most queries.
The reason is that the queries in this query load mostly use data from the quite structured \texttt{linkedTCGA} datasets which contain few similar characteristic sets, thus incurring a low number of relevant fragments per star pattern.
On the other hand, for \piqnic{} and \colchain{}, the fact that some star patterns with a low number of triple patterns include common predicates like \texttt{rdf:type} increases the number of relevant fragments.
The number of relevant nodes (Figure~\ref{subfig:nrn}) shows a similar trend to the number of relevant fragments since each fragment is replicated across 20 nodes; in some cases, however, where two relevant fragments are simultaneously replicated by some of the same nodes, the actual number of relevant nodes will be a bit lower than when the relevant nodes replicate exactly one relevant fragment.

Our results are similar for all query loads (figures provided on our website\footnotemark[3]) and show that even for the complex queries in query loads C and CH and the queries with a large number of intermediate results in query load L, \system{} presents a significant performance increase because it lowers the communication overhead.
For some queries, this is quite significant; for instance the queries C4 and S3 where \system{} increases execution time by up to two orders of magnitude.
Furthermore, some queries in the L and CH groups that timed out for \piqnic{} and \colchain{}, such as L3 and CH2, finished within the timeout of 1200 seconds for \system{}.
This is in line with the results presented in Section~\ref{subsec:scalability} and suggests that \system{} is able to complete more queries within the timeout than the state-of-the-art systems.

\subsection{Summary}\label{subsec:summary}
Our experimental evaluations show that \system{} significantly improves query performance while lowering the communication overhead compared to \piqnic{} and \colchain.
\system{} does so by distributing subqueries to other nodes such that the estimated network cost is limited as much as possible, and by processing entire star patterns over the individual fragments.
In doing so, \system{} decreases the network usage both in terms of the data transfer and number of requests, and increases performance by up to two orders of magnitude compared to the state of the art.
Moreover, \system{} does so while providing scalable performance under load; in fact, even when all nodes in the network issue queries concurrently, \system{} maintains efficient query processing.

%% file: conclusion.tex
In this paper, we proposed \system{} a novel query optimization approach for SPARQL queries over decentralized knowledge graphs.
%\system{} exploits exploits the fact that star-shaped subqueries can be processed relatively efficiently by the nodes thus decreasing the amount of intermediate results to be transferred over the network while not overleading singular nodes.
\system{} builds upon recent work on decentralized Peer-to-Peer (P2P) systems~\cite{DBLP:conf/esws/AebeloeMH19,DBLP:conf/www/AebeloeMH21} and introduces a novel fragmentation technique based on characteristic sets~\cite{DBLP:conf/icde/NeumannM11}, i.e., predicate families, as well as a novel indexing scheme that summarizes the sets of subjects and objects in a fragment using partitioned bitvectors.
Furthermore, \system{} proposes a query optimization strategy based on cardinality estimation, fragment compatibility, and data locality that is able to delegate the processing of (sub)queries to other, neighboring nodes in the network that hold relevant data.
We implemented our approach on top of two recent systems and evaluated \system's capabilities over well-known benchmarking suites containing real-world data and queries, as well as the performance of \system{} under load using large-scale synthetic datasets and stress-testing query templates.
The experimental results show that \system{} significantly reduces the network overhead when processing queries in a P2P network and, in doing so, increases performance by up to two orders of magnitude.

While we presented a novel distribution of the workload across nodes in a P2P network, \system{} also presents an opportunity to explore the effects of alternative strategies, e.g., for cost estimation, considering fragments optimized for object-object  joins (Figure~\ref{subfig:jvtype}), or alternative fragmentation and allocation strategies, e.g., based on SHACL/ShEx shapes~\cite{RabbaniLH21,RabbaniLH22}. 
Furthermore, we plan to expand the range of supported queries to include aggregation and analytical queries~\cite{IbragimovHPZ16,GalarragaJHP18} and to expand the framework with support of provenance both for data~\cite{HansenLGLTH20,AndersenGHJP14,rdf-star}, so that the system has information about the origin of the data it uses, as well as for queries~\cite{HernandezGH21} so that the system can explain how query answers were computed.